\documentclass[twoside,12pt]{article}
\usepackage{epsfig}
\usepackage[figuresright]{rotating}
\usepackage{epstopdf}

\usepackage{amssymb}
\usepackage[unicode=true, pdfusetitle,
 bookmarks=true,bookmarksnumbered=false,bookmarksopen=false,
 breaklinks=false,pdfborder={0 0 1},backref=false,colorlinks=false]
 {hyperref}

\def\Journal#1#2#3#4{{#1} {#2} (#4) #3 }

\def\NPB{{\em Nucl. Phys.} B}

\def\PLB{{\em Phys. Lett.} B}

\def\PRL{\em Phys. Rev. Lett.}
\def\PREV{\em Phys. Rev.}

\def\PRD{{\em Phys. Rev.} D}
\def\PRC{{\em Phys. Rev.} C}

\def\eps{\epsilon}
\def\r{\vec r}

\def\ss{\mbox{\boldmath $\sigma$}}
\def\mb{\,\mbox{mb}}
\def\fm{\,\mbox{fm}}
\def\MeV{\,\mbox{MeV}}
\def\GeV{\,\mbox{GeV}}
\def\TeV{\,\mbox{TeV}}
\def\Re{\,\mbox{Re}\,}
\def\Im{\,\mbox{Im}\,}
 \def\Pom{{ I\!\!P}}
 \def\Reg{{ I\!\!R}}
\def\st{\sigma_{tot}^{hN}}
\def\sel{\sigma_{el}^{hN}}
\def\sinhad{\sigma_{in}^{hN}}
\def\sdd{\sigma_{sd}^{hN}}

\def\sta{\sigma_{tot}^{hA}}
\def\sela{\sigma_{el}^{hA}}

\def\stn{\sigma_{tot}^{NN}}

\def\sq{\sigma_{\bar qq}}
\def\halftext{.471\textwidth}
\newcommand\la{\langle}
 \newcommand\ra{\rangle}
\def\lsim{\mathrel{\rlap{\lower4pt\hbox{\hskip1pt$\sim$}}
    \raise1pt\hbox{$<$}}}         
\def\gsim{\mathrel{\rlap{\lower4pt\hbox{\hskip1pt$\sim$}}
    \raise1pt\hbox{$>$}}}         

\newcommand{\beq}{\begin{equation}}
\newcommand{\eeq}{\end{equation}}
\newcommand{\beqn}{\begin{eqnarray}}
\newcommand{\eeqn}{\end{eqnarray}}
\newcommand{\be}{\begin{equation}}
\newcommand{\ee}{\end{equation}}
\newcommand{\bea}{\begin{eqnarray}}
\newcommand{\eea}{\end{eqnarray}}
\newcommand{\nn}{\nonumber}
\begin{document}

\begin{flushright} 

FERMILAB-PUB-12-487-PPD

USM-TH-303

\end{flushright}

\title{ \vspace{1cm} Nuclear Shadowing  in Electro-Weak Interactions}
\author{Boris  Z. Kopeliovich,$^{1}$ Jorge G.~Morf\'{\i}n,$^2$ Iv\'an~Schmidt$^1$\\
\\
$^1$Departamento de F\'{\i}sica,
Universidad T\'ecnica Federico Santa Mar\'{\i}a; and\\
Instituto de Estudios Avanzados en Ciencias e Ingenier\'ia; and\\
Centro Cient\'ifico-Tecnol\'ogico de Valpara\'iso;\\
Casilla 110-V, Valpara\'iso, Chile\\
$^2$Fermilab, Batavia, Illinois 60510, USA\\
}

\begingroup
\let\newpage\relax
\maketitle
\endgroup

\vspace{1 cm}

\begin{center}
To be published in `` Progress in Particle and Nuclear Physics 2012 ''.
\end{center}

\vspace{2 cm}

\begin{abstract} 
Shadowing is a quantum phenomenon leading to a non-additivity of electroweak
cross sections on nucleons bound in a nucleus. It occurs due to destructive interference of
amplitudes on different nucleons. Although the current experimental evidence for shadowing is dominated by charged-lepton nucleus scattering, studies of neutrino nucleus scattering have recently begun and revealed unexpected results.
\end{abstract}

\vfil\eject
\tableofcontents

\part{Theory of Nuclear Shadowing}
\section{Introduction}
\label{sec:introduction}

The term shadowing was naturally incorporated into quantum mechanics as a tag for one of the basic quantum  phenomena related to the destructive interference between elastic amplitudes on different scattering centers. The popular observable sensitive to
shadowing in a process, is the normalized ratio of the cross sections of this process on nuclear and nucleon targets, 
\beq
R_{A/N}=\frac{\sigma_{A}}{A\,\sigma_{N}}.
\label{1.20}
\eeq
Shadowing manifests itself in a suppressed value of this ratio, $R_{A/N}<1$, i.e. the nuclear cross section is less than the sum of the cross sections on all bound nucleons\footnote{Hereafter we neglect the isospin corrections, i.e. do not make difference between target protons and neutrons, unless specified. We also neglect the effect of binding, unless it is important.}. Intuitively, the source of the suppression is   
the survival probability 
for the projectile particle or its fluctuations to pass through  the nuclear medium and reach
a  bound nucleon deep inside the nucleus. In other words, the interaction with a given bound nucleon is shaded by the probability of having preceding interactions with other nucleons. This is easy to see on the example of the total hadron-nucleus interaction cross section.
It is clear that a strongly interacting particle has no chance to pass through a heavy nucleus without interaction. This means that the partial elastic amplitude at impact parameter less than the nuclear radius, $b<R_A$, nearly saturates unitarity bound, $\Im f_{el}^{hA}(b)\leqslant1$, i.e. the $\sigma_{tot}^{hA}\approx2\pi R_A^2$, or $R_{A/N}\propto A^{-1/3}$,
because the nuclear radius $R_A\propto A^{1/3}$. In the following sections we calculate these quantities more accurately within the Glauber-Gribov approach.

Notice that in exclusive reactions one should discriminate between shadowing and final state absorption. Both effects lead to a suppression of the ratio (\ref{1.20}), but have quite different origins.

\section{Shadowing in soft interactions}
\label{sec:soft}
\subsection{Shadowing in the Glauber model}
\label{sec:glauber}

The Glauber model~\cite{glauber} was the first theoretical approach that correctly calculated the effects of shadowing in hadron-nucleus interactions. The probability for a hadron to interact with the nucleus is one minus the probability of no interaction with any of the bound nucleons. So the $hA$ elastic amplitude at impact parameter $b$ has the eikonal 
form,
 \beq
 \Gamma^{hA}(\vec b;\{\vec s_j,z_j\}) =
1 - \prod_{k=1}^A\left[1-
 \Gamma^{hN}(\vec b-\vec s_k)\right]\ ,
 \label{1.40}
 \eeq
 where $\{\vec s_j,z_j\}$ denote the coordinates of the target nucleon
$N_j$. $i\Gamma^{hN}$ is the elastic scattering amplitude on a nucleon
normalized as,
 \beqn
\st &=& 2\int d^2b\,\Re\Gamma^{hN}(b);\nonumber\\
\sel&=& \int d^2b\, |\Gamma^{hN}(b)|^2\ .
\label{1.60}
 \eeqn
 
In the approximation of single particle nuclear density one can calculate
a matrix element between the nuclear ground states.
 \beqn
\left\la0\Bigl|\Gamma^{hA}(\vec b;\{\vec s_j,z_j\})
\Bigr|0\right\ra =
1-\left[1-{1\over A}\int d^2s\,
\Gamma^{hN}(s)\int\limits_{-\infty}^\infty dz\,
\rho_A(\vec b-\vec s,z)\right]^A\ ,
\label{1.80}
 \eeqn
 where 
 \beq
\rho_A(\vec b_1,z_1) = \int\prod_{i=2}^A
d^3r_i\,
|\Psi_A(\{\vec r_j\})|^2\ ,
\label{1.100}
 \eeq
 is the nuclear single particle density.
 
 The magnitude of shadowing depends on the process. Here we calculate the shadowing effects within the Glauber model for several basic processes.\\

\noindent
{\em Total cross section}.\\ 
The result Eq.~(\ref{1.80}) is related via unitarity to the total $hA$ 
cross section,
 \beqn
\sta&=&2\Re\int d^2b\,\left\{1 -
\left[1-{1\over A}\int d^2s\,
\Gamma^{hN}(s)\,T_A(\vec b-\vec s)\right]^A\right\}
\nonumber\\ &\approx&
2\int d^2b\, \left\{1-
\exp\left[-{1\over2}\,\st\,(1-i\rho_{pp})\,
T^h_A(b)\right]\right\}\ ,
\label{1.120}
 \eeqn
 where $\rho_{pp}$ is the ratio of the real to imaginary parts
of the forward $pp$ elastic amplitude;
 \beq
 T^h_A(b)= \frac{2}{\st}\int d^2s\, 
\Re\Gamma^{hN}(s)\,T_A(\vec b-\vec s)\ ;
\label{1.140} 
 \eeq
 and
 \beq
T_A(b) = \int_{-\infty}^\infty dz\,\rho_A(b,z)\ ,
\label{1.160}
 \eeq
 is the nuclear thickness function. 
 We use exponential form of $\Gamma^{hN}(s)$ throughout the paper,
 \beq
\Re \Gamma^{hN}(s) =
\frac{\st}{4\pi B_{hN}}\,
\exp\left(\frac{-s^2}{2B_{hN}}\right)\ ,
\label{1.180}
 \eeq
 where $B_{hN}$ is the slope of the differential $hN$ elastic cross
section. Note that the accuracy of the optical approximation (the second line in
(\ref{1.120})) is quite high for gold, $\sim 10^{-3}$. We use the optical form
throughout the paper for the sake of simplicity, although for numerical evaluations always rely on the accurate expression (the first line in (\ref{1.120})). 
The effective nuclear thickness Eq.~(\ref{1.140}), implicitly contains energy dependence, which is extremely weak.

In what follows we also neglect the real part of the elastic amplitude, unless specified, since it gives a vanishing correction  $\sim 
\rho_{pp}^2/A^{2/3}$.

In Table~\ref{tab:lhc}  the result of the calculation of the shadowing factor $R_{A/N}^{tot}$
for the total proton-lead cross section at $\sqrt{s}=5.5\TeV$ is shown.
 \begin{table}[h]
\begin{minipage}[t]{16.5 cm}
 \caption{
 The shadowing suppression factor Eq.~(\ref{1.20}) for the total, elastic, inelastic, quasielastic and production cross sections.}
\label{tab:lhc}
\end{minipage}
 \begin{center}
\begin{tabular}{|c|c|c|c|c|}
 \hline
 \vphantom{\bigg\vert}
   Model & 
$R_{A/N}^{tot}$
  & $R_{A/N}^{el}$
  & $R_{A/N}^{in}$
  & $R_{A/N}^{qel}$
   \\
\hline &&&\\[-2mm]
Glauber& 0.22 & 0.4 & 0.17& 0.03 \\
\hline   
\end{tabular}
\end{center}
 \end{table}
We use $\stn=93.1\mb$ and elastic slope $B^{NN}_{el}=20\GeV^{-2}$, which agree well with the recent measurements of the TOTEM experiment \cite{totem} at $\sqrt{s}=7\TeV$. We see how strong the shadowing effect for the total cross section is.\\

\noindent
{\em Elastic cross section}.\\
As far as the partial elastic amplitude is known, the elastic cross 
section reads,
 \beq
\sela=\int d^2b\, \left|1-
\exp\left[-{1\over2}\,\st\,T^h_A(b)\right]\right|^2\ .
\label{1.200}
 \eeq\\
From the numerical result depicted in Table~\ref{tab:lhc} we see that the elastic cross section is less suppressed by shadowing than the total one. This shows that even at the energy of LHC the proton-lead interaction does not reach yet the unitarity bound, where
the shadowing ratios Eq.~(\ref{1.20}) are expected to be alike for both processes.\\

\noindent
{\em Total inelastic cross section.}\\
 $\sigma_{in}^{hA}$ is given by the
difference between the total and elastic cross sections,
\beq
\sigma^{hA}_{in} = \sta-\sela=
\int
d^2b\,\left\{1-\exp\left[-
\sigma_{tot}^{hN}\,T^h_A(b)\right]\right\}\ .
\label{1.220}
 \eeq
 This cross section covers all inelastic channels, where either the hadron or the
nucleus (or both) break up. According to Table~\ref{tab:lhc} inelastic interactions are shadowed  stronger than the total cross section.
\\

\noindent
{\em Quasielastic cross section}.\\
As a result of the collision the nucleus can be excited to a state 
$|F\ra$. Summing over final states of the nucleus and applying the 
condition of completeness, one gets the quasielastic cross section,
 \beqn
\sigma^{hA}_{qel} &=& \sum\limits_F
\int d^2b\ \left[\left\la0\left|
\Gamma^{hA}(b)\right|F\right\ra^\dagger
\left\la F\left|\Gamma^{hA}(b)\right|0\right\ra  -
 \left |\left\la 0\left|\Gamma^{hA}(b)
\right|0\right\ra\right|^2\right]\nonumber\\
&=& 
\int d^2b\ \left[\left\la0\left|
\Bigl|\Gamma^{hA}(b)\Bigr|^2
\right|0\right\ra
- \left |\left\la 0\left|\Gamma^{hA}(b)
\right|0\right\ra\right|^2\right]\ .
\label{1.240}
 \eeqn
 Here we extracted the cross section of elastic scattering when the
nucleus remains intact.

Then in the fist term of this expression we  make use of the relation,
\beq
\Re\int d^2s\,\frac{T^h_A(\vec b-\vec s)}
{A}\left\{1-2\Gamma^{hN}(s)+
\left[\Gamma^{hN}(s)\right]^2\right\}\approx
1-{1\over A}\,T^h_A(b)(\st-\sel)\ ,
\label{1.260}
 \eeq
 and arrive at,
 \beq
\sigma^{hA}_{qel} =
\int d^2b\,\left\{
\exp\left[-\sinhad\,T^h_A(b)\right]-
\exp\left[-\st\,T^h_A(b)\right]\right\}\ .
\label{1.280}
 \eeq
 The large "volume" terms proportional to $A$ cancel in this expression, so one should expect a very strong
 nuclear suppression, as is confirmed by the result of evaluation presented in 
 Table~\ref{tab:lhc}. Strictly speaking, this suppression is due to a combination of
 shadowing in the initial state and the survival probability of the scattered hadron in the nuclear matter.
 Final state interactions are particularly important for exclusive channels
of hadron production. In the case of electro-weak interactions final state
interactions can imitate shadowing even when it is absent, e.g. when the
coherence time is short.\\

{\em Production cross section}.\\
Quasielastic scattering is a part of the inelastic cross section Eq.~(\ref{1.220}), and the only process with no production of new particles. Therefore it should be subtracted to get the total production cross section. The result is rather simple, 
 \beq 
\sigma^{hA}_{prod} = \sta-\sela-\sigma^{hA}_{qel}=
\int d^2b\,
\left\{1-\exp\left[-\sinhad\,T^h_A(b)\right]\right\}\ . 
\label{1.300}
 \eeq

\subsection{Inelastic shadowing corrections}\label{in-corr}
\label{sec:inelastic}

\subsubsection{Intermediate state diffractive excitations}\label{sect-kk}
\label{sec:intermediate}

\begin{figure}[htb]
\parbox{\halftext}{
\centerline{\includegraphics[width= 8.5 cm]{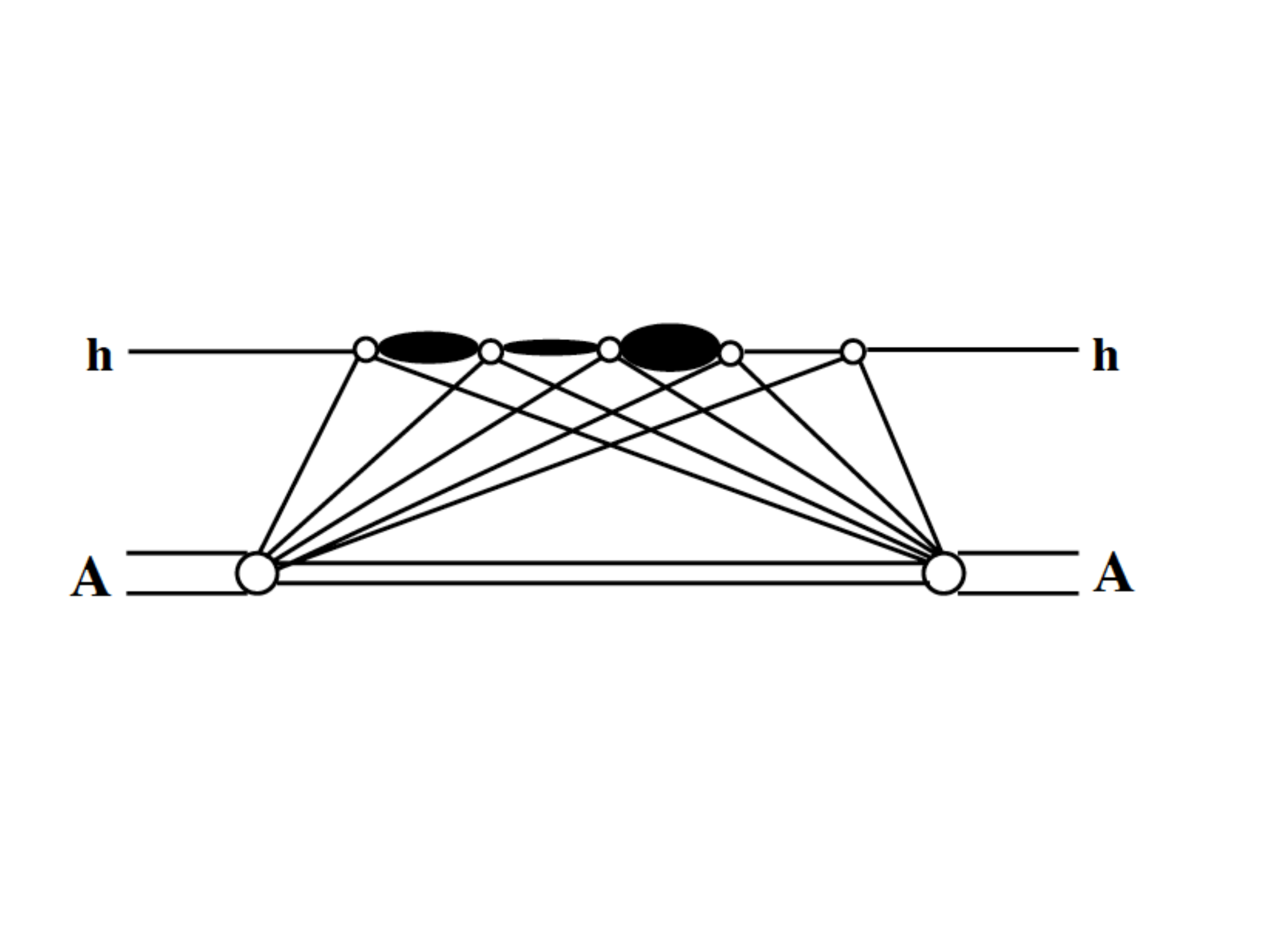}}
\caption{Diagonal and off-diagonal diffractive multiple excitations of the
projectile hadron in intermediate state}
\label{fig:diff}}
\hfill
\parbox{\halftext}{
\centerline{\includegraphics[width=8.5 cm]{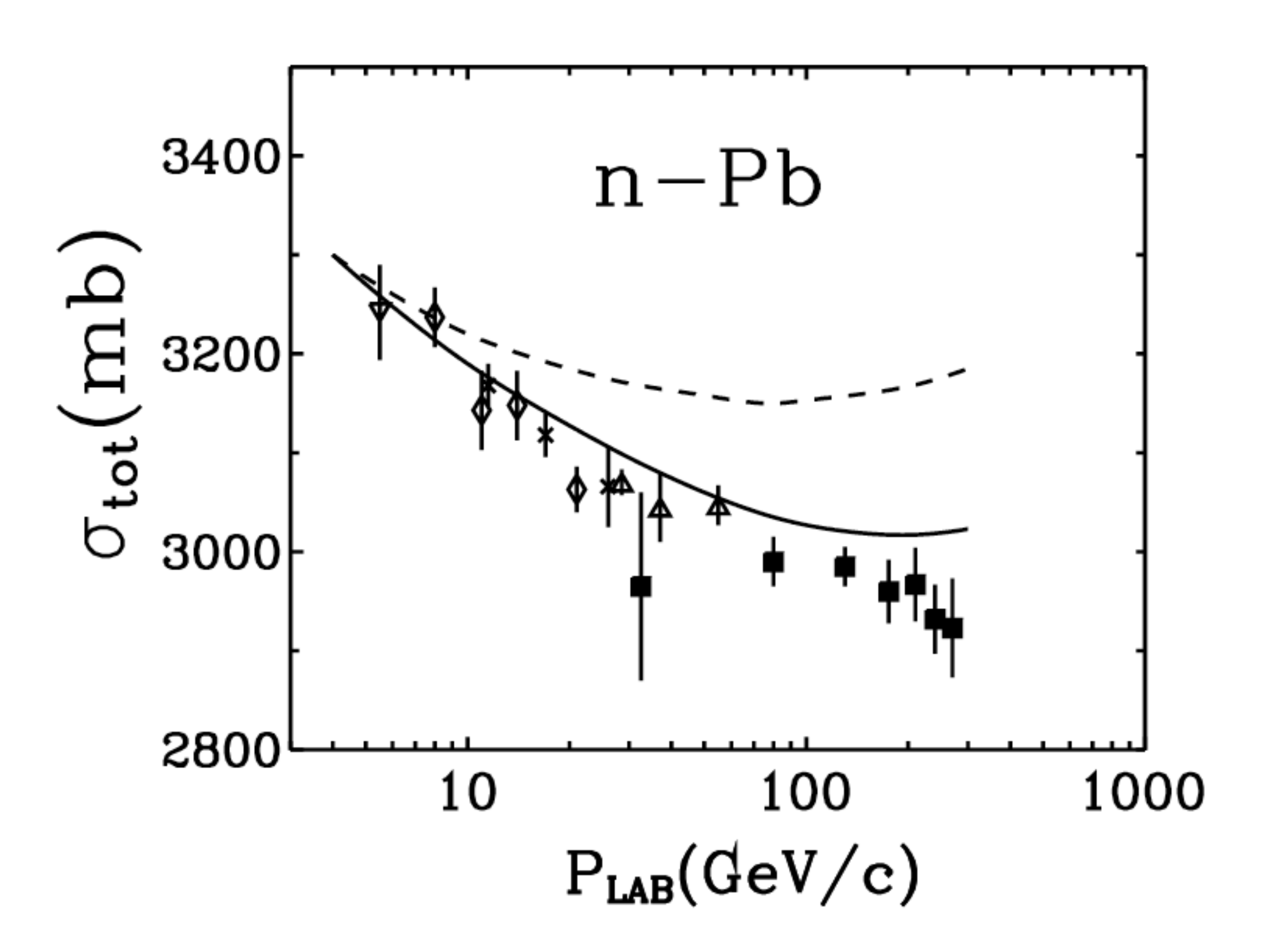}}
\caption{ Data and calculations \cite{murthy} for the total neutron-lead
cross section. The dashed and solid curves correspond to the
Glauber model and  corrected for Gribov shadowing respectively.}\label{fig:murthy}}
\end{figure}

The Glauber model is a single-channel approximation, therefore it misses the
possibility of diffractive excitation of the projectile in the intermediate
state as is  illustrated in Fig.~\ref{fig:diff}.
 These corrections called inelastic shadowing were introduced by
Gribov back in 1969 \cite{gribov}. The formula for the inelastic
corrections to the total hadron-nucleus cross section was suggested in
\cite{kk},
 \beqn
\Delta\sigma^{hA}_{tot} = 
- 8\pi\int d^2b\,
e^{-{1\over2}\sigma^{hN}_{tot}T_A(b)}\!\!\!
\int\limits_{M_{min}^2}\!\! dM^2
\left.\frac{d\sigma_{sd}^{hN}}
{dM^2\,dp_T^2}\right|_{p_T=0}
\int\limits_{-\infty}^{\infty}dz_1\,
\rho_A(b,z_1)
\int\limits_{z_1}^{\infty}dz_2\,
\rho_A(b,z_1)\,e^{iq_L(z_2-z_1)}\ ,
\label{1.320}
 \eeqn
 where $\sigma_{sd}^{hN}$ is the cross section of single diffractive 
dissociation $hN\to XN$ with longitudinal momentum transfer
 \beq
q_L=\frac{M^2-m_h^2}{2E_h}\ .
\label{1.340}
 \eeq

This correction makes shadowing stronger and nuclei more transparent \cite{kn}, because it is added with positive sign to the probability amplitude of having no interaction, $\exp(-\sigma_{hN}^{tot}T_A/2)$. 
It  takes care of the onset of
inelastic shadowing via phase shifts controlled by $q_L$ and does a good job describing data at low energies \cite{murthy,gsponer}, as one can also see
in Fig.~\ref{fig:murthy}.  Notice that the
higher order off-diagonal transitions neglected in (\ref{1.320}), including the diagonal transitions (or
absorption of the excited state), are important, but unknown. Indeed, the
intermediate state $X$ has definite mass $M$, but no definite
cross section. It was fixed in (\ref{1.320}) at $\sigma_{tot}^{hN}$ with no justification.
We will come back to this problem later.

There is, however, one case which is free of these problems,
shadowing in hadron-deuteron interactions. In this case no interaction in the intermediate
state is possible, and knowledge of diffractive cross section $hN\to XN$
is sufficient for calculations of the inelastic correction with no further
assumptions. In this case Eq.~(\ref{1.320}) takes the simple form
\cite{gribov,pr},
 \beq
\Delta\sigma^{hd}_{tot} = - 
2\int dM^2\int dp_T^2\,
\frac{d\sdd}{dM^2dp_T^2}\,
F_d(t)\ .
\label{1.360}
 \eeq

\subsubsection{Inelastic shadowing in the eigenstate representation}\label{eigen}
\label{sec:eigenstate}

If a hadron were an eigenstate of interaction, it could experience only 
elastic scattering (as a shadow of inelastic channels) and no diffractive
excitation were possible. In this case the Glauber formula would be exact and no
inelastic shadowing corrections were needed. This simple observation
suggests to switch from the basis of physical hadronic
states, which are the eigenstates of the mass operator, to the basis of a complete set of mutually orthogonal
states which are eigenstates of the scattering amplitude operator. This
was the driving idea of description of diffraction in terms of elastic
amplitudes \cite{pom,gw}, and becomes a powerful tool for calculation of
inelastic shadowing corrections in all orders of multiple interactions
\cite{kl}. Hadronic states (including leptons and photons) can be decomposed
into a complete set of such eigenstates $|k\ra$,
 \beq
|h\ra=\sum\limits_{k}\,\Psi^h_k\,|k\ra\ , 
\label{1.380}
 \eeq 
 where $\Psi^h_k$ are hadronic wave functions in the form of Fock state
decomposition. They obey the orthogonality conditions,
 \beqn
\sum\limits_{k}\,\left(\Psi^{h'}_k\right)^{\dagger}\,\Psi^h_k
&=&\delta_{h\,h'}\ ;
\nonumber\\
\sum\limits_{h}\,\left(\Psi^{h}_l\right)^{\dagger}\,\Psi^h_k
&=&\delta_{lk}\ .
\label{1.400}
 \eeqn

We denote by $f_{el}^{kN}=i\,\sigma_{tot}^{kN}/2$ the eigenvalues of the
elastic amplitude operator $\hat f$ neglecting its real part. We assume
that the amplitude is integrated over impact parameter, {\it i.e.} that
the forward elastic amplitude is normalized as
$|f_{el}^{kN}|^2=4\,\pi\,d\sigma_{el}^{kN}/dt|_{t=0}$. We can
express the elastic $f_{el}$ and off diagonal single-diffractive
$f_{sd}$ amplitudes as,
 \beq
f_{el}^{hN}=2i\,\sum\limits_k\,\left|
\Psi^h_k\right|^2\,\sigma_{tot}^{kN}
\equiv 2i\,\la\sigma\ra\ ;
\label{1.420}
 \eeq
 \beq
f_{sd}^{hN}(h\to h')=
2i\,\sum\limits_k\, (\Psi^{h'}_k)^{\dagger}\,
\Psi^h_k\,\sigma_{tot}^{kN}\ . 
\label{1.440}
 \eeq
 Notice that if all the eigen amplitudes were equal, the diffractive
amplitude (\ref{1.440}) would vanish due to the orthogonality relation,
(\ref{1.400}). The physical reason is obvious. If all the $f_{el}^{kN}$ are
equal, the interaction does not affect the coherence between the
different eigen components $|k\ra$ of the projectile hadron $|h\ra$.
Therefore, the off-diagonal transitions are possible only due to the differences
between the eigen amplitudes.

Summing all final states 
and making use of the completeness condition (\ref{1.400}),
then, excluding the elastic channels one arrives at \cite{kl,mp,zkl},
 \beq
16\pi\,\frac{d\sigma^{hN}_{sd}}{dt}\biggr|_{t=0}=
\sum\limits_i \left|\Psi^h_i\right|^2
\left(\sigma^{iN}_{tot}\right)^2
- \biggl(\sum\limits_i 
\left|\Psi^h_i\right|^2\sigma^{iN}_{tot}\biggr)^2
\equiv \la\sigma_{tot}^2\ra - \la\sigma_{tot}\ra^2\ .
\label{1.460}
 \eeq

If the lifetimes of the different eigenstate components of the hadron are sufficiently long, so that they do not mix with each other during propagation through the nucleus, the cross sections for different processes on nuclei can be written as,
 \beq
\sigma^{hA}_{tot} = 2\int d^2b\,\left\{1 -
\left\la\exp\left[-{1\over2}\,\sigma^{iN}_{tot}\,T^h_A(b)\right]
\right\ra\right\}
\label{1.480}
 \eeq
 \beq
\sigma^{hA}_{el} = \int d^2b\,\left|1 -
\left\la\exp\left[-{1\over2}\,\sigma^{iN}_{tot}\,T^h_A(b)\right]
\right\ra\right|^2
\label{1.500}
 \eeq
 \beq
\sigma^{hA}_{in} = \int d^2b\,\left\{1 -
\left\la\exp\Bigl[-\sigma^{iN}_{in}\,T^h_A(b)\Bigr]
\right\ra\right\}
\label{1.520}
 \eeq

Notice that the last expression for $\sigma^{hA}_{in}$ is already
free from the diffraction contribution. Although only elastic and
quasielastic cross sections were subtracted from $\sigma^{hA}_{tot}$
in the Glauber model, Eq.~(\ref{1.300}), after averaging over eigenstates
it turns out that diffraction is subtracted as well \cite{mine}.

The difference between the cross section
Eq.~(\ref{1.480}) and the Glauber approximation
Eq.~(\ref{1.120}), is in the way of averaging. In the former case the whole exponential is averaged,
while in the Glauber approximation only the exponent is averaged. The difference should correspond to the Gribov corrections summed in all orders. Indeed, the first order terms in the expansion of (\ref{1.480}) and (\ref{1.120}) cancel and in the second order using the relation (\ref{1.460}) we get,
 \beq
 \Delta\sta \approx
 \int d^2b\, {1\over4}\,\Bigl[\la\sigma^{iN}_{tot}\ra^2 -
\la(\sigma^{iN}_{tot})^2\ra\Bigr]\,T^h_A(b)^2
=- 4\pi\int d^2b\,T^h_A(b)^2\int dM^2\,
\frac{d\sigma^h_{sd}}{dM^2dt}\biggr|_{t=0}\ .
\label{1.580}
 \eeq
 This result is identical to Eq.~(\ref{1.320}), if we neglect there the 
phase shift vanishing at high energies, and also expand the 
exponential.

Note that since the inelastic nuclear cross section in the form
Eq.~(\ref{1.300}) is correct for eigenstates, one may think that
averaging this expression would give the correct answer. However, such a
procedure includes the possibility of excitation of the projectile and
disintegration of the nucleus to nucleons, but misses the possibility of
diffractive excitation of bound nucleons which is not a small correction.
We introduce a corresponding correction in the next section.

\subsubsection{Dipole description of shadowing}\label{dipole-soft}
\label{sec:dipole}

The light-cone dipole representation was proposed in \cite{zkl} as an
effective tool for calculation of hadronic cross sections and nuclear
shadowing, relying on the observation that color dipoles are the
eigenstates of hadronic interactions at high energies, and the eigenstate 
method \cite{kl} allows to sum up the Gribov inelastic 
corrections to all orders.

The key ingredient of this approach is the cross section of the
dipole-nucleon interaction, $\sq(r_T,s)$, which is an universal and flavor
independent function, depending on transverse separation $r_T$ and
energy. One cannot calculate it reliably, since that would involve nonperturbative effects,
but should fit it to data.  Once this cross section is known from data on a proton target, the nuclear effects can be predicted. Still, the results of calculations remain model dependent, because the hadronic wave functions participating in averaging Eq.~(\ref{1.460}) are poorly known. 

Applications of the dipole formalism to nuclei are especially
simple, if the energy is sufficiently high to freeze the fluctuations of
the dipole size during its propagation through the nucleus.
Otherwise one should rely on the path-integral technique
\cite{kz91,krt1,krt2}, which takes care of these fluctuations (see below, Sect.~\ref{sec:path}).

Due to color screening a colorless point-like dipole cannot interact with an
external color field. Since the underlying theory is non-abelian, the
interaction cross section for such dipoles vanishes at $r_T\to 0$ as
$\sq(r_T)\propto r_T^2$ \cite{zkl}, a phenomenon called color
transparency\footnote{Actually, the cross section behaves as $r_T^2\ln(r_T)$
\cite{zkl}, but with a good accuracy one can fix the logarithm at an effective
value of $r_T$ typical for the process under consideration.}. At high energies
nuclei are transparent for small-size fluctuations of the incoming hadron,
therefore the strong exponential attenuation suggested by the eikonal Glauber formula
cannot be correct and the nuclear medium should be more transparent, as was already mentioned above. For a dipole of a fixed size $r_T$ the eikonal form of Eqs.~(\ref{1.480})-(\ref{1.520}) is
exact. In this case the role of the eigenvalues of the cross section $\sigma^{iN}_{tot}$ is played by the dipole 
one $\sq(r_T)$, and the total hadron-nucleus cross section has the form \cite{zkl},
\beq
\sigma^{hA}_{tot}=2\int d^2b
\int d^2r_T\,\left|\Psi_h(r_T)\right|^2 \left[1-e^{-{1\over2}\sq(r_T)\,T_A(b)} \right].
\label{1.590}
\eeq

It is interesting to point out that although nuclear
transparency, which is the probability of no-interaction, exponentially falls with nuclear thickness, after averaging over dipole
sizes for a long path in the nuclear medium the resulting transparency  drastically changes \cite{zkl},
 \beq
e^{-\sq(r_T) T_A}\Rightarrow \int d^2r_T\,\left|\Psi_h(r_T)\right|^2 e^{-\sq(r_T)\,T_A} 
\propto \frac{1}{T_A}\ .
\label{1.600}
 \eeq
 Here we assumed a Gaussian shape for the $r_T$-dependence of the hadronic wave function $\Psi_h(r_T)$,
 and small-$r_T$ regime for the dipole cross section, $\sq(r_T)\propto r_T^2$, which is justified for a large nuclear thickness $T_A$.
 
 The result (\ref{1.600}) should be compared with the exponential attenuation in the
Glauber model, given by $\exp(-\st\,T_A)$. Apparently, the difference cannot originate from
the lowest order inelastic correction Eq.~(\ref{1.320}),  which has the same
exponential dependence on $T_A$. This color transparency effect
includes all the higher order inelastic corrections.

The Gribov inelastic shadowing corrections were calculated in \cite{mine,xsection,ciofi1,ciofi2} for various soft interaction processes. Although the nuclear transparency significantly increases for heavy nuclei, the related variation of  shadowing is pretty mild. This happens because for heavy nuclei the transparency term in the cross section 
(the second term in Eq.~(\ref{1.120})) is a small correction to the big first (volume) term.
Even a considerable variation of this small correction does not affect much the total cross section, i.e. shadowing. Indeed, the Gribov correction evaluated in Fig.~\ref{fig:murthy}
at low energy is several percent. With rising energy the hadron-nucleus amplitude 
approaches the black-disc limit (unitarity bound), where Gribov corrections vanish, because all eigen amplitudes become equal. We found that at the energy of LHC the Gribov corrections to the numbers in Table~\ref{tab:lhc} are only minus one percent.
So we skip this comparison here, but one can find the details of the calculations in \cite{mine,xsection,ciofi1}.

\subsection{Shadowing in photo-nuclear reactions}\label{sec:photons}

\begin{figure}[htb]
\epsfysize=9.0cm
\hspace*{3cm}
\begin{minipage}[t]{8 cm}
\epsfig{file=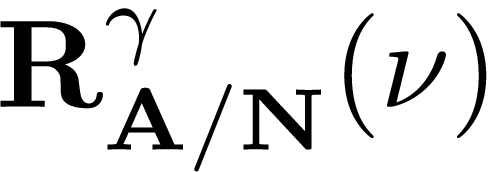,scale=0.4,angle=90}
\epsfig{file=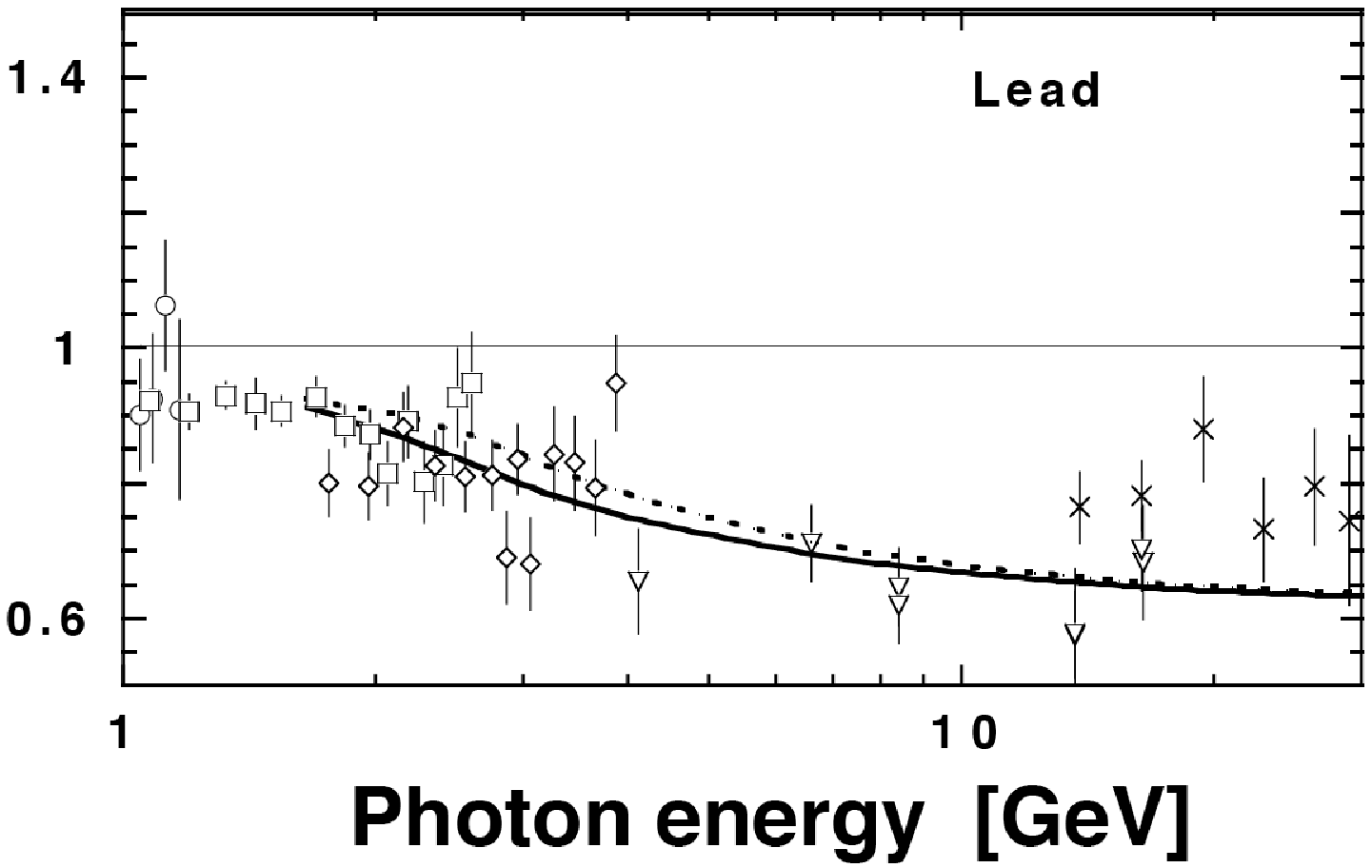,scale=0.7}
\end{minipage}
\vspace*{-5mm}
\begin{center}
\begin{minipage}[t]{16.5 cm}
\caption{Nuclear ratio $R^\gamma_{A/N}$ for lead as function of photon energy.
The curves correspond to a fit and model calculations \cite{muccifora}.\label{fig:photoabs}}
\end{minipage}
\end{center}
\end{figure}

At first glance the weakly interacting particles, like photons or leptons, should interact with nuclei with no shadowing. Indeed, applying the Glauber model expression 
(\ref{1.120}) to the photoabsorption cross section $\sigma^{\gamma A}_{tot}$ one
gets a vanishingly small shadowing effect. However, data presented in Fig.~\ref{fig:photoabs} clearly show that the photoabsorption cross section is strongly shadowed as function of energy. 
The nuclear ratio 
\beq
R^\gamma_{A/N}=\frac{\sigma^{\gamma A}_{tot}}
{Z\sigma^{\gamma p}_{tot}+(A-Z)\sigma^{\gamma n}_{tot}},
\label{1.620}
\eeq
is significantly suppressed by nuclear effects, which should be interpreted as shadowing. 

\begin{figure}[htb]
\epsfysize=9.0cm
\begin{center}
\epsfig{file=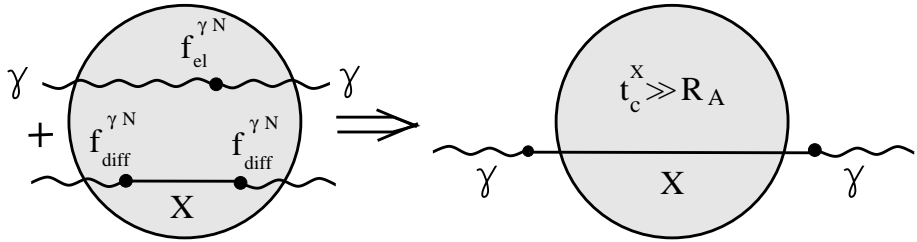,scale=0.4}
\begin{minipage}[t]{16.5 cm}
\caption{The sum of the two possibilities of either elastic $\gamma N$ scattering on a bound nucleon, or with preceding diffractive production of an intermediate state $X$ (left),
at high energies is equivalent to photon interaction via a hadronic fluctuation $X$ (right).
\label{fig:vdm}}
\end{minipage}
\end{center}
\end{figure}

The mechanisms of shadowing for photon interactions are well explained in the comprehensive review \cite{bauer}. The observed shadowing suppression   
is a manifestation of the hadronic properties of the photon. Namely, the photon
interacts with hadrons via its hadronic fluctuations. The observed smallness of the photoabsorption cross section is related to the smallness of the fluctuation probability (it is proportional to $\alpha_{em}$), while the hadronic cross section is large.
Although the fluctuation life time
$t\sim 1/m_h$ is quite short, it is subject to Lorentz time dilation, and at sufficiently high energies it maybe much longer than the nuclear size, as is illustrated in Fig.~\ref{fig:vdm}. 
In this regime the photon-nucleus interaction is indeed shadowed as much as in a hadron-nucleus collision.
The transition region between the short and long $t_c^X$ regimes can be described adding up the two contributions from the single- and double-step interactions, as is illustrated in the left part of Fig.~\ref{fig:vdm}. In the case of vector meson dominance, i.e. dominance of one pole in the dispersion relation in $Q^2$ for the amplitude, the sum of the two contributions to the photon nucleus amplitude, pictorially shown in the left part of  Fig.~\ref{fig:vdm}, reads \cite{bauer},
\beq
R^\gamma_{A/N}=1-\frac{8\pi}{A\sigma^{\gamma N}_{tot}}
\left.\frac{d\sigma^{\gamma\to V}_{diff}}{dp_T^2}\right|_{p_T=0}
\int d^2b\int\limits_{-\infty}^\infty dz_1\,\rho_A(b,z_1)
\int\limits_{-\infty}^{z_1} dz_2\,\rho_A(b,z_2)
e^{-iq_L^V(z_2-z_1)}\,e^{-{1\over2}\sigma^{V N}_{tot}T_A(b,z_2,z_1)},
\label{1.640}
\eeq
where $q_L^V=m_V^2/2E_\gamma$;
$T_A(b,z_2,z_1)=\int_{z_2}^{z_1} dz\,\rho_A(b,z)$. Here we neglected the real parts of the elastic and diffractive amplitudes.

This mechanism of shadowing is a particular case of Gribov corrections, and its evaluation is suffering of similar problems. First of all, even in the lowest order correction, the second term in Eq.~(\ref{1.640}), the value of $\sigma^{V N}_{tot}$ is unknown and is fitted to data.  The higher order multiple scattering corrections contain off-diagonal diffractive transitions $\gamma\to X\to X'\to X''...$, which are also unknown.
These higher order corrections become more important with increasing photon virtuality.
Indeed, the size of the $\bar qq$ fluctuation of the photon decreases as $\la r_T^2\ra\sim 1/Q^2$ and the magnitude of shadowing should diminish accordingly. In hadronic representation  it does not look that simple and can be achieved only by a specific tuning of diagonal and off-diagonal diffractive amplitude, which have to cancel each other.
This is usually modeled within generalized vector dominance models, but the predictive power of such approaches is very low.

The alternative color dipole approach naturally incorporated the color transparency features, however the $\bar qq$ distribution amplitudes cannot be calculated perturbatively
at low $Q^2$ and can only be modeled. These amplitudes were calculated within the instanton vacuum model in \cite{marat1} and the cross section of Compton scattering on protons and nuclei was evaluated  within the dipole formalism in \cite{marat2}.

\subsection{PCAC and shadowing of soft neutrino interactions}
\label{sec:PCAC}

Neutrinos are known as an significant source for the axial current, which is subject to nuclear shadowing, similar to that for the vector current, discussed in the previous section.
There are, however, specific  features of shadowing of the axial current, related to partial conservation of axial current (PCAC), which are absent in the case of vector current.
Indeed, the vector current is "trivially" conserved, $\partial_\mu\,J_\mu^V\propto
(m_n-m_p)\approx0$, since the proton and neutron masses cancel.
At the same time conservation of the axial current seems to be heavily broken, 
$\partial_\mu\,J_\mu^A\propto m_n+m_p$. In order to get the axial current conserved one has to introduce into $J_\mu^A$ additional terms besides $\gamma_5\gamma_\mu$, and this is how the massless Goldstone pseudo-scalar pion appears. In reality the pion has a small mass, this is why the current is conserved partially,
\beq
\partial_\mu J^A_\mu=m_\pi^2\,f_\pi\,\phi_\pi,
\label{2.20} 
\eeq 
where $m_\pi$ and $f_\pi\approx 0.93 m_\pi$ are the
pion mass and decay coupling, and $\phi_\pi$ is the pion field.

A beautiful manifestation of PCAC is the Goldberger-Treiman relation
\cite{treiman}, which bridges weak and strong interactions. It surprisingsly
connects the pion decay constant with the
pion-nucleon coupling, which seem to have very little in common.
Indeed, the former depends on the pion wave function, while the latter
is controlled by the wave function of the nucleon.  Nevertheless,
data on $\beta$-decay and muon capture confirm this
relation between very different physical quantities.

In the case of high energy neutrino
interactions PCAC leads to the Adler relation (AR)
between the cross sections of processes initiated by neutrinos and
pions \cite{adler},
\beq \left.\frac{d^2\sigma(\nu p\to
l\,X)}{dQ^2\,d\nu}\right|_{Q^2=0} = 
\frac{G^2}{2\pi^2}\,f_\pi^2\,\frac{E-\nu}{E\nu}
\,\sigma(\pi p\to X). 
\label{2.40} 
\eeq 
where $E$ is the neutrino energy;  $G=1.166\times 10^{-5}\,GeV^{-2}$ is
the electro-weak Fermi coupling; $Q^2=-q_\mu^2$, where
$q_\mu=k_\mu-k^\prime_\mu$ and $\nu=E-E^\prime$ are the 4-momentum
and energy transfer in the $\nu\to l$ transition (the same notation
as for neutrinos should not cause confusion). 

The Adler relation expresses the axial current interaction with that of the pion,
similar to the vector dominance model, which relates the interactions of the vector current and $\rho$-meson. This make it tempting to interpret the Adler relation as pion dominance.
 It turns out, however,  that a neutrino cannot fluctuate to a pion, $\nu\to\pi
l$, because the pion pole in the dispersion relation in $Q^2$ for
the axial current does not contribute to the interaction of the
neutrino at high energies \cite{Bell:1995nj,p-s,km}. 
Indeed, the axial current $J^A_\mu(Q^2)$
can be presented as,
\beq
J^A_\mu(Q^2)=\frac{q_\mu\,f_\pi}{Q^2+m_\pi^2}\,T(\pi p\to X) 
+ \frac{f_{a}}{Q^2+m_{a}^2}\,M_\mu(a p\to X)\,+\,...
\label{2.60}
 \eeq
Here the second and following terms represent the
contributions of the $a_1$ meson and other heavier
axial-vector states like $\rho-\pi$ and other multi-particle states.

The first term in (\ref{2.60}), corresponding to the pion pole,
contains the factor $q_\mu$, which then terminates its contribution
to the cross section, Eq.~(\ref{2.40}). Indeed, the amplitude of the
reaction is
\beq A(\nu\,p\to l\,X) \propto L_\mu\,J^A_\mu,
\label{2.80} 
\eeq
where $L_\mu=\bar
l(k^\prime)\gamma_\mu(1+\gamma_5)\nu(k)$ is the lepton current,
which is transverse, i.e.  $q_\mu\,L_\mu=0$ (up to the lepton mass,
assumed to be zero in what follows, unless specified). Therefore,
the pion term in (\ref{2.60}) does not contribute to the amplitude
Eq.~(\ref{2.80}), and this is true at any $Q^2$.

Thus, PCAC does not mean pion pole dominance,  but connects the contribution of heavy axial
states  (the second term in Eq.~(\ref{2.60})) with the nonexistent
pion contribution 
\cite{Bell:1995nj,p-s,km}. One can see that in the
 $Q^2$-dependence 
of the neutrino cross section at small $Q^2$.
Fitting the measured $Q^2$-dependence with the parametrization 
$(Q^2+M_{eff}^2)^{-2}$, one gets $M_{eff}\approx 1.1\GeV$ \cite{km},
which is a clear evidence of the dominance of heavy states in the axial current.

\subsubsection{Shadowing in the total neutrino cross section}
\label{sec:total}

Nuclear shadowing of neutrinos was predicted a long time ago \cite{bell-shad}
based on the simple observation that shadowing in the pion-nucleus cross section
immediately means that neutrino cross section is shadowed as well, because both
are connected by the Adler relation. However, this connection may be affected by coherence phenomena, which are discussed below.

As far as the effective mass of a typical hadronic fluctuation of a
neutrino, $\nu\to l+a$, is large, quite a high energy, $\nu\gsim 10\GeV$,
is needed to make the fluctuation lifetime,
 \beq
t_{fluct}=\frac{2\,E}{M_{eff}^2},
\label{2.100}
 \eeq
comparable with
the radii of heavy nuclei. Then one might jump to the conclusion that there
should be no shadowing at low energies.

\begin{figure}[htb]
\epsfysize=9.0cm
\begin{center}
\begin{minipage}[t]{8 cm}
\epsfig{file=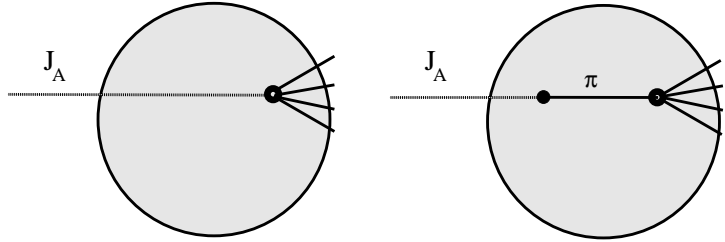,scale=0.4}
\end{minipage}
\begin{minipage}[t]{16.5 cm}
\caption{{\em Left:} Direct inelastic interaction of the axial current with a bound nucleon.
{\em Right:} Diffractive pion production preceding the inelastic collision. \label{fig:lc-shad}}
\end{minipage}
\end{center}
\end{figure}

However, this conclusion is not correct. It is based on the usual wisdom
that the fluctuation lifetime and the coherence time are equivalent
quantities, which is usually correct, but not in this case. 
The amplitude of inelastic neutrino-nucleus collision is shown schematically in Fig.~\ref{fig:lc-shad}.
The left picture corresponds to a direct inelastic interaction of the projectile axial current with a bound nucleon, while in the right picture the diffractively produced pion
experiences an inelastic collision. This looks similar to the calculation of shadowing for the photoabsorption cross section done in Sect.~\ref{sec:photons} in the case of vector dominance. In fact, it seems to be even better justified, since the pion pole is much closer to the physical region, than the $\rho$ pole. However, as we have just learned, it is not the pion pole which is behind the Adler relation, but heavier axial-vector singularities, which conspire
together and imitate the pion pole. Nevertheless, the diffractively produced pion
is on mass shell, but the details of the dynamics are hidden in the amplitude of diffractive neutrino-production of pion.

Since both contributions depicted in Fig.~\ref{fig:lc-shad} lead to the same final state, they can interfere. Squaring the inelastic amplitude we get the following expression for the ratio (\ref{1.20}) for the total neutrino-nucleus cross section \cite{nu-tot},
\beq
R^\nu_{A/N}=1 - \frac{8\pi}{A\sigma^{\nu N}_{tot}}
\left.\frac{d\sigma^{\nu\to\pi}_{diff}}{dp_T^2}\right|_{p_T=0}
\int d^2b\int\limits_{-\infty}^\infty dz_1\,\rho_A(b,z_1)
\int\limits_{-\infty}^{z_1} dz_2\,\rho_A(b,z_2)
e^{-iq_L^\pi(z_2-z_1)}\,e^{-{1\over2}\sigma^{\pi N}_{tot}T_A(b,z_2,z_1)},
\label{2.120}
\eeq
where $T_A(b,z_2,z_1)$ was defined in Eq.~(\ref{1.640}).
We neglected here the higher order terms in multiple interactions. 

At very low energy where $q_L^\pi$ is large, the second term in (\ref{2.120}) is suppressed, and the first term, which corresponds to a nuclear cross section proportional to $A$, dominates. At high energies $q_L^\pi\ll1$ can be neglected and the integrations over $z_{1,2}$ can be performed analytically \cite{nu-tot}. Due to the Adler relation the first term in (\ref{2.120}) and the volume part of the second term cancel, and the rest is the "surface" term $\propto A^{2/3}$.

It turns out that the high-energy regime actually starts at rather low energies if $Q^2\lsim
m_\pi^2$. Indeed, for elastic
neutrino-production of pions, $\nu p \to l\,\pi N$, the longitudinal
momentum transfer, $q_L^\pi=(m_\pi^2+Q^2)/2\nu$, i.e. the coherence time, 
\beq
t_c^\pi=\frac{1}{q_L^\pi}=\frac{2\nu}{m_\pi^2+Q^2},
\label{2.140}
\eeq
 is very long even at low
energy of few hundred $\MeV$. This is actually what matters for the onset of shadowing.
As for the fluctuation lifetime Eq.~(\ref{2.100}), it is indeed much shorter.

This is a result of the nontrivial origin of PCAC. The impossibility for the axial current to fluctuate into a
pion leads to the dominance of off-diagonal
processes, like $\nu\to \mu a_1$ and $a_1N\to \pi N$. The same happens for the
vector current, if one considers, for example, $\rho$ photoproduction via
intermediate excitation $\rho'$: $\gamma\to\rho'$ and $\rho' N\to\rho N$.
Such an off-diagonal contribution is negligibly small for the vector
current, but is a dominant one for neutrinos.  Only for diagonal
transitions the fluctuation lifetime and the coherence time are equal,
$t_{fluct}=t_c$. For neutrino interactions the former controls the $Q^2$
dependence of the cross section, while the latter governs shadowing.

\begin{figure}[htb]
\parbox{\halftext}{
\centerline{\includegraphics[width=7.5 cm]{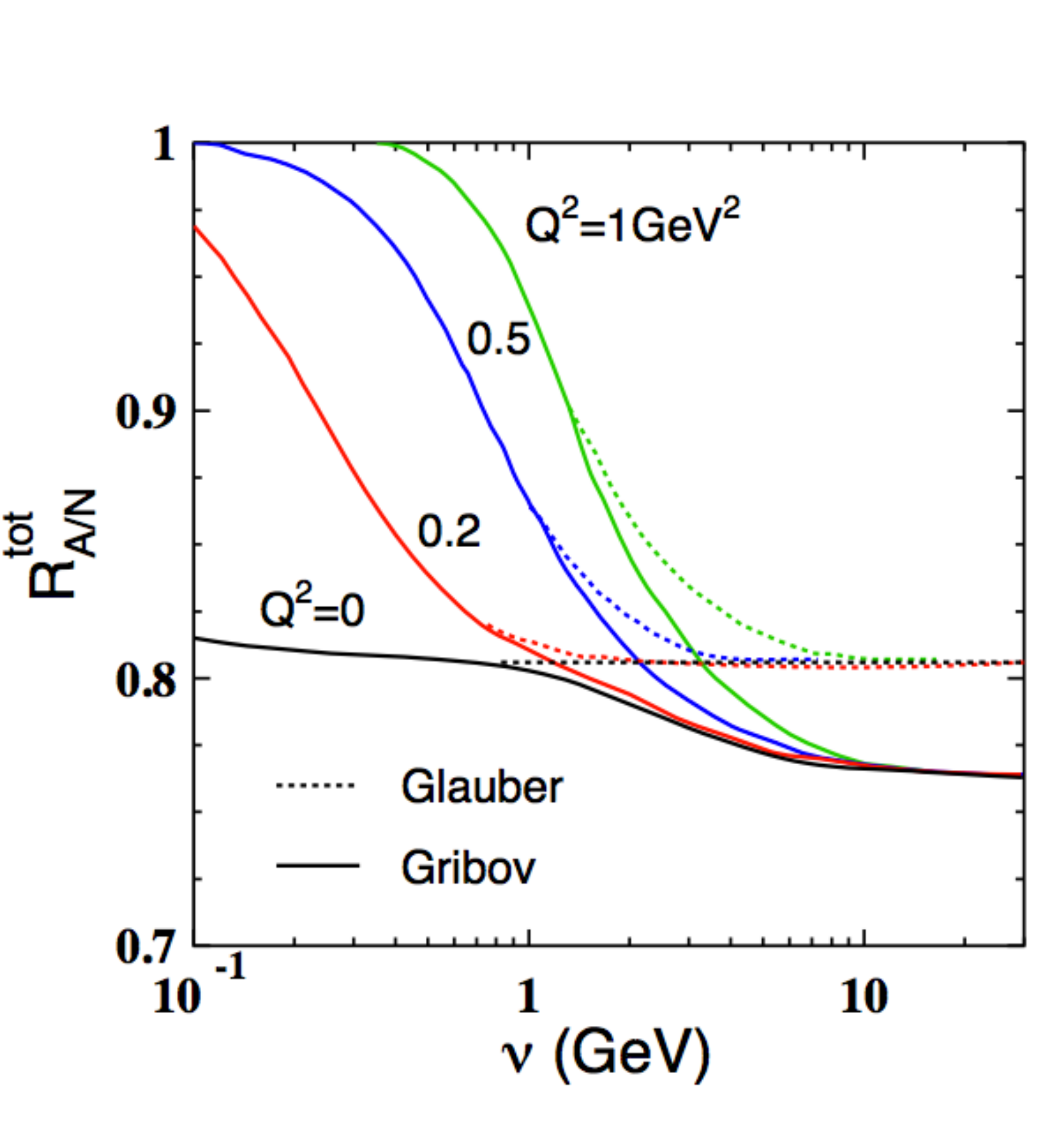}}
 \caption{The neon to nucleon ratio of total neutrino cross sections
 at different $Q^2$ \cite{nu-tot,gransasso}. Dashed and solid curves
correspond to the Glauber and Gribov corrected calculations. }
 \label{fig:shad-1}}
\hfill
\parbox{\halftext}{
\centerline{\includegraphics[width=7.5 cm]{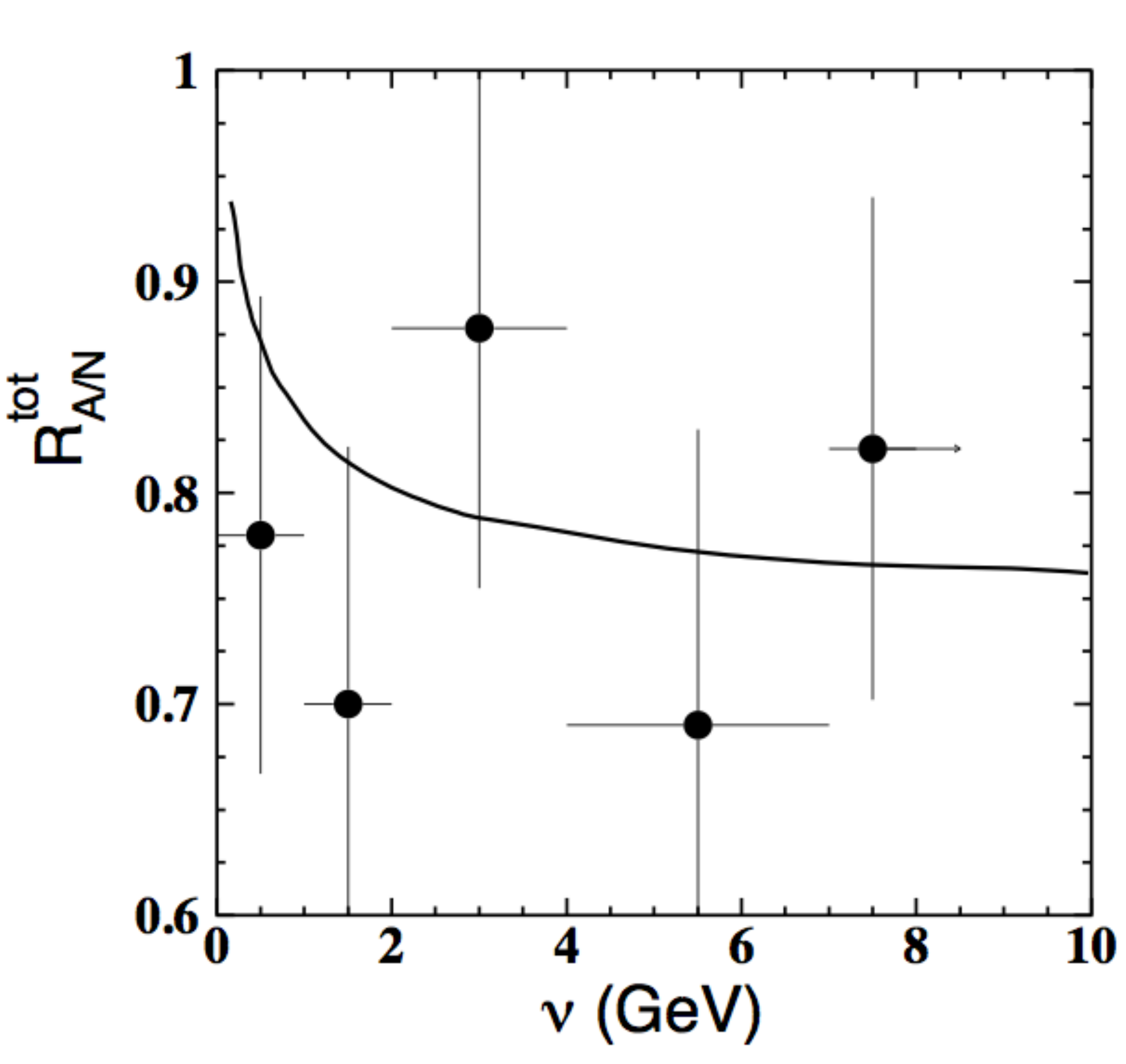}}
\caption{ The neon to proton ratio of the total neutrino cross sections,
calculated in \cite{nu-tot,gransasso} for $x<0.2$ and $Q^2<0.2\GeV^2$. The data present 
the BEBC results \cite{wa59}.}\label{fig:shad-2}}
 \end{figure}

The results of numerical evaluation of the shadowing effect Eq.~(\ref{2.120})
for neon are plotted in Fig.~\ref{fig:shad-1}.
The calculations \cite{nu-tot} done within 
the Glauber approximation, Eq.~(\ref{2.120}), and also including the Gribov's inelastic
corrections (important at high energies $\nu>m_{a_1}^2R_A$) are plotted in Fig.~\ref{fig:shad-1} by solid curves as function of energy, for different $Q^2$. 

The calculated shadowing effects are compared with BEBC data \cite{wa59} in Fig.~\ref{fig:shad-2}.
 As was anticipated, the shadowing exposes an early onset, and a significant suppression occurs at small $Q^2$ in
the low energy range of hundreds MeV. This is an outstanding feature of
the axial current. This seems to be supported by data, although 
 with rather poor statistics.

\subsubsection{Diffractive neutrino interactions: Breakdown of PCAC by shadowing}
\label{sec:breakdown}

The process $\nu+p\to l+\pi+p$ offers probably the most stringent test of PCAC in
neutrino interactions. Indeed, the analysis performed by Piketty and
Stodolsky \cite{p-s} revealed a potential problem related to the
dispersion representation, Eq.~(\ref{2.60}). Assuming the dominance of
the $a_1$ pole in the dispersion relation  they
arrived at an equation connecting the elastic and diffractive
pion-nucleon cross section, $\sigma(\pi p\to a_1 p)=\sigma(\pi
p\to\pi p)$. This relation strongly contradicts data: diffractive
production of $a_1$ meson is more than an order of magnitude
suppressed compared with the elastic cross section.

This puzzle was relaxed in \cite{belkov,km} by pointing out its shaky point,
namely, the $a_1$ pole cannot dominate in the axial current, since
it is quite a weak singularity compared to the $\rho$ pole in the
vector current case. In fact, the main contribution to the expansion
Eq.~(\ref{2.60}) comes from the $\rho$-$\pi$ and $3\pi$ cuts, related
to diffractive pion excitations. The invariant mass distribution
for diffractive pion excitations peaks at $M\approx 1.1\GeV$ \cite{transversity} and is
well explained by the so called Deck mechanism \cite{deck} of
diffractive excitation $\pi\to\rho\pi$. The interpretation of this peak has been a long
standing controversy, until a phase-shift amplitude analysis (see
references in \cite{pdg}) eventually revealed the presence of the
weak $a_1$ resonance having a similar mass. Summing up all
diffractive excitations (excluding large invariant masses
corresponding to the triple-Pomeron term), one concludes that the magnitudes of
single-diffractive and elastic pion-proton cross section are indeed
similar. This helps to resolve the Piketty-Stodolsky puzzle. 

Nevertheless, it was demonstrated that absorptive corrections break the relation between the heavy states and pion in the dispersion relation (\ref{2.60}) imposed by PCAC.
The deviation from the Adler relation was estimated in \cite{dispersion,marat3,marat4} at about $30\%$.

Even more dramatic breakdown of PCAC caused by nuclear shadowing was found in \cite{dispersion} for diffractive neutrino-production of pions on nuclei. These processes are usually classified as coherent or incoherent, which
according to the conventional terminology correspond to processes which leave the nucleus intact or the nucleus breaks up into fragments, respectively.

In what follows we assume the validity of the Adler relation for a nucleon target, in order to identify the net nuclear effects.
In the coherent pion production process the amplitudes on different
nucleons interfere, and the interference is enhanced by the condition that the nucleus remains intact.
The coherence effects can lead to substantial deviations from the AR and from simplified expectations, as is demonstrated below.

In addition to the pion coherence time, Eq.~(\ref{2.140}), another time scale related to heavy axial-vector state is important,
\beq
t_c^a=\frac{1}{q_L^a}=\frac{2\nu}{m_a^2+Q^2},
\label{2.160}
\eeq
The two scales control the amplitude of coherent neutrino-production \cite{dispersion},
whose imaginary part has the form,
 \beq
 M_{\nu A\to l\pi A}(\nu,Q^2,b)=M_1(\nu,Q^2,b)-M_2(\nu,Q^2,b),
 \label{2.180}
 \eeq
 where
 \beqn
 M_1(\nu,Q^2,b)&=&M_{\nu N\to l\pi N}
(\nu,Q^2) \int\limits_{-\infty}^\infty dz\,
e^{iq_L^\pi z}\,
  \rho_A(z,b)\,e^{-{1\over2}\sigma_{tot}^{\pi N}T_A(b,z)};
 \label{2.200}\\
 M_2(\nu,Q^2,b) &=&
 M_{\nu N\to laN}(\nu,Q^2)\,
 M_{aN\to \pi N}(\nu)
 \int\limits_{-\infty}^\infty dz\,
 e^{i(q_L^\pi-q_L^a) z}\,
  \rho_A(z,b)\,e^{-{1\over2}\sigma_{tot}^{\pi N}T_A(b,z)}
  \nonumber \\&\times&
 \int\limits_{-\infty}^z dz_1 e^{iq_L^a z_1}\,\rho_A(z_1,b)\,
e^{-{1\over2}\sigma_{tot}^{a N}[T_A(b,z_1)-T_A(b,z)]}.
 \label{2.220}
 \eeqn

The structure of the amplitude is similar to the one in Eq.~(\ref{2.120}). While the first term, $M_1$, corresponds to pion production by the neutrino without any preceding interaction, the second term $M_2$ corresponds to diffractive production of the heavy state $a$ preceding the pion production.
This is the first order Gribov inelastic shadowing correction \cite{gribov} to the coherent pion production amplitude.

\begin{figure}[htb]
\parbox{\halftext}{
\centerline{\includegraphics[width=7.5 cm]{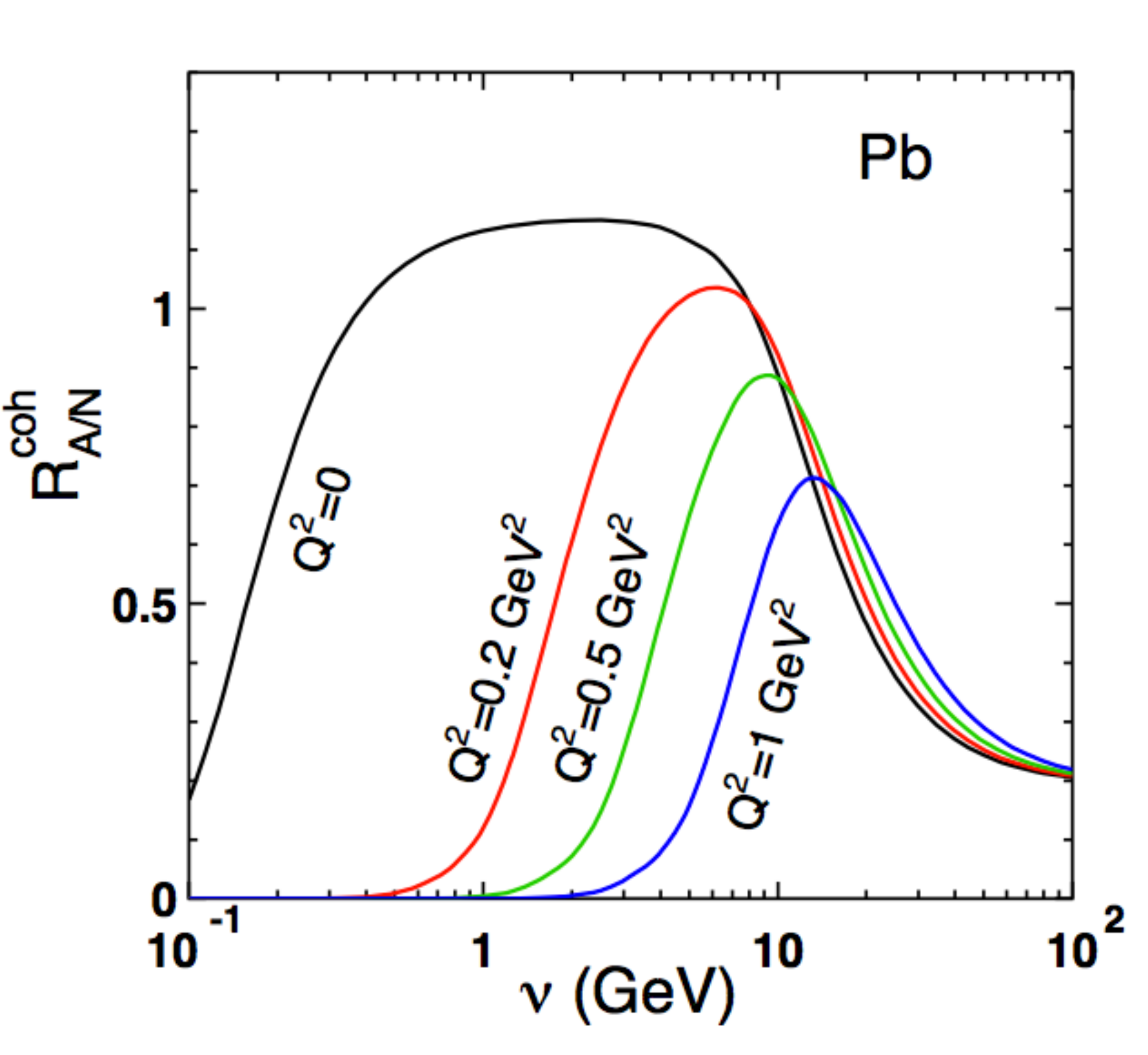}}
 \caption{Nuclear ratio $R^{coh}_{A/N}(\nu,Q^2)$ of $p_T$-integrated cross sections of coherent neutrino-production of pions on lead at different $Q^2$. }
 \label{fig:coh-Qdep}}
\hfill
\parbox{\halftext}{
\centerline{\includegraphics[width=7.5 cm]{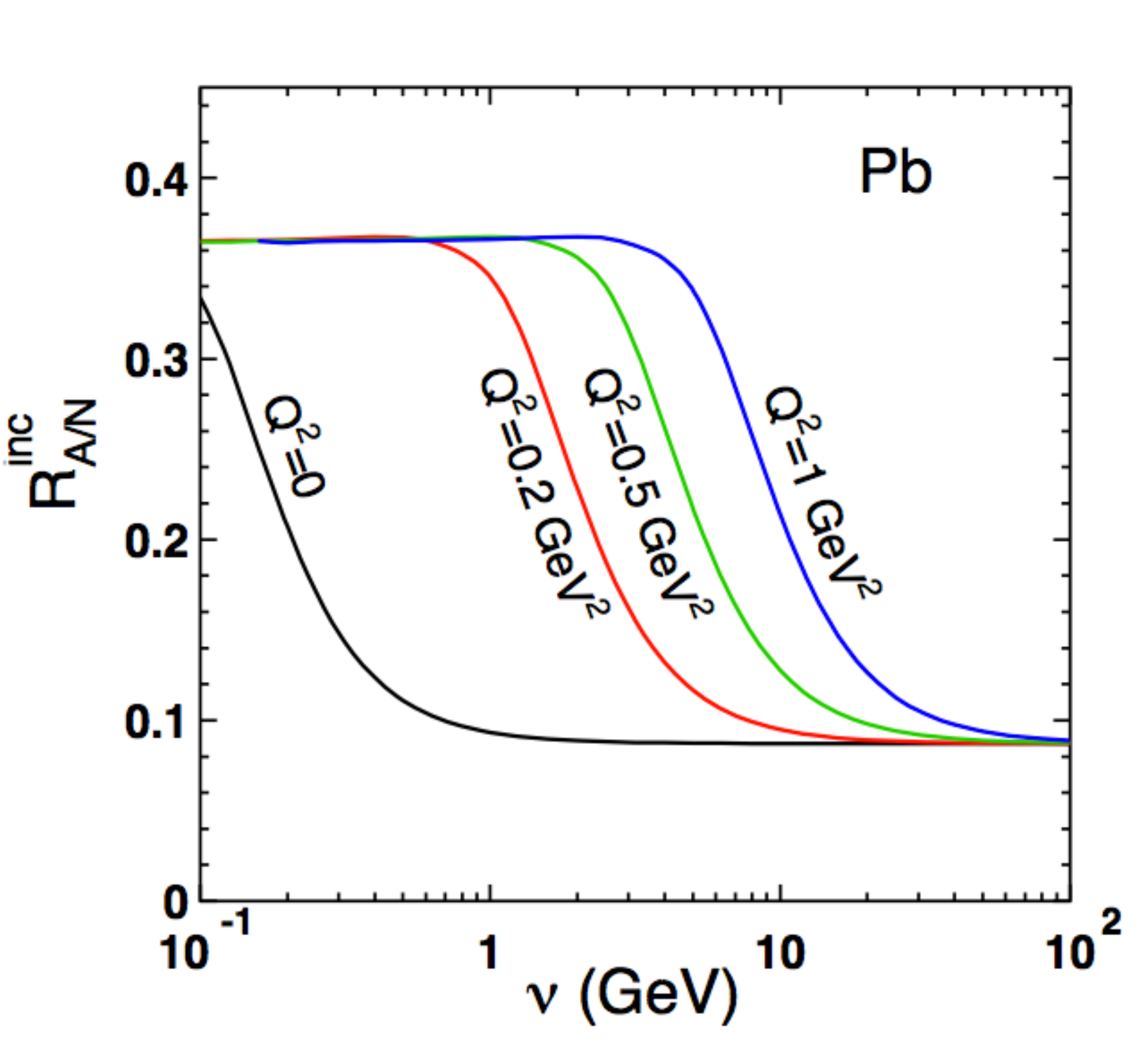}}
\caption{ The same as in Fig.~\ref{fig:coh-Qdep} for incoherent production on lead  at $Q^2=0,\ 0.2,\ 0.5$ and $1\GeV^2$.}
\label{fig:inc-Qdep}}
 \end{figure}

The results for coherent production of pions on lead are depicted in Fig.~\ref{fig:coh-Qdep}
as function of transferred energy $\nu$, for several values of $Q^2$.
The suppression predicted at low energies is related to the shortness of $t_c^\pi\ll R_A$ and lack of coherence. At higher energies the nuclear ratio at $Q^2=0$ forms a plateau in the energy interval $0.5\lsim\nu\lsim5\GeV$, which corresponds to the condition $t_c^\pi\gg R_A$, but $t_c^a\ll R_A$.
The height of the plateau corresponds to the Adler relation, which holds in this regime of short $l_c^a$. Remarkably, at higher energies $R^{coh}_{A/N}(\nu,Q^2=0)$ steeply falls from the $A$-dependence $A^{2/3}$ down to $A^{1/3}$, exposing a dramatic violation of the Adler relation. The source of the breakdown is shadowing, which begins at long $t_c^a$.

As function of $Q^2$ the plateau considerably shrinks leaving almost no room for 
the Adler relation (extrapolated to nonzero $Q^2$) to hold, as one can see from the
few examples depicted in Fig.~\ref{fig:coh-Qdep}.

A similar behavior is expected for incoherent production of pions $\nu+A\to l+\pi+A^*$,
except at the low energy range where no suppression is expected, because pions are produced on different nucleons incoherently. Therefore the plateau corresponding to the Adler relation starts at very low energies, as is demonstrated in Fig.~\ref{fig:inc-Qdep} for several values of $Q^2$. At higher energies, when $t_C^a\lsim R_A$ initial state shadowing causes an additional suppression, and similarly to the coherent case the $A$ dependence drops from $A^{2/3}$ down to $A^{1/3}$ as is demonstrated in Fig.~\ref{fig:inc-Qdep} (the details of calculations can be found in \cite{dispersion}).
Again, shadowing led to a dramatic breakdown of PCAC.

\section{Shadowing of deeply virtual photons}
\label{sec:virtual}

The sizable shadowing effects observed above were caused  by the large cross sections
of soft interactions of light hadrons, or soft hadronic fluctuations of photons and leptons at a low virtuality. It has been observed, however, that considerable shadowing 
also suppresses the cross section of hard reactions on nuclei, like deep-inelastic lepton scattering (DIS), Drell-Yan reaction, etc. In what follows we demonstrate that the dominant contribution to the shadowing effects comes from the soft component of hard processes, while the hard component is not shadowed.

The cross section of deep inelastic lepton scattering $l+N\to l'+X$ has the form,
\beq
\frac{d^2\sigma}{dxdQ^2}=\frac{4\pi\alpha_{em}^2}{Q^4}\left\{
\left(1-y-\frac{x^2y^2m_N^2}{Q^2}\right)\frac{F_2(x,Q^2)}{x}
+y^2F_1(x,Q^2)\right\},
\label{3.20}
\eeq
where the invariant structure functions can be expressed in terms of the total cross sections $\sigma_T(x,Q^2)$ and $\sigma_L(x,Q^2)$of transversely and longitudinally polarized virtual photons,
\beqn
F_2(x,Q^2) &=&
\frac{Q^2(1-x)}{4\pi^2\alpha_{em}}\,\left(\sigma_T+\sigma_L\right);
\nn\\
F_1(x,Q^2) &=&
\frac{Q^2(1-x)}{8\pi^2\alpha_{em}\,x}\,\sigma_T.
\label{3.40}
\eeqn
We use the standard notations, $\sqrt{s}$ and $W$ are the c.m. energies; $x=Q^2/(2qP)$, $y=Q^2/(2kP)$, where $k$, $q$ and $P$ are the 4-momenta of the initial lepton, virtual photon and the target nucleon, $Q^2=-q^2$.

Notice that the nuclear ratio of the structure functions,
\beq
\frac{F_2^A}{F_2^N}=\frac{\sigma_T^{\gamma^*A}+\sigma_L^{\gamma^*A}}
{\sigma_T^{\gamma^*N}+\sigma_L^{\gamma^*N}},
\label{3.44}
\eeq
which is independent of the lepton energy, is not the same as 
the measured ratio of the lepton-target cross sections,  
\beq
\frac{\sigma^{lA}_{DIS}}{\sigma^{lN}_{DIS}}=
\frac{F_2^A}{F_2^N}\frac{(1+\xi R_{L/T}^A)
(1+R_{L/T}^N)}{(1+R_{L/T}^A)(1+\xi R_{L/T}^N)},
\label{3.48}
\eeq
where
\beq
\xi=\frac{4(1-y)-\frac{4m_N^2x^2}{Q^2}}
{4(1-y)+2y^2+\frac{4m_N^2x^2}{Q^2}}
\label{3.52}
\eeq
is the photon polarization parameter, and
\beq
R_{L/T}^{A(N)}=\frac{\left(1+\frac{4m_N^2x^2}{Q^2}\right)\sigma_L^{\gamma^*A(N)}
+\frac{4m_N^2x^2}{Q^2}\sigma_T^{\gamma^*A(N)}}{\sigma_T^{\gamma^*A(N)}}
\approx\frac{\sigma_L^{\gamma^*A(N)}}
{\sigma_T^{\gamma^*A(N)}}.
\label{3.54}
\eeq
According to (\ref{3.48}) the lepton nucleon cross section ratio is only equal to the structure function
ratio, if there are no nuclear effects in $R_{L/T}^A$ or $\xi=1$. This is assumed in all NMC data
and supported by experimental observations. In the kinematical region relevant
for NMC, data for $R_{L/T}^A$ show practically no $A$ dependence \cite{Muecklich}. 
Nevertheless this problem needs further study and more precise data.

Already the first  DIS measurements at SLAC
\cite{SLAC} showed that the structure function $F_2(x,Q^2)$ 
is nearly constant as
function of $Q^2$ at fixed $x$. An explanation of this phenomenon was given by Bjorken \cite{Bjorken} and by
Feynman \cite{Feynman}. If the transverse momenta of the partons are
neglected, then the cross sections for transverse and longitudinal
photons scattering off spin-$1/2$ partons, i.\ e.\ quarks, are given by
\beqn
\sigma_T&=&\frac{4\pi^2\alpha_{em}Z_f^2}{Q^2(1-x)}\,
\delta\left(1-\frac{x}{x_q}\right)\nn\\ 
\sigma_L&=&0,
\label{3.60}
\eeqn
where $x_q$ is the momentum fraction of the proton carried by the struck quark
and $Z_f$ is the flavor charge in units of the elementary charge.
The $\delta$-function arises from momentum conservation and 
gives a physical meaning to the Bjorken variable. In the Breit frame \cite{Povh},
$x$ is the momentum
fraction of the proton carried by the struck quark.  
For massless quarks, 
the longitudinal cross section is zero due to helicity conservation.
Introducing the density $q_f(x)$ 
of quarks of flavor $f$, inside the
proton, one obtains a simple partonic interpretation of the structure
functions,
\beqn
F_2(x)&=&x\sum\limits_f^{}Z_f^2( q_f(x)+\bar q_f(x)),\\
F_L&=&0.
\label{3.80}
\eeqn
In the naive parton model, the structure functions depend only on $x$ and
not on $Q^2$, the longitudinal structure function vanishes and one obtains the
Callan-Gross relation \cite{CallanGross},
\beq
F_2-2xF_1=0.
\label{3.100}
\eeq
These equations are the basic results of the parton model and they are
approximately confirmed by experiment. 

One of the great achievements 
of QCD is the successful description of the deviations
from the naive parton model seen in experiment. 
In particular at low $x$
deviations from Bjorken scaling become quite pronounced.
In  the QCD improved parton
model, perturbation theory is applied to calculation of the corrections to the parton
model predictions. Increasing the photon virtuality one enhances the resolution of the parton probe, and one sees more partons. New partons appearing at higher resolution are generated by the splitting processes $G\to 2G$, $G\to q\bar q$, $q\to qG$.
The evolution of the structure function with $Q^2$ is described by the DGLAP
equations \cite{GL,Lipatov,AP,Dok}, which 
for the singlet parton densities read,
\beq
Q^2\frac{d}{dQ^2}
\left(
\begin{array}{c}
q_f(x,Q^2)\\[1em]
G(x,Q^2)
\end{array}
\right)
=
\frac{\alpha_s}{2\pi}\int\limits_{x}^1 \frac{dx_1}{x_1}
\left(
\begin{array}{cc}
P_{ff}\left(\frac{x}{x_1}\right)
&P_{fG}\left(\frac{x}{x_1}\right)\\[1em]
P_{Gf}\left(\frac{x}{x_1}\right)&P_{GG}\left(\frac{x}{x_1}\right)
\end{array}\right)
\left(
\begin{array}{c}
q_f(x_{1},Q^2)\\[1em]
G(x_{1},Q^2)
\end{array}
\right),
\label{3.120}
\eeq
where $q_f$ is a quark of a given flavor, and the splitting functions $P(x)$ are calculated perturbatively \cite{Ellis}.

All the soft physics is contained in the parton distributions. These are essentially contaminated by 
nonperturbative interactions and have to be parametrized at some input scale $Q_0^2$.
Different parametrizations have been provided by several collaborations \cite{GRV,MRST,CTEQ} performing global analyses of data in leading and next to leading orders. With the parton distributions as input, one can then 
calculate $F_2$ at a higher value of
$Q^2$. 
The other essential property is that the parton distributions are universal, i.e.
they do not depend on the process under consideration, but only on the hadron
state. This is based on QCD factorization proven for some processes and  up to higher twist effects \cite{css}.

\subsection{Phenomenology of shadowing in DIS}
\label{sec:DIS}

The space-time picture of an interaction within the parton model varies significantly  depending on the reference frame.
Only observables are Lorentz invariant, but not our theoretical ideas about the dynamics of the interaction.
In particular, what looks like absorption of the virtual photon by a parton in the nucleon in the Breit frame, looks very different in the rest frame of the proton. Here the same process looks like fluctuation of a high-energy virtual photon into a colorless $\bar qq$ dipole, with a subsequent interaction of the dipole with the target via gluonic exchanges. 

\begin{figure}[htb]
\hspace*{3cm}
{\includegraphics[width=5 cm]{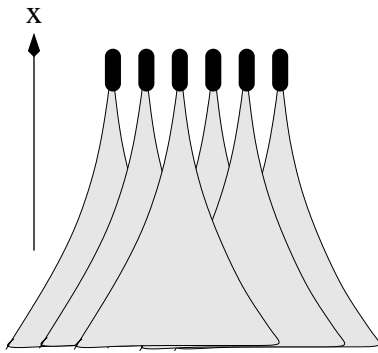}}\\[-3cm]
\hspace*{8cm}
\parbox{9cm}{ \caption{Overlap of parton clouds originated from different nucleons in a fast nucleus due to the reduction 
 of the Lorentz factor at small $x$  }
 \label{fig:clouds}}
 \end{figure}

Correspondingly, shadowing also looks quite different. In the infinite momentum frame of the nucleus both the nucleus and each bound nucleon are Lorentz contracted by the $\gamma$-factor.  A Lorentz-boosted nucleus looks like a pancake, as well as the bound nucleons. So if the nucleons do not overlap in the nuclear rest frame, they are still separated in the Breit frame also. However, the Lorentz $\gamma$-factor of the small-$x$ partons is reduced by the factor $x$, so they contract much less and start overlapping, like is illustrated in Fig.~\ref{fig:clouds}.
\vspace*{1.5cm}
This leads to a fusion of partonic chains originated from different bound nucleons \cite{kancheli}, and to a reduction of parton densities, and this produces shadowing.

In the nuclear rest frame the physics of shadowing is more intuitive and corresponds better to its optical analogy.
In this reference frame the $\bar qq$-dipole fluctuation of the photon propagates through the nucleus and experiences multiple interactions, which lead to a reduction of the photon flux, and eventually to a suppression of the cross section.
This is the same phenomenon as parton fusion, but seen from different reference frames.

The results of DIS measurements on nuclear targets are usually presented in the form of a ratio, Eq.~(\ref{1.20}), for the structure functions 
\beq
R_{A/N}(x,Q^2)=\frac{F^A_2(x,Q^2)}
{A\,F_2^N(x,Q^2)}.
\label{3.140}
\eeq
The shadowing suppression factor $R_{A/N}(x,Q^2)<1$ can be estimated using the same
two-step correction as for real photons, as is illustrated in Fig.~\ref{fig:vdm}.
It can be calculated with the same formula (\ref{1.320}) replacing $\sigma_{sd}^{hN}\Rightarrow\sigma_{sd}^{\gamma^*N}$, although the value of the absorption cross section remains problematic.

Shadowing in DIS on nuclei begins at sufficiently high energies $W^2\gg Q^2$, i.e. at small $x\ll1$. In this case the coherence time, also called Ioffe time, in the target rest frame reads,
\beq
t_c={1\over q_L}=\frac{2\nu}{Q^2+M_X^2}\approx \frac{1}{2m_N x}.
\label{3.160}
\eeq
One should expect the onset of shadowing at $t_c\gsim2\fm$, i.e. at $x<0.05$, and saturation at $t_c\gg R_A$. Actually the saturation is not exact due to the  contribution of large masses, increasing with $1/x$.
Expression (\ref{1.480}) written in the "frozen" approximation, i.e. with no mixing of different eigen components during propagation through the nucleus, includes all the inelastic shadowing corrections and can be applied to DIS. Expanding the exponential one gets in the lowest order in multiple interactions  \cite{povh1,povh2},
\beq
R_{A/N}(x,Q^2)=
1-{1\over4}\sigma_{eff}\la T_A\ra +...\,,
\label{3.210}
\eeq
where $\la T_A\ra = (1/A)\int d^2b\,T_A^2(b)$ is the mean value of the nuclear thickness function, and
\beq
\sigma_{eff}=\frac{\la(\sigma^{iN}_{tot})^2\ra}
{\la \sigma^{iN}_{tot}\ra}.
\label{3.215}
\eeq
It is interesting to notice that in this ratio the numerator and denominator are controlled by different scales.
Let us classify the hadronic fluctuations of a highly virtual photon as either hard or soft, as is presented in Table~\ref{tab:interplay}.
 \begin{table}[h]
\begin{minipage}[t]{16.5 cm}
 \caption{
 Contributions of soft and  hard fluctuations
of a virtual photon to the DIS cross section and to nuclear shadowing.}
\label{tab:interplay}
\end{minipage}
\begin{center}
\begin{tabular}{|c|l|l|l|l|c|}
 \hline 
Fluctuation & weight: $ W_i$ & $\sigma^{i}_{tot}$ & 
$W_i\,\sigma^{i}_{tot}$ &
 $W_i\,(\sigma^{i}_{tot})^2$\\
 \hline Hard & $\sim 1$ & $\sim 1/Q^2$ 
& $\sim 1/Q^2$  & $\sim 1/Q^4$\\
 \hline
 Soft & $\sim\mu^2/Q^2$ &$\sim1/\mu^2$& 
$\sim 1/Q^2$ & $\sim 1/\mu^2Q^2$\\
\hline
\end{tabular}
\end{center}
 \end{table}
Naturally, hard eigenstate fluctuations interact weekly with the cross $\sigma^i_{tot}\sim1/Q^2$, while soft ones have a large cross section $\sigma^i_{tot}\sim1/\mu^2$,
where $\mu$ is a soft hadronic scale. At the same time, hard fluctuations dominate
in a highly virtual photon, while soft ones appear rarely with a probability suppressed at least as $1/Q^2$ ($1/Q^4$ for longitudinally polarized photons). 
So smallness of the hard cross section is compensated by the large weight, and vice versa.
Thus, both contributions, hard and soft, behave like $1/Q^2$, and 
as Table~\ref{tab:interplay} shows, the DIS cross section $\la \sigma^i_{tot}\ra$
gets a finite contribution from the soft interactions at any high $Q^2$. However the shadowing term in (\ref{3.210}), as well as diffraction, turn out to be dominated by soft interactions, whose contribution scales as $1/Q^2$ similar to the DIS inclusive cross section.
Therefore the $x$-dependence of $\sigma_{eff}(x,Q^2)$ should be pretty mild,
$(1/x)^\delta$, where $\delta=2\Delta_{\Pom}(\mu^2)-\Delta_{\Pom}(Q^2)$.
The Pomeron intercept $\alpha_{\Pom}(0)=1+\Delta_{\Pom}(Q^2)$ rises with $Q^2$,
so the two terms in $\delta$ essentially compensate.

\begin{figure}[htb]
\parbox{\halftext}{
\centerline{\includegraphics[width=8 cm]{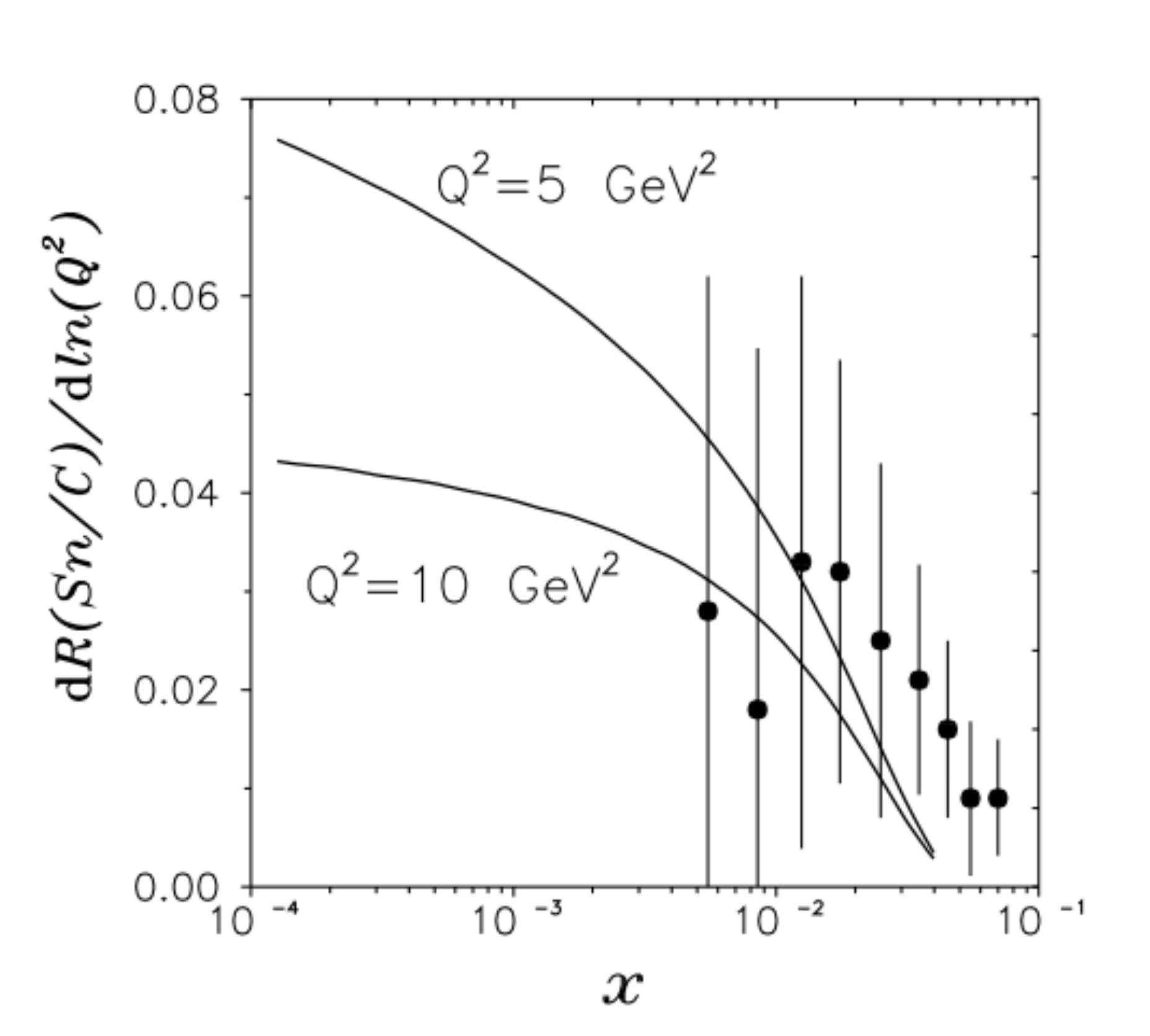}}
 \caption{Logarithmic $Q^2$-derivative of the ratio of structure
functions for tin to carbon. The data are from \cite{Muecklich}.}
 \label{fig:q-derivative}}
\hfill
\parbox{\halftext}{
\centerline{\includegraphics[width=9 cm]{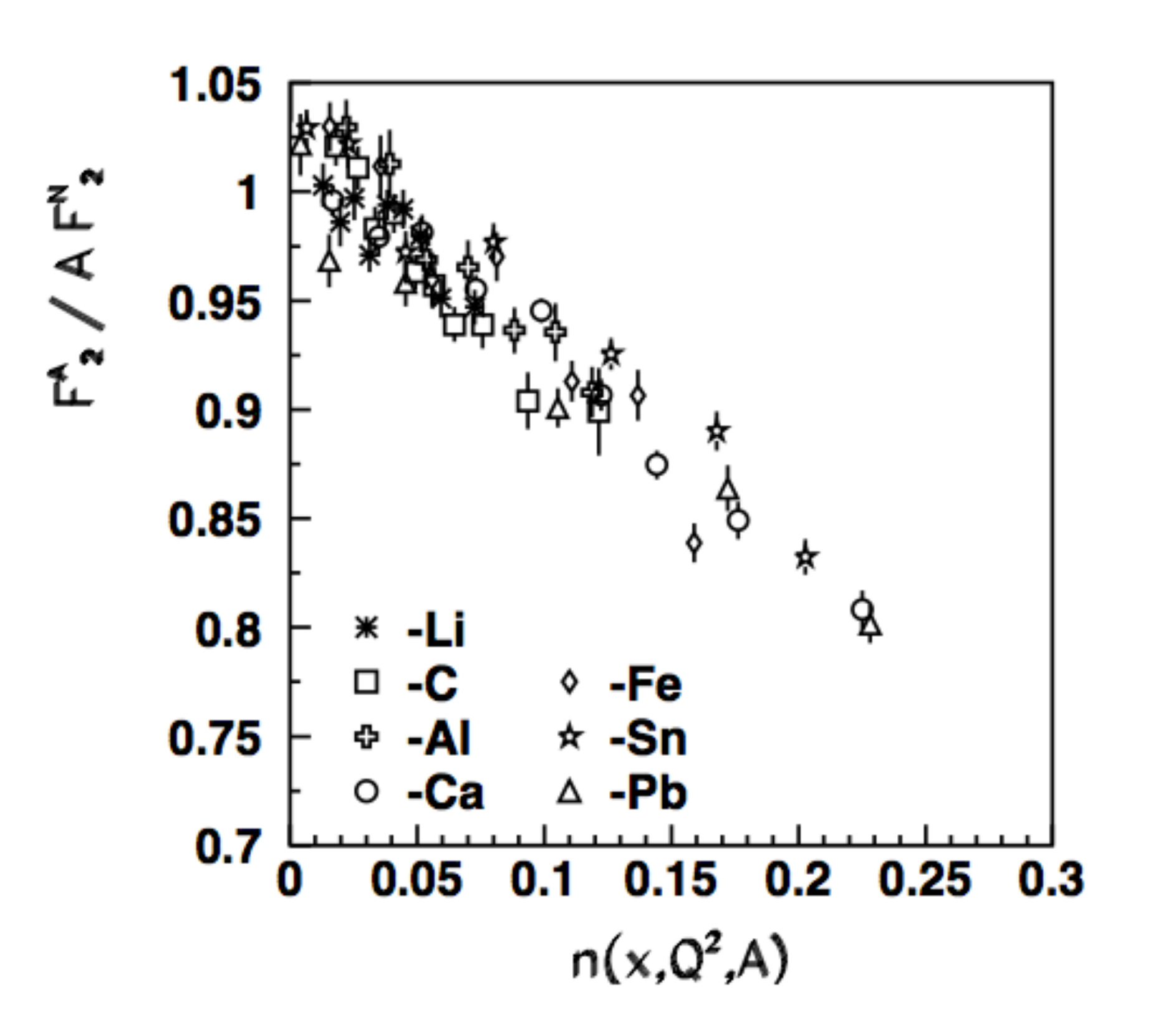}}
\caption{Nuclear shadowing versus scaling variable $n(x,Q^2,A)$ (see text).
The data for Li, C and Ca are from \cite{nmc-1,nmc2}.}
\label{fig:scaling}}
 \end{figure}

Another interesting consequence of the presence of two scales in the nuclear shadowing term in (\ref{3.210})-(\ref{3.215}) is the absence of $Q^2$ dependence in the nucleus-related denominator, $\la\sigma^i_{tot}\ra$. The $Q^2$ dependence of $R_{A/N}(x,Q^2)$ comes from the nucleon structure function $F_2^N(x,Q^2)$ in the denominator. Comparison with NMC data presented in Fig.~\ref{fig:q-derivative} demonstrated good agreement. 
This fact makes questionable the possibility of extracting the nuclear gluon density from $Q^2$
dependence of $R_{A/N}(x,Q^2)$ via a DGLAP based analysis. Indeed, the DGLAP equations relate the logarithmic $Q^2$ derivative of the structure function at high $Q^2$ to the gluon density \cite{prytz}. Correspondingly,
one can extract gluon shadowing from the $Q^2$ variation of $R_{A/N}(x,Q^2)$ as was done in \cite{pirner}.
However, data presented in Fig.~\ref{fig:q-derivative} can be explained with the lowest $\bar qq$ Fock component
of the virtual photon, containing no gluons.

Furthermore, one can calculate the effective cross section Eq.~(\ref{3.215}) since the numerator   $\la(\sigma^i_{tot})^2\ra$  according to Eqs.~(\ref{1.420}) and (\ref{1.580}) is related to diffraction,
\beq
\sigma_{eff}=\frac{16\pi}{\sigma^{\gamma^*N}_{tot}(x,Q^2)}
\left.\int dM_X^2\,\frac{d\sigma^{\gamma^* N}_{sd}(x,Q^2)}
{dM_X^2\,dp_T^2}\right|_{p_T=0},
\label{3.217}
\eeq

So far we neglected in (\ref{3.210}) only the higher order multiple interactions, hoping that $\sigma_{eff}\la T_A\ra$  is small. If, however, $x$ is not sufficiently small for the "frozen" approximation, the phase shifts diminish the shadowing effect. These corrections should be done similar to Eq.~{\ref{1.320}, and another simplifying approximation is the one made in (\ref{3.160}), where we fixed $M_X^2=Q^2$. In this case Eq.~(\ref{3.210}) is
generalized for the regime of shadowing onset as \cite{povh1,povh2},
\beq
R_{A/N}(x,Q^2)=
\frac{\sigma^{\gamma^*A}_{tot}(x,Q^2)}
{\sigma^{\gamma^*N}_{tot}(x,Q^2)} \approx
1-{1\over 4}\sigma_{eff}\,
\la T_A\ra\ F_A^2(q_L) \equiv1-n(x,Q^2,A),
\label{3.220}
\eeq
where the longitudinal nuclear form factor 
\beq
F^2_A(q_L) = 
\frac{1}{A\la T_A\ra}
\int d^2b\,\left|\int\limits_{-\infty}^{\infty}
dz\ \rho_A(b,z)\,e^{iq_L z}\right|^2
\label{3.240}
\eeq
takes into account the 
effects of the finite coherence time $t_c\approx 1/q_L$,
Eq.~(\ref{3.160}). At large $q_L > 1/R_A$, the nuclear 
form factor (\ref{3.240}) vanishes and suppresses the shadowing
term (\ref{3.220}). This is easily interpreted: for large $q_L$
the fluctuation lifetime and its path 
in nuclear medium are short, and shadowing is reduced.

In expression (\ref{3.220}) all ingredients are calculable for different nuclei, and as functions of $x$ and $Q^2$. This invites the introduction of a scaling variable $n(x,Q^2,A)$ defined in (59) \cite{povh1,povh2}, which should make shadowing universal for all nuclei, and values of $x$ and $Q^2$. In Fig.~\ref{fig:scaling}
NMC data points \cite{nmc-1,nmc2} for $R_{A/N}(x,Q^2)$, each having specific values of $x$ and $Q^2$ are plotted against the variable $n(x,Q^2,A)$. Data confirm the predicted scaling with good accuracy.

In terms of parton distribution functions (PDF) one can disentangle shadowing for quark and gluons, based on the triple-Regge phenomenology.
The cross section of the single-diffractive process, $\gamma^*+N \to X +N$, can be 
expressed in terms of the triple-Regge graphs. Indeed, summing up all final 
state excitations $X$, one can apply the unitarity relation to the 
$\gamma^*$-$\Pom$ amplitude, or to the total cross section $\sigma^{\gamma^*\Pom}_{tot}$, as is shown in Fig.~\ref{fig:3R-gamma}.
\begin{figure}[htb]
\parbox{\halftext}{
\centerline{\includegraphics[width=8 cm]{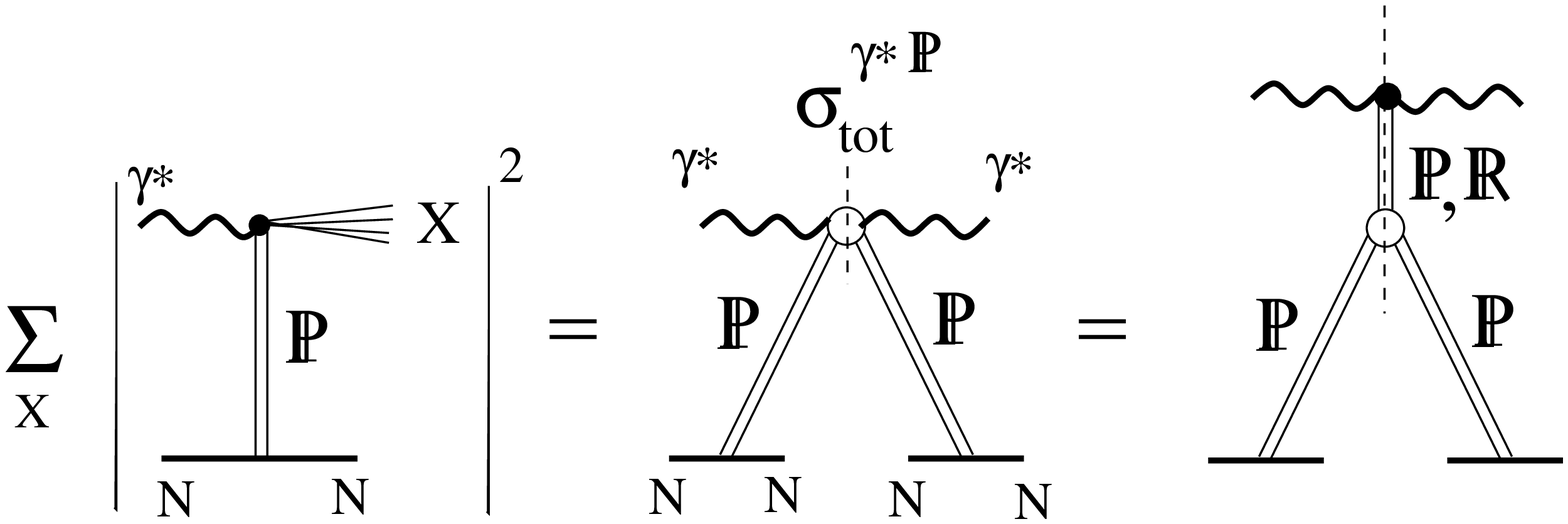}}
 \caption{The cross section of single diffraction, $\gamma^*N\to XN$ summed 
over all excitation channels at fixed effective mass $M_X$. 
The $\gamma^*$-$\Pom$ total cross section at high collision energies is a sum of Reggeon and Pomeron exchanges.}
 \label{fig:3R-gamma}}
\hfill
\parbox{\halftext}{
\centerline{\includegraphics[width=8 cm]{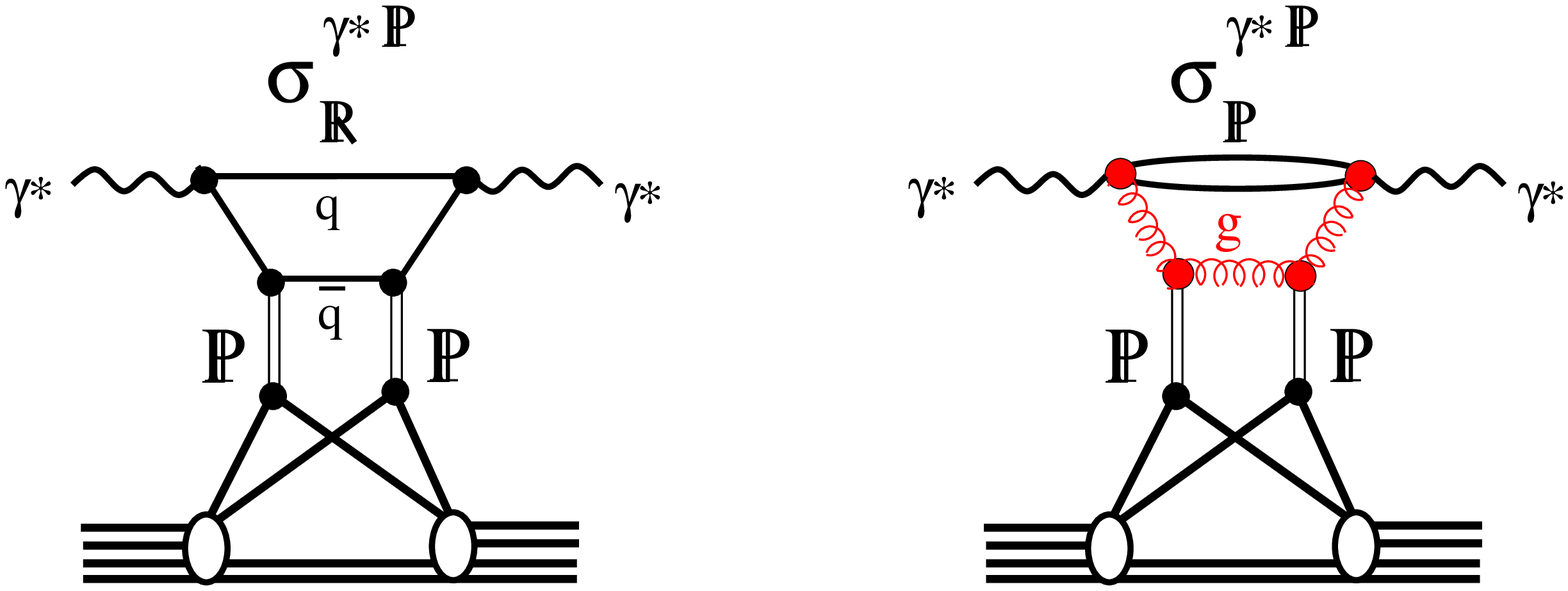}}
\caption{ The Gribov correction with $\bar qq$ (left) and $\bar qq+g$ (right) intermediate states with large invariant mass, which is related to small fractional momenta either of a quark, or the gluon respectively. }
\label{fig:pom-reg}}
 \end{figure}
The latter, according to (\ref{3.80}), is proportional to the structure function of the target, i.e. the Pomeron, so one can say that this way one can measure the PDF of the Pomeron \cite{schlein}.

 Provided that the effective mass of the excitation is large (but not too
much), $s_0\ll M_X^2\ll s$, one can describe the Pomeron-hadron elastic
amplitude via exchange of the Pomeron or secondary Reggeons in the
$t$-channel. Then one arrives at the triple-Regge graphs, which lead
to the cross section \cite{kklp},
 \beq
\frac{d\sigma_{sd}^{\gamma^*N\to XN}}{dx_F\,dt} =
\sum\limits_{i=\Pom,\Reg} G_{\Pom\Pom i}(t)
(1-x_F)^{\alpha_i(0)-2\alpha_\Pom(t)}
\left(\frac{s}{s_0}\right)^{\alpha_i(0)-1}.
\label{3.180}
 \eeq
 Here $x_F$ is the Feynman variable for the recoil nucleon in the c.m. of collision,
$x_F \approx 1-M_X^2/s$; $G_{\Pom\Pom\Pom}(t)$ and $G_{\Pom\Pom\Reg}(t)$ are phenomenological triple-Regge vertices.

In terms of quark-gluon intermediate states the Reggeon and Pomeron exchanges correspond to the photon fluctuating either to heavy $\bar qq$ or $\bar qqg$ fluctuations
respectively, as is depicted in Fig.~\ref{fig:pom-reg}. In the latter case what is important is that the large invariant mass $M_x$ is due to a large difference between the light-cone momenta of the $\bar qq$ and the $g$.
In terms of PDFs, the Reggeon and Pomeron parts of $\sigma^{\gamma^*\Pom}_{tot}$ correspond to measurements of the quark, or gluon distribution functions.

In the case of Gribov shadowing corrections the two Pomerons couple to different bound nucleons. Notice that  the Gribov correction in the elastic $\gamma^*$-$A$ amplitude, Fig.~\ref{fig:vdm},
comes from different unitarity cuts \cite{agk} corresponding to single and double inelastic interactions in the nucleus, or single diffraction, as shown in Fig.~\ref{fig:3R-gamma}.

\subsubsection{Models for Gribov corrections}
The Gribov correction to the nuclear PDF for the parton $j$ was calculated within the usual approach \cite{kk}, but formula (\ref{1.320}) was presented in the form \cite{fgs,fgs2011,kaidalov},
\beqn
\delta f_{j/A}(x,Q^2)&=&
-8\pi\,\Re\,\int d^2b
\int\limits_{-\infty}^{\infty} dz_1\,\rho_A(b,z_1)
\int\limits_{z_1}^\infty dz_2\,\rho_A(b,z_2)
e^{-{1\over2}\sigma_{eff}^{hN}T_A(b,z_1,z_2)}
\nonumber\\ &\times&
\int_x^{x_{\Pom,0}} dx_{\Pom}\,
e^{i x_{\Pom} m_N(z_2-z_1)}
f^{D(4)}_{j/N}(\beta,Q^2,x_{\Pom},p_T)_{p_T=0},
\label{3.200}
\eeqn
where $x_{\Pom}=(M_X^2+Q^2)/s$; $M_X$ is the invariant mass of the diffractive excitation; $\beta=x/x_{\Pom}$; $f^{D(4)}_{j/N}$ is the
diffraction PDF differential in four variables.
The small real part of the amplitude and terms $\sim1/A$ are neglected.

This approach suffers from the same problems as Eq.~(\ref{1.320}), namely,
the sum over the higher order Gribov corrections, depicted in Fig.~\ref{fig:diff}, is replaced by the eikonal attenuation exponential with  the effective cross section \cite{fgs} written in the same way as Eq.~(\ref{3.217}),
\beq
\sigma_{eff}=\frac{16\pi}{f_{j/N}(x,Q^2)}
\int_x^{x_{\Pom,0}} dx_{\Pom}\,
f^{D(4)}_{j/N}(\beta,Q^2,x_{\Pom},t_{min})\,
\label{3.205}
\eeq
There are, however, several concerns related to this assumption:
\begin{itemize}

\item Within the range of Bjorken $x$ where data on DIS on nuclei are available, this cross section is significantly varying during propagation through the nucleus. Only at very small $x\lsim10^{-3}$ hadronic fluctuations of a photon are "frozen" by Lorentz time dilation. Still, this never is precise enough for the fluctuations containing gluons, which are a source of gluon shadowing.

\item
Even if a fluctuation is "frozen", it is not correct to use the averaged cross section in the exponent, the whole exponential should be averaged. 
The example given above in Eq.~(\ref{1.600}) shows that the nucleus becomes much more transparent compared to the exponential attenuation assumed in (\ref{3.200}).
The difference is exactly the Gribov corrections. 
Notice that a better result was achieved in \cite{kaidalov}, based on the Schwimmer model \cite{schwimmer}.
The nuclear transparency was found to depend on nuclear thickness as in (\ref{1.600}).
However the way how the phase shift was introduced in the calculations had no justification, and the validity of the Schwimmer model at rather large $x$, where data are available, is doubtful.

\item Gluon shadowing is self-quenching. Once it becomes strong, it affects and suppresses the value of $\sigma_{eff}$, which is defined in (\ref{3.205}) on a free nucleon. Correspondingly, gluon shadowing gets reduced. Such a self-quenching process leads to a nonlinear equation
derived in \cite{b,k} in  the frozen dipole approach.

\end{itemize}

The calculations with Eq.~(\ref{3.200}) and data for the diffractive structure function led to a magnitude of gluon shadowing, which considerably exceeds the results of other approaches \cite{tony,weise,badelek,shaw,stan,capella-shad} and that of the dipole approach described below.

A much more simplified modeling of the Gribov corrections was performed in \cite{kp}. Although this model is outdated compared to the contemporary state of art in the field, we highlight here several features of this description of shadowing, because the model is frequently used by experimental groups analyzing their data. The model exhibits all the problems listed above, and creates even more troubles.  

\begin{itemize}

\item
The lowest order Gribov correction is calculated in \cite{kp} in analogy to the vector dominance model \cite{bauer}.
Namely, the invariant mass distribution of the diffractively produced intermediate states is replaced by a delta function,
$\delta(M^2-Q^2)$, reducing the Gribov corrections to a single-channel problem. On the contrary, the above discussed analyses perform mass integration, Eqs.~(\ref{1.320}),  (\ref{3.200}), in accordance with the measured mass-dependence of the diffractive cross section.

\item
The cross sections of diffractive processes $hp\to Xp$ in (\ref{1.320}), or $\gamma^*p\to Xp$ in (\ref{3.200}), are well measured and fitted to data. Instead, in \cite{kp} this cross section is fitted to data on shadowing, which have
incomparably lower statistics than in diffraction. 

\item
The falling $Q^2$ dependence of the effective absorption cross section is motivated by the dominant contribution of
heavy intermediate states of mass $M^2=Q^2$, which have smaller cross section. That is not correct, since the heavy hadrons have larger radius and larger cross section. This is the basis of the so called Bjorken puzzle \cite{ikl},
which is solved by a nontrivial cancellation between diagonal and off-diagonal diffractive amplitudes \cite{jennings}.

\item
Shadowing at large $Q^2$ is known to be dominated by the aligned-jet hadronic components of the virtual photon
\cite{aligned-jet} (see below Sect.~3.2.3), which lead to a logarithmic, rather than power \cite{kp}, $Q^2$-dependence of shadowing
\cite{mq,qiu}. 

\end{itemize}

We conclude that this model is oversimplified and has a very low predictive power compared with other approaches.

\subsection{Dipole representation}
\label{sec:dipolerep}

The problems of Eq.~(\ref{3.200}) listed above are typical for a description based on the hadronic representation,  since hadrons
are the eigenstates of the mass matrix, but they do not have a certain size, controlling  the nuclear transparency \cite{zkl}. 
Nevertheless, this description has the advantage of having in Eq.~(\ref{3.200})  definite phase shifts, controlled by $x_\Pom$ related to $M_X$. On the other hand, switching to the dipole representation, one gets certainty with the dipole sizes, but the mass and phase shifts become uncertain. Later on, in Sect.~\ref{sec:path} we will explain how to deal with the phase shifts within the dipole approach, but here we start assuming that the energy is high enough to neglect the phase shifts.

The dipole description of \cite{zkl} was applied to DIS in \cite{nz91}. The inclusive DIS cross section of a proton
gets the form,
\beq
\sigma_{T,L}^{\gamma^*p}(x,Q^2)=\int\limits_0^1d\alpha\int d^2r_T
\left|\Psi_{q\bar q}^{T,L}(\alpha,r_T,Q^2)\right|^2\sigma_{q\bar q}(r_T,x),
\label{4.70}
\eeq
where the $\Psi_{q\bar q}^{T,L}(\alpha,r_T,Q^2)$ are the light-cone (LC) wave functions
for the transition $\gamma^*\to q\bar q$. 
The LC wave functions can be calculated in perturbation
theory and read in first order in the fine structure constant $\alpha_{em}$  \cite{ks,bks,nz91},
\beq
\Psi^{T,L}_{\bar qq}(\vec r_T,\alpha,Q^2)=
\frac{\sqrt{\alpha_{em}}}{2\,\pi}\,
\bar\chi\,\widehat {\cal  O}^{T,L}\,\chi\,K_0(\epsilon r_T),
\label{4.80}
\eeq
where $\chi$ and $\bar\chi$ are the spinors of the quark
and antiquark respectively. Here $\alpha$ is the
light-cone momentum of the photon carried by the quark;
\beq
\epsilon^2 = \alpha(1-\alpha)Q^2 + m_q^2;
\label{4.90}
\eeq
$K_0(\epsilon r_T)$ is the modified Bessel function.
The operators $\widehat {\cal  O}^{T,L}$ for transversely and longitudinally polarized photons have the form,
\beq
\widehat {\cal  O}^{T}=m_q\,\vec\sigma\cdot\vec e +
i(1-2\alpha)\,(\vec\sigma\cdot\vec n)\,
(\vec {e}\cdot \vec\nabla_{T})
+ (\vec\sigma\times\vec e)\cdot\vec\nabla_{T}\ ,
\label{4.100}
\eeq
\beq
\widehat {\cal  O}^{L}= 2\,Q\,\alpha(1-\alpha)\,\vec\sigma\cdot\vec n\ ,
\label{4.120}
\eeq
where the two-dimensional operator $\vec\nabla_{T}$
acts on the transverse coordinate $\vec r_T$;
$\vec n=\vec q/q$ is a unit vector parallel to
the photon momentum; $\vec e$ is the polarization vector
of the photon.

For the universal total cross section of dipole-proton interaction we rely on the parametrization~\cite{gbw}
fitted to data for the proton structure function $F_2(x,Q^2)$ measured at high energies (small $x$) at HERA,
 \beq
\sigma_{q\bar q}(r_T,x)=
\sigma_0\left[1-\exp\left(-\frac{r_T^2Q_0^2}{4(x/x_0)^\lambda}\right)\right],
\label{4.125}
\eeq
where
$Q_0=1$ GeV and the three fitted parameters are 
$\sigma_0=23.03$ mb, $x_0=0.0003$, and $\lambda=0.288$.
This dipole cross section vanishes $\propto r_T^2$ at small distances, as
implied by color transparency~\cite{zkl}, and levels off exponentially at large separations,
which reminds eikonalization.
 Although this parameterization  might be unrealistic at large separations
(see discussion in \cite{kst2}), and does not comply with the DGLAP evolution (see improvements in \cite{bartels})
we will use it, because of its simplicity, and because DIS and Drell-Yan (see Sect.~\ref{sec:dy}) data are all available in the kinematical range
where (\ref{4.125}) works rather well. So in what follows we employ this parametrization for numerical calculations (unless specified).

\subsubsection{Shadowing and the lifetime of $\bar qq$ fluctuations of the photon}\label{sec:tc-dis}
\label{sec:lifetime}

Nuclear shadowing is controlled by the interplay between
two fundamental quantities.

\begin{itemize}
\item{The lifetime of photon fluctuations, or coherence time.
Namely, shadowing is possible only if the coherence 
time exceeds the mean inter-nucleon spacing in nuclei,
and shadowing saturates (for a given Fock component)
if the coherence time substantially exceeds the nuclear 
radius.}
\item{Equally important for shadowing is the transverse
separation of the $\bar qq$. In order to be shadowed the 
$\bar qq$-fluctuation of the photon has to interact with
a large cross section. 
As a result of color transparency \cite{zkl,bm,bbgg}, small size
dipoles interact only weakly and are therefore less shadowed. The dominant
contribution to shadowing comes from  the large
aligned jet configurations
\cite{ajm,fs} of the pair.}
\end{itemize}

The lifetime of a hadronic fluctuation given by Eq.~(\ref{3.160}) can be presented as

\beq
t_c= 
\frac{P}{x\,m_N} =
P\,t_c^{max}\ ,
\label{4.20}
\eeq
where $P=(1+M^2/Q^2)^{-1}$, and $t_c^{max}=1/m_N\,x$. 
The usual approximation is to assume that
$M^2 \approx Q^2$ since $Q^2$ is the only large 
dimensional scale available. In this case $P=1/2$.

For a noninteracting $\bar qq$ the coefficient  $P$ has a simple form,
\beq
P(k_T,\alpha)=\frac{Q^2\,\alpha\,
(1-\alpha)}{k_T^2+\epsilon^2}\ ,
\label{4.40}
\eeq
where $m_q$ and $k_T$ are the mass and transverse
momentum of the quark respectively. To find the mean value of the fluctuation lifetime in vacuum 
one should average (\ref{4.40}) over $k_T$ and $\alpha$ 
weighted with the wave function squared of
the fluctuation,
\beq
\la P\ra_{vac}=
\frac{\Bigl\la \Psi^{\gamma^*}_{\bar qq}
\Bigl|P(k_T,\alpha)\Bigr|
\Psi^{\gamma^*}_{\bar qq}\Bigr\ra}
{\Bigl\la \Psi^{\gamma^*}_{\bar qq}\Bigl|
\Psi^{\gamma^*}_{\bar qq}\Bigr\ra}\ .
\label{4.60}
\eeq

The normalization integral in the denominator of (\ref{4.60})
diverges at $r_T\to0$ for transversely polarized photons, 
therefore we arrive at the unexpected result
$\la P^T\ra_{vac}=0$. This means that in vacuum a transverse photon fluctuates mainly to heavy  $\bar qq$ pairs with large $k_T$, which have a vanishingly short lifetime.
However, they also have a vanishing small
transverse size $r_T\sim 1/k_T$ and interaction cross section.
Therefore, such fluctuation cannot be resolved by the interaction and do
not contribute to the DIS cross section.
To get a sensible result one should properly define the averaging
procedure. We are interested in the fluctuations which contribute
to nuclear shadowing, i.e. they have to interact at least twice.
Correspondingly, the averaging procedure 
has to be redefined as,
\beq
\la P\ra_{shad}=
\frac{\Bigl\la f_{sd}(\gamma^*\to\bar qq)
\Bigl|P(k_T,\alpha)\Bigr|
f_{sd}(\gamma^*\to\bar qq)\Bigr\ra}
{\Bigl\la f_{sd}(\gamma^*\to\bar qq)\Bigl|
f_{sd}(\gamma^*\to\bar qq)\Bigr\ra}\ ,
\label{4.140}
\eeq
where $f_{sd}(\gamma^*\to\bar qq)$ is the amplitude of diffractive
dissociation of the virtual photon on a nucleon 
$\gamma^*\,N\to\bar qq\,N$.

Thus, one should include in the weight  factor for the averaging the 
interaction cross section squared $\sigma^2_{\bar qq}(r_T,s)$, 
where $s=2m_N\nu-Q^2+m_N^2$.
Then, the mean value of the function $P(\alpha,k_T)$ reads,
\beq
\left\la P^{T,L}\right\ra=\frac{\int_0^1d\alpha\int d^2r_1d^2r_2
\left[\Psi_{q\bar q}^{T,L}\left(\vec r_2,\alpha\right)\right]^*
\sigma^N_{q\bar q}\left(r_2,s\right)
\widetilde P\left(\vec r_2-\vec r_1,\alpha\right)
\Psi_{q\bar q}^{T,L}\left(\vec r_1,\alpha\right)
\sigma^N_{q\bar q}\left(r_1,s\right)}
{\int_0^1d\alpha\int d^2r\,
\left|\Psi_{q\bar q}^{T,L}\left(\vec r_,\alpha\right)
\sigma^N_{q\bar q}\left(r,s\right)\right|^2}
\label{4.160}
\eeq
with
\beq
\widetilde P\left(\vec r_2-\vec r_1,\alpha\right)=
\int\frac{d^2k_T}{\left(2\pi\right)^2}\,
{\exp\left(-i\,\vec k_T\cdot\left( \vec r_2-\vec r_1\right)\right)}
{P\left(\alpha,k_T\right)} =
\frac{Q^2\alpha\left(1-\alpha\right)}
{2\pi}K_0\left(\varepsilon\left|\vec r_2
-\vec r_1\right|\right),
\label{4.180}
\eeq
which is a Fourier transform of  Eq.~(\ref{4.40}).

The results of numerical calculations for $\la P\ra$, for transversely and longitudinally polarized photons, are depicted in Fig.~\ref{fig:tc-q} as function of $Q^2$, and in Fig.~\ref{fig:tc-x} versus Bjorken $x$.
\begin{figure}[htb]
\parbox{\halftext}{
\centerline{\includegraphics[width=7.5 cm]{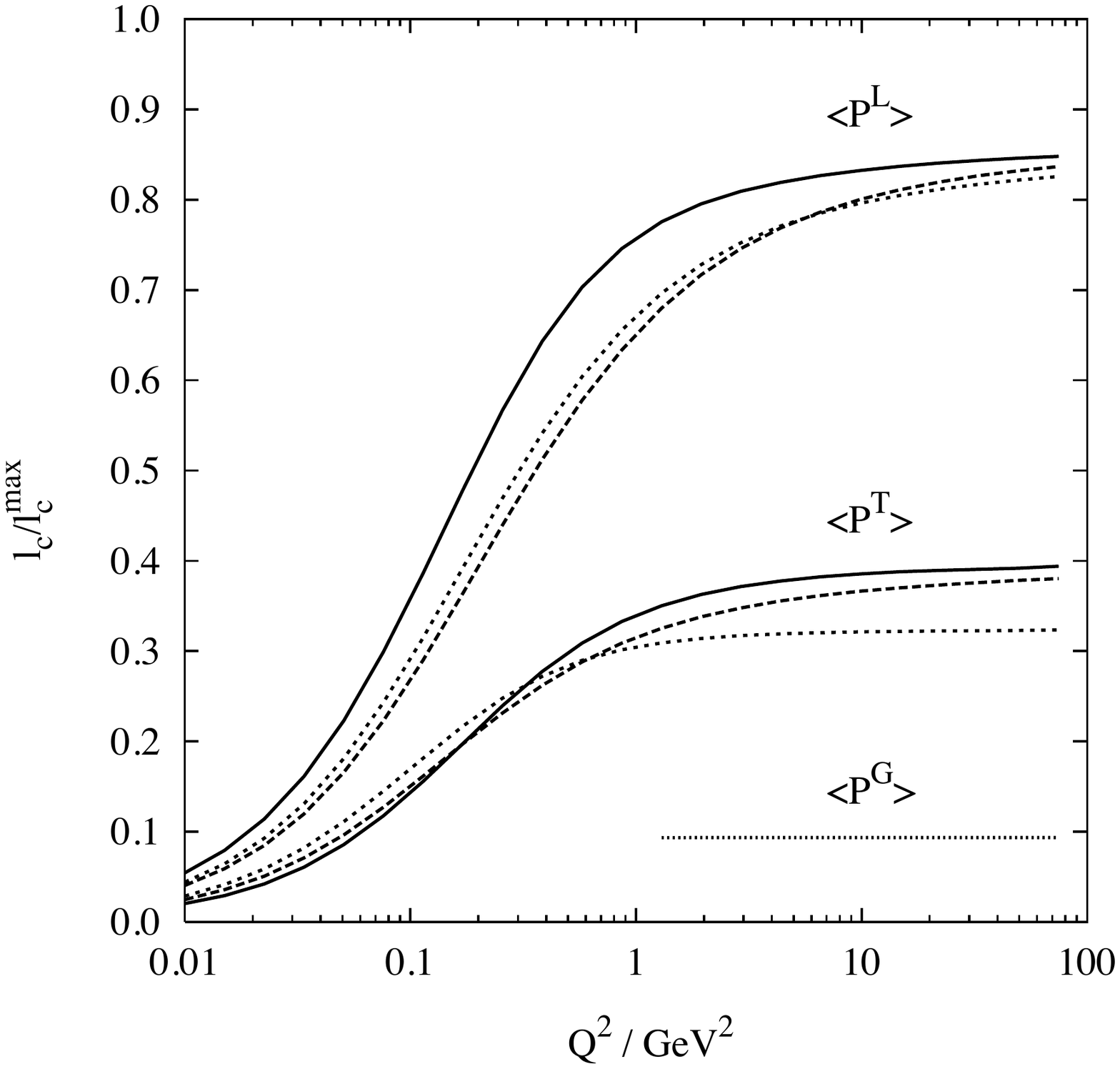}}
 \caption{$Q^2$ dependence of the factor $\left\la P\right\ra$ defined in (\ref{4.20}), at $x=0.01$, 
for the $\bar qq$ fluctuations of transverse 
and longitudinal
photons, and for $\bar qqG$ fluctuation, from top to 
bottom. Dotted and dashed and solid curves correspond to
different forms of the dipole cross section and distribution  functions
(see details in \cite{krt2}). }
 \label{fig:tc-q}}
\hfill
\parbox{\halftext}{
\centerline{\includegraphics[width=7.5 cm]{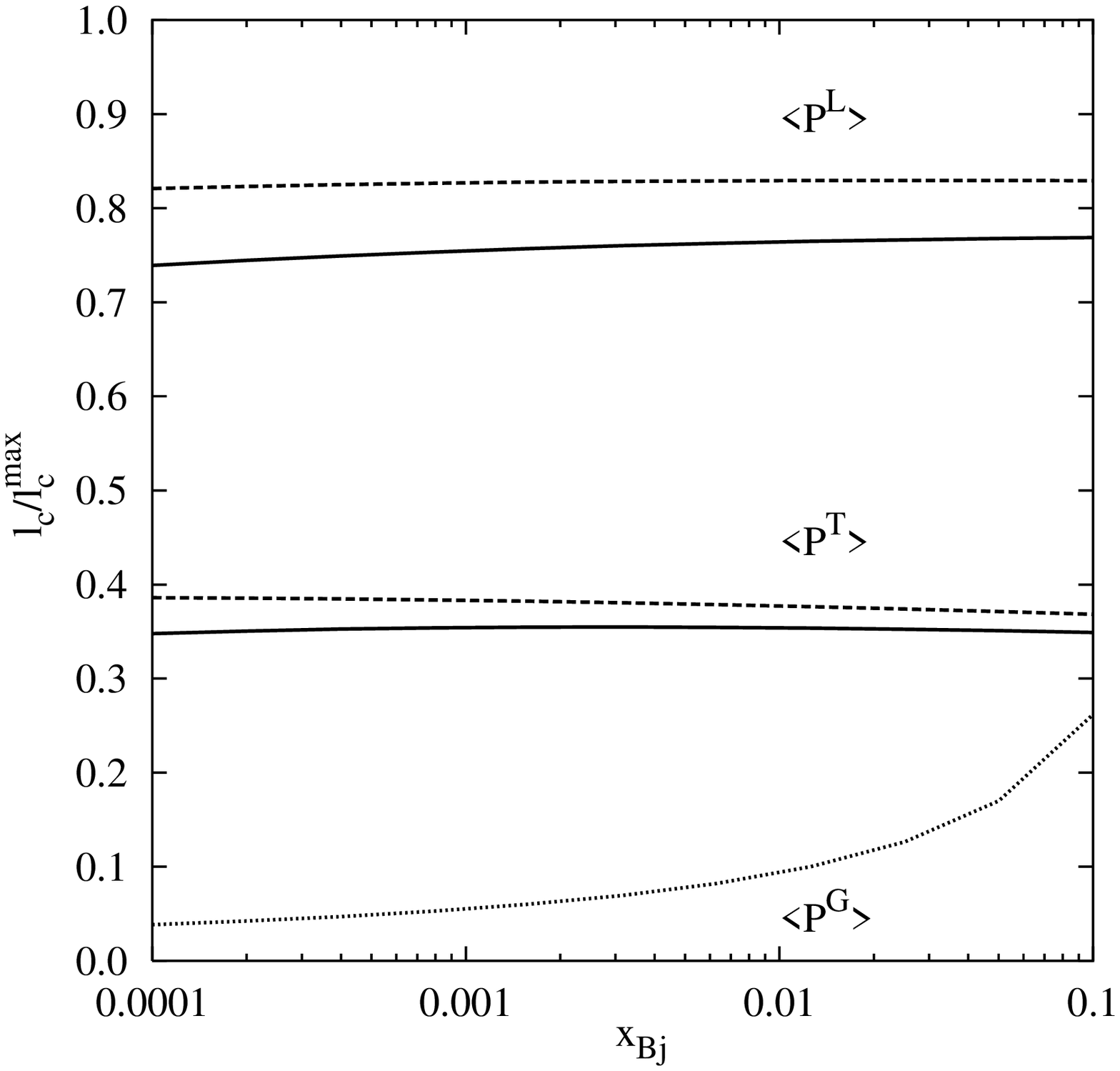}}
\caption{ Bjorken $x$ dependence of the factors $\left\la P^{T,L}\right\ra$
and $\left\la P^{G}\right\ra$ 
defined in (\ref{4.20}) corresponding to the coherence length 
for shadowing of transverse and longitudinal
photons and gluon shadowing, respectively. Solid and dashed curves
correspond to $Q^2=4$ and $40\,GeV^2$.}
\label{fig:tc-x}}
 \end{figure}
One can see that $\la P^L\ra>\la P^T\ra$, which means that a longitudinal photon produces lighter $\bar qq$ fluctuations than a transverse one. This is related to the suppression of the very asymmetric $\bar qq$ pairs with $\alpha\to0$ or $\alpha\to1$ in the distribution amplitude of a longitudinal photon. Such asymmetric pairs have the largest invariant mass.

\subsubsection{The path integral technique}\label{sec:path}

As far as the lifetime of partonic fluctuations of a photon significantly exceeds the nuclear size, the dipole approach is very suitable and easy tool in order to calculate the shadowing effects in DIS. Indeed, in this case one can rely on the ``frozen" approximation Eq.~(\ref{1.480}), which for interaction of a virtual photon has the form,
\beq
\sigma_{T,L}^{\gamma^*A}(x,Q^2)=2\int d^2b\int\limits_0^1d\alpha\int d^2r_T
\left|\Psi_{q\bar q}^{T,L}(\alpha,r_T,Q^2)\right|^2
\left[1-e^{-{1\over2}\sigma_{q\bar q}(r_T,x)T_A(b)}\right],
\label{4.200}
\eeq
where the $\Psi_{q\bar q}^{T,L}(\alpha,r_T)$ are given by Eq.~(\ref{4.80}), and in the quadratic form read,
\beqn
\left|\Psi_{q\bar q}^{T}(\alpha,r_T)\right|^2&=&
\frac{2N_c\alpha_{em}}{(2\pi)^2}\sum\limits_{f=1}^{N_f}Z_f^2
\left\{\left[1-2\alpha(1-\alpha)\right]\eps^2{\rm K}^2_1(\eps r_T)
+m_f^2{\rm K}^2_0(\eps r_T)\right\},
\label{4.210}\\
\label{psil}
\left|\Psi_{q\bar q}^{L}(\alpha,r_T)\right|^2&=&
\frac{8N_c\alpha_{em}}{(2\pi)^2}\sum\limits_{f=1}^{N_f}Z_f^2
Q^2\alpha^2(1-\alpha)^2{\rm K}^2_0(\eps r_T),
\label{4.220}
\eeqn

The advantage of the dipole description is clear,
Eq.~(\ref{4.200}) includes Gribov inelastic shadowing corrections to all orders multiple interactions~\cite{zkl},
what is hardly possible in the hadronic representation.
On the other hand, the dipoles having a definite size, do not have any definite mass, therefore the phase shifts between amplitudes on different nucleons cannot be calculated as simple as in Eq.~(\ref{1.320}). A solution for this problem was proposed in \cite{krt1}. If $t_c\lsim R_A$ the "frozen approximation is not appropriate and one should correct for the dipole size fluctuations during propagation through the nucleus. This can be done with the path integral technique \cite{feynman}, which sums up different propagation paths of the partons.

For a $\bar qq$ component of the photon Eq.~(\ref{4.200}) should be replaced by,
\beqn
&&\Bigl(\sigma^{\gamma^*A}_{tot}\Bigr)^{T,L} = 
A\,\Bigl(\sigma^{\gamma^*N}_{tot}\Bigr)^{T,L}
\,-\, \frac{1}{2} Re\int d^2b
\int\limits_0^1 d\alpha
\int\limits_{-\infty}^{\infty} dz_1 \int\limits_{z_1}^{\infty} dz_2
\int d^2r_1\int d^2r_2\,
\\ & \times &
\Bigl[\Psi^{T,L}_{\bar qq}\left(\varepsilon,\r_2\right)\Bigr]^*\,
\rho_A\left(b,z_2\right)\sigma_{q\bar{q}}^N\left(s,\r_2\right)
G\left(\vec r_2,z_2\,|\,\vec r_1,z_1\right)\,
\rho_A\left(b,z_1\right)\sigma_{q\bar{q}}^N\left(s,\r_1\right)
\Psi^{T,L}_{\bar qq}\left(\varepsilon,\r_1\right), \nonumber
\label{4.240}
\eeqn
The Green's function
$G\left(\vec r_2,z_2\,|\,\vec r_1,z_1\right)$ describes propagation 
of a  $\bar qq$ pair  in an
absorptive medium, having initial separation $\r_1$ at the initial position $z_1$, up to the point $z_2$, where it gets separation $\r_2$, as is illustrated in Fig.~\ref{fig:GF}. 
\begin{figure}[htb]
\parbox{\halftext}{
\centerline{\includegraphics[width=6 cm]{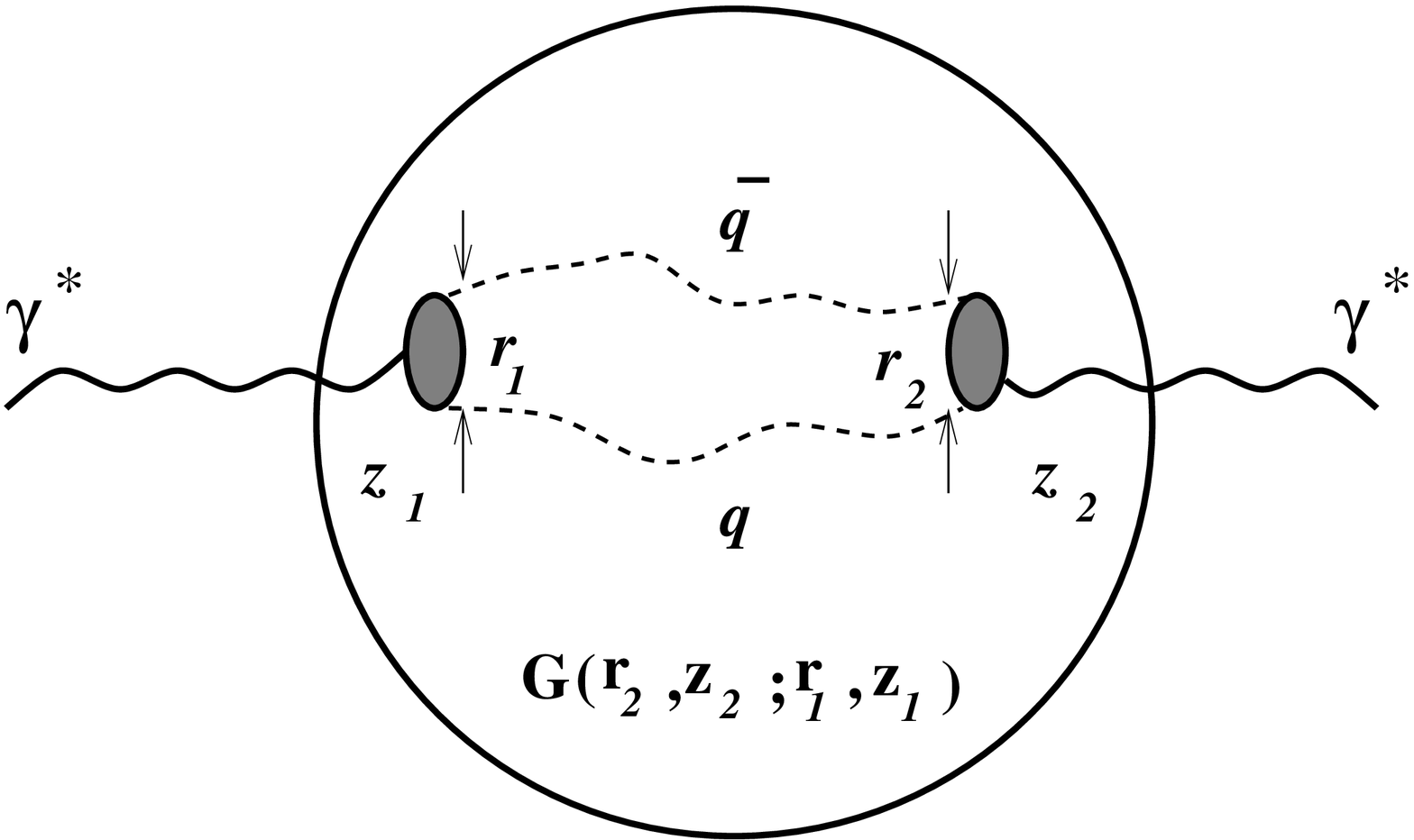}}
 \caption{Propagation of a $q\bar q$-pair through a nucleus between points with longitudinal coordinates $z_1$ and $z_2$.  The evolution of the $\bar qq$ separation from the initial, $\r_1$,  up to the final, $\r_2$, due to  the transverse motion of the quarks, is described by the
Green's function $G\left(\vec\rho_2,z_2;\vec\rho_1,z_1\right)$. }
 \label{fig:GF}}
\hfill
\parbox{\halftext}{
\centerline{\includegraphics[width=6 cm]{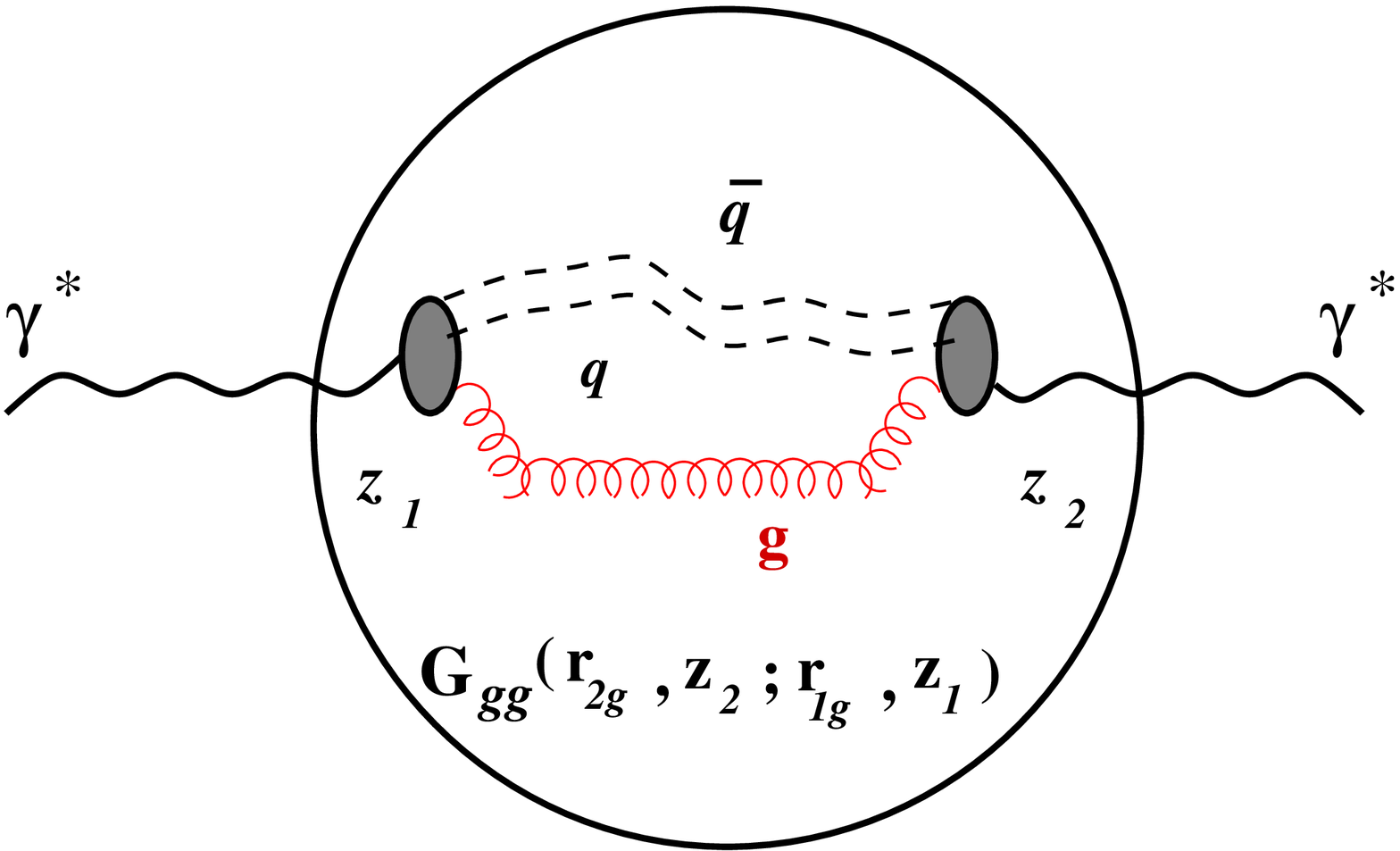}}
\caption{Propagation through a nucleus of the $q\bar q-g$ fluctuation of a longitudinally polarized photon.  
Neglecting the small, $\sim1/Q^2$ size of the color-octet $\bar qq$ pair, the effective octet-octet dipole propagation is described by the Green's function $G_{gg}(\r_{2g},z_2;\r_{1g},z_1)$. 
}
\label{fig:GF-g}}
 \end{figure}
It satisfies the evolution equation,
\beq
\left[i\frac{\partial}{\partial z_2}
+\frac{\Delta_\perp\left(r_2\right)-\varepsilon^2}
{2\nu\alpha\left(1-\alpha\right)}
+U(r_2,z_2)
\right]
G\left(\vec r_2,z_2\,|\,\vec r_1,z_1\right)
=
i\delta\left(z_2-z_1\right)
\delta^{\left(2\right)}\left(\vec r_2-\vec r_1\right).
\label{4.260}
\eeq
The light-cone potential term in the left-hand side (l.h.s.) of
this equation describes nonperturbative interactions within the dipole, and its absorption in the medium.
The real part the potential responsible for nonperturbative quark interactions was modeled and fitted to data of $F_2^p$ in \cite{kst2}. Here we fix $\Re U(r_2,z_2)=0$, and treat quarks as free particles for the sake of simplicity.
The imaginary part of the potential describes the attenuation of the dipole in the medium,
\beq
\Im U(r,z)=-{1\over2}\,\sigma_{\bar qq}(r)\,\rho_A(b,z).
\label{4.270}
\eeq

At small $x$ when the coherence length substantially
exceeds, $l_c^{T,L}\gg (z_2-z_1)$ (the nuclear radius) the solution of
Eq.~(\ref{4.240}) much simplifies, $G(\vec r_2,z_2\,|\,\vec r_1,z_1)
\propto \delta^{\left(2\right)}\left(\vec r_2-\vec r_1\right)$,
 so Lorentz time dilation ``freezes'' the variation
of transverse $\bar qq$ separation.
Correspondingly, the total cross
section gets the simple form of Eq.~(\ref{4.200}).

Eq.~(\ref{4.260}) can be solved
analytically if the medium density is constant, $\rho_A=\rho_0$, and then the dipole cross section has the simple form,
$\sigma_{q\bar q}(r)=Cr^2$.  The solution is the harmonic
oscillator Green function with a complex frequency \cite{kz91},  
\begin{equation} 
G\left(\r_2,z_2;\r_1,z_1\right)
=\frac{\gamma}{2\pi\sinh\left(\omega\Delta z\right)}
\exp\left\{-\frac{\gamma}{2}\left[\left(r_2^2+r_1^2\right)
\coth\left(\omega\Delta z\right)-\frac{2\r_2\cdot\r_1} 
{\sinh\left(\omega\Delta z\right)} \right]\right\},
\label{4.280}
\end{equation}  
where  
\beqn
\Delta z& = & z_2-z_1, \\ 
\omega^2 & =&
\frac{i\,C\rho_0}{\nu\alpha\left(1-\alpha\right)},\\ 
\gamma^2 & = & -i\,C\rho_0\nu\alpha\left(1-\alpha\right).
\label{4.300}
\eeqn
This formal solution and Eq.~(\ref{4.240}) properly account for all
multiple scatterings and for the finite lifetime of the hadronic
fluctuations of the photon, as well as varying  transverse separation of the $q\bar q$ pair
during propagation through the medium.
It effectively sums up all the Gribov corrections including
nonzero phase shifts between different amplitudes.

For practical applications these results can be extended to a realistic nuclear density varying with coordinates
as is described in \cite{kz91}.
Also the real part of the light-cone potential $U(r,z)$ in (\ref{4.240}) was modeled in \cite{kst2,krt2}
to incorporate nonperturbative effects. 

The best data available today for nuclear shadowing in DIS
are for the structure function  
ratio tin over carbon from NMC \cite{nmc}. These data are shown
in Figs.~\ref{fig:f2sncx} and \ref{fig:f2sncq} as function of Bjorken $x$ and $Q^2$ respectively.
\begin{figure}[htb]
\parbox{\halftext}{
\centerline{\includegraphics[width=9 cm]{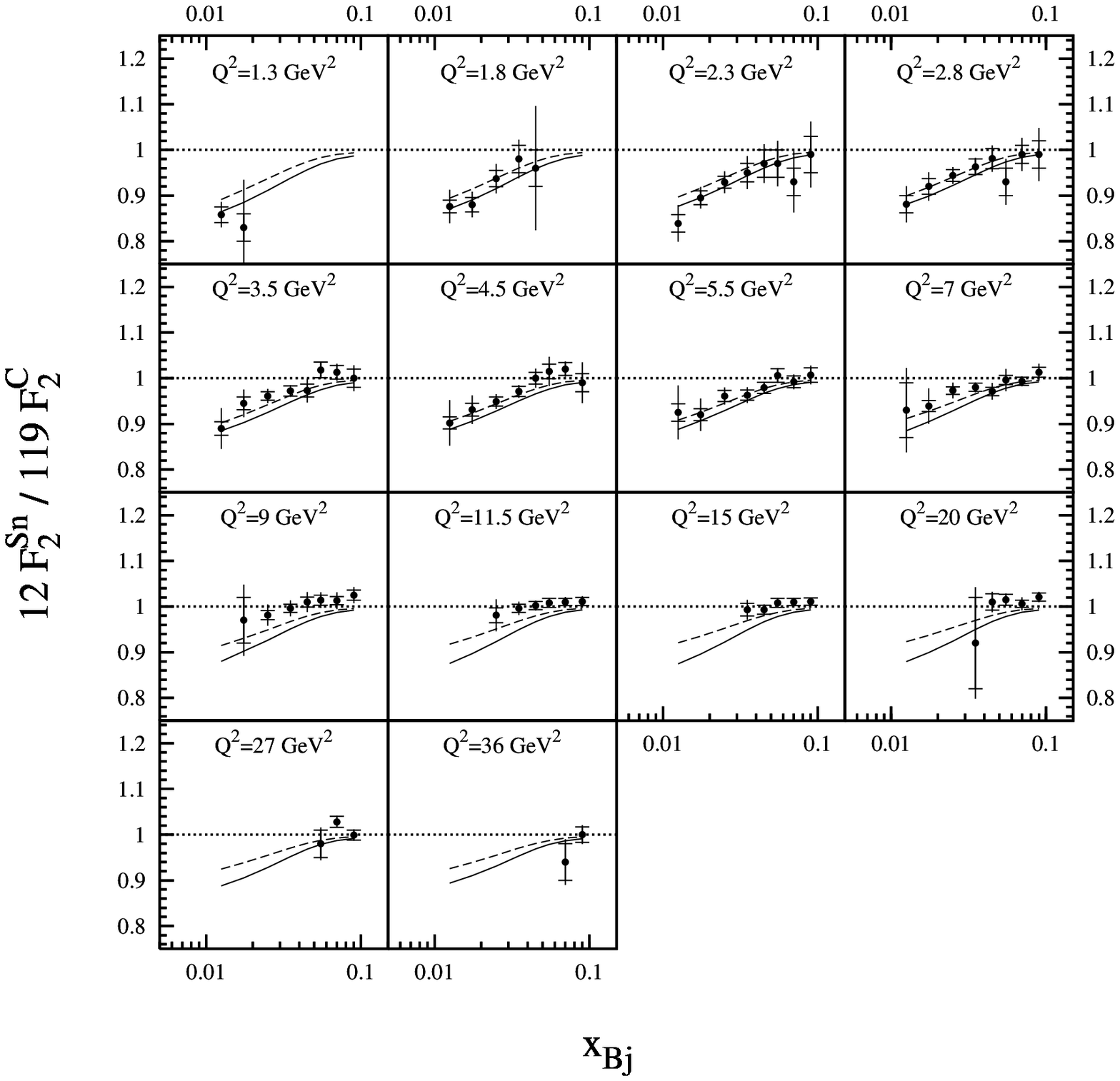}}
 \caption{The Bjorken-$x$ dependence of nuclear shadowing in DIS
      for the structure function ratio of tin relative to carbon.
      The data are from the NMC experiment \cite{nmc}.  
    The solid and dashed curves are calculated including or excluding the nonperturbative effects respectively.
 }
 \label{fig:f2sncx}}
\hfill
\parbox{\halftext}{
\centerline{\includegraphics[width=9 cm]{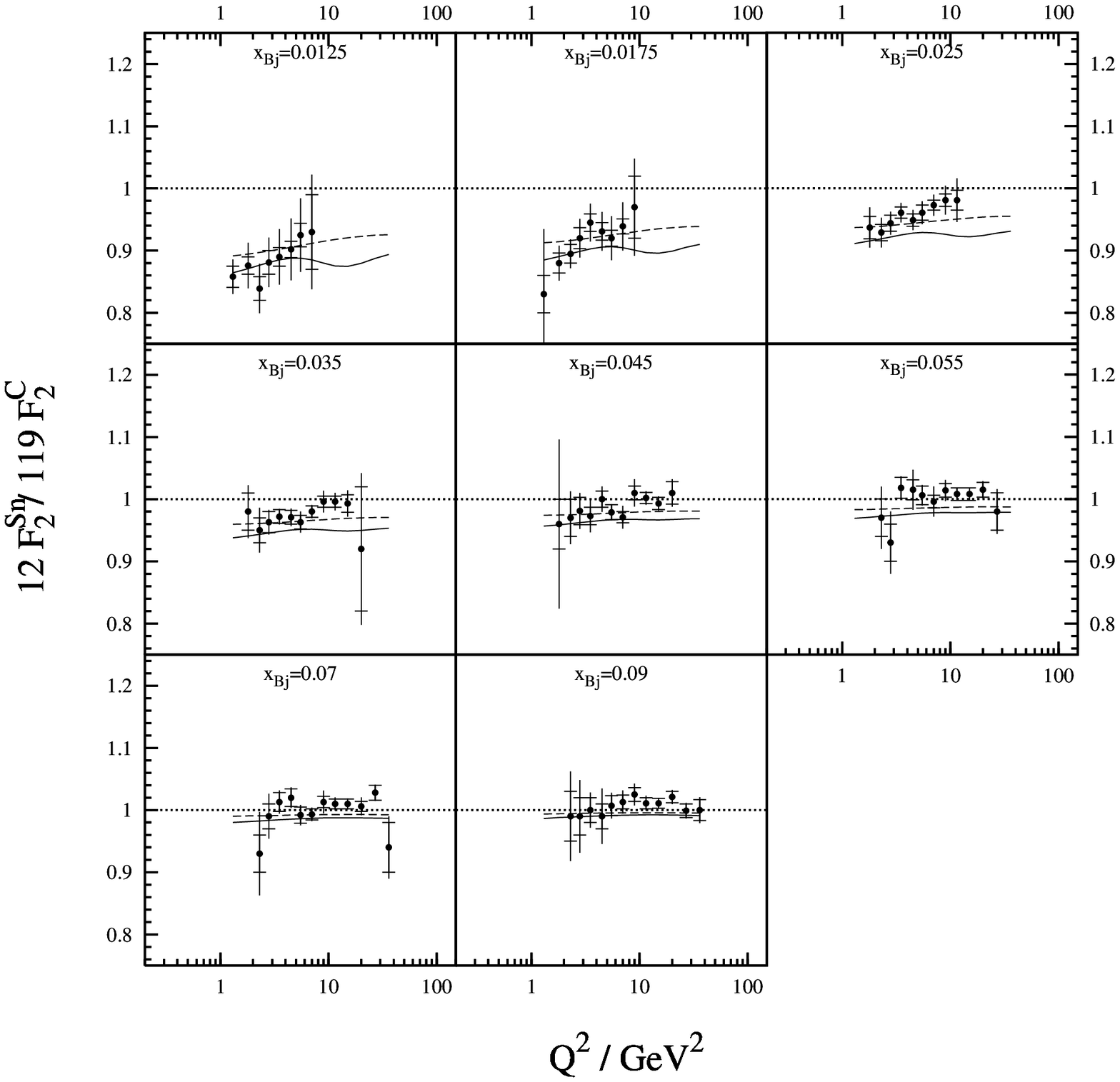}}
\caption{The $Q^2$ dependence of nuclear shadowing in DIS
      for the ratio of  the structure functions of tin to carbon.
      The data \cite{nmc}
      are the same as shown in fig.~\ref{fig:f2sncx}, and  
      the curves  have the same meaning.}
\label{fig:f2sncq}}
 \end{figure}
The numerical results of the calculations, which are performed either disregarding or including the nonperturbative effects,
are plotted in Fig.~\ref{fig:f2sncx} by dashed and solid curves respectively. One can see that inclusion of the nonperturbative effects does not lead to a significant change of the magnitude of shadowing. Comparison with NMC data \cite{nmc,nmc951} shows pretty good agreement.
Note however that inclusion of the
antishadowing effect might shift the curves upwards. Indeed, the low-$x$ tail of the EMC suppression 
observed at at large $x$ should be an enhancement, as is suggested by the momentum conservation sum rule.
Whether this will happen 
at all values of $x$ or only around $x\sim 0.1$, depends on the model.

As an example of the $A$ dependence of shadowing, the numerical results for the ratios of various nuclei to carbon are depicted in Fig.~\ref{fig:nmcc},
in comparison with NMC data \cite{nmc,nmc951}.
\begin{figure}[htb]
\parbox{\halftext}{
\centerline{\includegraphics[width=8cm]{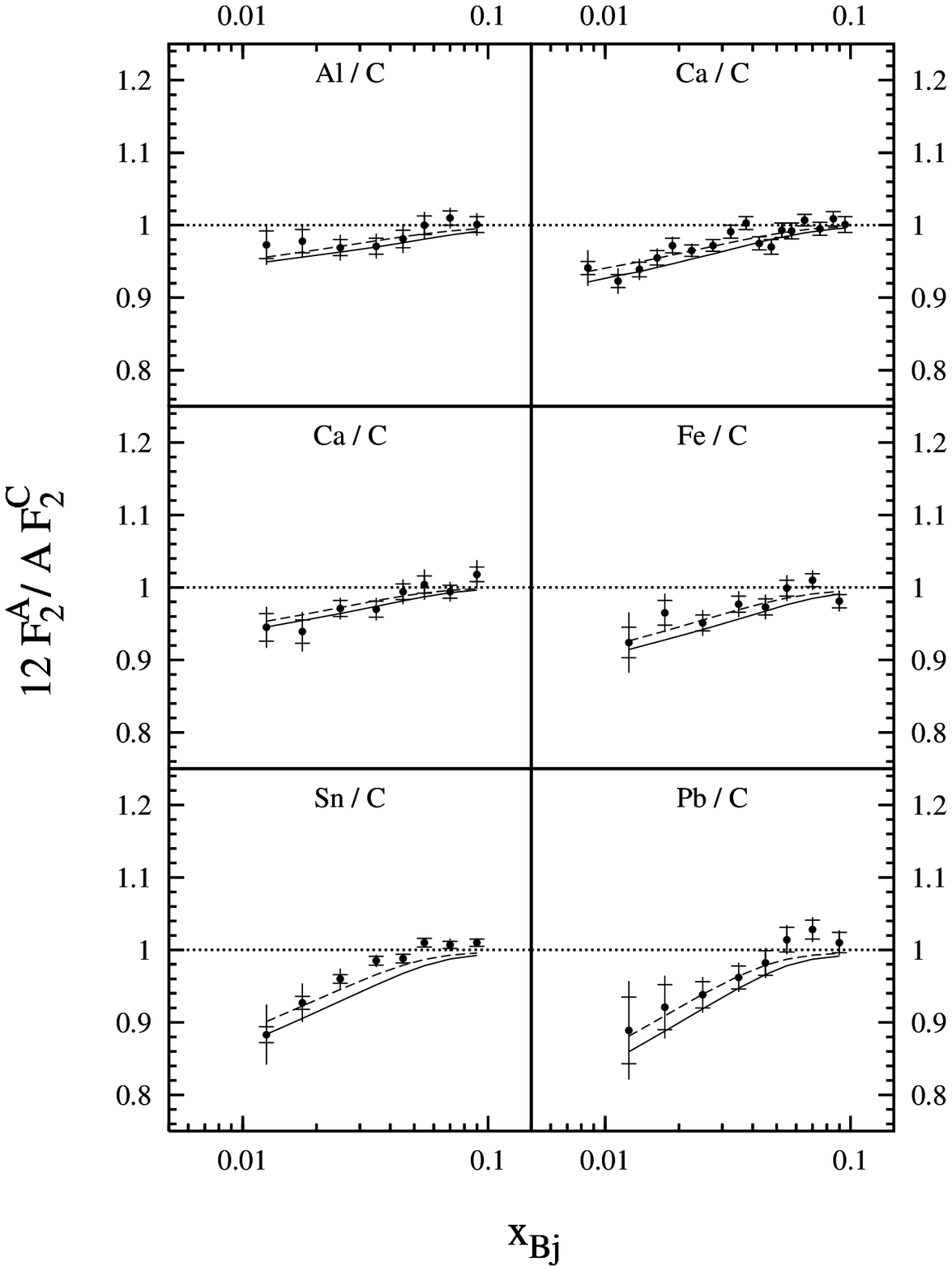}}
 \caption{Comparison between calculations for shadowing in DIS and experimental
      data from NMC \cite{nmc,nmc951}
      for the structure functions of
      different nuclei relative to carbon.  
$Q^2$ ranges within $3\le
        Q^2\le 17\GeV^2$.  The curves  
      have the same meaning as in Fig.~\ref{fig:f2sncx}
}
 \label{fig:nmcc}}
\hfill
\parbox{\halftext}{
\centerline{\includegraphics[width=8.5 cm]{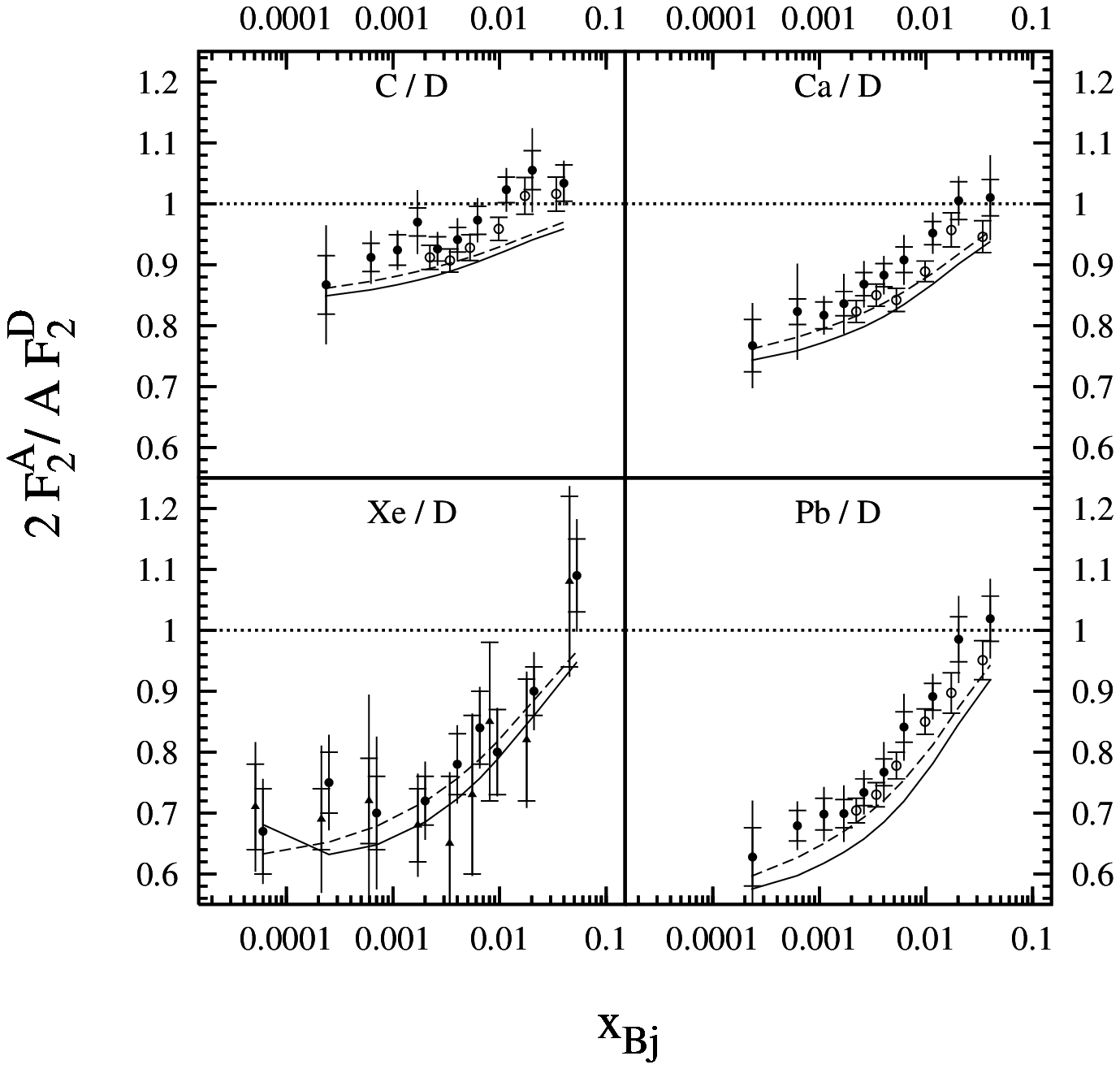}}
\caption{E665 data for shadowing in DIS for various nuclei \cite{e66592,e66595}.
      Full circles show data taken with electromagnetic calorimeter cuts, while
      the open circles result from the FERRAD radiative 
      correction
      code. 
      Hadronic cuts were applied for the triangles in the lower left plot. 
 $Q^2$ ranges within $0.15\le Q^2\le~22.5\GeV^2$, except for the xenon data,  
        $0.03\le Q^2\le 17.9\GeV^2$.      The curves  
      have the same meaning as in Fig.~\ref{fig:f2sncx}.  }
\label{fig:e665}}
 \end{figure}
 Calculations including no interaction within the $\bar qq$ pair
are shown by dashed curves. The calculations are parameter free and agree with data rather well. 
Introduction of a nonperturbative interaction between $q$ and $\bar q$ described by the real part of the  light-cone potential $\Re U(r,z)$, parametrized in the oscillator form (see details in \cite{kst2,krt2}), does not produce any significant change. These results, plotted in Fig.~\ref{fig:nmcc} by solid curves, also agree with the data.

We also compare our calculations with data from the E665 experiment
\cite{e66592,e66595}, which covered much lower values of $x$, Fig.~\ref{fig:e665}.  
 The agreement with E665 data is not as good as with the NMC data. Indeed, the two datasets seem to be somewhat inconsistent \cite{nmc,Muecklich}, so it looks challenging to reproduce both. 

The disagreement between NMC and E665 vanishes however, in the
$F_2$ ratio relative to carbon. This observation might give a clue to the origin of the ``disagreement" between the results of the two experiment. The modification of the properties of bound nucleons compared with free ones is a popular interpretation of the EMC effect, which is the observed suppression of nuclear $F_2(x,Q^2)$ at large $x$.
This effect should propagate down to small $x$ as an enhancement (momentum conservation sum rule), which is indeed observed at $x\sim 0.1-0.2$. Such a few percent enhancement should have similar magnitudes for carbon and heavier nuclei, because the nuclear density does not vary much. However for deuterons the EMC effect and related small-$x$ enhancement are much weaker. Therefore the ratios $A/D$ should be shifted by few percent upwards  
compared with $A/C$ ratios. This is what happens and can be seen in Figs.~\ref{fig:nmcc} and \ref{fig:e665}.
Our calculations of shadowing have not been corrected for the EMC effect of antishadowing, and this is probably why they agree with the NMC data for $A/C$ but underestimate the E665 data for $A/D$.

Notice that the measurement of nuclear shadowing in inclusive DIS, where only the final lepton is detected,
is a very difficult task
because of large radiative corrections \cite{joerg-phd}. These corrections  occur, e.g. when
the incident lepton reduces its momentum due to electromagnetic bremsstrahlung and does not
produce a deep inelastic event \cite{Levy}.
In the older publications \cite{nmc-1,nmcold2}, the 
NMC collaboration calculated the
radiative corrections employing the computer program FERRAD, which relies on the
theoretical analysis \cite{Mo}.
For a reevaluation of the shadowing data \cite{nmc951} three different codes
were tested, FERRAD, an improved version of FERRAD and
TERAD. The last code relies on calculations 
\cite{B1,B2,B3} and was used for the published NMC result. 
A more detailed discussion on radiative corrections for NMC and
HERMES can be found in \cite{Jeroen}. 
One may say that the whole
shadowing effect in data  is calculated. Without
radiative corrections, the cross section ratio would be around
unity \cite{Jeroen}. However, the correctness of the calculation was
checked by comparing to the "hadron tagged" data, making sure that really a deep
inelastic event was measured.

A different way to identify the deep inelastic events was chosen by the E665
collaboration \cite{e66592,e66595}, 
where cuts in the electromagnetic calorimeter were applied. 
In \cite{e66592} the xenon 
data obtained in this way were compared to an evaluation
with hadron tagging, see Fig.~\ref{fig:e665}. The two methods give consistent
results.
In \cite{e66595}, the E665 collaboration also applied the FERRAD code for 
a comparison with NMC.
The results are depicted by open circles in Fig.~\ref{fig:e665}.  One recognizes a
systematic discrepancy between the two evaluation methods. The radiative
corrected  ratios are lower by $\sim$5\% than the points from the
calorimeter analysis. 

\subsubsection{Shadowing of longitudinal vs transverse photons}\label{sec:L/T}

Eqs.~(\ref{4.70}) and (\ref{4.240}) allow to calculate the absolute values of $\sigma_L$ and $\sigma_T$, as well as the magnitude of nuclear shadowing for each of them. The results for $R_{L/T}^p=\sigma_L^p/\sigma_T^p$ calculated at $x=0.01$ are plotted as function of $Q^2$ in Fig.~\ref{fig:lt}. 
\begin{figure}[htb]
\parbox{\halftext}{
\centerline{\includegraphics[width=8 cm]{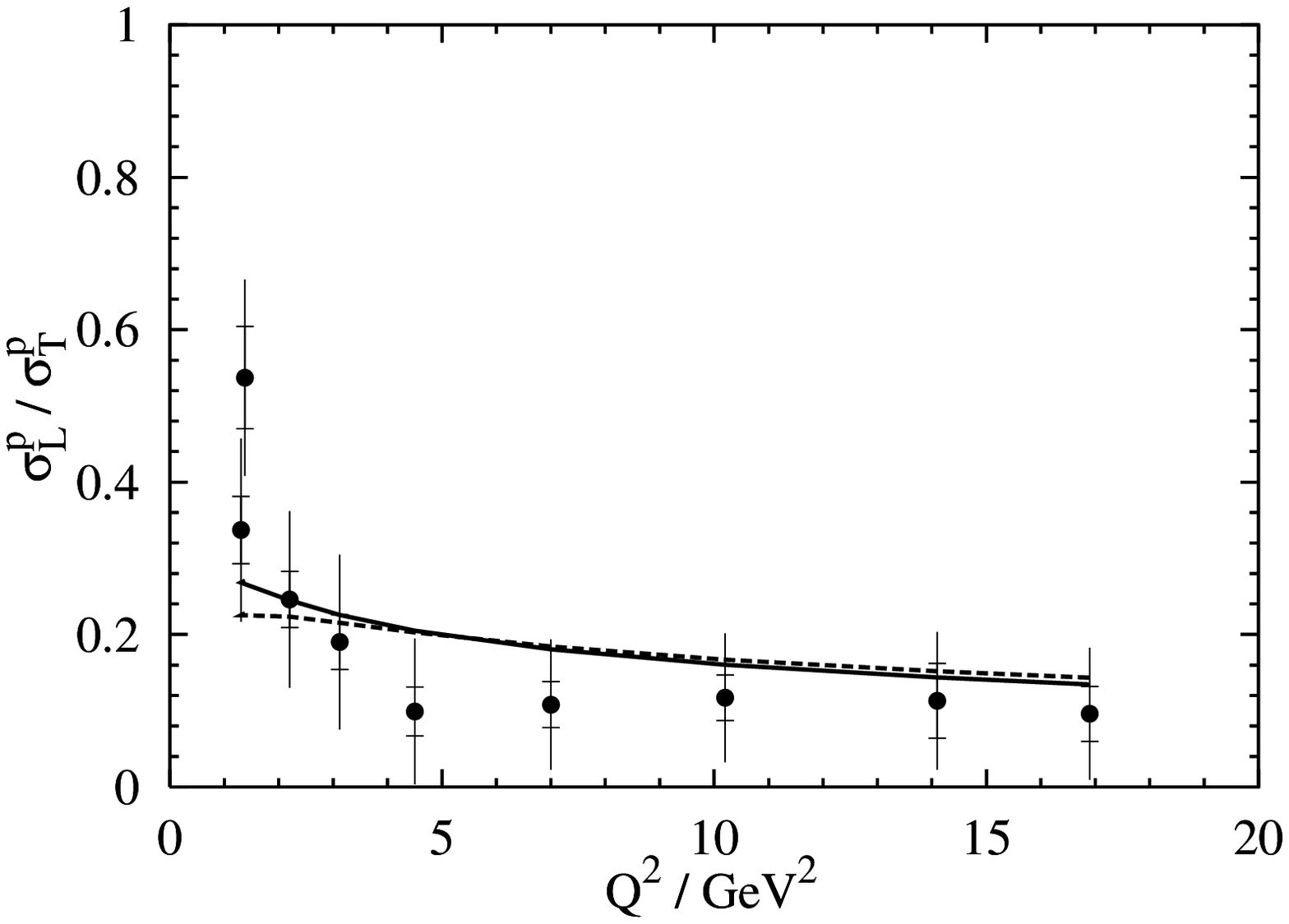}}
 \caption{Ratio $R^p_{L/T}$ on a proton as function of $Q^2$ calculated with Eqs.~(\ref{4.200})-(\ref{4.220}).
 For the curve notations see Fig.~\ref{fig:f2sncx}.
     The NMC data
      \cite{nmc1} are from the left to the right for
      $x=0.0080$, $0.0045$, $0.0125$, 
      $0.0175$, $0.025$, $0.035$, $0.05$, $0.07$, and $0.09$.}
 \label{fig:lt}}
\hfill
\parbox{\halftext}{
\centerline{\includegraphics[width=7 cm]{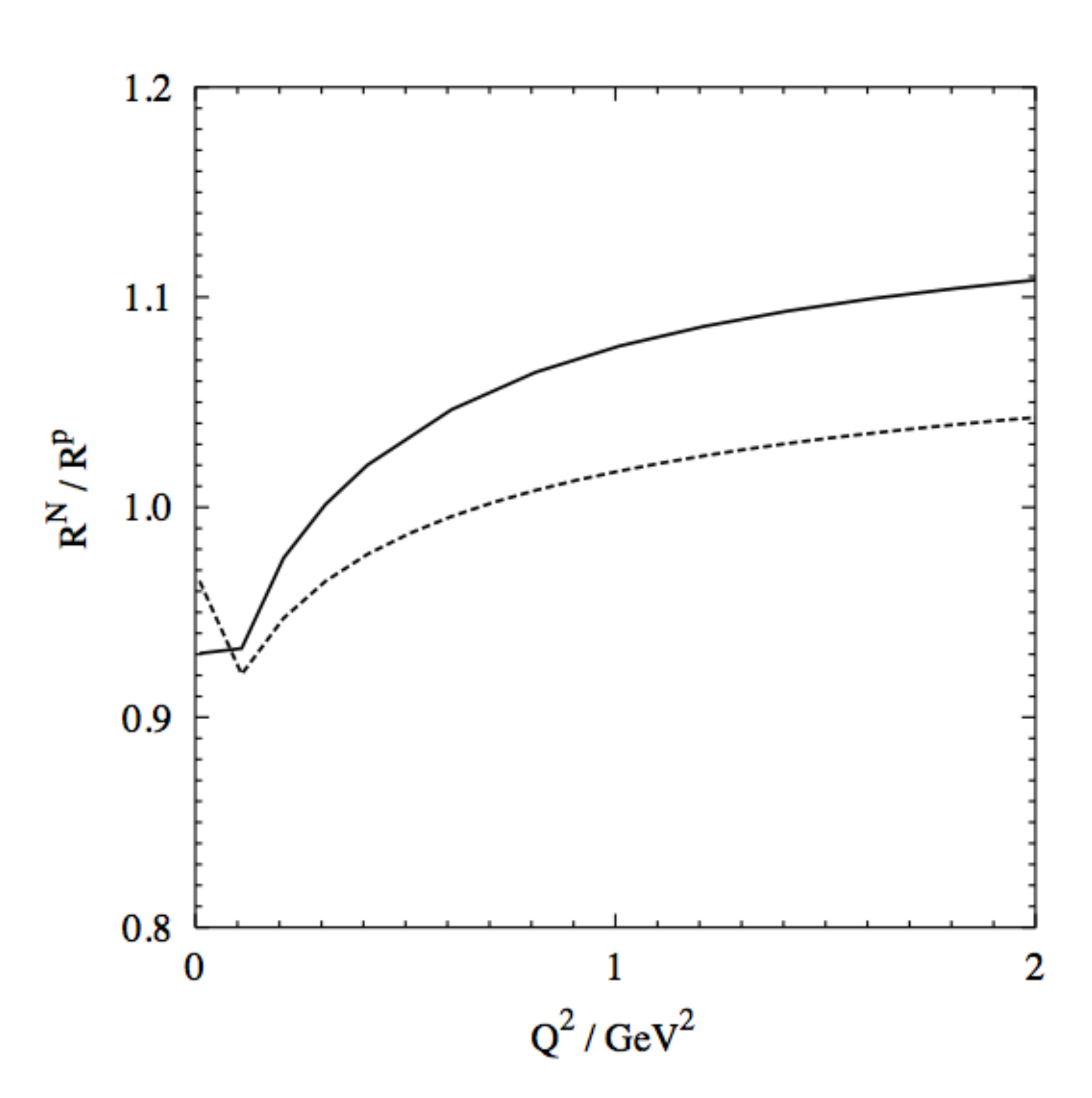}}
\caption{Ratio of longitudinal to transverse 
cross sections $R_{L/T}=\sigma_L/\sigma_T$ calculated for nitrogen $^{14}N$ divided by the same ratio for a proton, $R_{L/T}^{^{14}N}/R_{L/T}^p$. For the curve notations see Fig.~\ref{fig:f2sncx}.
}
\label{fig:lt-n}}
 \end{figure}
The solid and dashed curves, like previously, correspond to calculations with or without inclusion of the nonperturbative effects in the $\bar qq$ distribution amplitude of the photon. Theory agrees well with NMC data
\cite{nmc1} measured at close values of Bjorken $x$.

The ratio turns out to be quite small, $R_{L/T}^p\lsim 0.2$. This is understandable intuitively as a result of a smaller transverse size of $\bar qq$ fluctuations in longitudinally polarized photons. Indeed, the mean transverse separation squared,
\beq
\la r_T^2\ra=\frac{1}{Q^2\,\alpha(1-\alpha)+m_q^2},
\label{4.302}
\eeq
is minimal for the symmetric fluctuations, $\alpha\sim1/2$, but reach the maximal size of the order of the confinement radius for very asymmetric pairs at $\alpha\to 0,\ 1$. The latter configurations are suppressed by the distribution function of longitudinal photons, Eq.~(\ref{4.220}), so the $\bar qq$ fluctuations are more symmetric,
i.e. have smaller size, while fluctuations of transverse photons get a considerable distribution from asymmetric, large size fluctuations.

This observation has important consequences for the relation Eq.~(\ref{3.48}) connecting the shadowing effects in the nuclear structure function $F_2^A$ and nuclear DIS cross section $\sigma^{lA}_{DIS}$.
The observed nuclear ratio (\ref{1.20}) for the DIS cross section equals to that for the structure functions $F_2$ only if (i) either $R_{L/T}\ll1$ and can be neglected in (\ref{3.48}); (ii) or $\xi=1$; (iii) or  $R_{L/T}$ has no nuclear dependence, $R^A_{L/T}=R^N_{L/T}$.

Regarding the condition (i), indeed $R_{L/T}^p$ is rather small, but not that small to neglect it in (\ref{3.48}).
Regarding (ii), the value of $\xi$, according to (\ref{3.52}), at small $x$ is mainly controlled by the transferred fractional energy $y=\nu/E_l$. Apparently, at the same of $x$ and $Q^2$ the value of $\nu$ is also the same in high-energy (NMC, E665) and low-energy (JLAB, HERMES) experiments. Therefore the values of $y$ and $\xi$ are very different. The low-energy experiments at small $x\sim0.1$ are mainly done with large $y\to1$, i.e. small $\xi\ll1$, while at high energies $y\ll1$ and $\xi\sim1$.  In the latter case, according to (\ref{3.48}) one can safely treat the ratio of the DIS cross sections as that for $F_2$. 

However, at smallest $x$ even in large energy experiments the transferred fractional energy reaches large values, so $\xi$ is small, and one should have a look at the $A$-dependence of $R_{L/T}^A$. This ratio was calculated in \cite{krt2} for nitrogen $^{14}N$ and hydrogen at $x=0.01$, and the ratio of the results $R_{L/T}^{^{14}N}/R_{L/T}^p$ is plotted in Fig.~\ref{fig:lt-n} as function of $Q^2$. We see that the difference between $R_{L/T}$ on  $^{14}N$ and $p$ ranges within about $10\%$. This correction is diluted by the small value of $R_{L/T}\approx0.2$ down to about $2\%$ even in the case of $\xi\to0$ where the nuclear corrections are maximal.

Notice that the weak nuclear dependence of $R_{L/T}$ predicted in \cite{krt2}, was at the time of publication in dramatic disagreement with HERMES data \cite{hermes}, which showed a nuclear enhancement of $R_{L/T}$ by a factor of several. However, few years later the HERMES Collaboration discovered a mistake in their measurements \cite{hermes-wrong}.

\subsubsection{Nuclear shadowing for valence quarks}\label{sec:valence}

Nuclear shadowing for valence quarks 
is usually believed to be small \cite{ekr}, if it occurs at all.  It was
demonstrated in \cite{krtj}, however, that shadowing for valence quarks is quite sizable, 
even might be stronger than the shadowing of sea quarks.  
 Notice in this regard that the nuclear structure function 
$F_2(x)$ is different from the quark distribution function in an essential
way; namely, the former contains shadowing effects and therefore the baryon 
number sum rule is not applicable to it \cite{Brodsky}, 
a difference that might explain the 
discrepancy of present results compared to Ref.~\cite{ekr}.

Note that we relate the nuclear cross section  Eq.~(\ref{4.200}) to shadowing for sea quarks
because the dipole cross section $\sigma^N_{q\bar q}(r_T,x)$ was fitted to HERA data at very low $x<0.01$,
so it includes only
the part corresponding to gluonic exchanges in the
cross-channel. Therefore, this is the part of the sea generated via gluons
(there are also other sources of the sea, for instance the meson cloud of
the nucleon, but they steeply vanish with $x$).  The fact that
the color-dipole cross section includes only the part generated by gluons 
is the reason why it
should not be used at larger $x$.  This part
of the dipole cross section can be called the Pomeron in terms of Regge
phenomenology. In the same framework, one can relate the valence quark
distribution in the proton to the Reggeon part of the dipole cross section,
which has been neglected so far. 
So, to include valence quarks in the dipole formulation of DIS, one
should replace
 \beq
\sigma_{q\bar q}(r_T,x) \Rightarrow
\sigma^{\Pom}_{q\bar q}(r_T,x) + 
\sigma^{\Reg}_{q\bar q}(r_T,x)\ ,
\label{4.305}
 \eeq
where the first (Pomeron) term corresponds to the gluonic part of the cross
section, responsible for the sea quarks in the nucleon structure function.
The second (Reggeon) term must reproduce the distribution of valence quarks
in the nucleon; this condition constraints its behavior at small $x$.
One can guess that it has the following form,
 \beq
\sigma^{\Reg}_{q\bar q}(r_T,x) = 
\tilde N\,r_T^2\,\sqrt{x}\ ,
\label{4.310}
 \eeq
where $\sqrt{x}$ should reproduce the known $x$ dependence of valence quark
distribution (as, in fact, motivated by Regge phenomenology), and the factor
$r_T^2$ is needed to respect the Bjorken scaling. The factor $\tilde N$ will
cancel in what follows.

We are now in a position to calculate shadowing for valence quarks by
inserting the cross section Eq.~(\ref{4.305}) into the eikonal
expression Eq.~(\ref{4.200}). If one
expands the numerator in powers of $\sigma^{\Reg}_{q\bar q}$ and picks
out the linear term\footnote{The small size of $\sigma^{\Reg}_{q\bar q}(r_T,x)$
at small $x$ motivates such an expansion; however, one should note
that it would not be proper to include the higher powers 
of the Reggeon cross section.  Indeed, the Reggeons correspond to
planar graphs.  These cannot be eikonalized since they lead to the so-called AFS
(Amati-Fubini-Stangelini)  planar graphs, which vanish at high energies
\cite{gribov}.}, then one arrives at the following expression for nuclear
shadowing of the valence quarks,
\beq
R_v(x,Q^2) =\frac{\int d^2b\, T_A(b)
\int_0^1 d\alpha\int d^2r_T
\Bigl|\Psi_{q\bar q}(r_T,\alpha,Q^2)\Bigr|^2
\sigma^{\Reg}_{q\bar q}(r_T,x)\,
e^{-{1\over2}\sigma^{\Pom}_{q\bar q}(r_T,x)
T_A(b)}}
{A\int_0^1 d\alpha\int d^2r_T\,
\Bigl|\Psi_{q\bar q}(r_T,\alpha,Q^2)\Bigr|^2\,
\sigma^{\Reg}_{q\bar q}(r_T,x)}.
\label{4.315}
 \eeq
The results of numerical calculations with this expression are compared with 
the results of the global fit \cite{ekr} in Fig.~\ref{fig:valence}.
\begin{figure}[htb]
\hspace*{0.7cm}
  \scalebox{0.7}{\includegraphics*{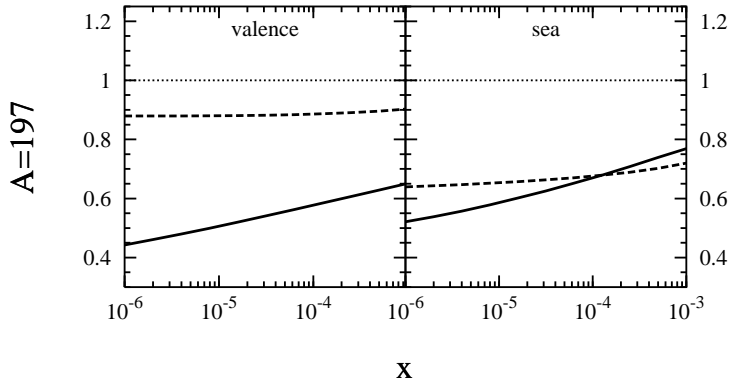}}\hfill
  \raise1.7cm\hbox{\parbox[b]{2.5in}{
    \caption{
      \label{fig:valence}
     Shadowing for sea and valence $u$-quarks in DIS off gold at $Q^2=20\GeV^2$. 
Solid lines are calculated
from Eq.~(\ref{4.315}), 
while dashed curves show the EKS98 parameterization 
\cite{ekr}.
    }
  }
}
\end{figure}
We see that shadowing for valence quarks is 
stronger than for sea quarks, and is much stronger 
than in the parameterization of \cite{ekr}. 
Note that the global fit of \cite{ekr} assumes that the nuclear valence quark 
distribution has to satisfy the baryon number sum rule, which takes it for granted
that the structure function is a measure of the number of quarks in the target.
However, the very meaning of shadowing is a reduced access of the probe (the virtual photon) to some of the bound nucleons. So the reduction of the effective number of valence quarks in the target, does not mean that baryon number is not conserved, it
only shows that the probe is not sufficiently hard. Increasing $Q^2$ one does not get rid of the soft component of DIS \cite{povh2}.

Unfortunately, it will be 
impossible to extract the low-$x$ valence quark distribution of a nucleus
from DY experiments, because the nuclear structure function is 
dominated by sea quarks. Maybe neutrino-nucleus scattering experiments and accurate measurement of
shadowing for $F_3^{\nu N}+F_3^{\bar\nu N}$ (see Sect.~\ref{sec:nu-dis}) can provide information on shadowing of valence quarks.

\subsubsection{``Antishadowing"}
\label{sec:anti}

Shadowing is a quantum-mechanical phenomenon, which is impossible  for hard reactions in classical physics, since multiple interactions have too small cross sections to shadow each other. Only if a hadronic quantum fluctuation has a sufficiently long lifetime to undergo multiple interactions (at least twice), can two amplitudes on two different nucleons interfere, either destructively (shadowing), or constructively (antishadowing). This is why
the term antishadowing is labelled by quotation marks in the title of this section, because this small enhancement of the nuclear
ratio $F_2^A(x,Q^2)/F_2^N(x,Q^2)$, which was observed \cite{nmc-rev} at $x\sim0.1-0.2$, is well outside of the coherence region. Therefore no coherence phenomena, either shadowing, or antishadowing are possible.
Indeed, the fluctuation lifetime given by the Ioffe time, Eq.~(\ref{3.160}), for this $x$-interval is very short $0.5-1\fm$ compared to the mean nucleon spacing in nuclei, $\sim 2\fm$. Although the observed nuclear enhancement cannot be related to antishadowing, this terminology is widely adopted so we will use it.

No consensus has been reached so far regarding the mechanisms responsible for this effect, as well as for the EMC effect, which is a nuclear suppression at larger $x\sim 0.2-0.7$ \cite{nmc-rev}. As far as both of these collective effects cannot be caused by coherence, they may be only related to a modification of the bound nucleons compared with free ones. This subject goes far beyond the scope of the present paper focused on coherence phenomena, but a detailed discussion of the medium effects can be found in the comprehensive review \cite{emc-rev}.
However, those effects might propagate down to smaller $x$ and also contribute to the observed nuclear suppression, therefore we briefly discuss here possible implications at small $x$.

Suppression of the quark PDF of a medium-modified bound nucleon  at large $x$, should cause an enhancement 
at smaller $x$ due to the baryon number (number of valence quarks) conservation sum rule. This enhancement would propagate down to small $x$, however shadowing onsets at $x\lesssim0.1$ and overtakes the enhancement.
In this qualitative picture the nuclear enhancement observed in the interval $0.1\lesssim x\lesssim0.2$ is a result of the interplay of two phenomena, which have quite different origin. One, the EMC effect, has no relation to coherence,
and results from medium modification of the properties of bound nucleons. Another, the shadowing suppression,
is caused by destructive interference of the DIS amplitudes on different bound nucleons, regardless whether these nucleons are, or are not modified by the medium. A quantitative analysis based on the model of "swelling nucleons" was performed in \cite{barone1} in good agreement with data.

\subsubsection{Coherence time for gluon radiation and gluon shadowing}
\label{sec:glue-shad}

Shadowing in the nuclear gluon distributing function at small 
$x$, which looks like gluon fusion $gg\to g$ in the infinite 
momentum frame of the nucleus, 
should be treated in the rest frame of the nucleus as
shadowing for the Fock components of the photon 
containing gluons. Indeed, the first shadowing term contains
double scattering of the projectile gluon via exchange of two 
$t$-channel gluons, which is the same Feynman graph as
gluon fusion. Besides, both of them correspond to the triple-Pomeron
term in diffraction which controls shadowing (see Eq.~(\ref{3.180}) and Fig.~\ref{fig:3R-gamma}).

The lowest Fock component containing a gluon is the $|\bar qqg\ra$. 
The coherence time controlling 
shadowing depends on the effective mass
of the $|\bar qqg\ra$, which should be expected to be 
heavier than that for a $|\bar qq\ra$. Correspondingly,
the coherence time $\la t^g_c\ra$ should be
shorter and the onset of gluon shadowing  
is expected to start at smaller $x$.

For the coherence time one can rely on 
the same Eq.~(\ref{3.160}), but with the invariant mass of the fluctuation,
\beq
M^2_{\bar qqg}=\frac{k_T^2}{\alpha_g(1-\alpha_g)}+
\frac{M_{\bar qq}^2}{1-\alpha_g}\ ,
\label{4.320}
\eeq
where $\alpha_g$ is the fraction of the photon momentum 
carried by the gluon, and
$M_{\bar qq}$ is the effective mass of the $\bar qq$ pair.
This formula is, however, valid only in the perturbative limit.
It is apparently affected by the nonperturbative interaction of gluons,
which was
found in \cite{kst2} to be much stronger than that for a $\bar qq$.
Since this interaction may substantially modify
the effective mass $M_{\bar qqg}$, we switch to
the formalism of Green's function described above, which
recovers Eq.~(\ref{4.320}) in the limit of high $Q^2$.

We treat gluons as massless and transverse. For the factor $P$ defined in
(\ref{4.20}), in the case of gluon shadowing one can write,
\beq\label{gluonl}
\Bigl\la P^{g}\Bigr\ra = 
\frac{N^g}{D^g}\ ,
\label{4.340}
\eeq
where 
\beqn\nonumber
N^{g}&=& m_N\,x\,
\int d^2r_{1g}\,d^2r_{1q\bar q}\,d^2r_{2g}\,
d^2r_{2q\bar q}\,d\alpha_q
\,d\ln(\alpha_g)\,
\widetilde \Psi^{\dagger}_{\bar qqg}
\left(\vec r_{2g},\vec r_{2q\bar q},
\alpha_q,\alpha_g\right)\\ & \times &
\left(\int\limits_{z_1}^\infty dz_2\, G_{\bar qqg}
\left(\vec r_{2G},\vec r_{2q\bar q},z_2;
\vec r_{1g},\vec r_{1q\bar q},z_1\right)
\right)\,
\widetilde \Psi_{\bar qqg}\left(\vec r_{1g},\vec r_{1q\bar q},
\alpha_q,\alpha_g\right)
\label{4.360}
\eeqn
\beqn\nonumber
D^{g}&=&\int d^2r_{1g}\,d^2r_{1q\bar q}\,d^2r_{2g}\,
d^2r_{2q\bar q}\,d\alpha_q
\,d\ln(\alpha_g)\,
\widetilde \Psi^{\dagger}_{\bar qqg}
\left(\vec r_{2g},\vec r_{2q\bar q},
\alpha_q,\alpha_g\right)\\ & \times &
\delta^{\left(2\right)}\left(\vec r_{2g}-\vec r_{1g}\right)\,
\delta^{\left(2\right)}
\left(\vec r_{2q\bar q}-\vec r_{1q\bar q}\right)\,
\widetilde \Psi_{\bar qqg}\left(\vec r_{1g},\vec r_{1q\bar q},
\alpha_q,\alpha_g\right)
\label{4.380}
\eeqn
Here we have introduced the Jacobi variables, 
$\vec r_{q\bar q}=\!\vec R_{\bar q}-\vec R_{q}$ and
$\vec r_{g}=\!\vec R_g-(\alpha_{\bar q}\vec R_{\bar q}+\alpha_q\vec R_{q})
/(\alpha_{\bar q}+\alpha_q)$. $\vec R_{g,q,\bar q}$ are the position vectors of
the gluon, the quark and the antiquark in the transverse plane and
$\alpha_{g,q,\bar q}$ are the longitudinal momentum fractions.

Differently from the case of a $|\bar qq\ra$ Fock state,
where we found that  at high $Q^2$ perturbative QCD can be safely used for shadowing calculations,
the nonperturbative effects remain important for the
$|\bar q\,q\,g\ra$ component even for highly virtual photons.
High $Q^2$ squeezes the $\bar qq$ pair down to a size $\sim 1/Q$,
while the mean quark-gluon separation at $\alpha_g \ll 1$ 
depends on the strength of gluon interaction which
is characterized in this limit by the parameter $b_0\approx 0.65\GeV$ \cite{kst2}.
The presence of such a semi-hard scale, which considerably exceeds $\Lambda_{QCD}$, is confirmed
by various experimental observations \cite{two-scales}, in particular by the observed strong suppression of the 
diffractive gluon radiation \cite{kst2}. In nonperturvative QCD models this scale is related to the instanton size \cite{shuryak,zahed}, $r_0\sim 1/b_0\approx 0.3\fm$.

For $Q^2\gg b_0^2$ the $\bar qq$ is small,
$r^2_{\bar qq} \ll r_g^2$, and one can treat the $\bar qqg$ system
as a color octet-octet dipole, as is illustrated in Fig.~\ref{fig:GF-g}.
Then the three-body Green's function $G_{q\bar qg}$ factorizes,
\beq
G_{q\bar qg}\left(\vec r_{2g},
\vec r_{2q\bar q},z_2;\vec r_{1g},\vec r_{1q\bar q},z_1
\right)\ \Rightarrow\ 
G_{q\bar q}\left(\vec r_{2q\bar q},z_2;
\vec r_{1q\bar q},z_1\right)\,
G_{gg}\left(\vec r_{2g},z_2;\vec r_{1g},z_1\right)\ .
\label{4.400}
\eeq
The color octet-octet  Green's function $G_{gg}$, describing the propagation of a glue-glue dipole through the medium, satisfies the simplified  evolution equation \cite{kst2},
\beq
\left[\frac{\partial}{\partial z_2}-\frac{Q^2}{2\nu}
+\frac{\Delta_\perp\left(r_{2g}\right)}
{2\nu\alpha_g\left(1-\alpha_g\right)}
-\frac{b_0^4\,r_{2g}^2}
{2\nu\alpha_g\left(1-\alpha_g\right)}
\right]
G_{gg}\left(\vec r_{2g},z_2;\vec r_{1g},z_1\right)
=\delta\left(z_2-z_1\right)
\delta^{\left(2\right)}\left(\vec r_{2g}-\vec r_{1g}\right)
\label{4.420}
\eeq

Correspondingly, the modified $\bar qqg$ wave function
simplifies too,
\beq\widetilde \Psi_{\bar qqg}
\left(\vec r_{g},\vec r_{q\bar q},
\alpha_q,\alpha_g\right) \Rightarrow
-\,\Psi^L_{\bar qq}\left(\vec r_{q\bar q},
\alpha_q\right)\ \vec r_{g}\cdot\vec\nabla\,
\Psi_{qg}\left(\vec r_{g}\right)\ 
\sigma_{gg}^N\left(x,r_{g}\right)\ ,
\label{4.440}
\eeq
where the nonperturbative quark-gluon wave function 
has a form \cite{kst2},
\beq
\Psi_{qg}\left(\vec r_{g}\right)=
\lim_{\alpha_g\to 0}\Psi_{qg}\left(\alpha_g,r_{g}\right)=
- \frac{2i}{\pi}\,
\sqrt\frac{\alpha_s}{3}\ 
\frac{\vec e\cdot\vec r_{g}}{r^2_{g}}\,
\exp\left(-\frac{b_0^2}{2}\,r^2_{g}\right)\ ,
\label{4.460}
\eeq
and the color-octet dipole cross section reads,
\beq
\sigma_{gg}^N\left(x,r_{g}\right)
=\frac{9}{4}\,
\sigma_{q\bar q}^N\left(x,r_{g}\right)\ .
\label{4.480}
\eeq

Within these  approximations  we can evaluate the factor $\la P^G\ra = \la l_c^g\ra/l_c^{max}$ given by
Eq.~(\ref{4.340}).
The results are depicted in Figs.~\ref{fig:tc-q} and \ref{fig:tc-x}. With the approximations made above,
the calculations  cannot cover the low $Q^2$ region and are perform
at $Q^2>1\GeV^2$. The gluon coherence length turns out to be much shorter than 
both $l_c^T$ and $l_c^L$ for $|\bar qq\ra$ fluctuations. This
observation corresponds to delayed onset of gluon shadowing,
shifted to smaller $x$ compared with quark shadowing, as was predicted in \cite{kst2}.

Although gluon shadowing is related to the higher Fock component of the photon, $\gamma^*\to\bar qqg$,
shadowing might be related also to a large $\bar qq$ separation, or $g-\bar qq$. The former is related to quark shadowing, which is a higher twist effect, but even increasing $Q^2$ one cannot get rid of this contribution.
As was demonstrated above, the main contribution to shadowing of transversely polarized photons comes from very asymmetric sharing of light-cone momentum, $\alpha\sim m_q^2/Q^2$. To get a net effect of gluon shadowing one should suppress this term by taking either heavy flavors \cite{ihkt-psi,kt-charm}, or DIS with longitudinally polarized photons \cite{kst2}. Here we rely on the latter.

Longitudinal photons can serve to measure the gluon density because they
effectively couple to color-octet-octet dipoles. This can be understood in
the following way: the light-cone wave function for the transition
$\gamma^*_L\to q\bar q$ does not allow for large, aligned jet configurations \cite{bks}. 
Thus, unlike the transverse case, all $q\bar q$ dipoles from longitudinal
photons have size $1/Q^2$ and the double-scattering term vanishes like
$\propto 1/Q^4$. The leading-twist contribution for the shadowing of 
longitudinal
photons arises from the $|q\bar q g\ra$ Fock state of the photon. Here again,
the distance between the $q$ and the $\bar q$ is of order $1/Q^2$, but the
gluon can propagate relatively far from the $q\bar q$-pair. In addition,
after the emission of the gluon, the pair is in an octet state. Therefore, the
entire $|q\bar qg\ra$-system appears as a $gg$-dipole, and the shadowing
correction to the longitudinal cross section is just the gluon shadowing we
want to calculate.

One can also see that from the expression for the cross section
of a small size dipole \cite{fs,barone2},
\beq
\sigma_{\bar qq}^{A,N}(r_T,x) \approx
\frac{\pi^2}{3}\,\alpha_s(Q^2)\,
x\,g_N(x,Q^2)\ ,
\label{4.500}
\eeq
where $g_N(x,Q^2)$ is the gluon density at
$Q^2\sim 1/r_T^2$. Thus, we expect nearly the same nuclear shadowing
at large $Q^2$ for the longitudinal photoabsorption cross section and
for the gluon distribution,
\beq
\frac{\sigma^L_A(x,Q^2)}{\sigma^L_N(x,Q^2)}\approx
\frac{g_A(x,Q^2)}{g_N(x,Q^2)}
\label{4.520}
\eeq

The shadowing 
correction to $\sigma^L_A(x,Q^2)$ has the form (compare with (\ref{4.240})),
\beqn
\Delta\sigma^L_A(x,Q^2) \!\!&=&\!\!-\, 
\Re\int d^2b
\int\limits_{-\infty}^{\infty} dz_1
\int\limits_{z_1}^{\infty} dz_2\,
\rho_A(b,z_1)\rho_A(b,z_2)
\int d^2r_{2g}\,d^2r_{2\bar qq}\,d^2r_{1g}\,d^2r_{1\bar qq}
\int d\alpha_q\,d{\rm ln}(\alpha_g)
\nonumber\\ &\times& 
F^{\dagger}_{\gamma^*\to\bar qqg}
(\vec r_{2g},\vec r_{2\bar qq},\alpha_q,\alpha_g)\, 
G_{\bar qqg}(\vec r_{2g},\vec r_{2\bar qq},z_2;\vec r_{1g},\vec r_{1\bar qq},z_1)\, 
F_{\gamma^*\to\bar qqg}(\vec r_{1g},\vec r_{1\bar qq},\alpha_q,\alpha_g)
\label{4.540}
\eeqn
Assuming $Q^2\gg b_0^2$ we can neglect $r_{\bar qq}\ll r_g$. The net diffractive amplitude $F_{\gamma^*\to\bar qqg}(\vec r_{1g},\vec r_{1\bar qq},\alpha_q,\alpha_g)$ takes the form of Eq.~(\ref{4.440}), and we can rely on the factorized relation (\ref{4.400}) for the 3-body Green's function, with equation (\ref{4.420}) for the evolution of the gluonic dipole. The latter has the solution,
\beq
 G_{gg}(\vec r_{2g},z_2;\vec r_{1g},z_1) =
\frac{\Gamma}{2\pi\,{\rm \sinh}(\Omega\,\Delta z)}
\exp\left\{-\frac{\Gamma}{2}\,
\left[( r_{1g}^2+ r_{2g}^2)\,{\rm coth}(\Omega\,\Delta z) -
\frac{2\vec r_{1g}\cdot\vec r_{2g}}{{\rm sinh}(\Omega\,\Delta z)}
\right]\right\}\ ,
\label{4.560}
\eeq
where
\beqn
\Gamma&=&\sqrt{{b_0}^4-i\,\alpha_g(1-\alpha_g)\,\nu\,C_{gg}\,
\rho_0}\nonumber\\
\Omega &=& \frac{i\,\Gamma}
{\alpha_g(1-\alpha_g)\,\nu},
\label{4.580}
\eeqn
and other notations are the same as in Eq.~(\ref{4.280}), except $C_{gg}={9\over4}C$ according to (\ref{4.480}).

The results of numerical calculation of (\ref{4.540}) for the ratio
\beq
R_g(x,Q^2)=\frac{g_A(x,Q^2)}{A\,g_N(x,Q^2)},
\label{4.600}
\eeq
are depicted in Fig.~\ref{fig:glue} as function of Bjorken $x$ for $Q^2=4$ and $40\GeV^2$.
\begin{figure}[htb]
\parbox{\halftext}{
\centerline{\includegraphics[width=8 cm]{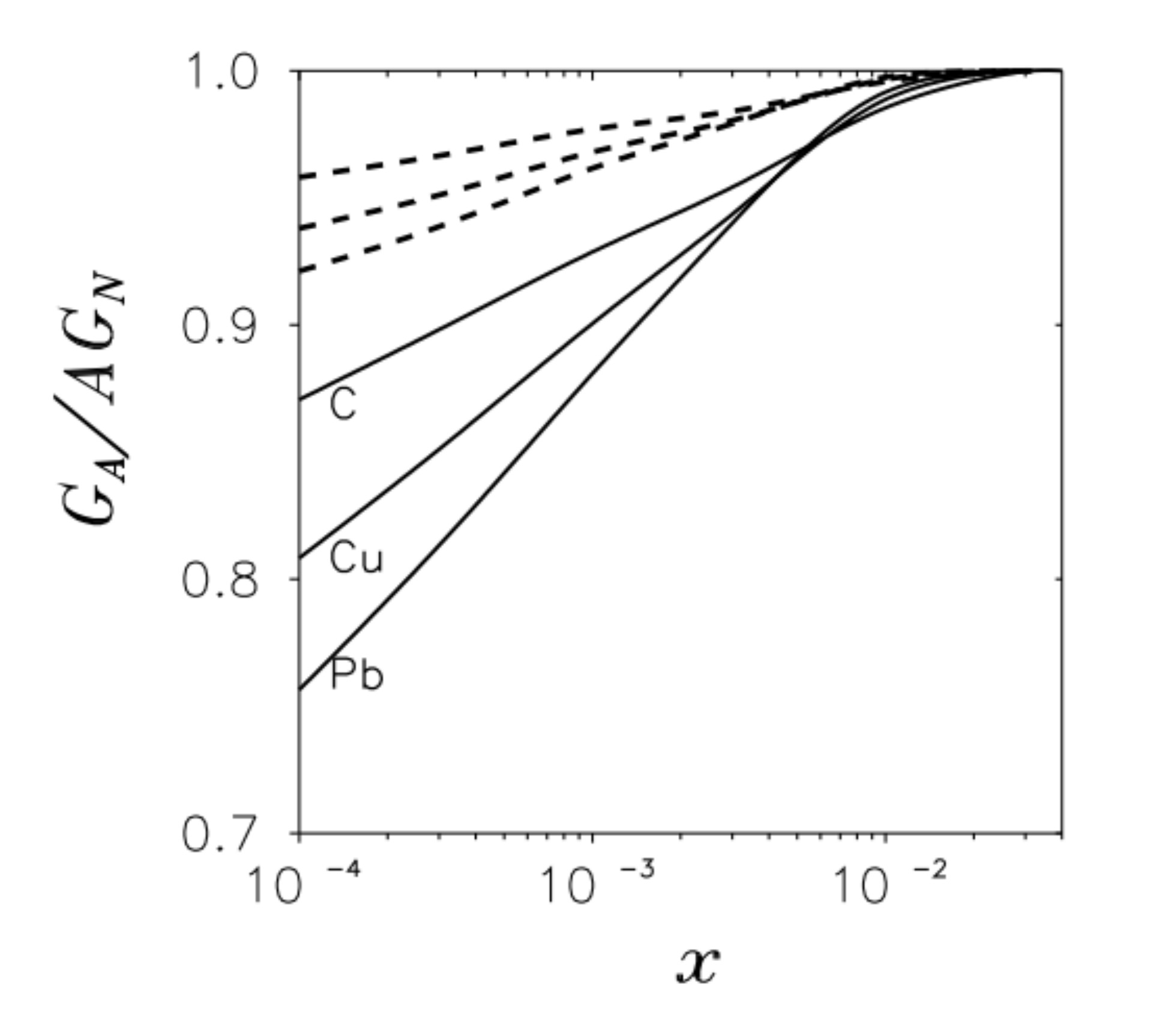}}
 \caption{Ratio of the gluon distribution functions in nuclei (carbon, copper and lead) 
 and nucleons at small Bjorken $x$ and $Q^2 = 4\GeV^2$ (solid curves) and $40\GeV^2$ (dashed curves).
 }
 \label{fig:glue}}
\hfill
\parbox{\halftext}{
\centerline{\includegraphics[width=9 cm]{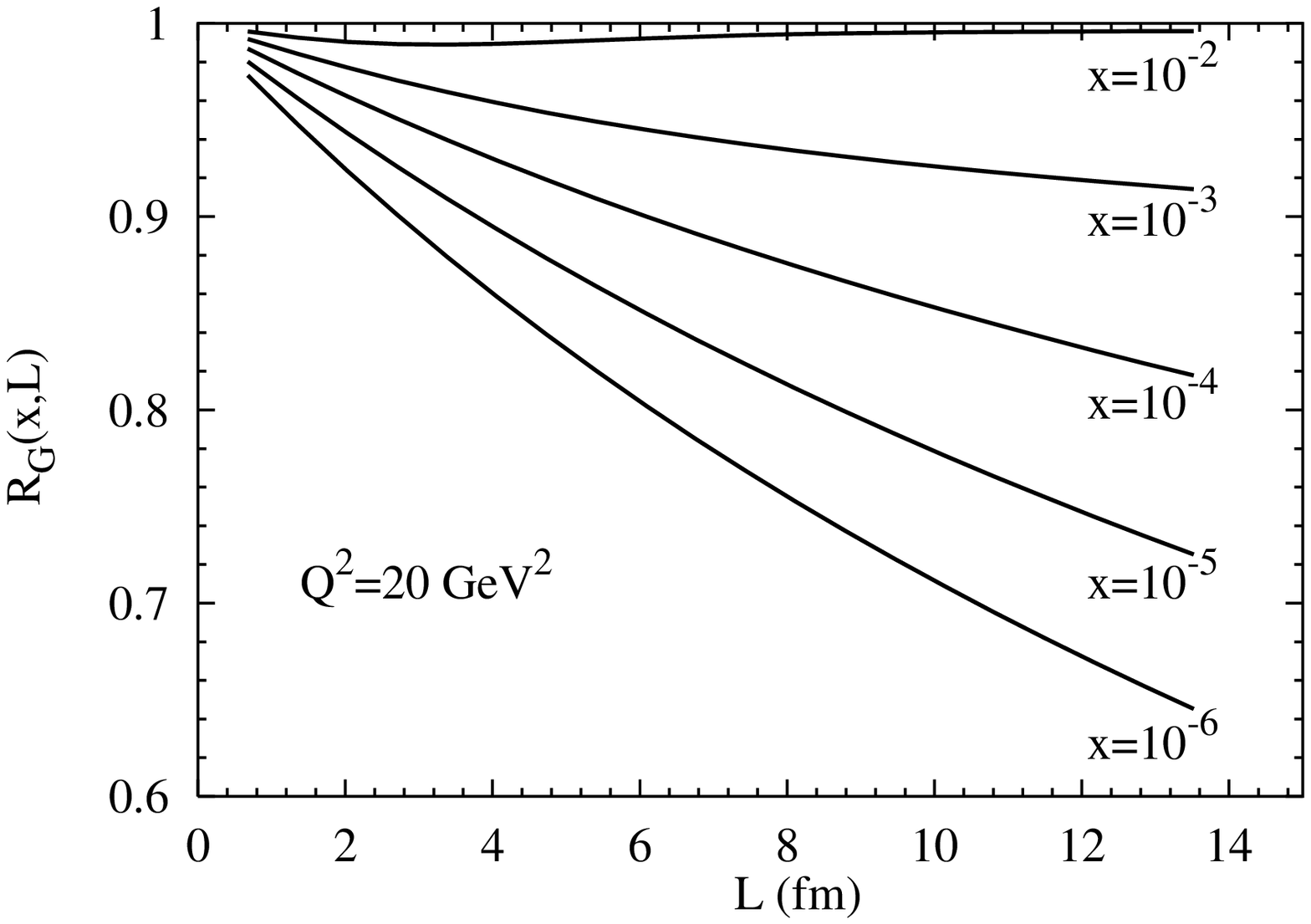}}
\caption{Gluon shadowing vs.\ the length of the nuclear medium $L=2\sqrt{R_A^2-b^2}$. 
All curves are for $Q^2=20$ GeV$^2$ but for different values of~$x$.}
\label{fig:gshadvsl}}
 \end{figure}

Our results for gluon shadowing as a function of the length of 
the nuclear medium at impact parameter $b$ are shown in 
Fig.~\ref{fig:gshadvsl}. 
The calculations are performed for lead with a uniform nuclear density of
$\rho_A=0.16\,\fm^{-3}$.  The 
small size of the $gg$ dipole leads to a rather weak gluon shadowing. For most
values of $x$, gluon shadowing increases as a function of $L$ as one would
expect.  At the largest value of $x=0.01$, however, gluon shadowing becomes
smaller as $L$ increases, and $R_g$ approaches $1$. Although this behavior
seems to be counterintuitive, it can be easily understood by noting that at
$x=0.01$ the coherence length of the $|q\bar qg\ra$-Fock state becomes very
small and the form factor of the nucleus suppresses shadowing \cite{krt2}. 

\section{Drell-Yan process}\label{sec:dy}

The Drell-Yan (DY) process,
\beq
h_1+h_2\to \bar ll+X,
\label{5.20}
\eeq
 in the kinematical region where the invariant dilepton mass $M$ is
small compared to the center of mass energy $\sqrt{s}$
 is of similar theoretical interest as DIS at low Bjorken $x$. Moreover, the cross sections of these processes are related by the factorization theorem \cite{css}.
In contrast to DIS, where only the total cross section can be measured, there is a variety of observables which can be measured in the DY process, such as the transverse momentum distribution or
the angular distribution of the lepton pair.

The fractional light cone momenta of the dilepton relative the colliding hadrons,
\beq
x_1= \frac{p^+_{\bar ll}}{p^+_{h_1}};\ \ \ \ \ 
x_2 = \frac{p^-_{\bar ll}}{p^-_{h_2}},
\label{5.40}
\eeq
satisfy the relations
\beqn
x_1x_2&=&\frac{M^2+p_T^2}{s};
\nonumber\\
x_1-x_2&=&x_F,
\label{5.60}
\eeqn
where $M$ is the dilepton invariant mass, and $x_F$ is the Feynman variable.

Within the collinear approximation of the lowest order parton model $p_T=0$, and the fractional light-cone momenta
Eq.~(\ref{5.60}) coincide with the standard Bjorken $x_{1,2}$ of the $q$ and $\bar q$ annihilating into the heavy dilepton, $\bar qq\to\bar ll$. In this lowest order approximation the cross section of the process (\ref{5.20}) 
is expressed in terms of quark and antiquark densities,
\beq
\frac{d\sigma}{d\tau}=\frac{4\pi\alpha^2_{em}}{3N_c\,M^2}
\int_0^1dx_1\sum_f Z_f^2
\left\{q_f(x_1)\bar q_{f}(\tau/x_1)+(1\leftrightarrow 2)\right\}.
\label{5.80}
\eeq
Here $\tau=M^2/s$, and the factor $N_c$ (number of colors) appears in the denominator,
because quark and antiquark must have the same color in order to annihilate.

Some features of dilepton production
cannot be understood in the lowest order approximation. 
\begin{itemize}
\item{Cross sections calculated straightforwardly from (\ref{5.80}) are too small
by a factor of $2-3$ compared to the measured value. This discrepancy is usually
treated by the introducing of a so called $K$-factor. The $K$-factor is approximately
independent of $M^2$, but it is process dependent.}
\item{Dileptons with large transverse momenta, of the order of few GeV, are observed in
experiment. There are however no transverse momenta in the naive parton model, which treats the process in the collinear approximation, Eq.~(\ref{5.80}).
Phenomenologically, one can introduce a primordial momentum distribution of the
quarks. However, the
width necessary to describe data is much larger than
what one would expect from Fermi motion of the quarks.}
\end{itemize}

These problems can be settled by taking into account the first order QCD
corrections. Due to the
radiation of the gluon, the quark acquires a transverse momentum. 
In this way, the
pQCD corrections provides the missing mechanism for the production of lepton
pairs with large transverse momentum $p_T$ \cite{fritzsch,transal,transal2}. 

The next-to-leading (NLO) correction solves most of the problems of the naive parton
model. It explains how large $p_T$-dileptons are produced and account for almost
all of the $K$-factor \cite{dyreport}. 
Yet, not all the problems are settled by this correction.
Since it is numerically large, one has to investigate how much higher order
corrections affect the result. Furthermore, the transverse momentum spectrum
is not well described even qualitatively. The theoretical result agrees with data only at
$p_T^2\sim M^2$, and even diverges at $p_T\to 0$:
$d\sigma/dp_T^2\propto p_T^{-4}$, while the observed cross section is of course finite.
The reason for this behavior is that 
large logarithms
$\ln({M^2}/{p_T^2})$
occur in higher order corrections
and one has to re-sum all these terms.
This is possible within pQCD \cite{resumal,resumcol,Pirner}, by a re-summing
soft gluons radiated from the quark or the antiquark.

\subsection{Dipole description of heavy dilepton production}
\label{sec:heavy}

Although cross sections and all observables are Lorentz invariant, the partonic interpretation of
the process depends on the reference frame. We have seen this in
the case of DIS, and similar considerations can be
made for DY. It was pointed out in \cite{hir} that in the target
rest frame, DY dilepton production looks like bremsstrahlung, rather than parton
annihilation.  A quark (or an antiquark)
from the projectile hadron radiates a virtual
photon on impact on the target, as is illustrated in Fig.~\ref{fig:dyfig}.
\begin{figure}[htb]
\parbox{\halftext}{
\centerline{\includegraphics[width=7 cm]{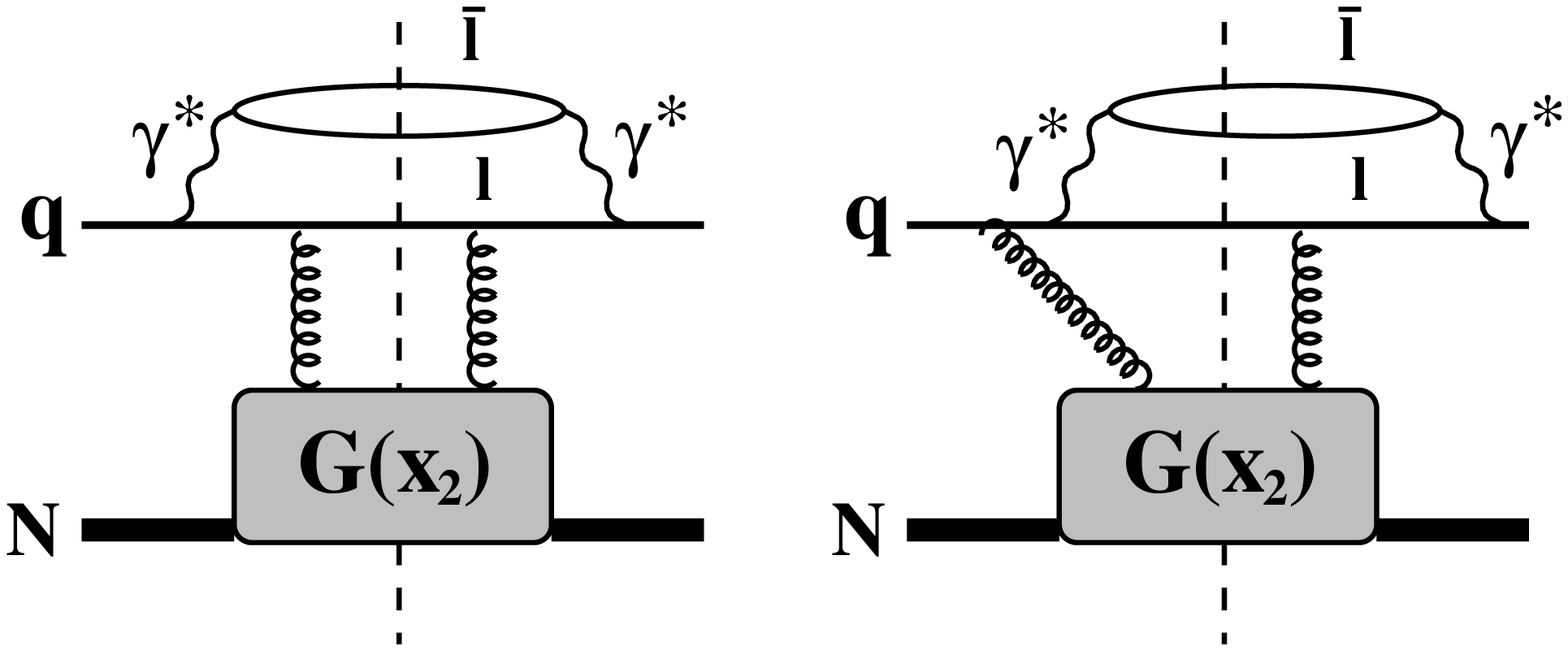}}
 \caption{Feynman diagrams squared for the cross section of di-lepton production in 
a $q$-$N$ collision.  In the target rest frame the
      quark (or antiquark) scatters off the target color field and radiates a
      massive photon, which decays into the lepton pair. The photon
      can be radiated before or after the quark hits the target.
      $G(x)=xg(x)$ is the gluon distribution function in the proton. }
 \label{fig:dyfig}}
\hfill
\parbox{\halftext}{
\centerline{\includegraphics[width=7 cm]{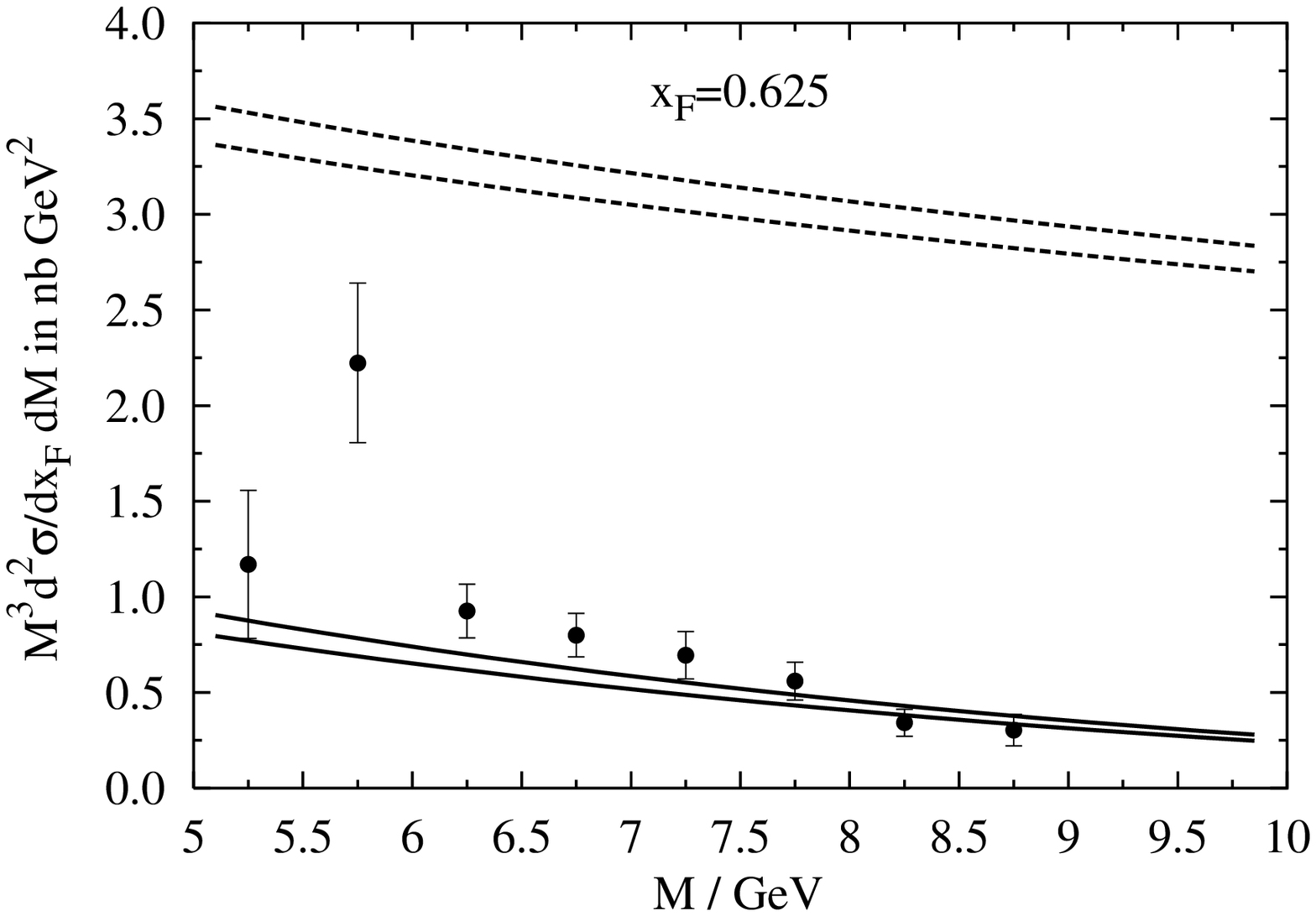}}
\caption{Data  from E772 \cite{dydata} and solid curves are for the DY cross section in $pD$ collisions
 at $\sqrt{s}=38.8$ GeV, dashed curves are for $\sqrt{s}=500\GeV$. 
  For each energy, the upper and lower curves correspond to massless quarks, or $m_f=200\MeV$.
}
\label{fig:dy-data}}
 \end{figure}
The radiation can occur before (not shown) or after
the quark scatters off the target. 
The sum of the two contributions gets the form in impact parameter representation similar to the dipole interaction in DIS. Such a dipole description of the DY process was proposed in \cite{hir} and developed in many publications including \cite{kst1,bhq,krt3,krtj,brazil}.

The cross section for radiation of a virtual photon from a quark after
scattering on a proton, can be written in factorized light-cone form,
\beq
\frac{d\sigma(qp\to \gamma^*X)}{d\ln\alpha}
=\int d^2\rho\, |\Psi^{T,L}_{\gamma^* q}(\alpha,\rho)|^2
    \sigma_{q\bar q}(x_2,\alpha\rho),
    \label{5.100}
\eeq
similar to the case of DIS.
Here, $\sigma_{q\bar q}$ is the dipole-proton cross section, which depends on the $q\bar q$ separation 
$\alpha\rho$,
where $\rho$ is  the photon-quark transverse separation and $\alpha$ 
is the fraction of 
the light-cone momentum of the initial quark taken away by the photon.
In (\ref{5.100}) $T$ and $L$ stand for radiation of transverse and longitudinal photons respectively.

An interesting feature of (\ref{5.100}) is the appearance of the dipole 
cross section, although there is no physical $q\bar q$-dipole in Fig.~\ref{fig:dyfig}.
The physical interpretation of (\ref{5.100}) is similar to DIS. The projectile quark is expanded in the
interaction eigenstates. We keep only the first two,
\beq
|q\ra=\sqrt{Z_2}|q_{bare}\ra+\Psi^{T,L}_{\gamma^* q}|q\gamma^*\ra+\dots,
\eeq
where $Z_2$ is the wave function renormalization constant for fermions.
In order to produce a new state the interaction must resolve between the two Fock 
states, {\it i.e.} they have to interact differently. Since only the bare quarks
interact in both Fock components the difference arises from the relative displacement 
in transverse plane of the quark radiating the photon.  If $\rho$ is the transverse separation between the
quark and the photon, the $\gamma^*q$ fluctuation has a center of gravity in the
transverse plane which coincides with the impact parameter of the parent quark.
The transverse separation between the photon and the center of gravity is
$(1-\alpha)\rho$ and the distance between the quark and the center of gravity is
correspondingly $\alpha\rho$. A displacement in coordinate space corresponds to
a relative phase factor in momentum space, 
$-\exp({\rm i}\alpha\vec\rho\cdot\vec k_\perp)$, which produces the color 
screening factor exactly like in the dipole cross section.

The LC distribution functions for the $|q\gamma^*\ra$ component can be written in a form similar to DIS,
\beq
\Psi^{T,L}_{\gamma^* q}(\alpha,\vec\rho)=\frac{\sqrt{\alpha_{em}}}
{2\pi}
\left(\bar\chi_q\widehat {\cal  O}_{\gamma^*q}^{T,L}\chi_{q}\right)
{\rm K}_0(\eta\rho),
\label{5.120}
\eeq
where $\chi$ are two component spinors, and
\beq
\eta^2 = m_f^2 \alpha^2 + M^2 \left(1-\alpha\right). 
\label{5.140}
\eeq
The operators $\widehat {\cal  O}_{\gamma^*q}^{T,L}$ \cite{kst1} look similar to, but are different from the analogous operators Eqs.~(\ref{4.100}) and (\ref{4.120}) in the case of DIS,
\beqn
\widehat {\cal  O}_{\gamma^*q}^T&=&
{\rm i}m_f\alpha^2\vec e\cdot(\vec n\times\vec\sigma)-
{\rm i}(2-\alpha)(\vec e\cdot\vec\nabla_\rho)
+\alpha\vec e\cdot(\vec\sigma\times\vec\nabla_\rho),\nonumber\\
\widehat{\cal  O}_{\gamma^*q}^L&=&
2M(1-\alpha).
\label{5.160}
\eeqn
The two dimensional gradient $\vec\nabla_\rho$ acts on the transverse coordinate
$\vec\rho$; $\vec n$ is the unit vector parallel to the momentum 
of the projectile quark; and
$\vec\sigma$ are the Pauli spin-matrices.
The LC distribution functions for the transition $q\to \gamma^* q$
read explicitly, for a given flavor of unit charge
\beqn
|\Psi_{\gamma^* q}(\alpha,\rho)|^2   &=&
    |\Psi^T_{\gamma^* q}(\alpha,\rho)|^2
  + |\Psi^L_{\gamma^* q}(\alpha,\rho)|^2, 
  \label{5.180}
  \\
|\Psi^T_{\gamma^* q}(\alpha,\rho)|^2 &=&
  \frac{\alpha_{em}}{\pi^2}\left\{
     m_f^2 \alpha^4 {\rm K}_0^2 \left(\eta\rho\right)
   + \left[1+\left(1-\alpha\right)^2\right]\eta^2
     {\rm K}_1^2 \left(\eta\rho\right)\right\}, 
     \label{5.200}\\
\label{dylcl}|\Psi^L_{\gamma^* q}(\alpha,\rho)|^2 &=&
  \frac{2\alpha_{em}}{\pi^2}
  M^2 \left(1-\alpha\right)^2 {\rm K}_0^2 \left(\eta\rho\right).
  \label{5.220}
\eeqn
Comparing
(\ref{5.200}) and (\ref{5.220}) with their DIS counterparts (\ref{4.210}) and
(\ref{4.220}) shows that the factor $N_c$ is no longer present 
in the $q\to \gamma^* q$ LC wavefunctions. This is due to the $N_c$ in the
denominator of (\ref{5.80}). Furthermore, we see that 
$|\Psi^T_{\gamma^* q}(\alpha,\rho)|^2$
 has an extra factor of $2$, because in the
DY process, one has to sum over the transverse polarizations
of the photon, rather than
average, as in DIS.

Eventually, we are in a position to calculate the DY cross section in hadron-proton collisions,
relying on the elementary cross section Eq.~(\ref{5.100}).
According to the definition (\ref{5.40}) the photon carries away the LC momentum fraction
$x_1$ from the projectile hadron, therefore the hadronic cross
section reads, 
\beqn
\frac{d\sigma(hN\to\gamma^*X)}{dM^2dx_F}&=&\frac{\alpha_{em}}{3\pi M^2}\,
\frac{x_1}{x_1+x_2}\int\limits_{x_1}^1\frac{d\alpha}{\alpha^2}
\sum_fZ_f^2\left\{q_f\left(\frac{x_1}{\alpha}\right)+
q_{\bar f}\left(\frac{x_1}{\alpha}\right)\right\}
\frac{d\sigma(qN\to \gamma^*X)}{d\ln\alpha}\nonumber\\
\nonumber\\
&=&\frac{\alpha_{em}}{3\pi M^2}\,
\frac{1}{x_1+x_2}\int\limits_{x_1}^1\frac{d\alpha}{\alpha}
F_2^h\left(\frac{x_1}{\alpha}\right)
\frac{d\sigma(qN\to \gamma^*X)}{d\ln\alpha},
\label{5.240}
\eeqn

Employing in (\ref{5.100}) the universal dipole cross section in the parametrization \cite{gbw},
which we have already used in DIS, one can perform a numerical evaluation in a parameter free way. 
As an example, we compare in Fig.~\ref{fig:dy-data}
the results of numerical evaluation of the DY cross section, plotted by solid curves, with data from the E772 experiment \cite{dydata} at $\sqrt{s}=38\GeV$. To show the sensitivity to the quark mass in (\ref{5.140})
the calculation was done with $m_f=0$ (upper curve) and $m_f=200\MeV$ (lower curve).
The dashed curves show predictions for the energy of RHIC $\sqrt{s}=500\GeV$.

Remarkably, the parameter free calculations explain well the observed cross section without any adjustment, and
without any $K$-factor. This is the advantage of the dipole approach: the phenomenological dipole cross section is fitted to DIS data, $l+p\to l'+X$, at small Bjorken $x$ which include all higher order effects, all possible gluon radiation. Therefore, one should not try to add to the above results higher order corrections, which would be double counting. This approach is also an alternative to the twist decomposition, since all higher twist effects are included by default.
A numerical test of the dipole approach for Drell-Yan reaction vs NLO
parton model calculations was performed in \cite{Raufeisen:2002zp}, and no visible
deviation was found.

As we already noticed, the DY process provides more observables than DIS, in particular the di-leptons are produced
with a certain Feynman $x_F$ and transverse momentum $q_T$. The $x_F$ distribution was  calculated above in (\ref{5.240}). The dipole approach also allows to calculate the differential DY cross section. The $q_T$ dependence comes from the interference of the DY amplitudes with different impact parameters and has the form~\cite{kst1,jkt},
 \beqn\nonumber
\frac{d\sigma(qN\to \gamma^*X)}{d\ln\alpha\, d^2q_{T}}
&=&\frac{1}{(2\pi)^2}
\int d^2\rho_1d^2\rho_2\, 
e^{{\rm i}\vec q_{T}\cdot(\vec\rho_1-\vec\rho_2)}
\Psi^*_{\gamma^* q}(\alpha,\vec\rho_1)
\Psi_{\gamma^* q}(\alpha,\vec\rho_2)\\
&\times&
\frac{1}{2}
\Bigl[\sigma^N_{q\bar q}(\alpha\rho_1,x_2)
+\sigma^N_{q\bar q}(\alpha\rho_2,x_2)
-\sigma^N_{q\bar q}(\alpha|\vec\rho_1-\vec\rho_2|,x_2)\Bigr]\ .
\label{5.260}
 \eeqn

Replacing the $q_T$-integrated quark-nucleon cross section $d\sigma(qp\to \gamma^*X)/d\ln\alpha$ in
(\ref{5.240}) by the differential cross section (\ref{5.260}), one gets the differential $q_T$-dependent 
DY cross section in $hN$ collisions.

\subsection{Drell-Yan process on nuclear targets}\label{sec:dy-nucl}

\subsubsection{Shadowing in di-lepton production}\label{sec:dy-shad}

The total cross section of the DY process on a nuclear target can be calculated with the same  equation (\ref{5.240})
with a replacement $d\sigma(qp\to \gamma^*X)/d\ln\alpha\Rightarrow d\sigma(qA\to \gamma^*X)/d\ln\alpha$.
The calculation of the quark-nucleus DY cross section is also rather straightforward. One can employ  Eq.~(\ref{4.200}), which was used for DIS on nuclei, replacing the photon distribution functions $\Psi^{T,L}_{\bar qq}(\alpha,r_T)$ by the $\gamma^*q$ distribution functions $\Psi^{T,L}_{\gamma^*q}(\alpha,r_T)$ given by Eq.~(\ref{5.120}). Examples of numerical results are presented in Fig.~\ref{fig:atotal}. The $A$-dependence of the $q_T$-integrated DY cross section is predicted \cite{krtj} for the energies of RHIC and LHC for proton and deuteron collisions with different nuclei. 
\begin{figure}[htb]
\parbox{\halftext}{
\centerline{\includegraphics[width=7.5 cm]{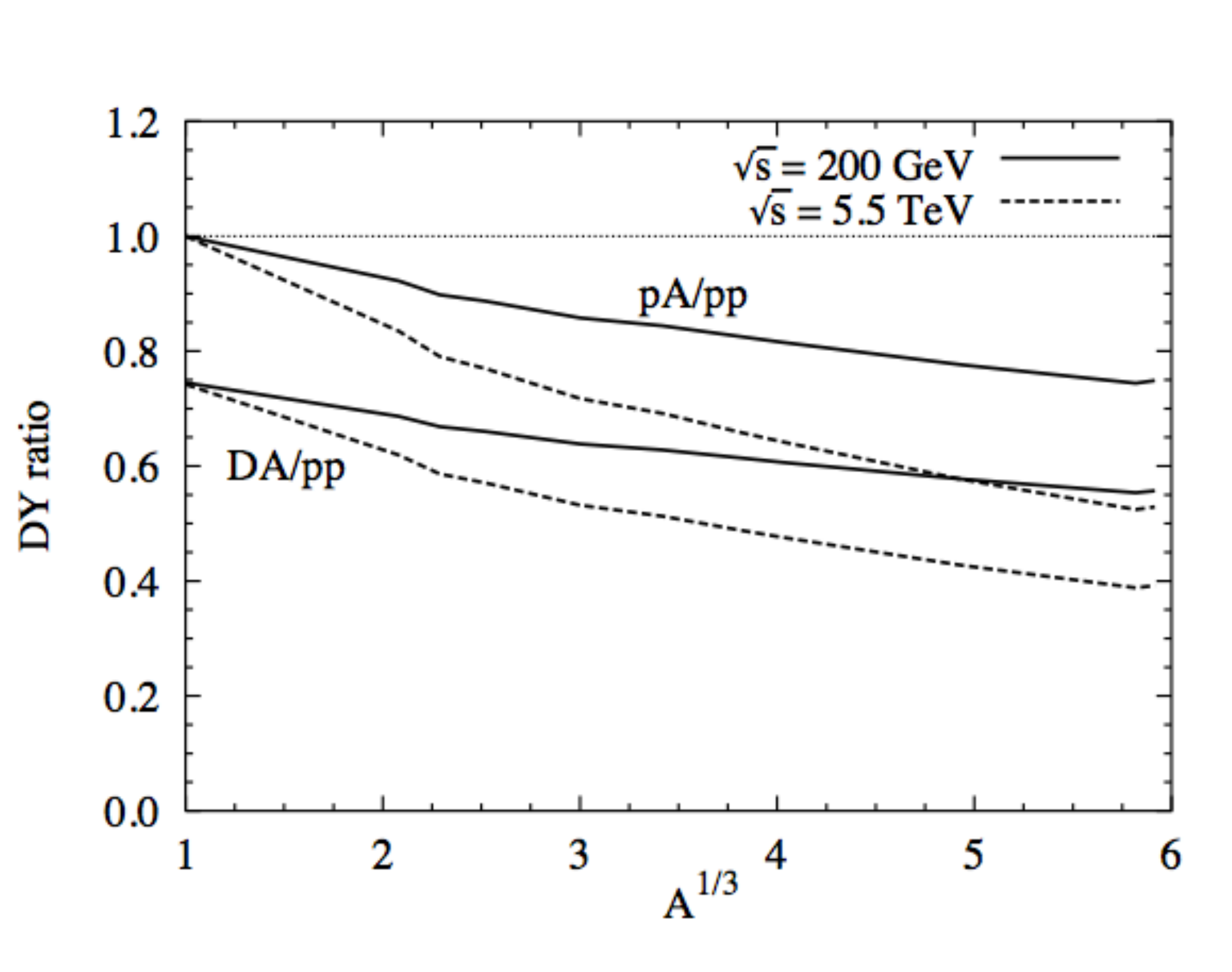}}
 \caption{Shadowing for the total DY cross section as function of $A^{1/3}$ at $M=4.5$ GeV and $x_F=0.5$. 
The lower pair of curves is for deuteron -- gold scattering
and the two upper curves are for proton -- gold collisions. }
 \label{fig:atotal}}
\hfill
\parbox{\halftext}{
\centerline{\includegraphics[width=7.5 cm]{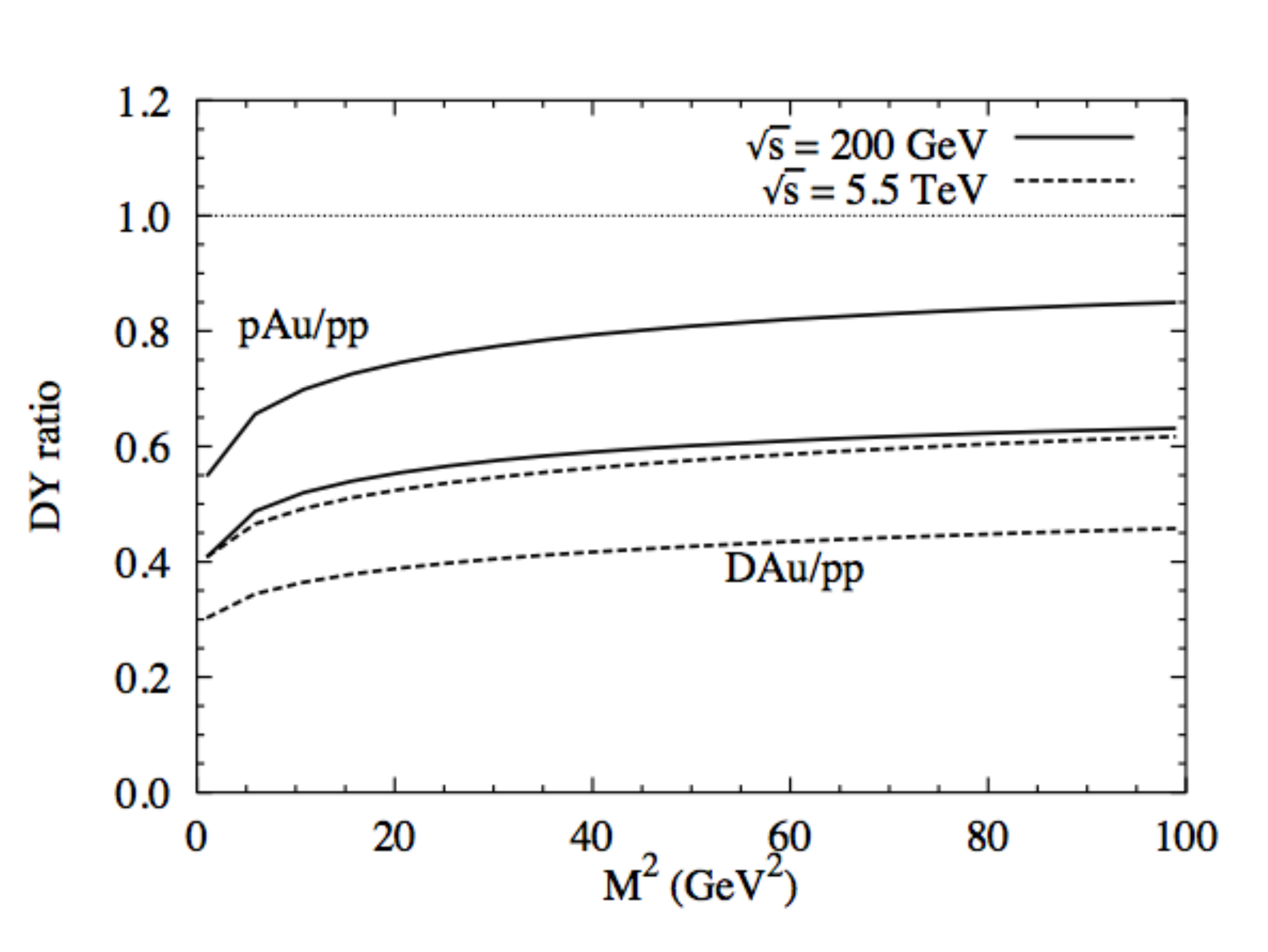}}
\caption{Shadowing for the total DY cross section 
in proton -- gold (upper curves) and deuterium -- gold 
(lower curves) collisions
at the energies of RHIC and LHC at $x_F=0.5$
as function of  di-lepton mass $M^2$.
}
\label{fig:mtotal}}
 \end{figure}
Calculations are done in the "frozen-dipole" approximation which is sufficiently accurate at these energies, and $x_F=0.5$. The scale dependence of these calculations is presented in \ref{fig:mtotal}. As in DIS, the magnitude of shadowing slowly decreases with rising scale. Notice that gluon shadowing is important at these high energies and was included in the calculations (see details in \cite{krtj}).

The results demonstrate a considerable difference between the nuclear effects for proton and deuteron beams, which is related to isotopic effects. The radiation of a heavy photon is sensitive to the quark charge.

\subsubsection{Nuclear modification of the transverse momentum distribution}\label{sec:qT-dependence}

The quark-nucleon differential DY cross section, Eq.~(\ref{5.260}) is easily converted to the quark-nucleus one, if the conditions for the "frozen" approximation are fulfilled.  Then, the cross section reads \cite{kst1,krtj},
 \beqn
\frac{d\sigma(qA\to \gamma^*X)}{d\ln\alpha d^2q_{T}}
&=&\frac{1}{(2\pi)^2}\int d^2b
\int d^2\rho_1d^2\rho_2\, 
e^{{\rm i}\vec q_{T}\cdot(\vec\rho_1-\vec\rho_2)}
\Psi^*_{\gamma^* q}(\alpha,\vec\rho_1)
\Psi_{\gamma^* q}(\alpha,\vec\rho_2)\nonumber\\
&\times&
\left[1-e^{-{1\over2}\sigma^N_{q\bar q}(\alpha\rho_1,x_2)T_A(b)}
-e^{-{1\over2}\sigma^N_{q\bar q}(\alpha\rho_2,x_2)T_A(b)}
+e^{-{1\over2}\sigma^N_{q\bar q}(\alpha|\vec\rho_1-\vec\rho_2|,x_2)T_A(b)}\right].
\label{5.280}
 \eeqn
 Convoluting this cross section with the quark distribution function of the beam hadron, one arrives at the differential hadron-nucleus DY cross section,
\beq
\frac{d\sigma(hp\to\gamma^*X)}{dM^2dx_Fd^2q_T}=\frac{\alpha_{em}}{3\pi M^2}\,
\frac{1}{x_1+x_2}\int\limits_{x_1}^1\frac{d\alpha}{\alpha}
F_2^h\left(\frac{x_1}{\alpha}\right)
\frac{d\sigma(qA\to \gamma^*X)}{d\ln\alpha\,d^2q_T}\,.
\label{5.300}
\eeq

The numerical results of calculations of the nuclear ratio (\ref{1.20}) with equations (\ref{5.240}) and (\ref{5.300}),
at the energies of RHIC and LHC are depicted by dashed curves in Fig.~\ref{fig:rg}.
\begin{figure}[htb]
\hspace*{0.5cm}
 \scalebox{0.55}{\includegraphics*{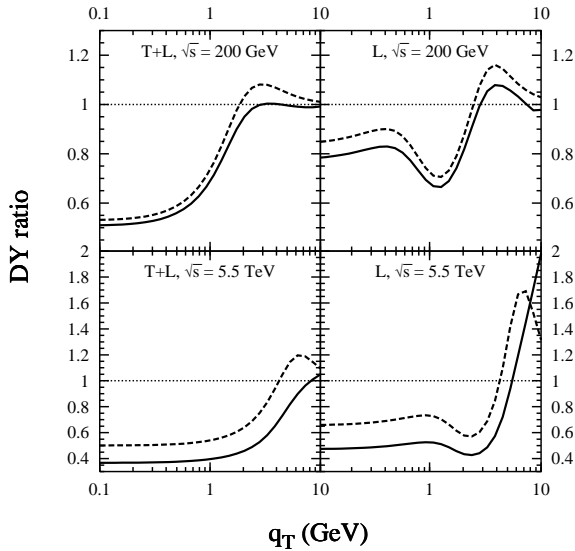}}\hfill
  \raise0.9cm\hbox{\parbox[b]{3in}{
    \caption{
      \label{fig:rg}
      The influence of gluon shadowing on the nuclear ratio Eq.~(\ref{1.20})
      of the DY differential cross sections. Dashed curves
are calculated without gluon shadowing, while solid curves include gluon shadowing.
The influence on the longitudinal DY cross section is shown separately
in the two left-hand 
plots $(L)$. The two plots on the right show the DY ratio for 
the sum of the transverse and longitudinal cross sections $(T+L)$. 
Calculations are done at $x_F = 0.625$, $M = 6.5\GeV$, and the energies of RHIC and LHC.}
  }
}
\end{figure}
The ratio exhibits the so called Cronin effect (e.g. see Cronin effect for hadrons \cite{knst-prl}), a nuclear enhancement at medium values of $q_T$. This can be interpreted as a result of increased mean transverse momentum of the di-lepton produced on a nucleus compared to a proton target. Intuitively is can be understood as a result of multiple interactions of the quark which "shakes off" a heavy photon with increase $q_T$ after getting a stronger kick from multiple rescatterings.

Gluon shadowing was also added in these calculations, and the results are plotted in Fig.~\ref{fig:rg} by solid curves.
One can rely on the small-$\rho$ approximation, $\sigma_{\bar qq}(\rho)\propto \rho^2$, which is quite accurate for DY processes at $q_T>1\GeV$ \cite{krt3}. In this approximation the dipole cross section is proportional to the gluon density \cite{FS},
 \beq
\sigma_{q\bar q}^N(\rho,x)\Bigr|_{\rho\to0}
=\frac{\pi^2}{3}\alpha_s\!
\left(\frac{\lambda}{\rho^2}\right)
\rho^2\,G_N\!\!\left(x,\frac{\lambda}{\rho^2}\right),
 \label{5320}
 \eeq
where $\lambda\approx10$ is a phenomenological parameter.
On a nuclear target one should replace $G_N\left(x,\frac{\lambda}{\rho^2}\right)\Rightarrow G_A\left(x,\frac{\lambda}{\rho^2}\right)/A$, and correspondingly in Eq.~(\ref{5.280}),
 \beq
\sigma^N_{q\bar q}(\rho,x)\Rightarrow
\sigma^N_{q\bar q}(\rho,x)R_G\!\left(x,\frac{\lambda}{\rho^2},b\right)\ ,
\label{5.340}
 \eeq
The gluon contribution is appreciable, especially at the energies of LHC, and it practically terminates the Cronin enhancement for radiation of transversely polarized photons.

\subsubsection{A word of caution: incoherent effects imitating coherence}
\label{sec:incoherent}

 In the case of DIS we saw above that the smaller $x$ is, the longer is the coherence time $t_c\sim(2xm_N)^{-1}$. Correspondingly, shadowing increases with $1/x$.
 This is not true anymore for DY processes. According to (\ref{5.60}) $x_2$ reaches the minimal value when $x_1\to1$. This is the kinematical limit and one needs to be cautious. 
 The coherence length (same as time) for a fluctuation of a quark $q\to \bar ll\,q$
is given by,
\beq
l_c=\frac{2E_q}{M^2_{\bar llq}-m_q^2}\ ,
\label{5.360}
\eeq
where
\beq
M^2_{\bar llq}=\frac{M^2}{1-\alpha}+\frac{m_q^2}{\alpha}
+ \frac{q_T^2}{\alpha\,(1-\alpha)}\ 
\label{5.380}
\eeq
Apparently $\alpha>x_1$, therefore when $x_1\to1$, then $\alpha\to1$ and $t_c\to0$, contrary to the naive expectation based on the experience with DIS. 

 The mean coherence length can be evaluated similar to how it was done in Sect.~\ref{sec:tc-dis} for DIS, i.e. by averaging $l_c$ weighted with the $\gamma^*q$ distribution functions and dipole cross sections. The results obtained in\cite{eloss1,eloss2}for $pp$ collisions at $800\GeV$ are plotted by solid curves in Fig.~\ref{fig:dy-lc}.
\begin{figure}[htb]
\parbox{6.5cm}{
\centerline{\includegraphics[width=6 cm]{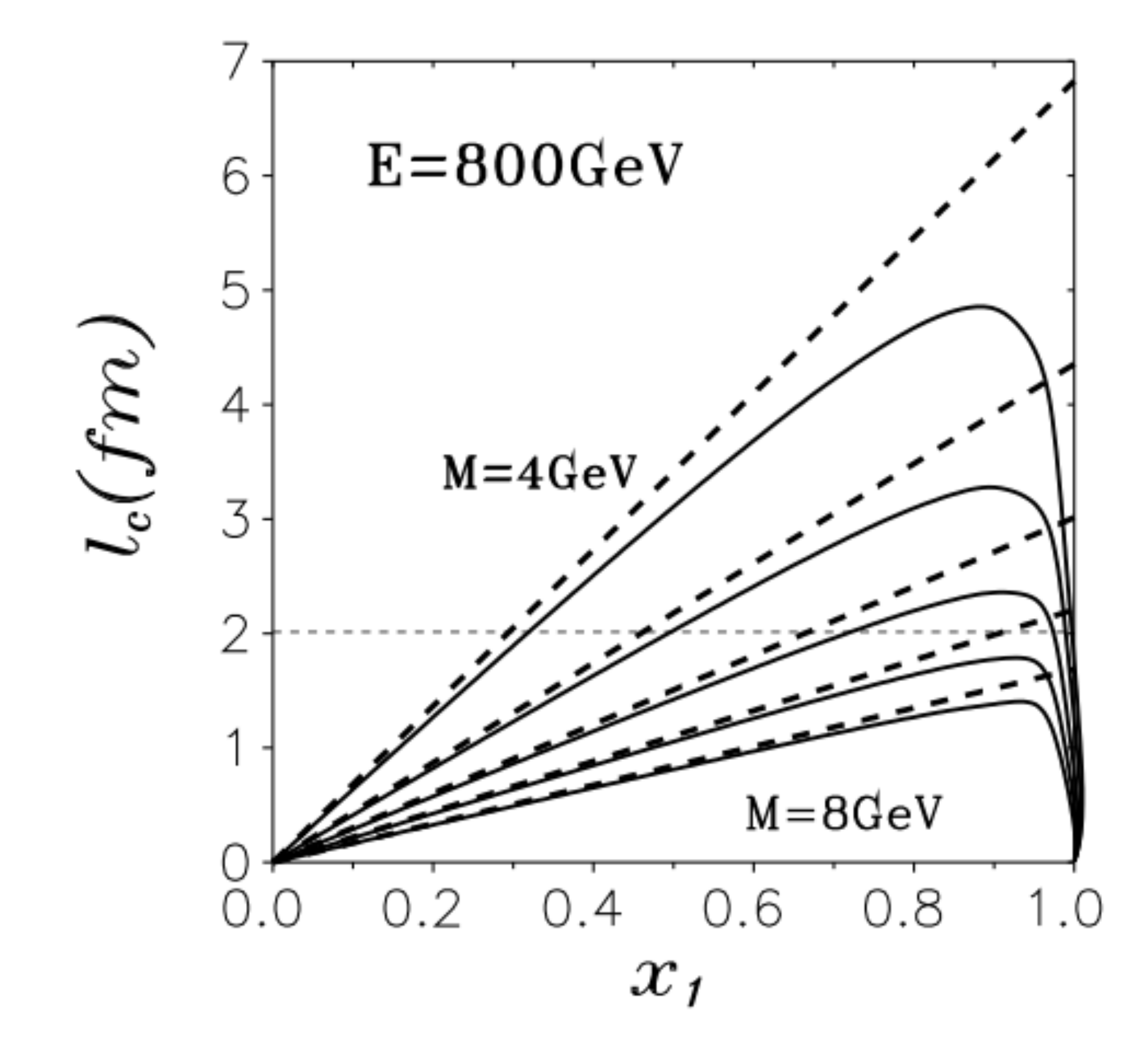}}
 \caption{The mean coherence length (\ref{5.360}) vs $x_1$
at different dimuon masses.  Dashed lines show the predictions of QCD 
factorization dependent only on $x_2$. }
 \label{fig:dy-lc}}
\hfill
\parbox{10.5cm}{
\centerline{
\includegraphics[width=6 cm]{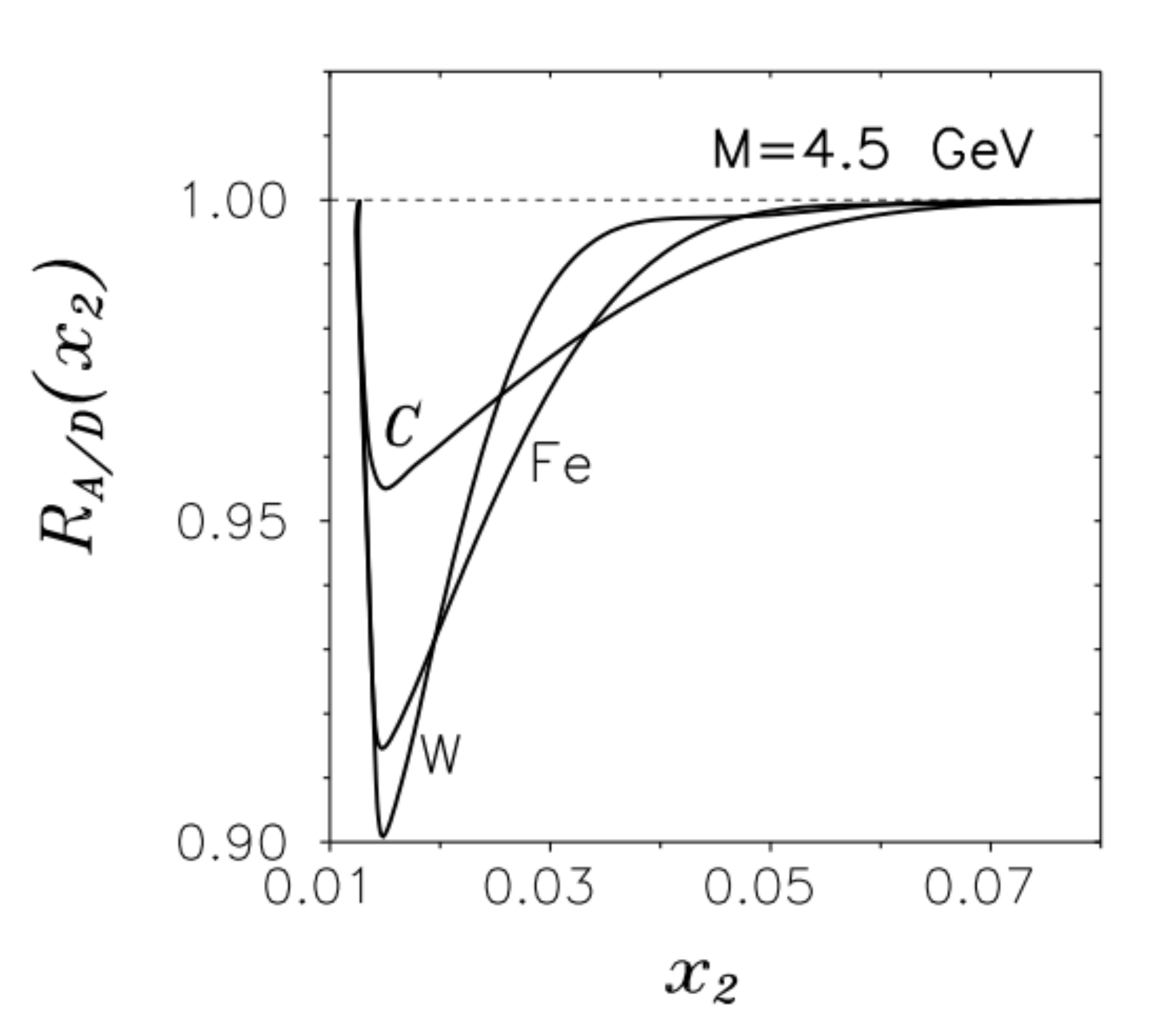}
\includegraphics[width=5.5 cm]{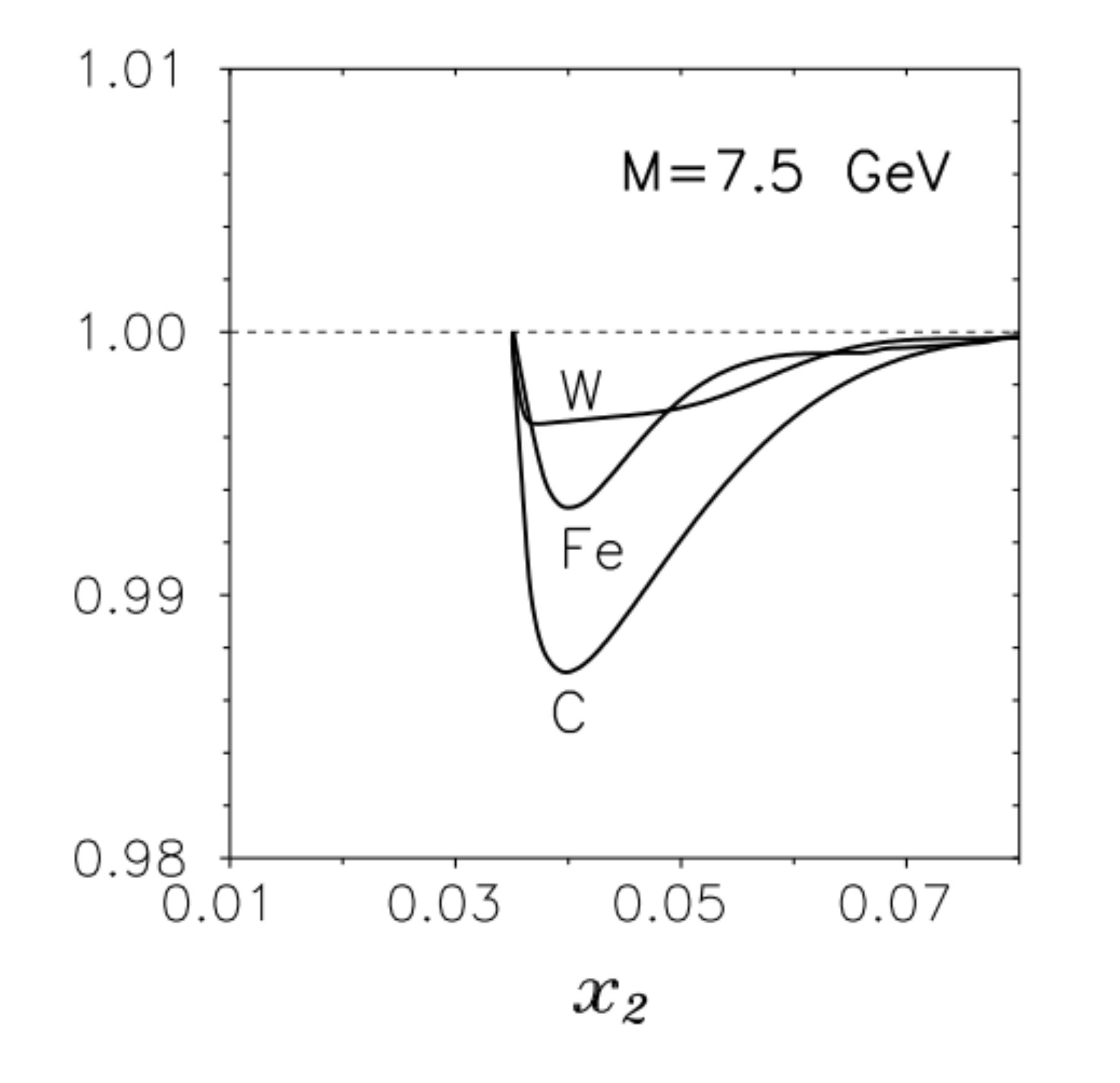}}
\caption{Shadowing in DY reaction on carbon, iron and tungsten as function
of $x_2$ for $M=4.5\,GeV$ (left) and $7.5\GeV$ (right). Nuclear shadowing disappears
at large and small $x_2$ because the coherence length, Eqs.~(\ref{5.360}) and 
(\ref{5.380}), vanishes in these limits.
}
\label{fig:dy-shad}}
 \end{figure}
Although at small $x_1$ $\la l^{DY}_c\ra$ rises, because $x_1$ is decreasing, upon approaching
the kinematic limit $x_1\to1$ is steeply falls down. Correspondingly, no shadowing is possible in this regime.
Dashed lines linearly rising with $1/x^2\propto x_1$ correspond to the usual behavior (\ref{4.20}) for DIS, and would be valid also for DY, if QCD factorization were correct.

As a result of these kinematical constraints the interval of $x_2$ available for shadowing in DY data of the E772/E866
experiments is rather narrow and the magnitude of shadowing \cite{eloss2} depicted in Fig.~\ref{fig:dy-shad} is considerably smaller that what one would expect based on QCD factorization.
 
 As long as shadowing in DY processes is weak at large $x_1$, one may wonder what causes the considerable nuclear suppression observes at $x_1\to1$ in DY data of the E772 and E866 experiments? An example of data is depicted in Fig.~\ref{fig:dy-e772}.
\begin{figure}[htb]
\hspace*{1.5cm}
 \scalebox{0.55}{\includegraphics*{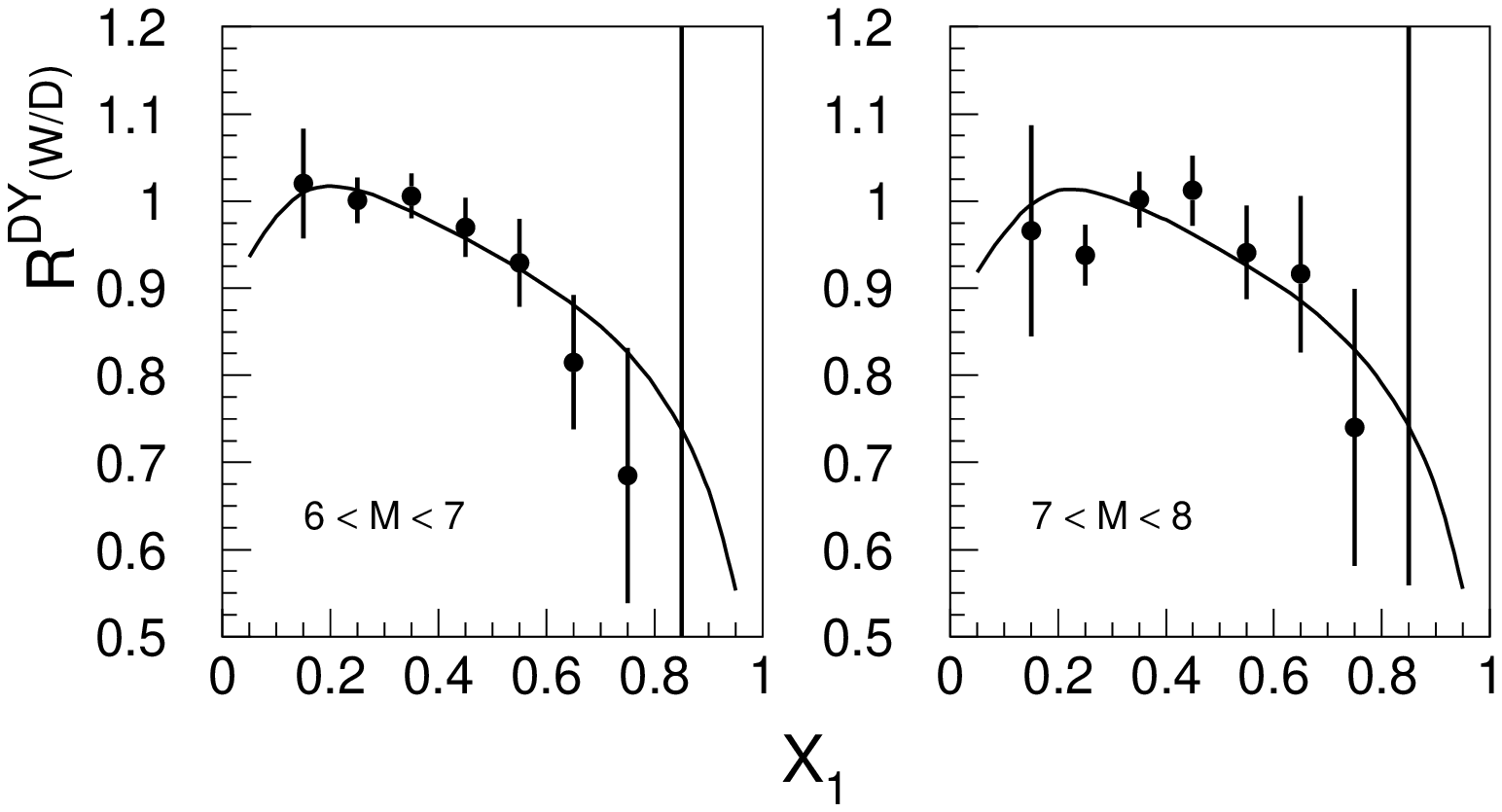}}\hfill
  \raise 0.8cm\hbox{\parbox[b]{3in}{
    \caption{
      \label{fig:dy-e772}
 Ratio of DY cross sections on tungsten to deuterium as a function of
$x_1$. Data \cite{e772} are selected to have large di-lepton masses to eliminate nuclear shadowing.     
The curves are calculated in \cite{forward} including energy loss effects.} 
}}
\end{figure}
The data are selected with large di-lepton invariant mass $M>6\GeV$. According to Fig.~\ref{fig:dy-lc}
the coherence length for such a large scale is shorter than the mean nucleon spacing $2\fm$ at all $x_1$,
therefore shadowing is impossible, while the data demonstrate a strong nuclear suppression steadily growing towards $x_1=1$. 

Qualitatively, one should expect any hadronic system propagating through a medium to dissipate energy \cite{eloss1,eloss2,forward}. Apparently, this makes it more difficult to produce any particle, in particular a di-lepton, carrying a large fractional LC momentum $x_1\to1$, i.e. one should expect a nuclear suppression of leading particles.
In terms of the Fock state decomposition this can be seen as nuclear enhancement of higher components containing more gluons \cite{forward}. Correspondingly, should expect an $x_1$-scaling for the nuclear suppression. The results on numerical evaluation \cite{forward} is plotted in Fig.~\ref{fig:dy-e772}.

Notice that these results of the E772 and E866 DY experiments are included in most of the global DGLAP based analyses 
 aiming at the extraction of the nuclear PDFs from data. The incorrect treatment of these DY data as shadowing certainly causes a bias in the results of such analyses.
 
 The effect of increase nuclear suppression at the edge of the kinematic range is a general phenomenon and has been also observed in hadron-nucleus collisions (see discussion and data in \cite{kn-iop}).

The differential cross section of the DY reaction was found above to be modified by the coherence effects on nuclei.
However, even if the coherence length is short, a similar modification is possible. Fig.~\ref{fig:qt-e866} presents data from the E866 experiment \cite{vasiliev,leitch}, which exhibit a pronounced Cronin effect, although the coherence length is too short to explain this by coherence.
\begin{figure}[htb]
\hspace*{1cm}
 \scalebox{0.75}{\includegraphics*{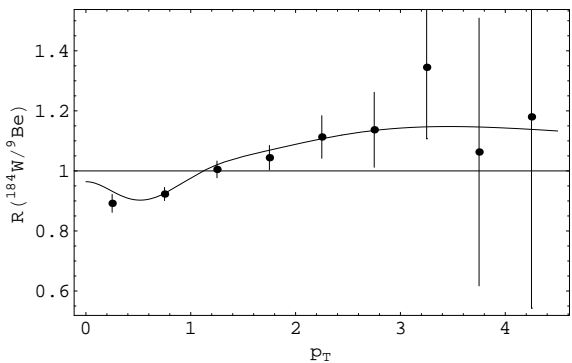}}\hfill
  \raise1cm\hbox{\parbox[b]{3in}{
    \caption{
      \label{fig:qt-e866}
     Comparison of theoretical prediction of $R_{W/Be}(p_T)$ vs. $p_T$
(in $\GeV/c$) to experiment for $x_2 = 0.05$.  Data are from the
FermiLab E772/E866 experiments~\cite{vasiliev,leitch}.  The curve shows the result of the calculation of broadening in the color dipole approach \cite{cronin-mikkel}. }
  }
}
\end{figure}
Even if $l_c$ is short, the projectile quark experiences multiple interactions in the nucleus and $p_T$-broadening prior radiation of the di-lepton. The increase $p_T$ of the quark is transferred to the radiated dilepton as $q_T=\alpha p_T$. The quark broadening can be also calculated in the dipole approach \cite{jkt,broadening}
\beq
\Delta p_T^2=2T_A(b)\,\frac{\partial\sigma_{\bar qq}(r_T)}{\partial r_T^2}\Biggr|_{r_T=0}
\label{5.400}
\eeq
The result of numerical evaluation \cite{cronin-mikkel} of the Cronin effect is compared with data in Fig.~\ref{fig:qt-e866}.

\section{DIS of neutrinos}\label{sec:nu-dis}

\subsection{Parton model}
\label{sec:parton}

The charge current (CC) neutrino interaction $\nu(\bar\nu)+N\to l(\bar l)+X$ probes three different types of structure functions \cite{handbook},
\beqn
\frac{d\sigma^{\nu(\bar\nu)}}{dx\,dy}&=&
\frac{\pi\alpha_{em}^2 m_NE_\nu}
{2\sin^4(\theta_W)(M_W^2+Q^2)^2}
\left[{y^2\over2}2xF_1^{\nu(\bar\nu)}(x,Q^2)+
\left(1-y-y\frac{xm_N}{2E_\nu}\right)F_2^{\nu(\bar\nu)}(x,Q^2)\right.
\nonumber\\ &\pm&\left.
\left(y-{y^2\over2}\right)xF_3^{\nu(\bar\nu)}(x,Q^2)\right],
\label{5.420}
\eeqn
where the sign $\pm$ corresponds to either neutrino or antineutrino beams, respectively.

The new interesting features comes with the structure function $F_3$, which in the leading order carries information about the number of valence quarks $N_v=3$ in the nucleon. Namely, measuring the difference between the neutrino and antineutrino cross sections one arrives at a relation given by  the Gross-Llewellyn Smith sum rule \cite{gls,handbook},
\beq
{1\over2}\int\limits_0^1 {dx\over x}\left( xF_3^{\nu N}+xF_3^{\bar\nu N}\right)=
N_v\left[1-\frac{\alpha_s(Q^2)}{\pi}-\frac{C}{Q^2}+{\cal O}(Q^{-4})\right]
\label{5.440}
\eeq

The current value of the left-hand side of Eq.~(\ref{5.440}) measured in the CCFR experiment \cite{CCFR1,CCFR2} is
$2.5\pm 0.18\pm 0.078$, which is 1.8 standard deviations below the expected value \cite{handbook}. Notice, however, that the CCFR data were taken on an iron target, therefore the structure function $F_3^A$ is subject to nuclear shadowing, reducing its value.

\subsection{Shadowing: dipole description}
\label{sec:shad-dipole}

As usual, the dipole representation is the most effective tool for prediction of nuclear shadowing.
The PDFs describing light and heavy flavors at small $x$ in the nucleon correspond to fluctuations of the $W$-boson into quark-antiquark dipoles, which interact with the target. The Cabibbo enhanced  transitions are to light quark dipoles, or strange-charm pairs. The perturbative light cone distribution functions, compared to Eqs.~(\ref{4.210})-(\ref{4.220}),
also include the axial parts, and read \cite{zoller},
\beqn
\left|\Psi^T_{\bar qQ}(\alpha,r_T)\right|^2&=& \frac{\sqrt{2}G_FN_c}{(2\pi)^3}(m_W^2+Q^2)
\left\{\left[(1-\alpha)^2m^2+\alpha^2\mu^2\right]
K_0^2(\tilde\epsilon r_T)
\right.\nonumber\\ &+& \left.
\left[\alpha^2+(1-\alpha)^2\right]\tilde\epsilon^2\,K_1^2(\tilde\epsilon r_T)\right\}
\label{5.460}
\eeqn
\beqn
\left|\Psi^L_{\bar qQ}(\alpha,r_T)\right|^2&=& \frac{G_FN_c}{\sqrt{2}(2\pi)^3}
\left(1+\frac{m_W^2}{Q^2}\right)
\left\{\left(\left[2\alpha(1-\alpha)Q^2+(m_Q-m_q)\left[(1-\alpha)m_Q-\alpha m_q\right]\right]^2
\right.\right.\nonumber\\ &+& \left.\left.
\left[2\alpha(1-\alpha)Q^2+(m_Q+m_q)\left[(1-\alpha)m_Q+\alpha m_q\right]\right]^2
\right)K_0^2(\tilde\epsilon r_T) \right.
\nonumber\\ &+&
2(m_Q^2+m_q^2)\tilde\epsilon^2 K_1^2(\tilde\epsilon r_T)\Biggr\},
\label{5.470}
\eeqn
where
\beq
\tilde\epsilon^2=\alpha(1-\alpha)Q^2+\alpha m_q^2+(1-\alpha)m_Q^2.
\label{5.480}
\eeq

Using these distribution functions and the eikonal form (Eq.~(\ref{1.590})) of the nuclear amplitude, the shadowing effects for neutrino-nucleus interactions were calculated in \cite{zoller,mochado}. The results of \cite{mochado} for 
$^{56}Fe$ are depicted in Figs.~\ref{fig:F2} and \ref{fig:F3} for the nuclear ratios of $F_2$ and $xF_3$ respectively, plotted vs $Q^2$ at fixed Bjorken $x$.
\begin{figure}[htb]
\parbox{\halftext}{
\centerline{\includegraphics[width=7.5 cm]{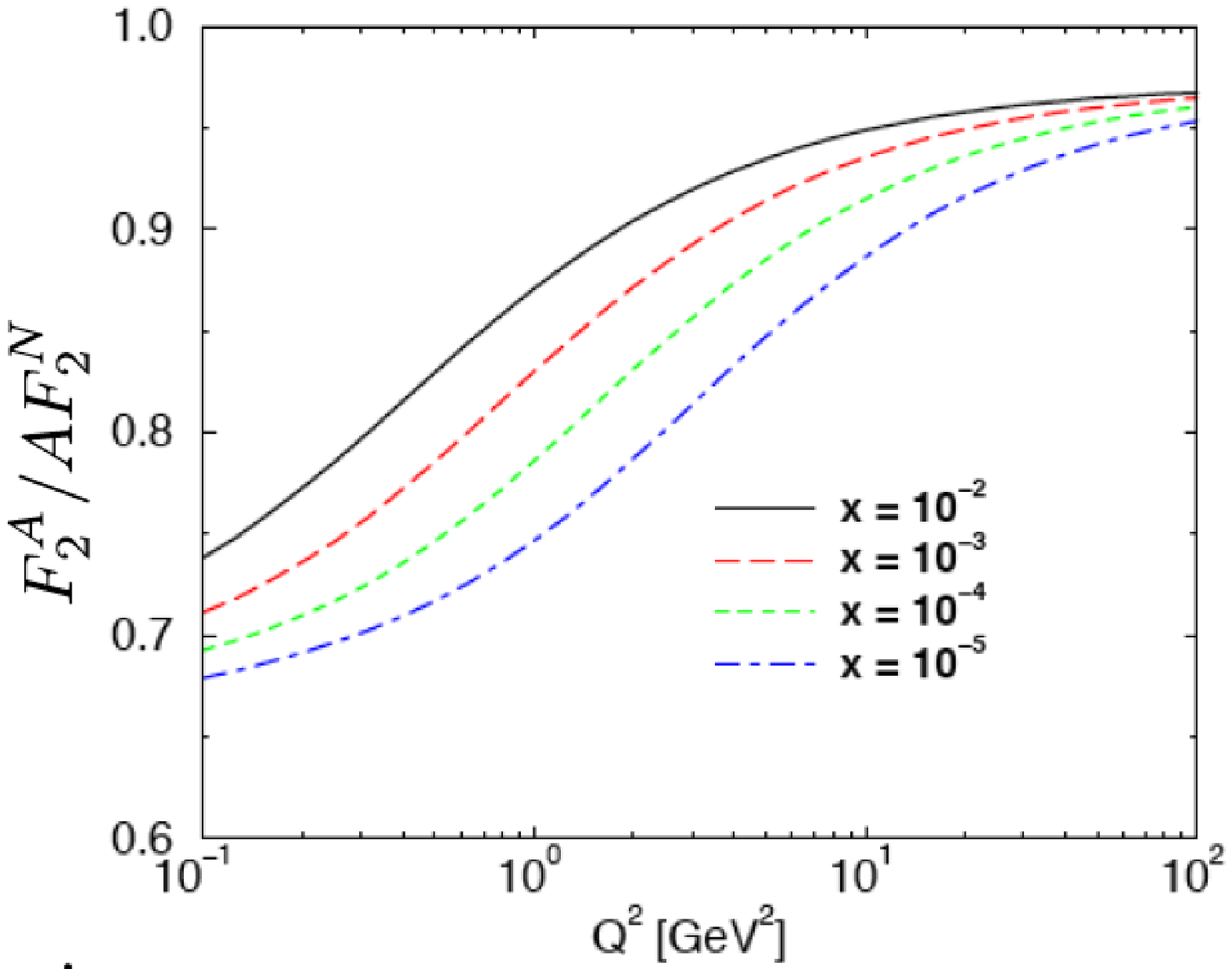}}
 \caption{The nuclear ratio $F_2^A /AF_2$ as function of  $Q^2$ at fixed $x$. }
 \label{fig:F2}}
\hfill
\parbox{\halftext}{
\centerline{\includegraphics[width=7.5 cm]{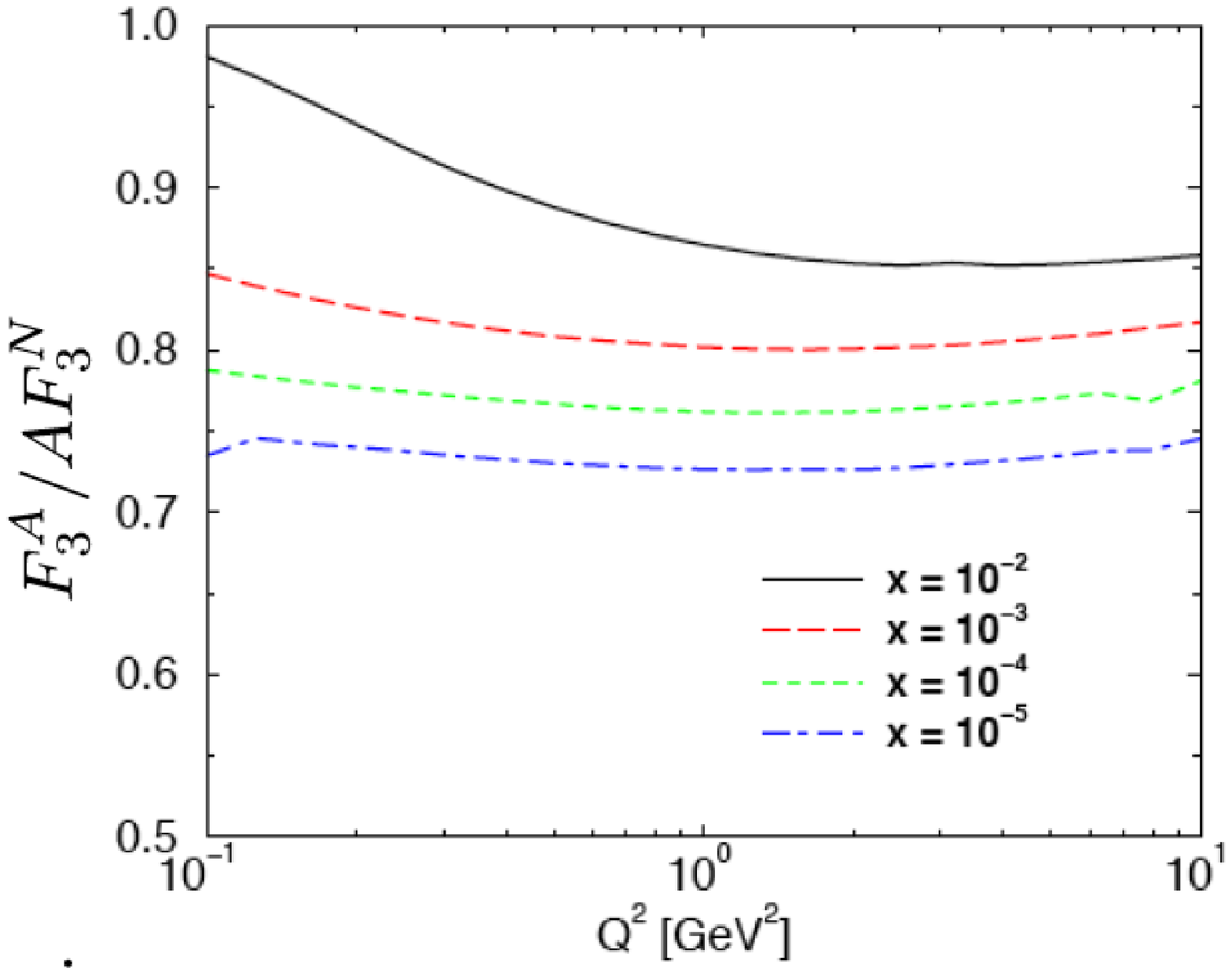}}
\caption{The nuclear ratio $xF_3^A /AxF_3$ as function of  $Q^2$ at fixed $x$.
}
\label{fig:F3}}
 \end{figure}
 Nuclear shadowing for $F_3$ demonstrates a remarkably weak scale dependence.

Unfortunately, none of these calculations is sufficiently realistic to be compared with data:
(i) At small $x<10^{-2}$ gluon shadowing provides substantial corrections. It reduces the cross section of neutrino interaction on bound nucleons, and at the same time make the nuclear medium more transparent. 
(ii) At larger $x\gsim10^{-2}$ gluon shadowing can be neglected (see Sect.~\ref{sec:glue-shad}), however the "frozen"
dipole size approximation employed in the above calculations is not valid, and should be replaced by the path integral technique (see Sect.~\ref{sec:path}). (iii) At low $Q^2\lesssim1\GeV^2$ the perturbative distribution functions (\ref{5.460})-(\ref{5.470}) are not to be trusted. Nonperturbative distribution functions were derived for vector \cite{marat1,marat2}
and axial \cite{marat3,marat4} currents in the instanton vacuum model.

Notice that the expected magnitude of nuclear suppression, $xF_3^A/AxF_3^N\approx 0.85$ explains well the observed deviation from the Gross-Llewellyn Smith sum rule, Eq.~(\ref{5.440}).
Power and quark mass corrections to this sum rule were discussed and evaluated in \cite{qiu-vitev}.
The dipole description is an alternative way of calculation, which is supposed to contain all such terms.

\part{Experimental Study of Shadowing in Lepton-Nucleus Scattering: Recent Developments}

\section{Introduction}
The experimental study of shadowing and anti-shadowing in lepton nucleus scattering has been dominated by charged-lepton nucleus scattering.  $\mu$-A scattering at CERN and Fermilab and e-A scattering at SLAC provided early evidence that the value of F$_2(x,Q^2)$ per nucleon in nuclei was measurably different than F$_2(x,Q^2)$ measured in deuterium in the low-x shadowing region. Most of these experiments were performed in the '80s and '90s, and there have been only a few such experiments, notably the HERMES experiment~\cite{Airapetian:2002fx}, since the shutdown of muon DIS experiments at CERN and Fermilab and e-A experiments at SLAC.  Several experiments at Jefferson Laboratory studied mid-to-high x nuclear effects in more detail, but did not contribute significantly to the study of shadowing.  With the upcoming availability of the Jefferson Laboratory 12 GeV program offering a wider kinematic range, it is expected that charged-lepton nucleus studies of shadowing and anti-shadowing will begin again.

On the other hand, studies of nuclear effects with neutrinos have only recently become possible with high-statistics DIS experiments that also provide detailed covariant error matrices, allowing the inclusion of correlated errors into the analysis.  Neutrino nucleus scattering introduces a new facet in the study of shadowing (and all x-dependent nuclear effects) with the presence of the axial-vector current.  As shown in the theoretical treatment of shadowing, differences between $\ell A$ and $\nu A$ scattering in the shadowing region are expected.  More detailed and varied programatic studies of nuclear effects with neutrinos are now being initiated with the availability of intense neutrino beams designed and commissioned for neutrino oscillation measurements.  The need for this study is further reinforced by the necessity of understanding these nuclear effects in detail, to better estimate the systematic errors on measured neutrino oscillation parameters.  These new studies of neutrino nucleus scattering will dominate this part of the review.

\section {Charged-lepton Nucleus scattering}
\label{sec-charged}
The present understanding of nuclear effects in lepton-nucleus scattering is mainly based on charged-lepton--nucleus ($\ell A$) DIS data. In the early 80s, the European Muon Collaboration (EMC) \cite{Aubert:1983xm} found that the per-nucleon structure functions $F_{2}$ for iron and deuterium differ as a function of x$_{Bjorken} \equiv x$. This intriguing result initiated a series
 of follow-up  experiments ~\cite{Bodek:1983qn, Bodek:1983ec, Arnold:1983mw, Dasu:1988ru, Bari:1985ga, Benvenuti:1987az, Ashman:1988bf, Arneodo:1989sy, Amaudruz:1991cc, nmcold2, Amaudruz:1992wn, Adams:1992vm, Adams:1992nf, Adams:1995is, Gomez:1993ri} 
to investigate the nuclear modifications of this ratio, $R[F_{2}^{\ell A}] =F_{2}^{\ell A} /(A\, F_{2}^{\ell N})$, over a wide range of nuclear targets with atomic number $A$.
These experiments established that when scattering off nucleons within a nucleus in the deep-inelastic region with
$Q^{2}\geq1\ {\rm GeV}^{2}$, the ratio of cross section per nucleon in nuclei to that in deuterium varies considerably in the kinematic range from relatively small Bjorken $x$ $\sim10^{-2}$ to large $x$ $\sim0.8$.  The behavior of the ratio $R[F_{2}^{\ell A}]$
can be divided into four regions: 

\begin{itemize}
\item the shadowing region - R $\leq$ 1 for x $\lesssim$ 0.1,
\item the antishadowing region - R $\geq$ 1 for 0.1 $\lesssim$ x $\lesssim$ 0.25 ,
\item the EMC effect - R $ \leq$ 1 for 0.25 $\lesssim$ x $\lesssim$ 0.7,
\item and the Fermi motion region - R $\geq$ 1 for x $\gtrsim$ 0.7.
\end{itemize}

There is no single inclusive model  explaining the nuclear modifications across the whole x region.
The shadowing suppression at small $x$ is the topic of this review, and the previous sections have summarized the current status of the theoretical understanding of shadowing.
The anti-shadowing region is theoretically less well understood but, as indicated, might be explained by
the application of momentum, charge, and/or baryon number sum rules.  
The modifications at medium $x$ (the so-called {}``EMC effect") are still lacking a convincing, community-accepted explanation, but are usually
described as nuclear binding and medium effects \cite{Geesaman:1995yd}.  It has also been shown~\cite{Arrington:2003} that the EMC effect persists at lower Q and W in the resonance/transition region.
The rapid rise of the ratio at large $x$ is attributed to the Fermi motion of the nucleons within the nucleus.

Notice that the currently available experimental data from charged lepton-nucleus DIS scattering have been obtained with fixed target experiments, where the values of $x$ and $Q^2$ are strongly correlated.  With the additional requirement that $Q^2$ be large enough in order to apply perturbative QCD (e.g.,~$Q^2 > 1$ GeV$^2$),  the majority of available experimental data covers a relatively limited region in x, where the effect of nuclear shadowing is far from saturation.  The review by  Geesaman, Saito and Thomas~\cite{Geesaman:1995yd}  summarized the experimental situation in 1995, and  since then there has been limited amounts of new data on shadowing from charged-lepton nucleus scattering in the DIS region.  
Although limited recent data from charged-lepton nucleus scattering on shadowing are available, recent analyses of the shadowing region to extract nuclear parton distributions have been performed.  Current work on this topic will be covered in Section~\ref{sec-npdf}

With these qualifications, the evidence for nuclear effects in charged-lepton nucleus scattering can be summarized in Figure~\ref{fig:slac}, which displays the $F_{2}^{Fe}/F_{2}^{D}$ structure function ratio, as measured by both the SLAC eA and the BCDMS $\mu$A collaborations. The SLAC/NMC curve is the result of an A-independent parametrization fit to calcium and iron charged-lepton DIS data \cite{Bodek:1983qn,Bari:1985ga,Benvenuti:1987az,Gomez:1993ri,Dasu:1993vk,Owens:2007kp}. This has often been used as a standard nuclear correction factor to convert data from a nuclear target to a free-nucleon target, for both charged-lepton and neutrino interactions. Although not displayed in this figure, there was also  available experimental evidence for shadowing that is shown in the following figures.  Notice also in those figures that, particularly in the shadowing region, a single A-independent curve does not reflect the experimental situation.


%

\begin{figure}[h]
\begin{center}
\includegraphics[width=0.45\textwidth]{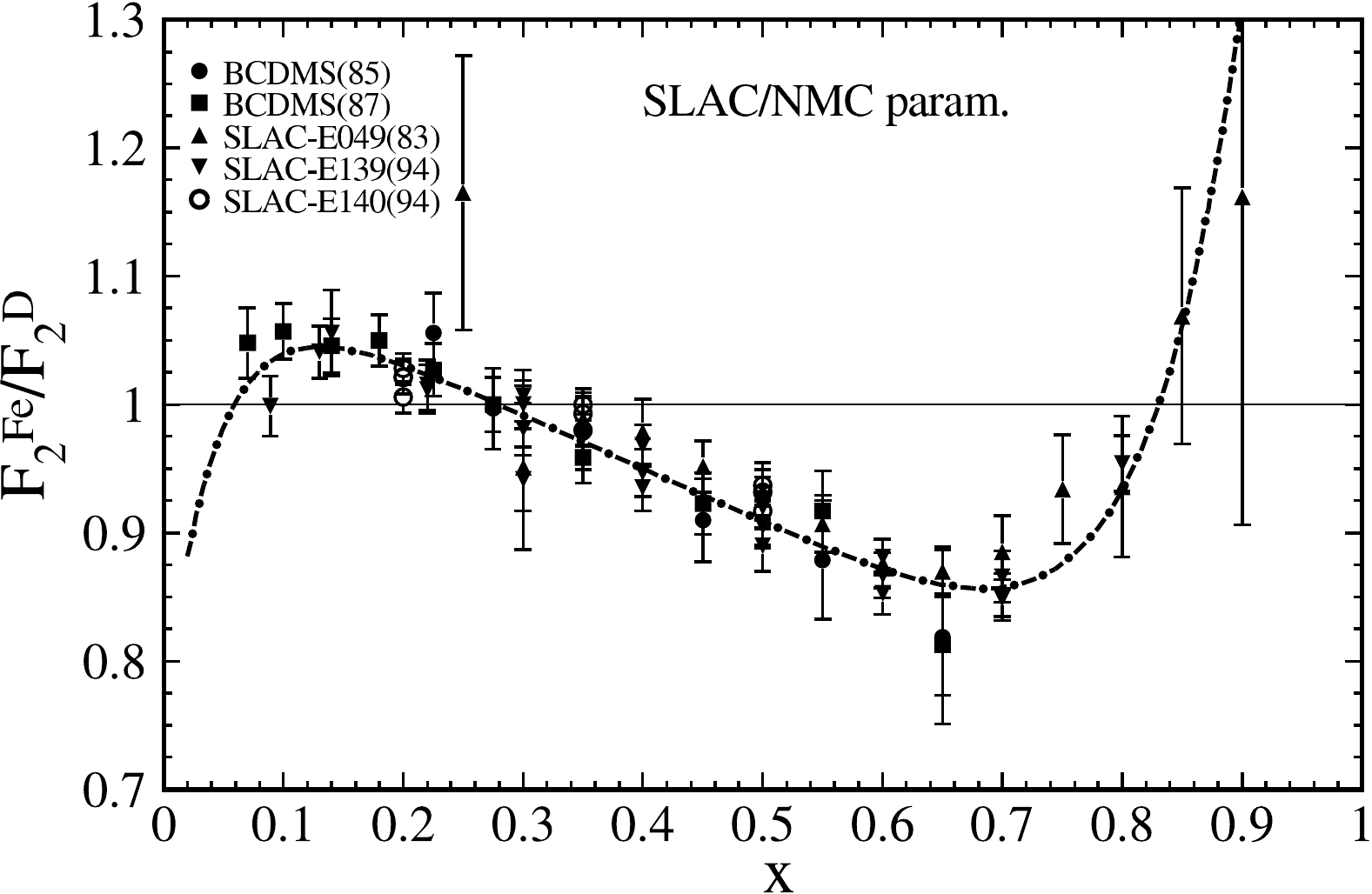}
\caption{Nuclear correction factor, $F_{2}^{Fe}/F_{2}^{D}$, as a function of
$x$. The parameterized curve is compared to SLAC and BCDMS data \cite{Bodek:1983qn,Bari:1985ga,Benvenuti:1987az,Gomez:1993ri,Dasu:1993vk}.  The details of the shadowing regime are shown in the following Figure ~\ref{fig:gst-3} and Figure ~\ref{fig:gst-4} for other nuclei since Fe data in the shadowing region did not survive other applied cuts.}
\label{fig:slac}
\end{center}
\end{figure}

A more detailed look at shadowing experimental results, for specific nuclei such as C, N, Ca, Ag, Sn and Xe, is shown in Figures~\ref{fig:gst-3} and ~\ref{fig:gst-4}.  These data have been used in the comparison with the parameter-free calculations shown in Figure~\ref{fig:e665} and also appear in Figure~\ref{fig:scaling}.  Even for deuterium itself, data from $\mu$-nucleus scattering in the experiments E665 at Fermilab and NMC at CERN show clear evidence of shadowing in the ratio of deuterium to hydrogen,~Figure~\ref{fig:gst-12} .

\begin{figure}[h]
\begin{center}
\includegraphics[width=0.45\textwidth]{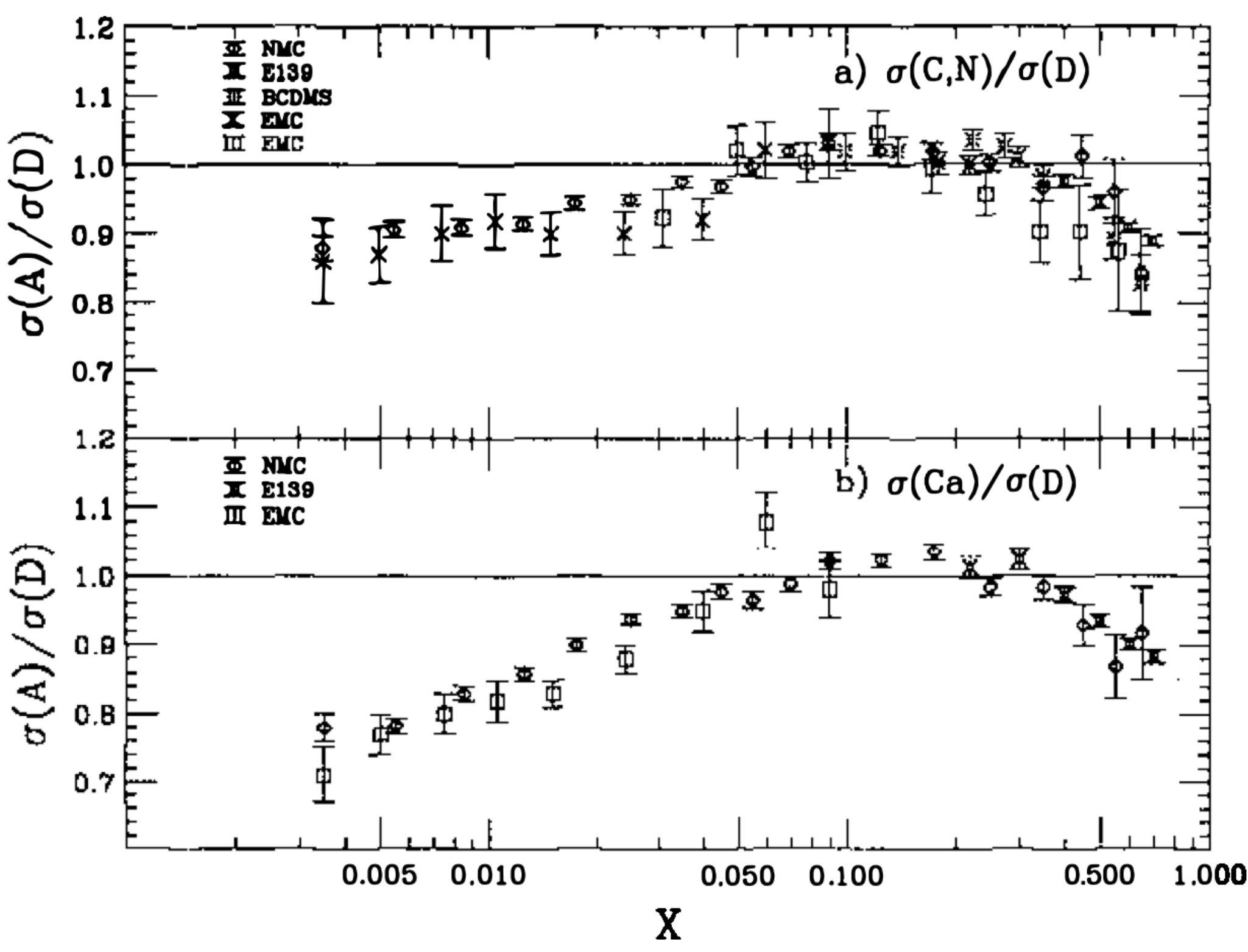}

\caption{Ratios of the deep inelastic cross section on targets of (a) carbon and nitrogen and (b) calcium to those of deuterium~\cite{Gomez:1993ri,Aubert:1983xm,Amaudruz:1991cc,nmc}}.
\label{fig:gst-3}
\end{center}
\end{figure}

\begin{figure}[h]
\begin{center}
\includegraphics[width=0.45\textwidth]{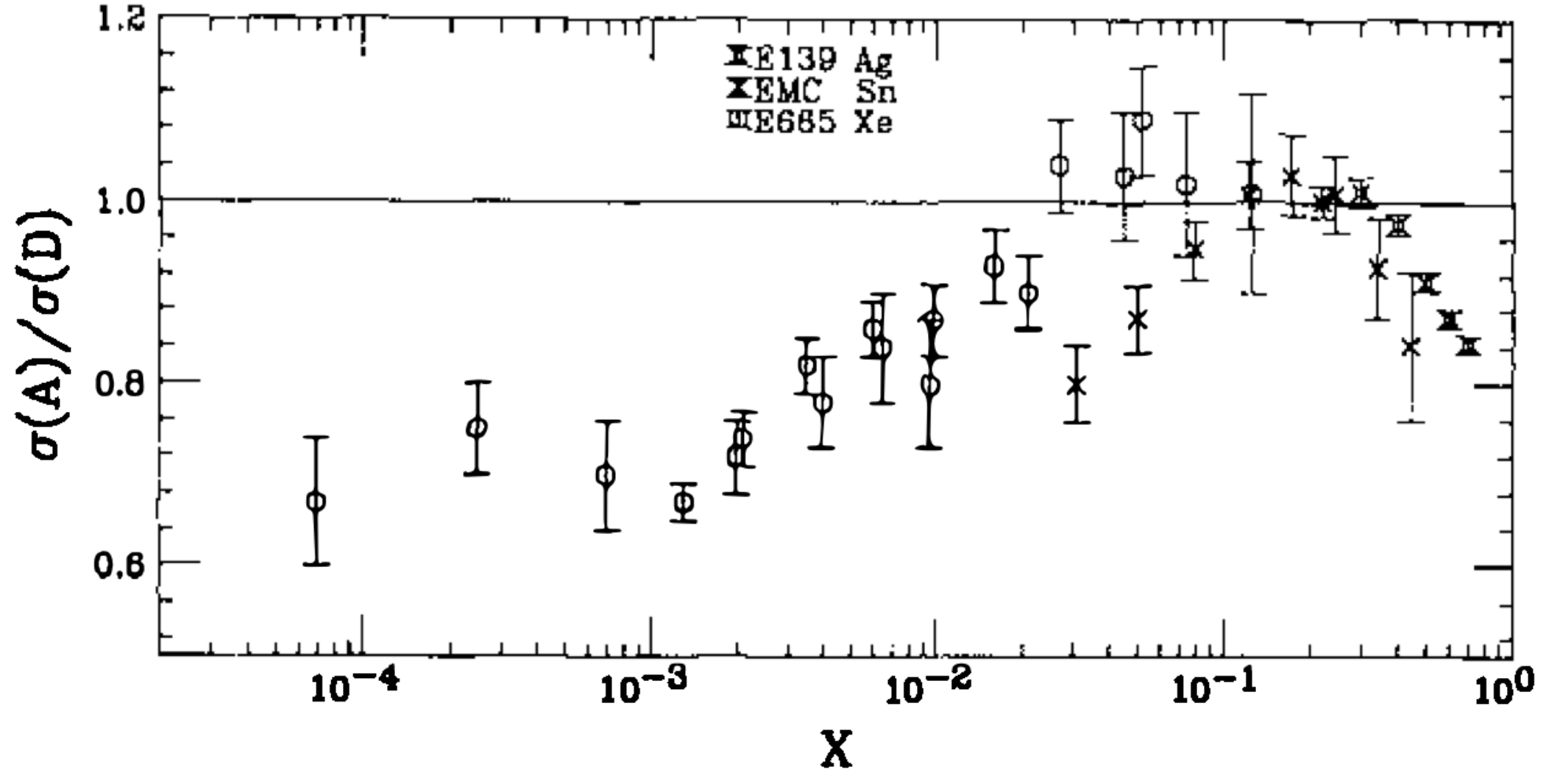}

\caption{Ratios of the deep-inelastic cross section on targets of tin,  xenon and silver to those of deuterium~\cite{Gomez:1993ri,Aubert:1983xm,Adams:1995is}}.
\label{fig:gst-4}
\end{center}
\end{figure}

\begin{figure}[h]
\begin{center}
\includegraphics[width=0.45\textwidth]{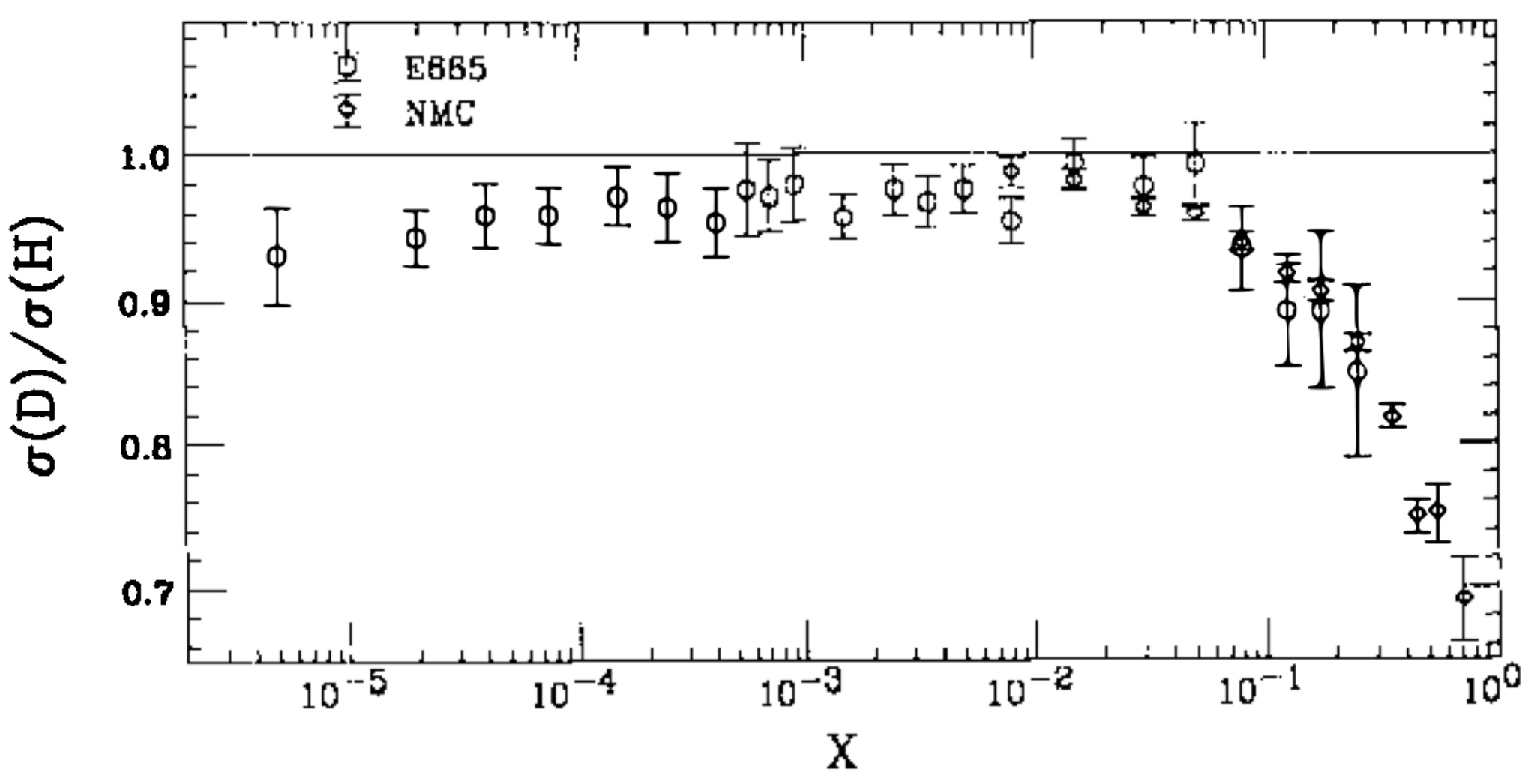}

\caption{Nuclear correction ratio, $F_{2}^{D}/F_{2}^{H}$, as a function of $x$~\cite{nmc,Adams:1995is}.}
\label{fig:gst-12}
\end{center}
\end{figure}

All experimental results in the shadowing region involving charged-lepton nucleus scattering are consistent with the onset of shadowing occurring at a value of the Ioffe time (\ref{3.160}) roughly equivalent to the separation between nucleons in a given nucleus.  Shadowing increases with increasing A and decreases with increasing $Q^2$.  The $Q^2$-dependence in the shadowing region, a small but non-zero slope in log $Q^2$, as seen by the NMC experiment, is consistent with descriptions~\cite{Melnitchouk:1995am} of shadowing prevalent some years ago, of low-$Q^2$ scattering being ascribed to vector meson dominance, as illustrated in Figure~\ref{fig:f2sncq}.  An understanding of the role of the longitudinal cross section in shadowing has been provided by the HERMES experiment~\cite{Airapetian:2002fx} that examined the ratio of $R_A/$ $R_D$ with R = $\sigma_L$ / $\sigma_T$ the longitudinal to transverse DIS cross sections.  They found that $R_A/$ $R_D$ is consistent with 1.0 for A up to $^{84}Kr$ and the kinematic range down to x = 0.01 and $Q^2$ = 0.5 $GeV^2$.  Furthermore, the HERMES data were quite consistent with earlier measurements by the NMC collaboration~\cite{Amaudruz:1992wn,Arneodo:1996rv}  accumulated at  considerably higher energies.  No significant  $Q^2$-dependence was observed over the wide $Q^2$ range covered by the combined HERMES and NMC experiments.


\section{Neutrino Nucleus Scattering}
The measurement of neutrino deeply inelastic scattering, besides providing valuable information on the weak current, can be a significant aid in determining the parton distribution functions (PDFs).  Neutrino and antineutrino DIS data directly measure specific quark flavors including the strange quark and strange anti-quark distributions.  Neutrino scattering provides the only current input to determine s and $\overline{s}$ in the proton via the strange-to-charm quark transition and the subsequent detection of a charmed particle decay.  Combining neutrino Charged Current (CC) $W^{\pm}$ interactions  with other probes, such as charged lepton  $(\gamma,Z)$, allows us to disentangle the separate PDF flavor components.

While for charged-lepton scattering the use of a nuclear target is a choice,  for high statistics neutrino scattering experiments it is a necessity, due to the weak nature of the $\{W^{\pm},Z\}$ interactions.  This complicates the interpretation of neutrino scattering data,  since these data are used for the extraction of free nucleon PDFs, which depend on nuclear correction factors to account for the nuclear effects of the nucleon being bound in a nucleus.   

As in charged-lepton nucleus scattering,  these nuclear correction factors essentially convert the experimentally-measured structure functions in a bound nucleon within a nucleus to the corresponding structure functions within a free nucleon.  For charged-lepton nucleus scattering these factors are usually given in terms of curves, such as in Figure~\ref{fig:slac}, measured for a given nucleus at the relevant Q.   In earlier QCD global fits that attempted to include neutrino DIS data, the charged-lepton nuclear correction factors were simply applied to neutrino nucleus scattering results.  From the theoretical side it was thought that these  factors would be the same for neutrino nucleus scattering as for charged-lepton nucleus scattering, with the possible exception of the shadowing/anti-shadowing region.  

Early attempts to include neutrino-nucleus DIS scattering data, corrected with charged-lepton nuclear correction factors, introduced such tension in the shadowing region at low-x in global QCD fits (described below) that the low-x neutrino data was excluded early-on in CTEQ nucleon PDF global analyses.  In a more recent examination of high-x parton distribution functions, carried out by the CTEQ collaboration~\cite{Owens:2007kp} and  \cite{Owens:2007zz}, indications began to accumulate that the nuclear correction factors for neutrino nucleus scattering could indeed be different than those for charged-lepton nucleus scattering, and {\em not only in the shadowing region}.  

Although we will be concentrating on the shadowing region, in this section we will describe current studies of nuclear effects in neutrino nucleus interactions across the full x range, since the behavior of the cross section as it approaches the shadowing region from higher x is important. Using recent neutrino-Fe and neutrino-Pb scattering data to directly extract the neutrino nucleus nuclear correction factors, we will compare them to those extracted from charged lepton scattering, as well as to specific model-dependent corrections provided by Kulagin and Petti~\cite{kp}.  The limitations of the Kulagin-Petti model have been emphasized in section~\ref{sec:DIS}.  However, it is the only ``model" that has been available for $\nu$ nucleus scattering, and therefore it has been heavily used by experimental and phenomenological analyses.  Finally, we will concentrate on what neutrino scattering data off nuclei is telling us about shadowing and anti-shadowing, in interactions involving the axial-vector current.

\subsection{Early Neutrino Nucleus Scattering Results}

Soon after the discovery that muon interactions off bound high-A nuclei exhibited a different x dependence than muon interactions off deuterium, this observation was tested with neutrino nucleus interactions.  The earliest tests  were dominated by bubble chamber results~\cite{Parker:1983yi,CooperSarkar:1984eb,Hanlon:1985yg,Kitagaki:1988wc,Guy:1989iz},  with one notable exception involving the placement of a hydrogen target upstream of the CDHS (mostly) iron detector~\cite{Abramowicz:1984yk}.

None of the bubble chamber experiments was able to find a statistically significant effect such as that found by the EMC collaboration.   The combined CERN WA25 and WA59 analysis of $\nu$ and $\overline{\nu}$ interactions off deuterium and neon ~\cite{CooperSarkar:1984eb} concluded that their results were consistent with the EMC higher $Q^2$ data for the x range between 0.3 and 0.6, and consistent with the SLAC lower $Q^2$  data~\cite{Arnold:1983mw,Bodek:1983ec} for all x. They also concluded that there was no $Q^2$-dependence in the low-x shadowing region, in the range of 0.25 to 14 $GeV^2$.  However, within the shadowing region they did see a depletion in the ratio of Ne to D, as indicated earlier with Figure~\ref{fig:shad-1}.  Nevertheless, the statistics are limited and the claim of an overall normalization uncertainty of 1\% between the two independent exposures of WA25 and WA59, based on the comparison of ``muon" distributions in the alcoves, may be underestimated based on more recent understanding of the signal vs background in these alcove measurements. The main challenge of these early $\nu$ A experiments was the accumulation of sufficient statistics with bubble chambers to make significant comparisons between cross sections of different nuclei.  An additional difficulty in the comparison across experiments was the $\geq$ 20\% uncertainty on the flux of incoming neutrinos for each independent experiment.  As mentioned above, even attempts at comparison of experiments in the same beam were hampered by uncertainties in signal to noise in the measurements of muon alcove intensities used for normalization.

A procedure was developed by the Fermilab E-745 heavy liquid bubble chamber collaboration\cite{Kitagaki:1988wc} to use the bubble-chamber's ability to isolate nucleon recoils (``dark tracks") as indication of interactions off tightly-bound nuclei.  By taking the ratio of events with and without these dark tracks, the collaboration found an x-dependence of this ratio similar to that found by the earlier-referenced EMC collaboration.  However, a follow-up analysis by a BEBC collaboration ~\cite{Guy:1989iz} showed that one could get a similar x-dependent ratio distribution by applying the same criteria to neutrino scattering off both deuterium and hydrogen, thus suggesting this criteria was not indicative of scattering off a tightly bound nucleus.

Although the conclusion of the CDHS study of the ratio of iron to hydrogen cross sections and structure functions was that there were no statistically significant differences indicative of nuclear binding effects in iron, their results, figure 1a of their publication~\cite{Abramowicz:1984yk}, certainly follow the trend displayed in the more statistically powerful CERN muon and SLAC electron results.



\subsection {Recent Neutrino Nucleus Scattering Data: The ratio of $\nu$-A  to $\nu$-Nucleon Cross sections}
The more contemporary study of  $\nu$ nucleus scattering using high-statistics experimental results with careful attention to multiple systematic errors began with the CDHSW, CCFR/NuTeV $\nu$ Fe, the NOMAD $\nu$ C and the CHORUS $\nu$ Pb experiments.  Whereas the NuTeV \cite{nutev} and CHORUS \cite{chorus} Collaborations have published their full data sets, NOMAD \cite{Altegoer:1997gv} has not yet done so.  Although NOMAD displayed a small part of their preliminary data in a 2004 publication  \cite{Petti:2004wy}, using their data would require extracting their measured points with errors from figures, an inaccurate procedure at best.   The more detailed analyses presented in the following will be made with mainly the CDHSW, CCFR and NuTeV $\nu$ Fe results and the CHORUS $\nu$ Pb results.

The most direct comparison between scattering neutrinos off nuclei and off nucleons is a comparison of the cross sections.  This involves none of the assumptions that will be needed to extract structure functions from these cross sections nor the procedures needed for global fits. Members of the CTEQ collaboration have studied this direct cross section comparison~\cite{Owens:2007kp, Owens:2007zz}.
The eventual goal of this study was to extract the nuclear correction factors from  $\nu$ nucleus scattering, in order to compare them with the similar nuclear correction factors from charged-lepton nucleus scattering, and to use them in incorporating $\nu$ nucleus data in global analyses.   This, in turn, requires taking a ratio of the measured $\nu$ A cross sections to the corresponding {}``free-nucleon" quantities.  For the charged-lepton nuclear correction factors, the measured per-nucleon cross sections or structure functions off nucleus A were compared directly to measured charged-lepton scattering results off deuterium.  For neutrino scattering, a similar ratio using scattering off deuterium is impossible, since there exist only low-statistics bubble chamber data (see~\cite{383021} and references therein).   The statistical and systematic accuracy of these early bubble chamber neutrino deuterium results is not sufficient for determining a neutrino nuclear correction factor. 

\paragraph{Determining {}``free" Cross sections and Structure Functions for Comparisons}
To address this, the nCTEQ collaboration decided to approximate the deuterium denominator in the neutrino correction factors by creating ``free" neutrons and protons from PDFs determined using a new reference fit that involved data off mainly hydrogen and deuterium targets.  
This reference fit starts with the CTEQ6.1M PDFs~\cite{CTEQ}
and removes any neutrino nucleus data, keeping data involving proton and deuterium targets.   Nuclear corrections for deuteron targets were included where appropriate. Details on the treatment of heavy quarks, which primarily affects the low-$x$ region for neutrino induced processes, and therefore are significant when examining the shadowing region, as well as high-x corrections, can be found in reference \cite{Owens:2007kp}.  The resulting data for the reference came from the BCDMS Collaboration \cite{bcdms} for $F_2^p {\ \rm and \ } F_2^d$, from the NMC Collaboration \cite{nmc1} for $F_2^p {\ \rm and \ } F_2^d/F_2^p$, from H1~\cite{H1} and ZEUS~\cite{ZEUS} for $F_2^P$, from CDF \cite{cdf} and D\O\ \cite{d0} for inclusive jet production, from CDF \cite{cdf_w} for the $W$ lepton asymmetry, from E-866 \cite{e866_ratio} for the ratio of lepton pair cross sections for $pd$ and $pp$ interactions, and from E-605 \cite{e605} for dimuon production in $pN$ interactions\footnote{Note that the E-605 data were taken on a copper target, but the nuclear corrections in the relevant kinematic region, outside the shadowing region, have been measured \cite{dimuon_nuc} to be consistent with $A^1$, so no nuclear corrections were included for these data}.

\paragraph{The CTEQ determination of the ratio of $\nu$-Fe and $\nu$-Pb to $\nu$-Nucleon Cross sections}
In their analysis of $\nu A$ scattering data, the CTEQ group used two of the recent sets of experimental results mentioned; the CHORUS $\nu$ Pb and the NuTeV $\nu$ Fe data sets, examining the behavior of the as measured cross sections displayed as a ratio to the ``free-nucleon" cross section.  Figure~\ref{no_shift_no_cor} shows the results with only statistical errors displayed on the measured points.  In general, there is agreement between the different data sets, which is somewhat surprising. Aside from a few low-statistics CHORUS $\overline \nu$ data points, in the approach to the shadowing region from high-x (0.6 to 0.08) there is quite good agreement between  $\nu$ and $\overline \nu$ and between Fe and Pb.   

The CHORUS $\overline \nu$ Pb data displays a very {}``traditional" x-behavior with (from high- to low-x) an antishadowing region starting at x = 0.3, with a maximum and turnover toward the shadowing region at x $\sim$ 0.15.  On the other hand, the CHORUS $\nu$ Pb data, with significantly smaller errors, have a much smaller antishadowing region shifted toward lower-x. 
Similarly, in approaching the shadowing region from high-x, the NuTeV $\overline \nu$ data follow the CHORUS $\nu$ Pb data closely.  The NuTeV $\nu$ data, again having much smaller errors than the  $\overline \nu$ data, are again consistent with both the CHORUS $\nu$ Pb and the NuTeV $\overline \nu$ data in the approach to the shadowing region.

\begin{figure}
\begin{center}
\includegraphics[height=3 in]{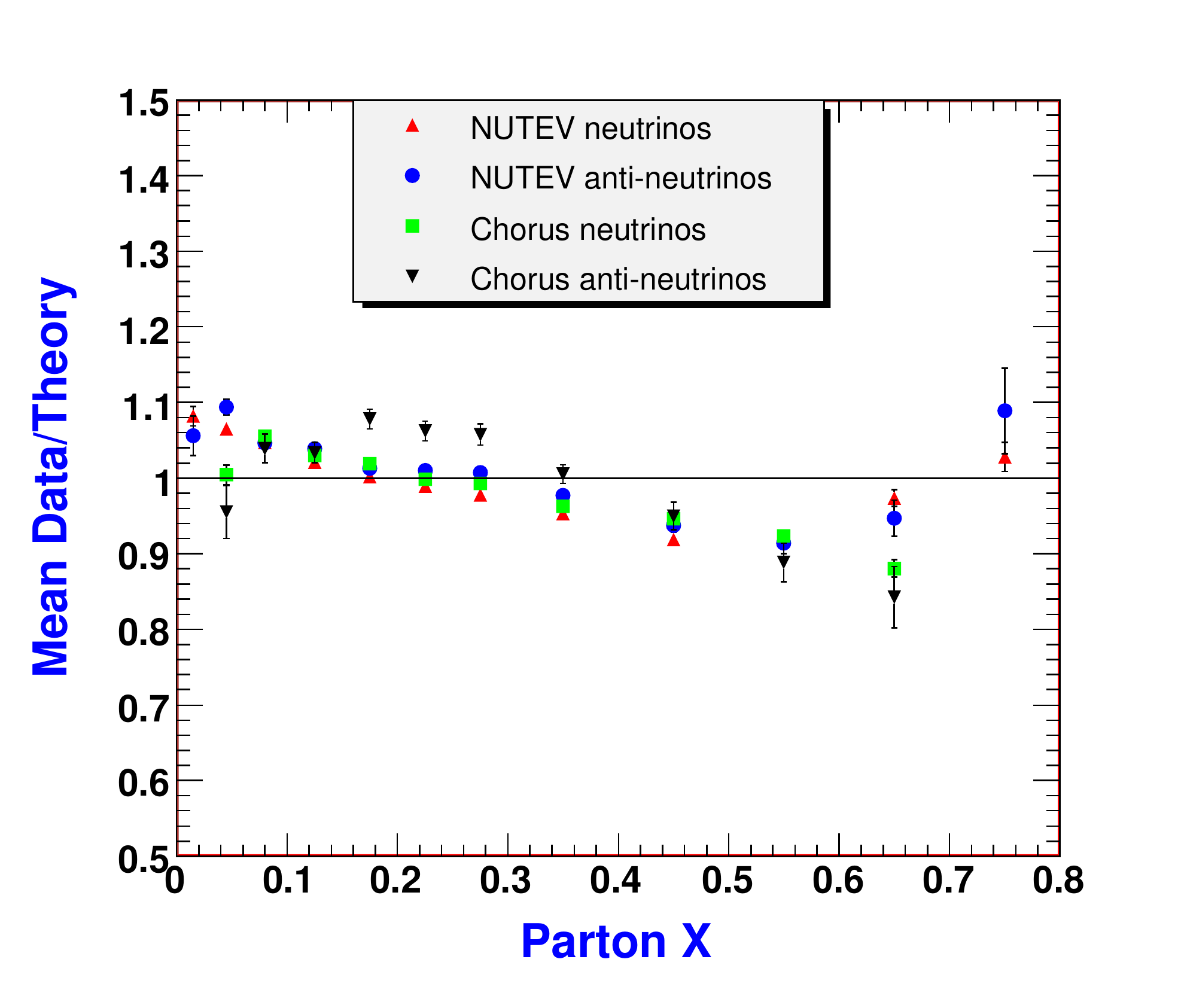}
\caption{Comparison between the as measured Chorus and
NuTeV neutrino cross section data off nuclei, without any nuclear corrections, and the reference free-nucleon fit ({}``Theory").}
\label{no_shift_no_cor}
\end{center}
\end{figure}

However, in the low-x shadowing region the results are quite different, as shown in Figure~\ref{Detail5-lowx} (low-x details of Figure~\ref{no_shift_no_cor}).  Only the lower-statistics CHORUS  $ \nu$ and $ \overline \nu$ cross section ratios turn over and approach the expected R $\leq$1.0 axis.  From the same figure, the NuTeV  data sets show a somewhat different behavior.  Only at the very lowest x = 0.015 point (with the largest error) does the NuTeV $\overline \nu$ data turn over toward the shadowing region.  The considerably more accurate NuTeV $ \nu$ data {\em does not exhibit a shadowing turn-over even at the lowest x = 0.015 point}.   
 
\begin{figure}
\begin{center}
\includegraphics[height=2.5 in]{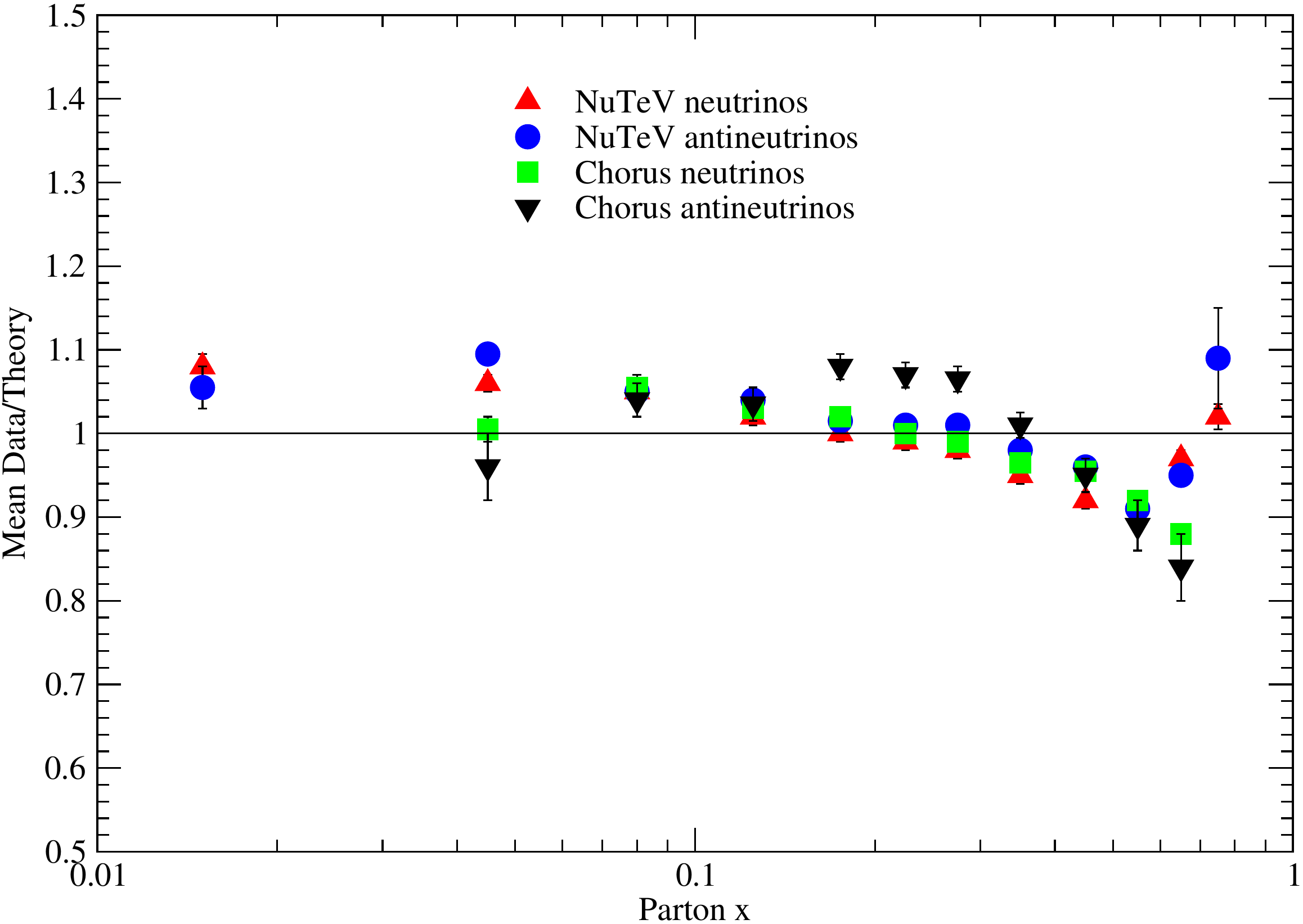}
\caption{Low-x details of the ratio of the Chorus and NuTeV measured cross sections, without any nuclear corrections, to the ``free-nucleon" reference cross section.}
\label{Detail5-lowx}
\end{center}
\end{figure}

It is reasonable to ask if this observed behavior of $\nu$ and $\overline \nu$ off Fe and Pb is consistent with expectations.  One of the few models available that has been specifically designed to predict the $\nu$ A nuclear corrections and is widely used by the experimental neutrino community is the model of S. Kulagin and R. Petti (K-P) \cite{Kulagin:2004ie, Kulagin:2007ha, Kulagin:2007ju}.  Their approach includes a QCD description of the nucleon structure functions at high x, as well as the treatment of Fermi motion and nuclear binding, off-shell correction to bound nucleon structure functions, nuclear pion excess and nuclear shadowing.  Their model involves several mechanisms in specific regions of  x$_{Bj}$, to enable  covering the entire x$_{Bj}$ range.  This model has limitations, particularly in the shadowing region, emphasized in~\ref{sec:DIS}. However, it predicts a shadowing turnover that is not seen in the NuTeV Fe data.  How significant is this deviation of the NuTeV measurements from expectations can be determined by carefully considering the errors of the measured points.

A way of understanding how flexible the CHORUS and NuTeV measured points are within their quoted errors is to apply a correction, like the K-P correction, to these cross section ratios, allowing the so-corrected central value to vary, with an applied penalty term for excursions, within the systematic errors.  This then also serves as a test of the K-P model's predicted nuclear correction factors for $\nu$ A scattering.
CTEQ corrected all four data sets with the Kulagin-Petti factors, while bringing in the systematic errors of each data set into the fit.  Within the larger systematic errors of the CHORUS data, the K-P nuclear correction factors successfully correct the CHORUS data, bringing the corrected data to the R = 1.0 axis (within errors), as shown in Figure~\ref{Detail6-lowx}.  On the other hand, contrary to expectations the uncorrected NuTeV data ratios are already at values of R $\geq$ 1.0 with small errors in the "shadowing" region.  The K-P correction factors, being modeled to raise the values of R (toward 1.0) in the shadowing region, force the NuTeV values even further away from 1.0 when applied to the NuTeV data points.  This is a very different result than the one obtained in $\ell Fe$ scattering.

\begin{figure}
\begin{center}
\includegraphics[height=2.5 in]{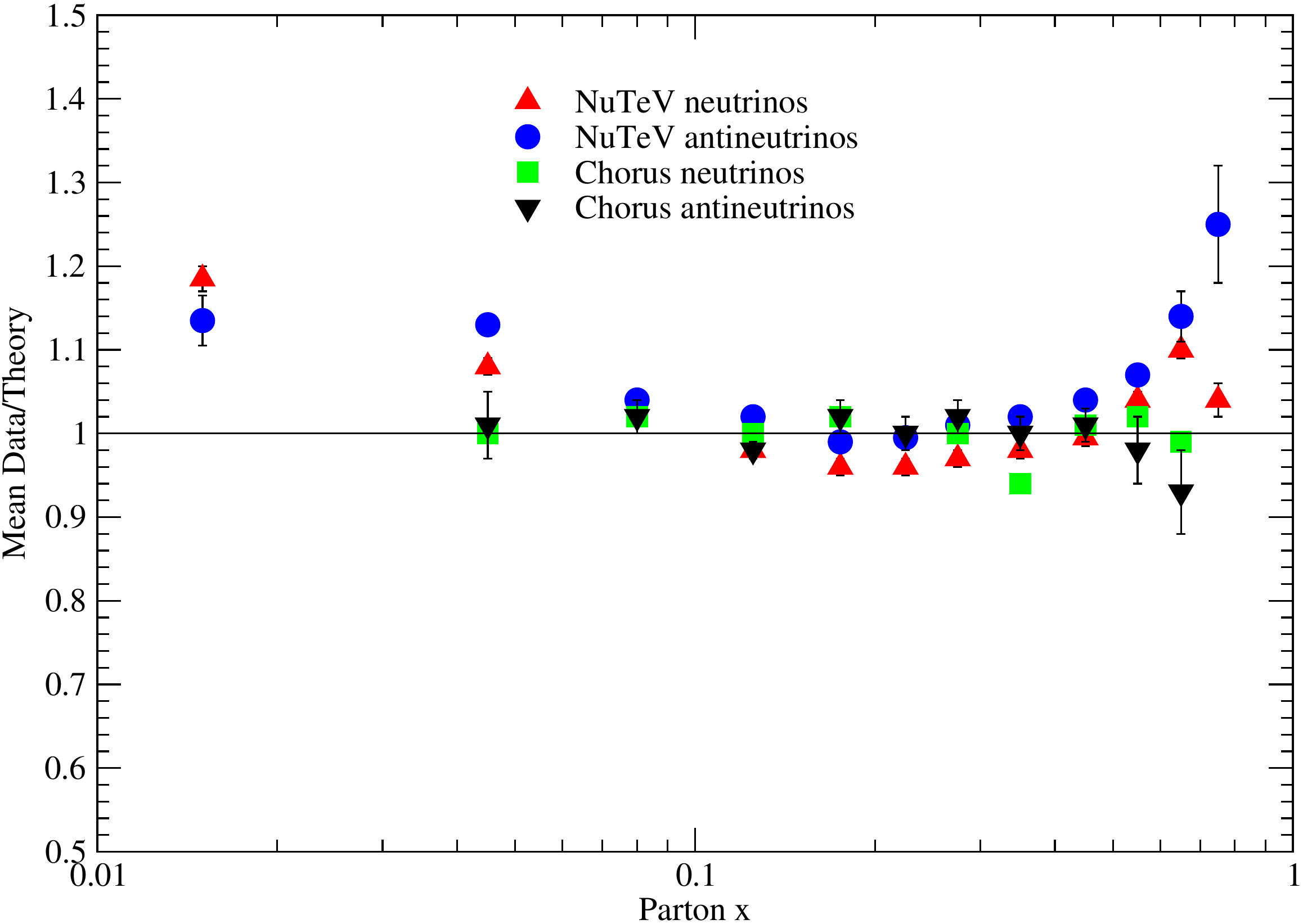}
\caption{As an indication of the size of their systematic errors and the effect of the K-P nuclear correction factors, the low-x details of the comparison between the Chorus and NuTeV neutrino cross section data and the free-nucleon reference cross sections are shown, after the K-P corrections have been applied and the measured points have been allowed to vary within the quoted systematic errors to reduce the $\chi^2$ (see text).}
\label{Detail6-lowx}
\end{center}
\end{figure}

\paragraph{Evidence for Shadowing from $\nu$ A Cross sections}
We then conclude that the CHORUS data, within their errors, are consistent with the predictions of a more conventional model in the shadowing/anti-shadowing region.  The NuTev data is certainly {\em NOT} consistent with conventional expectations in the shadowing/antishadowing region and the NuTeV $\nu$ cross section data do not show any shadowing behavior at all, down to the lowest x value of x = 0.015.  
	

\subsection {Recent Neutrino Nucleus Scattering Data: Global QCD Fits and Nuclear Parton Distribution Functions}
\label{sec-npdf}
To appreciate the more recent analyses of nuclear effects in neutrino nucleus scattering, it is necessary to understand the concept and application of global QCD analyses of various experimental data.  It has been through such global fits, attempts to find a single fit-solution describing multiple data sets, that tension between neutrino nucleus scattering and other data sets have suggested that nuclear correction factors applied to the neutrino scattering data need to be re-examined.

The QCD Parton Model provides a comprehensive framework for describing general high energy processes. In this framework, the cross-section for a hadron-hadron collision process is written as the convolution of a set of universal non-perturbative Parton Distribution Functions (PDFs $f_i(x,Q^2_f)$) and a perturbatively-calculable hard cross section $d\hat{\sigma}$.  

\begin{eqnarray}
d\sigma^{l + N \rightarrow l' + X} & = & \sum_{i=q,\overline{q}} f_i^N(Q^2) \otimes d\hat{\sigma}^{l + i\rightarrow l' + i'}(\mu^2,Q^2). 
\label{eq:fac}
\end{eqnarray}

\noindent
The parton distributions describe the structure of hadrons, and are the link between the physically measured cross-sections $d\sigma^{l + N \rightarrow l' + X}$ and the basic processes of the theory.  Obviously, the precise determination of these functions is of fundamental importance for the interpretation of experimental results within the Standard Model.  These parton distribution functions should be applicable to all processes.  A goal of global QCD fits is to extract a set of parton distribution functions, while testing this concept of universality.  It is this concept of universality that will be considered in the discussions that follow.

Although global analyses began in the late 1970's with the work of Gluck and Reya~\cite{Gluck:1977ah}, and continued with the analyses of Duke and Owens~\cite{Duke:1983gd} and Martin, Roberts and Stirling~\cite{Martin:1987bx}, it wasn't until  the global analyses of Morf\'{\i}n and Tung \cite{Morfin:1990ck}, which were initiated in the 1988 Snowmass Workshop\cite{Tung:1989cs}, that increased emphasis was placed on careful consideration of experimental systematic errors.   Since then there have been many subsequent global analyses, accommodating an increasing ensemble of types of data with increasingly sophisticated treatment of experimental errors.  The leading groups working on global analyses of free-nucleon PDFs today are the CTEQ (works described in this review and~\cite{Guzzi:2011sv}), the MSTW~\cite{arXiv:1006.2753}, and the NNPDF~\cite{942827} collaborations.  There has been no reason to suspect that independent global analyses could not be applied to experimental data off nuclear targets as well as nucleon targets.  The challenge has been to perform global fits to a {\em combination} of nucleon-target and nucleus-target data, which call for nuclear correction factors to compensate for nuclear effects.  The results of fits to nuclear targets and the determination of nuclear correction factors will be discussed below.

Global analyses adopt functional forms appropriate for parton distributions at all Q, so the evolved distributions can then be given in a simple analytic form, which varies smoothly over the entire range of x. The goal is to minimize the role of the chosen initial point of evolution.  A simple form that satisfies the above considerations is, for example:
\begin{eqnarray}
 xf(x,Q_0^2) = c_0x^{c_1} (1-x)^{c_2} e^{c_3x} (1+ e^{c_4}x)^{c_5}
\label{eq:input}
\end{eqnarray} 
where the c-coefficients are called the Òshape parametersÓ. An advantage of this parametrization for a study of shadowing is that it provides a way to emphasize the small-x behavior of the parton distributions via the value of $c_1$.	

Once the parameterization of the PDFs has been chosen at the initial scale $Q_0^2$ as indicated above, the procedure for a ``global fit" is then:
\begin{itemize}
\item Evolve the PDFs from the initial $Q_0^2$ to the relevant larger scales $Q^2 > Q_0^2$ of the experimental data by the DGLAP equations.
\item Compute the physical observables to be tested. 
\item Determine how well the computed observables match the experimental ones via a $\chi^2$-function that compares the measured experimental values and their associated statistical and systematic errors with the computed values. 
\item Vary the c-coefficients until the best agreement with the data is reached by observing a minimum in the $\chi^2$-function.  
\end{itemize}
The actual process of finding the minimum $\chi^2$ in practice is a non-trivial task, since the $\chi^2$ involves both statistical and systematic experimental errors. General-purpose minimization packages like \texttt{MINUIT} \cite{James:1975dr} can be used to initiate the process, but global fitting is also an {\em art} that requires additional considerations.

\subsubsection{Global Fits: Nuclear Structure Functions, Nuclear Parton Distributions and Correction Factors}

A comparison of results from neutrino scattering as well as charged-lepton scattering off nuclear targets can be done with global fits to extract {\em nuclear} parton distributions.  With these nPDFs, one can then study both the magnitude and the shape of the nuclear correction factors $R^A_i(x, Q_0)$, for both neutrino and charged-lepton scattering.  The experimental input can be, preferably, the measured differential cross sections or the extracted nucleon/nuclear structure functions. 

\paragraph{Using the ratio of extracted $\nu$-A  to $\nu$-Nucleon Structure Functions}
A possible experimental input to these global fits involves the extracted nuclear structure functions from the differential cross sections.  With sufficient statistics and control of systematic errors the structure functions $F_2(x,Q^2)$, $xF_3(x,Q^2)$ and, $R_L(x,Q^2)$ = $\sigma_L / \sigma_T$ can be determined independently for neutrino and antineutrino, by fitting the y-distribution of events in a given x-$Q^2$ kinematic {}``box".  This procedure leads to a total of {\em six} structure functions.  

Somewhat less demanding in statistics and control of systematic errors, the ``average" structure functions $F_2(x,Q^2)$ and $xF_3(x,Q^2)$ can be determined from fits to combinations of the neutrino and antineutrino differential cross sections, with several assumptions.  This is the procedure used by the NuTeV collaboration.  The sum of the $\nu$ and $\overline {\nu}$ differential cross sections, yielding $F_2$ can then be expressed as (see also \ref{5.420}):

\begin{eqnarray}
&&{\frac{d^2\sigma}{dx dy}}^{\nu}+{\frac{d^2\sigma}{dx dy}}^{\overline\nu}
={\frac{G_{F}^2
M E}{\pi}}\Big[2\Big(1-y-{\frac{M x y}{2E}} 
 +  {\frac{y^2}{2}{\frac{1+4 M^2 x^2/ Q^2}
         {1+R_{L}}}}\Big) F_{2} + {y} \Big(1 - {\frac{y}{2}}\Big)
         \Delta xF_{3}\Big].
\end{eqnarray}

\noindent
where now $F_{2}$ is the {\em average} of $F_{2}^{\nu}$ and $F_{2}^{\overline {\nu}}$, and the last term is proportional to the difference in $xF_3$ for neutrino and antineutrino probes, $\Delta xF_3=xF_3^{\nu}-xF_3^{\overline{\nu}}$.  At leading order, assuming symmetric $s$ and $c$ seas, this is $4x\left(s-c\right)$.  The cross sections are also corrected for the excess of neutrons over protons in the iron target (5.67\%), so that the presented structure functions are for an isoscalar target.   

A significant step in the determination of $F_2(x,Q^2)$ in this manner, that surely affects the low-x (shadowing) values, is the assumed $\Delta xF_3$ and $R_L(x,Q^2)$.  NuTeV uses a NLO QCD model as input (TRVFS \cite{trvfs}), and assumes an input value of $R_L(x,Q^2)$ that comes from a fit to the world's charged-lepton measurements \cite{rworld}.  
It is important to emphasize that the NuTeV $F_2(x,Q^2)$ extracted points in the shadowing region with x $\leq 0.1$ are extremely dependent on these two assumptions.  In addition, from the NuTeV publications it is difficult to determine if the assumed R and $\Delta xF_3$ have been corrected for nuclear effects up to Fe and, if so, how has this been done.  For this reason it is preferable to directly use the measured differential cross sections in these fits rather than the extracted average structure functions.

\paragraph{Determining the Nuclear Correction Factors}
There are two alternative approaches for determining these nuclear correction factors.  They both lead either directly or indirectly to the concept of nuclear parton distribution functions.  One approach that will be referred to as method 1, is to use the basic global analysis formalism described above, implemented in a global PDF fitting package of experimental results off nuclei, with {\em no} nuclear corrections applied to the analyzed data.   If the nuclear modifications of the $x$-shape in this analysis are modest, the initial parametrization of Equation~\ref{eq:input} should provide enough flexibility to also accommodate these now {\em nuclear} parton distribution functions (nPDF).  Notice that the resulting nPDFs are frequently referred to as the PDFs of a bound nucleon. Such an interpretation is reasonable at large $x$, because the lifetime of hadronic fluctuations is short, and the incoming lepton reaches a bound nucleon deep inside the nucleus without any preceding interactions. However, this is not true at small $x$ in the shadowing region. A long living hadronic fluctuation coherently interacts with a whole chain of nucleons, rather than with a separate bound one, therefore the DIS process probes the {\em collective} PDFs of the nucleus. These nPDFs are different from the ones of a free nucleon, and are subject to shadowing, i.e. they rise with the number of nucleons slower than linearly.  This can also be interpreted in the infinite momentum frame of the nucleus, where as is explained in Sect.~3.1 and illustrated in Fig.~\ref{fig:clouds}, the parton clouds originating from different nucleons overlap and can fuse at small $x$, where they form a common parton distribution for all involved nucleons. These collective nPDFs cannot and should not be identified with any individual bound nucleon.

Given then both nPDFs and free-nucleon PDFs, the nuclear correction factors can be determined for a given nucleus.  Within the parton model, the nuclear correction factor $R^A_i(x, Q_0)$ for an experimentally measurable quantity $M_i$ can be defined as follows: 

\begin{equation}
R^A_i(x, Q_0) = \frac{M^A_i(x,Q_0)}{ M_i^N (x,Q_0)}
\label{eq:R}
\end{equation}

\noindent
where $M^A_i(x,Q_0)$ represents the measurable quantity computed with nuclear PDFs, and $M_i^N (x,Q_0)$ is the same quantity constructed for the average nucleon, (n+p)/2, out of the free-nucleon PDFs.

An alternative approach, method 2, is to assume a set of free-nucleon PDFs, calculate $f_i^p (x,Q_0)$ and $f_i^n (x,Q_0)$, and then use global fitting techniques to fit the nuclear correction factors, also called weighting-factors that modify the free nucleon PDFs to provide agreement with the measured nuclear quantity.  This method also yields nPDFs as a by-product of the fit:

\begin{equation}
M_i^A (x,Q_0) = R^A_i(x, Q_0) \, 
    \frac{1}{A} \left[ Z\,f_i^{p} (x,Q_0) 
                + (A - Z) f_i^{n} (x,Q_0) \right] ,
\label{eq:R2}
\end{equation}


\noindent
In either method, $R^A_i(x,Q_0)$ depends on the observable under consideration since different observables may be sensitive to different  combinations of PDFs.  For example the nuclear correction factor $R^A_i(x,Q_0)$ for $F_2^A$ and $F_3^A$ will, in general,  be different.  Additionally, the nuclear correction factor for $F_2^A$ will yield different results for the $\nu$-$Fe$ process as compared with the charged-lepton $\ell^\pm$-$Fe$ process.  In both cases this is due to the different parton contributions to each process.

Contemporary attempts to use global fitting techniques to extract these nuclear parton distribution functions and/or the nuclear correction factors began with the pioneering work of Eskola and Paukkunen~\cite{Eskola:1998df,arXiv:0906.2529,Eskola:2007my} and later in collaboration with C. Salgado~\cite{Eskola:2010jh}. The study continued with the works of Hirai, Kumano and Nagai~\cite{Hirai:2001np,Hirai:2004wq,Hirai:2007sx} and deFlorian and Sassot~\cite{deFlorian:2003qf, deFlorian:2011fp}, all who used variations of method 2.  The latest group to compute nPDFs, the nCTEQ group, employs method 1~\cite{arXiv:1012.0286,arXiv:0907.2357,arXiv:0812.3370,arXiv:0710.4897}.

\subsubsection{Global Fits: The nCTEQ Nuclear Parton Distributions and Correction Factors}

In contrast to the procedure employed by other nPDF global analysis groups, the nCTEQ group\footnote{This section is a summary of work performed by a CTEQ/Grenoble/Karlsruhe collaboration consisting of: K.~Kovarik (Karlsruhe, Inst. Technology), I.~Schienbein (Joseph Fourier U.), J.~Y.~Yu (Southern Methodist U.), C.~Keppel (Hampton U), J.~G.~Morf\'{\i}n (Fermilab), F.~I.~Olness (Southern Methodist U.), J.~F.~Owens (Florida State U.) and T.~Stavreva (Joseph Fourier U.)}. employs method 1, that is to use the basic global analysis formalism, implemented in a global PDF fitting package of results off nuclei, with no nuclear corrections applied to the analyzed data and no assumed set of free-nucleon PDFs.   This then is a direct fit of the nPDFs themselves.


To confirm that this procedure that uses nuclear global fitting techniques and does {\em not assume} a set of nucleon PDFs to start with, is reasonable, the procedure is first applied to charged-lepton scattering results.  The goal of the nCTEQ group is to determine an independent evaluation of the charged-lepton nuclear correction factor, $F_{2}^{Fe}/F_{2}^{D}$, that is preferred by the data. It can than be directly compared with the SLAC-NMC factor and with the predictions of other models and global charged-lepton fits.   After cuts on Q ($\geq$ 2.0 GeV) and W ($\geq$ 3.5 GeV), the fit used over 700 data points over the complete x range, which  are listed in Tables I, II and III of publication~\cite{arXiv:0907.2357}.  The nuclei included in the fit ranged from He to Pb in ratios to deuterium, and Be to Pb in ratios to C.  In particular the data surviving the cuts in the ratios of C, Ca and Cu to deuterium were the main drivers to the turnover into the shadowing region at low x.  There were no Fe to deuterium ratios for x $\leq$ 0.7 (the shadowing region) that survived the cuts. 

The nCTEQ result is displayed in Figure~(\ref{fig:nucl+-}-a) for $Q^{2}=5$~GeV$^{2}$, and in Figure~\ref{fig:nucl+-}-b) for $Q^{2} =20$~GeV$^{2}$. The figures of the ratio $F_{2}^{Fe}/F_{2}^{D}$ are in good agreement with the SLAC/NMC parameterization of Figure~\ref{fig:slac} and with the fits from Hirai-Kumano-Nagai (HKN07) \cite{Hirai:2007sx}.  The data points displayed in Figure~\ref{fig:nucl+-} are the data available off Fe that satisfied the cuts.  The success of this comparison suggests that this method can also be successfully applied to $\nu$-A scattering results. 

\begin{figure*}
\begin{picture}(500,155)(0,0) 
\put(0,0){\includegraphics[width=0.45\textwidth]{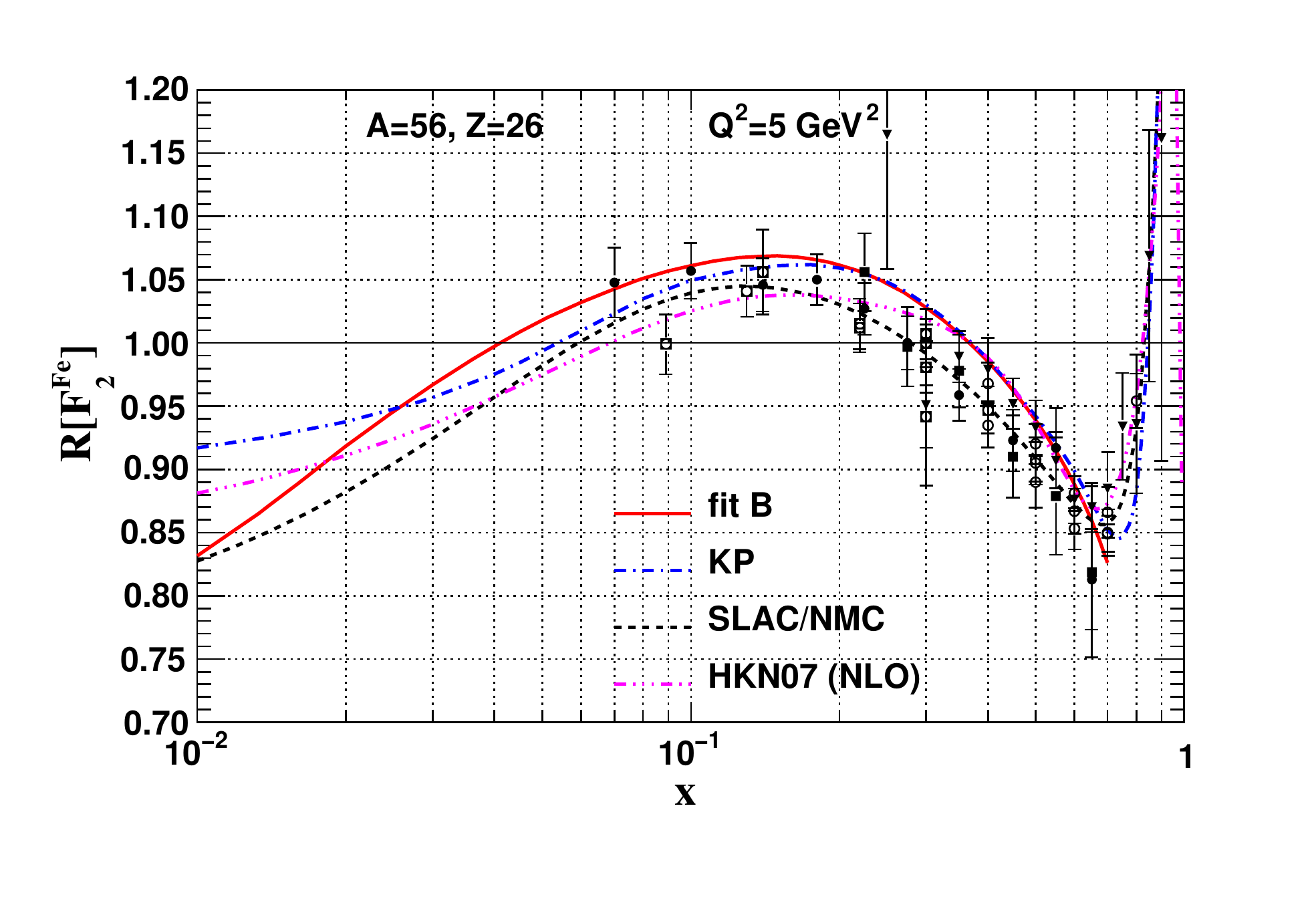}}
\put(260,0){\includegraphics[width=0.45\textwidth]{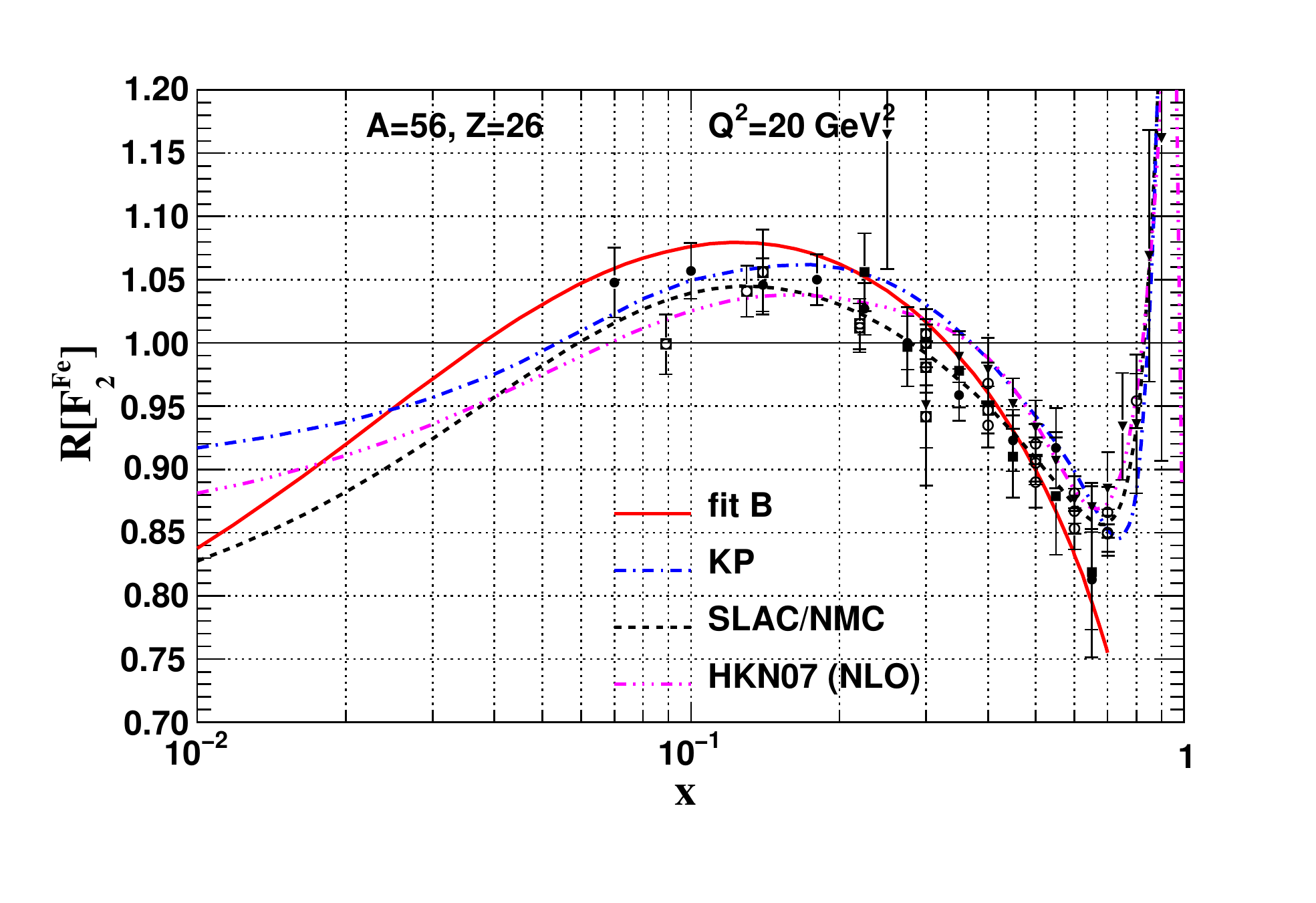}}
\put(114,0){$(a)$} 
\put(374,0){$(b)$}
\end{picture} 
\caption{Using a fit to all charged-lepton data ($\ell^{\pm}A$) and DY data, off a variety of nuclei, that passed their cuts, the nCTEQ computed nuclear correction factor, $F_{2}^{Fe}/F_{2}^{D}$ (solid line labeled fit B), as a function of $x$ is shown in Figure-(a) for $Q^{2}=5\,{\rm GeV}^{2}$ and Figure-(b) for $Q^{2}=20\,{\rm GeV}^{2}$. Both fits are compared with the SLAC/NMC parameterization, as well as the Hirai et \textit{al.} (HKN07), (Ref.~\cite{Hirai:2007sx}) fit to charged-lepton nucleus scattering data and the Kulagin-Petti model predictions. The data points displayed in the Figures are the measured Fe to D ratios that survived the cuts.}
\label{fig:nucl+-}
\end{figure*}


\paragraph{Nuclear Correction Factors for $F_2^\nu(x,Q^2)$ and  $F_2^{\bar\nu}(x,Q^2)$ }

A similar procedure is then used to extract the neutrino nuclear correction factor $F_{2}^{Fe}/F_{2}^{D}$.  To apply this procedure now to $\nu$-A scattering, there are several data sets that can be considered.  The earliest is the CDHSW $\nu$-Fe data. followed by the CCFR $\nu$-Fe data, the NuTeV $\nu$-Fe data and finally the CHORUS $\nu$-Pb data.  However, if one demands a full covariant error treatment of the published data, yielding maximal discriminatory power of the data, there is essentially only the NuTeV $\nu$-Fe and CHORUS  $\nu$-Pb scattering data for input.  This would also be a natural selection by the fit itself.  If the CDHSW and CCFR data, with their errors calculated via the sum of the squares of statistical and systematic errors, were added to the NuTeV and CHORUS data with their full covariant error matrix for the fit, the weight of the CDHSW and CCFR data points in the combined fit would be greatly reduced.  This was shown in the earlier treatment of neutrino cross sections, when the NuTeV Fe and the CHORUS Pb data were combined.  In this case, even though both data sets have full covariant error matrices, the relatively small NuTeV errors with respect to the CHORUS errors enable the NuTeV data points to dominate the combined fit.

An additional input to the fit is the NuTeV/CCFR di-muon data~\cite{Goncharov:2001qe}, which are sensitive to the strange quark content of the nucleon.  However, no other data such as charged-lepton--nucleus ($\ell^{\pm}A$) and DY data have been used.  Because the neutrino data alone do not have the power to constrain all of the PDF components, a minimal set of external constraints~\cite{arXiv:0710.4897} also has to be imposed and some of these external assumptions do indeed affect the behavior of the fit parton distributions at small x - the shadowing region.  These include the Callan-Gross relation ($F_2^{\nu A} = 2 x F_1^{\nu A}$) as well as use of the assumption s = $\overline s$ and c = $\overline c$.  
 
It is important to note that the nCTEQ fit is made directly to the double differential cross sections in order to extract the set of nPDF consistent with the NuTeV measured differential cross section points.  The fit does {\em not} use the extracted NuTeV results of the average value of $F_2(x, Q^2)$, which contains all the assumptions made by NuTeV to extract them.  The extracted nPDFs are then combined with the earlier-described free-nucleon PDFs to form the individual values of the nuclear correction factor R for a given x and $Q^2$, separately for neutrino and anti-neutrino - not the average of both - as shown in Figure~\ref{fig:fig6a} for $\nu$--$Fe$  and in Figure~\ref{fig:fig6b} for $\bar\nu$--$Fe$. Since the difference between $F_2(\nu A)$ and $F_2(\bar\nu A)$ is small, the consistency of these extracted values of $F_2(x, Q^2)$ with the measured average values from NuTeV can be shown in Figure~\ref{fig:fig6a}.

\begin{figure*}
\begin{picture}(500,155)(0,0) 
 \put(0,0){\includegraphics[width=0.45\textwidth]{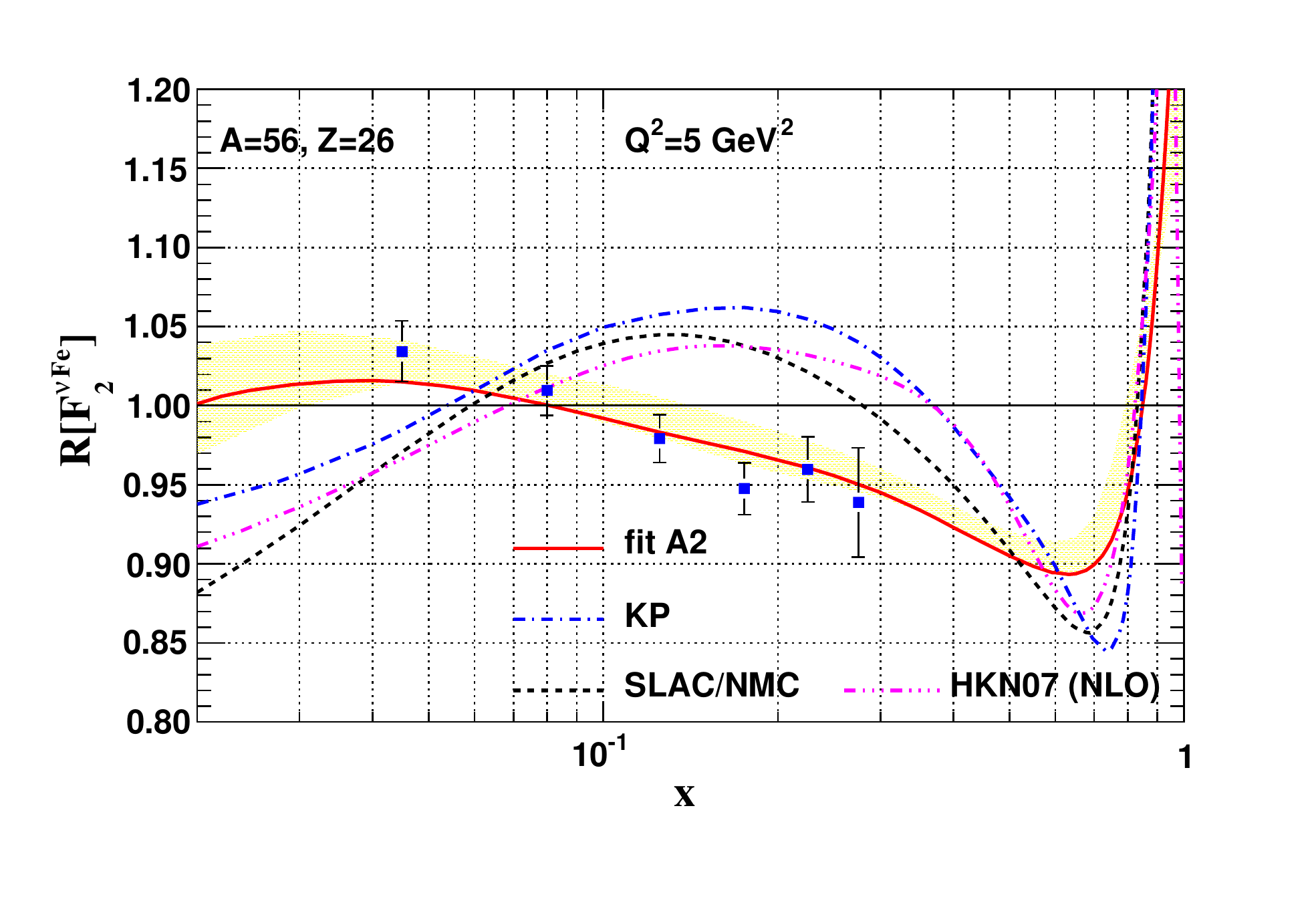}}
\put(260,0){\includegraphics[width=0.45\textwidth]{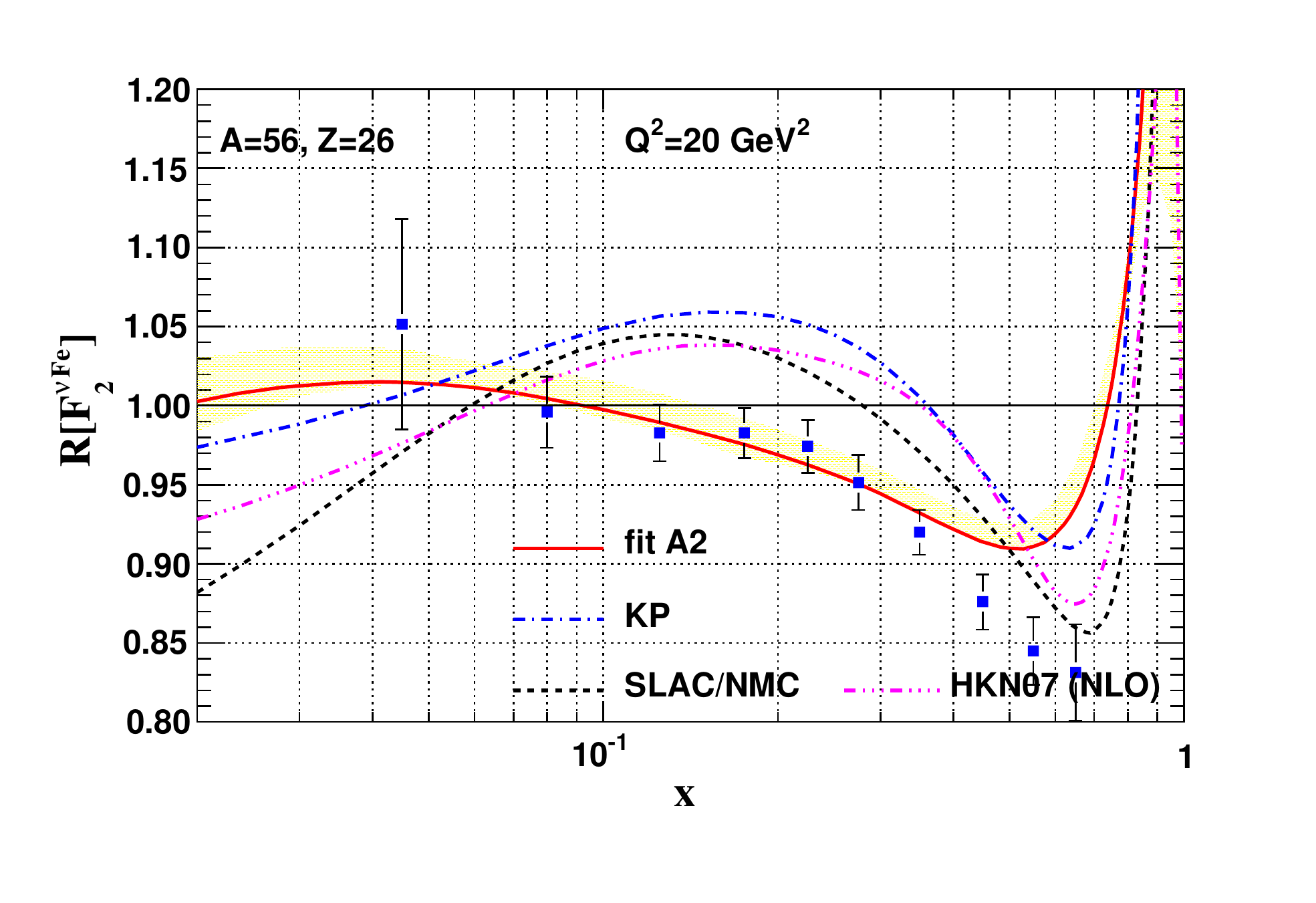}}
\put(114,0){$(a)$} \put(374,0){$(b)$} \end{picture} 
\caption{ 
Nuclear correction factor $R$ for the structure function $F_2$ in charged current $\nu Fe$ scattering at  a)~$Q^2=5~GeV^2$ and b)~$Q^2=20~GeV^2$.  The  solid curve shows the result of the nCTEQ analysis of NuTeV differential cross sections (fit A2), divided by the results obtained with the reference fit (free-proton) PDFs;  the uncertainty from the A2 fit is represented by the yellow band. Plotted also are NuTeV data points of the average $F_2$ (see text) to illustrate the consistency of the fit with the input points. For comparison the correction factor from the Kulagin--Petti model \protect\cite{kp} (dashed-dot line),  HKN07  \protect\cite{Hirai:2007sx} (dashed-dotted line), and the SLAC/NMC parametrization (dashed line) of the charged-lepton nuclear correction factor are also shown.
%
}
\label{fig:fig6a}
\end{figure*}

\begin{figure}
\begin{picture}(500,155)(0,0) 
 \put(0,0){\includegraphics[width=0.45\textwidth]{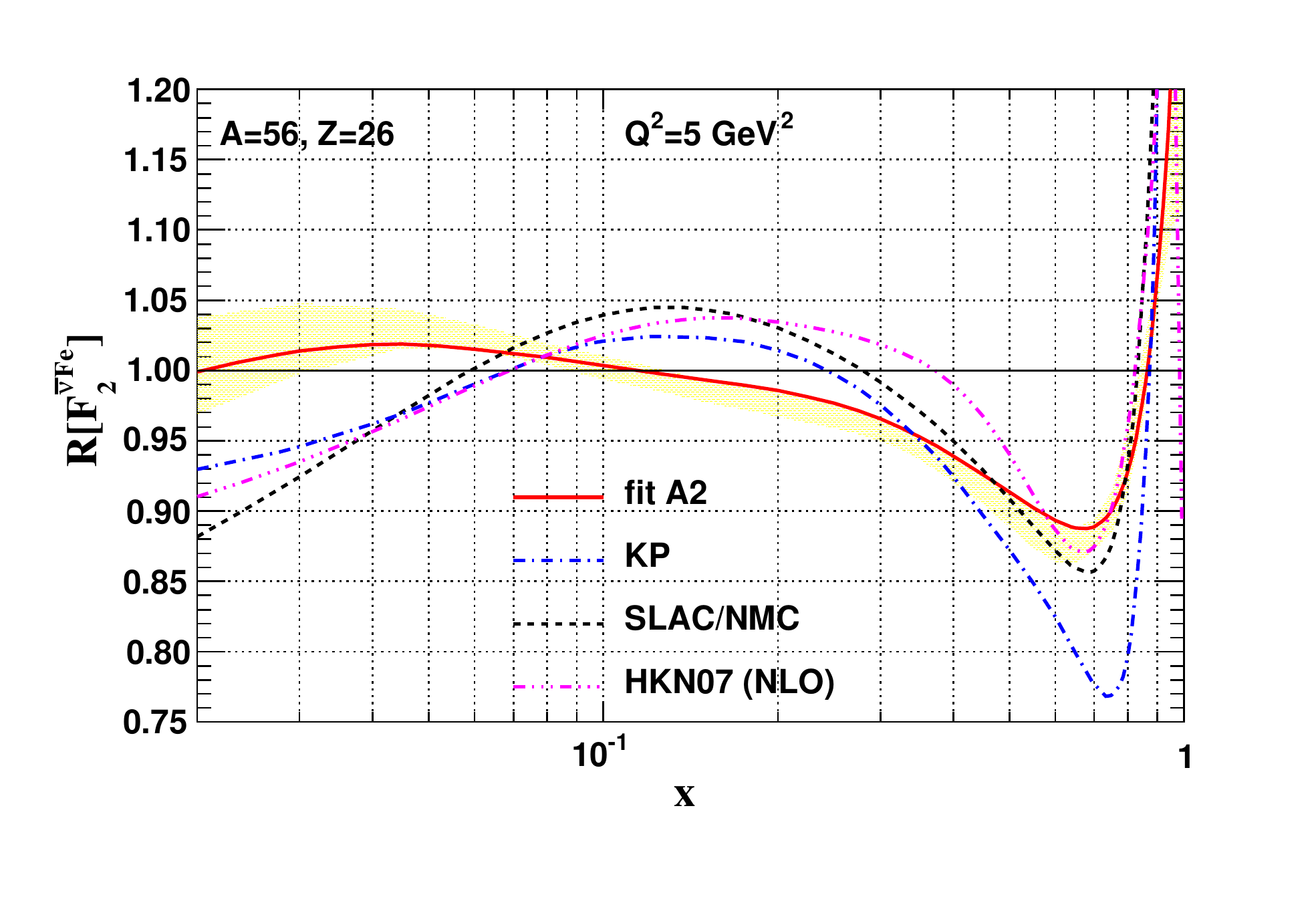}}
\put(260,0){\includegraphics[width=0.45\textwidth]{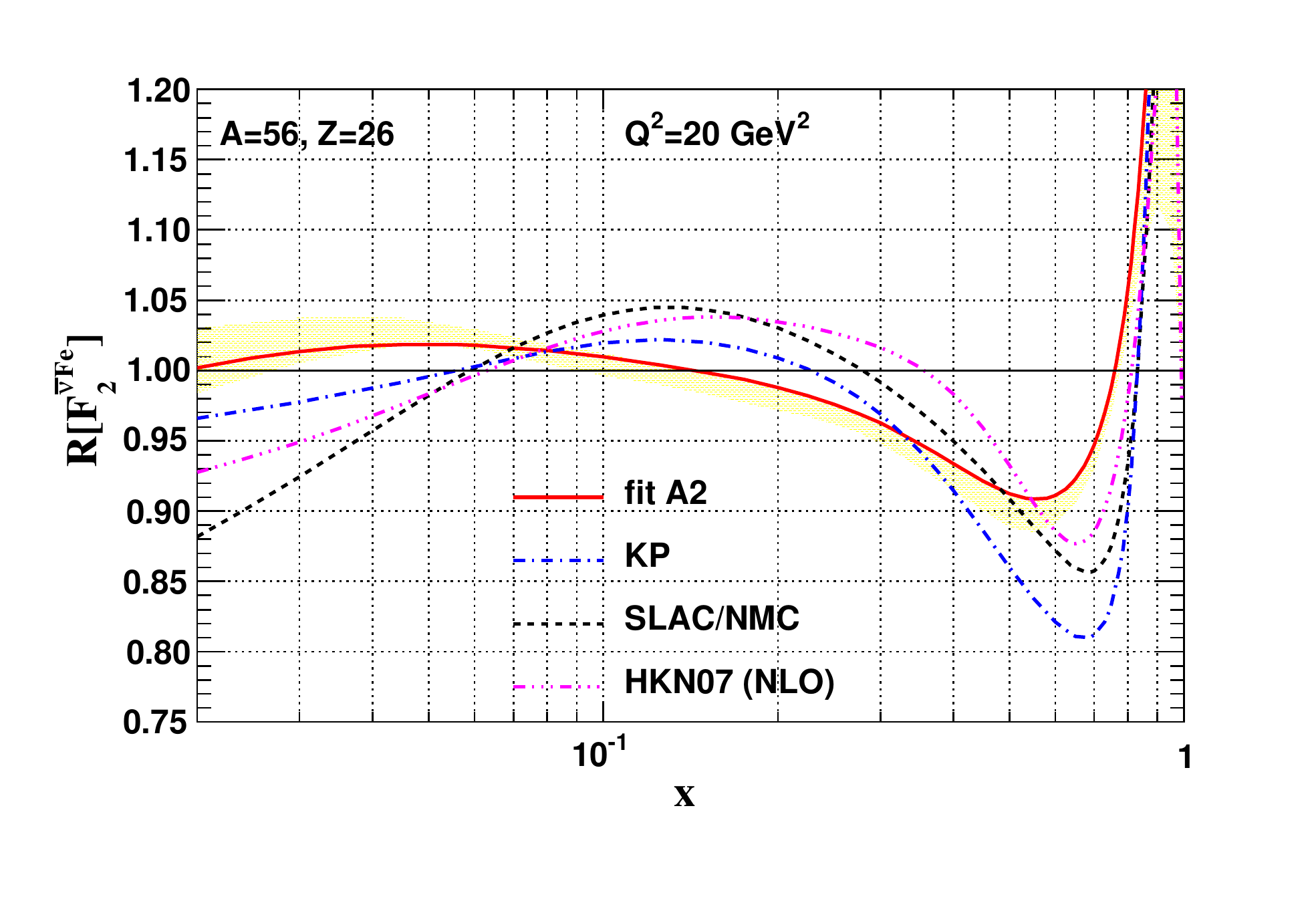}}
\put(114,0){$(a)$} \put(374,0){$(b)$} \end{picture} 
\caption{
The same as in Fig.~\protect\ref{fig:fig6a} for $\overline{\nu} Fe$
scattering.
}
\label{fig:fig6b}
\end{figure}

It is also of course possible to combine these fitted nPDFs to form the individual values of the average of $F_2(\nu A)$ and $F_2(\bar\nu A)$ for a given x and $Q^2$, to compare directly with the NuTeV published values of this quantity.  This was recently done and the nCTEQ preliminary results\footnote{We thank Karol Kovarik,(Karlsruhe, Inst. Technology) for allowing us to use these preliminary results presented at the 2012 DIS conference in Bonn~\cite{DIS12}} for low-$Q^2$ are shown in Figure~\ref{fig:F2lowQ}.

The comparison between the nCTEQ fit that passes through the NuTeV measured points, and the charged-lepton fit is very different in the lowest-x, lowest-Q region and gradually approaches the charged-lepton fit with increasing Q.  However, the slope of the fit approaching the shadowing region from higher x, where the NuTeV measured points and the nCTEQ fit are consistently below the charged-lepton A fit, make it difficult to reach the degree of shadowing evidenced in charged-lepton nucleus scattering at even higher $Q^2$ 

\begin{figure}[h]
\begin{center}
\includegraphics[width=0.75\textwidth]{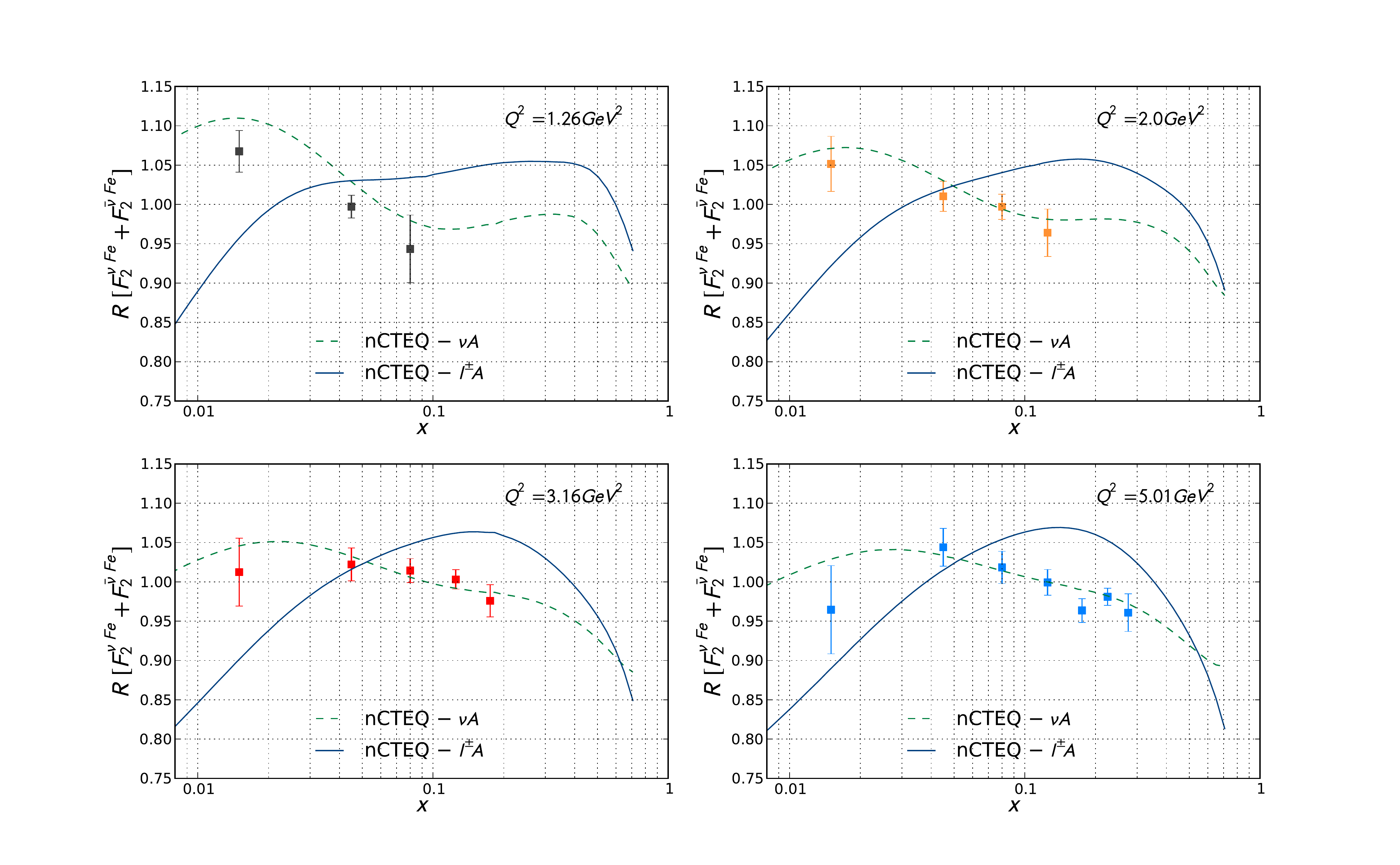}
\caption{ 
Nuclear correction factor $R$ for the average $F_2$ structure function in charged current $\nu Fe$ scattering at  $Q^2$ =1.2, 2.0, 3.2 and 5.0 $GeV^2$ compared to the measured NuTeV points.  The green dashed curve curve shows the result of the nCTEQ analysis of $\nu$ A (CHORUS, CCFR and NuTeV) differential cross sections plotted in terms of the average $F_2^{Fe}$ divided by the results obtained with the reference fit (free-proton) PDFs.  For comparison, the nCTEQ fit to the charged-lepton data from Figure~\ref{fig:nucl+-} is shown by the solid blue curve.}
\label{fig:F2lowQ}
\end{center}
\end{figure}

 %


\paragraph{nCTEQ: Shadowing in $\nu$-A Scattering}
As concluded above, the nCTEQ fit to $\ell^{\pm}A$ and DY data is generally consistent with the SLAC/NMC parameterization and HKN fits to calibrate and certify the nCTEQ method. However, the nCTEQ fits $\nu Fe$ scattering in Figures~\ref{fig:fig6a}, \ref{fig:fig6b} and \ref{fig:F2lowQ} do not agree with the charged-lepton nucleus fits.  Although there is a general tension over most of the x-range, the biggest difference is in the shadowing region, where the non-existent turn-over of the NuTeV $\nu Fe$ cross section data in the shadowing region at lower $Q^2$ is again evident.  Since lower $Q^2$ data dominates the low-x region, this also explains the lack of any shadowing turnover at all in the ratio of cross-sections in Figure`\ref{no_shift_no_cor}.

\subsubsection{Global Fits: The Eskola-Paukkunen-Salgado (EPS) Nuclear Parton Distributions and Correction Factors}
Although the discussion of nuclear parton distributions began in the late 70's\cite{Berlad:1979tc}, it was with the early work of K.~J.~Eskola and colleagues~\cite{Eskola:1998iy} that the more modern era of extracting nuclear parton distributions via global fits to a variety of lepton-nucleus data began.  Their systematic consideration of  $\nu A$ scattering was treated in their work of 2009, published in 2010 \cite{Paukkunen:2010hb}, and then revisited to specifically address the consistency of $\nu A$ DIS results and general nPDF results~\cite{Paukkunen:2010qj}.

Their analysis~\cite{Paukkunen:2010hb} uses method 2 with the CTEQ6.6 free-nucleon PDFs as their basis set.  It is important to note that instead of using the full covariant error matrix,  they have chosen to add the statistical and systematic uncertainties in quadrature when computing $\chi^2$.  This implies that the total errors they are using are {\em larger} than those used by nCTEQ and thus have less discriminatory power when compared to other results.  They form a ratio of data versus theory that is based on CTEQ6.6:
\begin{equation}
 R^{\rm CTEQ6.6} \equiv \frac{\sigma^{\nu,\overline\nu}\left({\rm Experimental} \right)}{\sigma^{\nu,\overline\nu}\left({\rm CTEQ6.6} \right)},
\end{equation}
When using  $\nu A$ as the experimental numerator and comparing it to a denominator calculated with free-nucleon PDFs (CTEQ6.6) a rather poor fit could be naturally expected.  Then they form a second purely theoretical ratio:
\begin{equation}
 R^{\rm CTEQ6.6 \times EPS09} \equiv \frac{\sigma^{\nu,\overline\nu}\left({\rm CTEQ6.6 \times EPS09} \right)}{\sigma^{\nu,\overline\nu}\left({\rm CTEQ6.6} \right)}, \label{eq:nuclear_mod_npdfs}
\end{equation}
using the nuclear correction factors from their EPS09~\cite{Eskola:2009uj} analysis and introducing CTEQ6.6 PDF uncertainties based in turn on upper and lower cross section uncertainties based on the PDF uncertainty analysis.  
If the nuclear correction factors determined by EPS09, based on charged-lepton nucleus scattering, are correct also for $\nu A$ scattering, then $R^{\rm CTEQ6.6}$ should be consistent with $R^{\rm CTEQ6.6 \times EPS09}$.  

They report their conclusions as a function of additional necessary radiative and target mass corrections that are made either separately, together or not at all.  An example, from Table 1 of their publication, is their findings for the $\chi^2/N$-values when making neither and then both the radiative and target-mass corrections:

\begin{table}[h]
\begin{center}
{\footnotesize
\begin{tabular}{ccc}
 Radiative and Target Mass corrections & CTEQ6.6 &  CTEQ6.6$\times$EPS09 \\
\hline
\hline
 NuTeV	& 1.51  & 1.05 \\
 CHORUS & 1.15  & 0.79 \\
 CDHSW  & 1.10  & 0.71 \\
 \\
No Radiative or Target Mass corrections & CTEQ6.6 &  CTEQ6.6$\times$EPS09 \\
\hline
\hline
 NuTeV	& 1.35  & 1.08 \\
 CHORUS & 1.23  & 1.09 \\
 CDHSW  & 0.96  & 0.86  \\
\end{tabular}
}
\caption[]{\small The $\chi^2/N$-values computed using CTEQ6.6 with and without nuclear
modification from EPS09. The numbers are given for calculations with and without the radiative
and the target mass corrections.}
\label{Table:chi2values}
\end{center}
\end{table}

From their comparison of the change in $\chi^2$ as they add these corrections, they conclude:

\begin{itemize}
 \item 
Whatever way they make the calculation, the one without
nuclear corrections from EPS09 gives substantially larger $\chi^2$.
 \item
For CHORUS and CDHSW data, the $\chi^2$s get consistently smaller as the radiative and
target mass are applied.
 \item
For NuTeV data, the $\chi^2$ remains practically unchanged whether radiative or
target mass corrections are applied or not.
\end{itemize}

They note that in their fit to the NuTeV data shown in Figure~\ref{Fig:Q_nu_average_NuTeV}, their predictions are in perfect agreement for some NuTeV energy ranges, whereas there are significant and inconsistent  deviations in other energy bins.

\begin{figure}[!htb]
\center
\hspace{-1.2cm}
\includegraphics[scale=0.65]{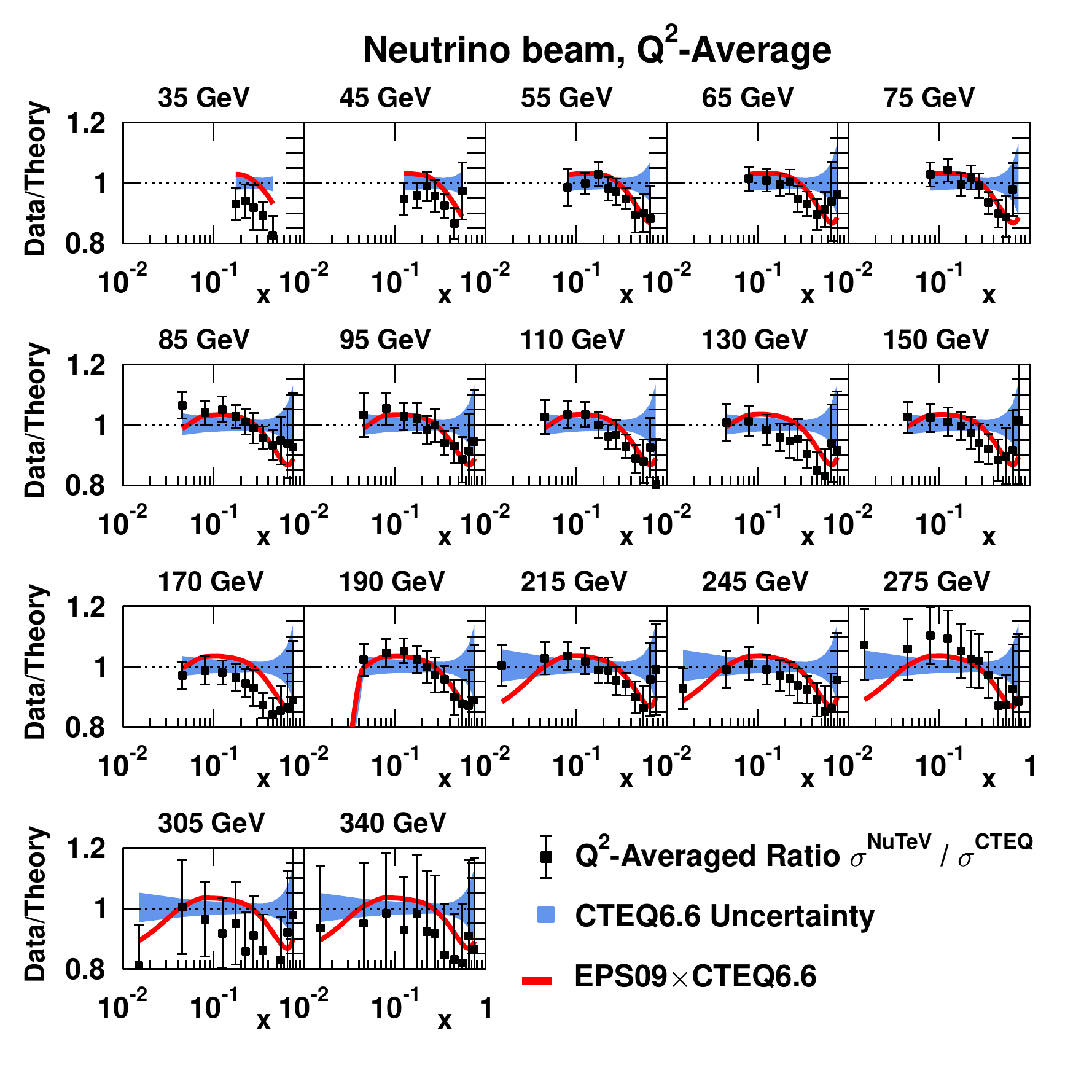}
\caption[]{\small The $Q^2$-averaged NuTeV neutrino data compared to the EPS09*CTEQ6.6 fit.}
\label{Fig:Q_nu_average_NuTeV}
\end{figure}

\paragraph{Eskola-Paukkunen-Salgado: Shadowing in $\nu$-A Scattering}
Their overall conclusion of the analysis of CDHSW, CHORUS and NuTeV data sets is that only the NuTeV neutrino data is inconsistent with nuclear PDFs determined from charged-lepton nucleus scattering.  They attribute the NuTeV discrepancy to an unexplained, neutrino energy-dependent fluctuations in the data and assert that the NuTeV data should not be used in making conclusions about the universality of nuclear PDFs.  Without the NuTeV data they conclude that there is no inconsistency between $\nu$-A and $\ell$-A scattering in any of the x-dependent nuclear effects regions - including the shadowing region.

\subsubsection{Global Fits: The Hirai-Kumano-Nagai Nuclear Parton Distributions and Correction Factors}

This group extracts nPDFs using method 2, by fitting weight factors.  They have a complete set of nPDFs for fits to data other than $\nu$-A scattering.  They do comment on nuclear effects in $\nu$-A scattering~\cite{Hirai:2009mq}, but do not make explicit fits including  $\nu$-A scattering results.

\subsubsection{Global Fits: The de Florian-Sassot (dFS) Nuclear Parton Distributions and Correction Factors}
The de Florian-Sassot global analyses yielding nPDFs began in 2004~\cite{deFlorian:2003qf}.  They were the first group to apply a NLO analysis to nuclear data.  In their initial analyses they used a convolution relation to relate the ``free-nucleon" PDFs to the bound nPDFs via a weight-function $W_i(y,A,Z)$:

\begin{equation}
f_i^A(x_N,Q_0^2) = \int_{x_N}^A\,\,  \frac{dy}{y} W_i(y,A,Z)\,\, f_i(\frac{x_N}{y},Q_0^2)
\label{eq:f2}
\end{equation}

\noindent
that allowed them to perform evolutions in Mellin space for increased accuracy.  They used the GRV98~\cite{GRV98} free- proton  PDFs as their basis set and fit the weight functions to a variety of charged-lepton DIS and Drell-Yan large-A data.  They performed no error analysis of the nPDF sets.  Their extracted values of $F_2^A/F_2^D$ = $R^A_{\nu}(f_i)$ are shown in Figure~\ref{fig:ad2}.  As can be seen they fit  quite well the shadowing region for all values of A.

\begin{figure}[h]
\begin{center}
\includegraphics[width=0.65\textwidth]{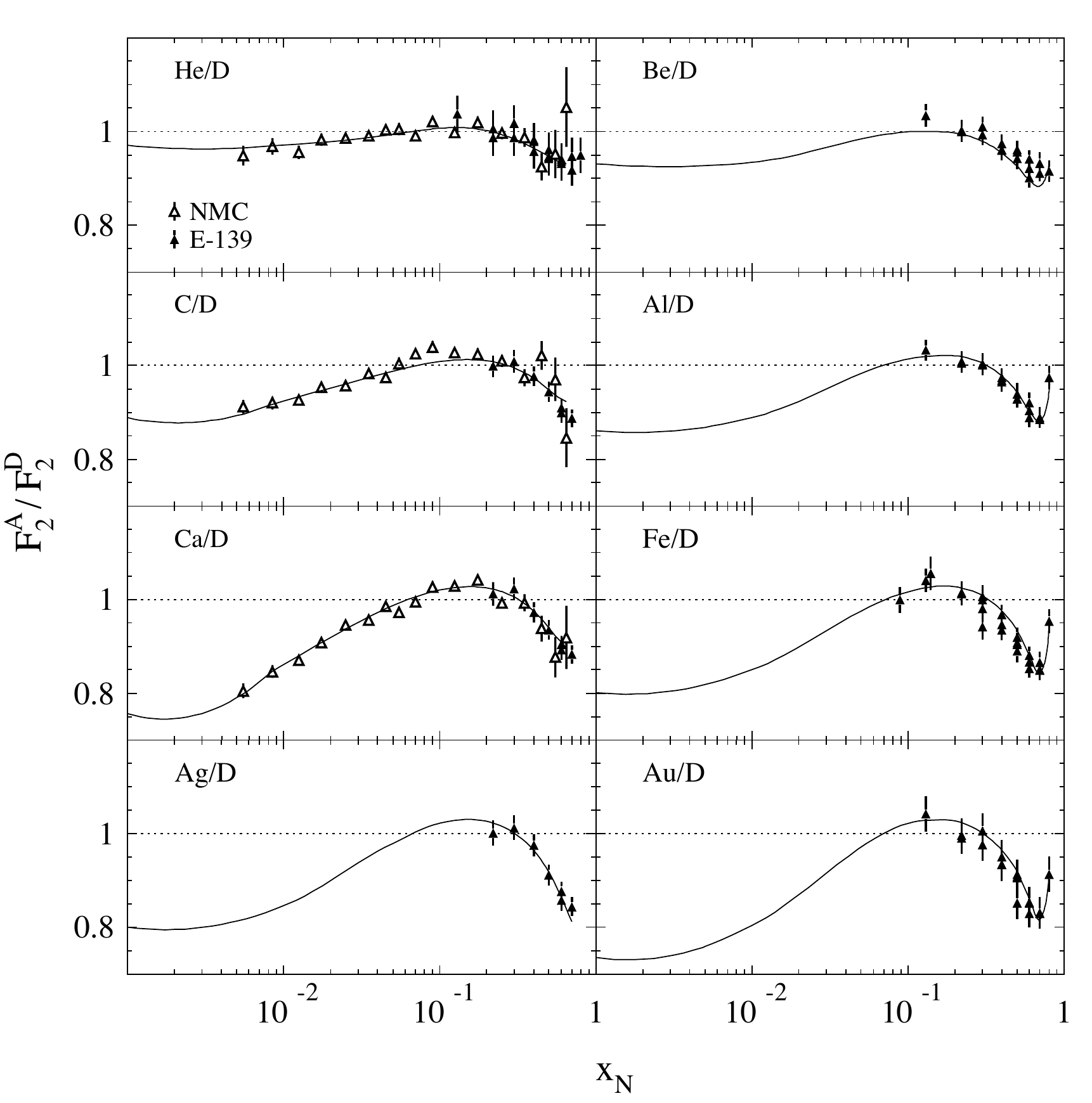}
\caption{$F_2^A/F_2^D$ data. The lines interpolate the
values obtained with the NLO nPDF set at the respective $Q^2$, and extrapolate 
to low $x_N$}.
\label{fig:ad2}
\end{center}
\end{figure}

The group latest global analysis~\cite{deFlorian:2011fp} now also includes P. Zurita and M. Stratmann as co-authirs.  There are considerable updates from the first version outlined above. They have replaced the procedure based on the convolution integral Eq.(\ref{eq:f2}) with the more conventional approach of a ``nuclear modification factor" (nuclear correction factor) given by Eq.(\ref{eq:R2}).  They also have chosen to use the MSTW~\cite{Martin:2009iq} NLO free-proton PDFs as their basis set.  Their strategy is to parametrize the $R^A_i(x_N,Q_0)$ in Eq.~(\ref{eq:R}) for individual quarks and gluons as, for example, the valence quarks, where both valence quark distributions are assigned the same nuclear modification factor $R^A_v)$, which they parametrize as:
\begin{eqnarray}
\nonumber
R^A_v(x,Q^2_0)  &=&  \epsilon_1 \, x^{\alpha_v} (1-x)^{\beta_1}  \times  \\ 
         & &   (1 + \epsilon_2 (1-x)^{\beta_2}) (1 + a_v  (1-x)^{\beta_3})\,,
\label{eq:rval}
\end{eqnarray}

They have expanded the data set used in their fit to include not only DIS and Drell-Yan off larger-A targets, but now also  $\nu$-A scattering and inclusive pion production in deuteron-gold collisions.  

Concentrating on their fit to  $\nu$-A results, they include the data sets from NuTeV (Fe), Chorus (Pb) and CDHSW (Fe).  They fit to the averaged structure functions $F_{2,3}\equiv (F_{2,3}^{\nu A} + F_{2,3}^{\bar{\nu}A})/2$ from appropriate linear combinations of neutrino and antineutrino CC DIS differential cross sections, as opposed to the differential cross section for $\nu$ and  $\bar{\nu}$ independently.  Their treatment of data point uncertainties uses the statistical and systematic uncertainties of the data added in quadrature as compared to the nCTEQ analysis that uses the full correlated covariant error matrix provided by the NuTeV collaboration.  Both of these choices make considerable difference when compared to the nCTEQ analysis. They conclude that the CC data for the averaged structure function $F_2$ are quite well reproduced  both in shape and in magnitude, {\em within the experimental uncertainties}, by their fit, as illustrated by Fig.~\ref{fig:f2neutrino}.

\begin{figure*}[tbh!]
\begin{center}
\vspace*{-0.6cm}
\epsfig{figure=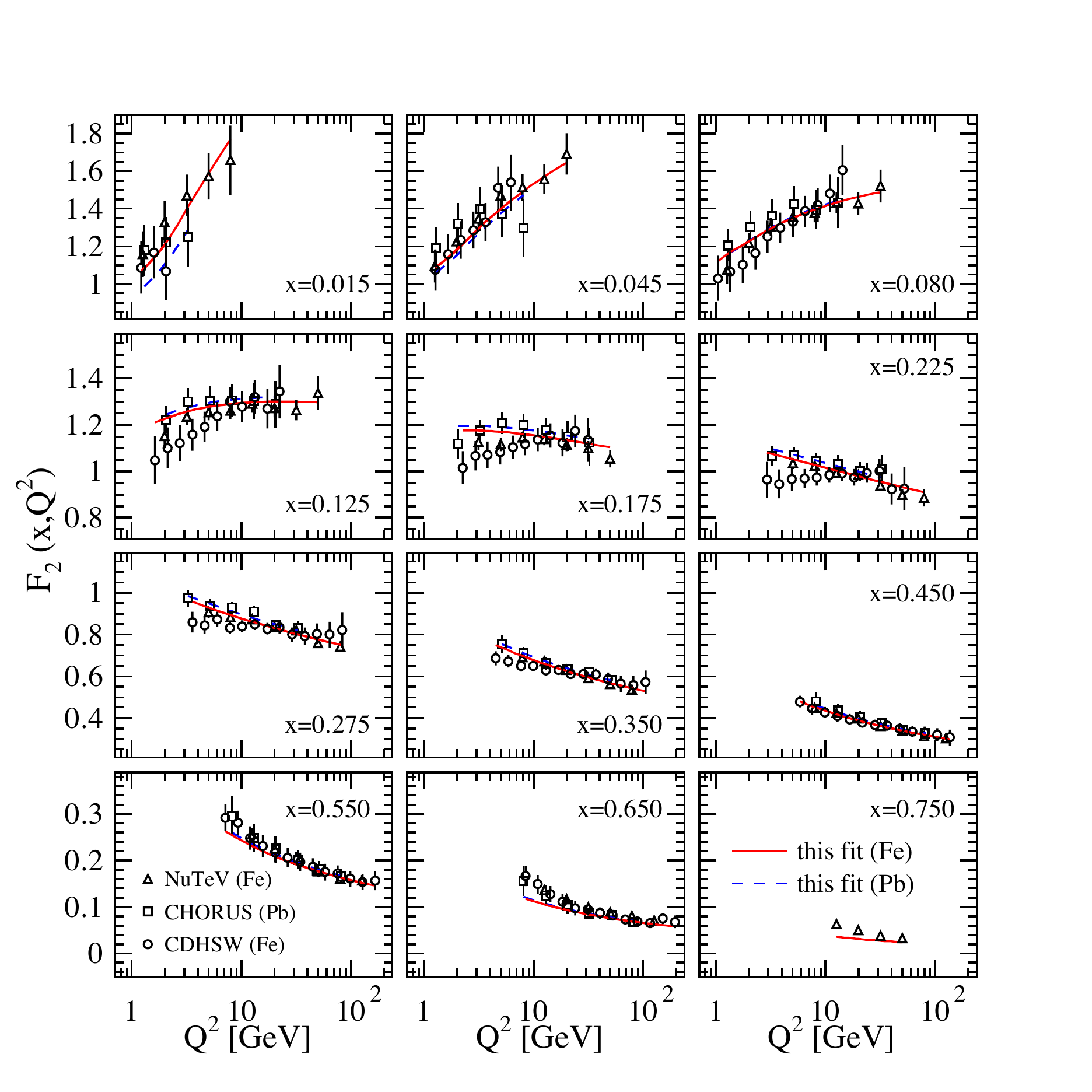,width=0.65\textwidth}
\end{center}
\vspace*{-0.5cm}
\caption{\label{fig:f2neutrino}
The de Florian et al.NLO fit compared to the averaged CC DIS structure function $F_2$ as a function of $Q^2$ in various bins of $x_N$ for NuTeV (Fe) and CHORUS (Pb) targets shown as solid and dashed lines, respectively. }
\end{figure*}

\paragraph{de Florian-Sassot: Shadowing in $\nu$-A Scattering}
On the basis of their analysis they conclude that there is no disagreement between neutrino and charged-lepton nuclear correction factors and that a common set of nPDFs fitting both neutrino and charged lepton data over the full x range, including the shadowing region, can be found.

\subsection{Comparison of the $\ell^{\pm}A$ and $\nu A$ Nuclear Correction Factors}
There is an obvious difference in the conclusions drawn by the nCTEQ group compared to those drawn by the EPS and dFS groups with respect to the $\ell^{\pm}A$ and $\nu A$ nuclear correction factors, particularly for the shadowing region.  

For the nCTEQ analysis, the contrast between the charged-lepton ($\ell^{\pm}A$) case and the neutrino ($\nu A$) case is striking.  While their charged-lepton results that fit to charged-lepton and Drell-Yan data generally align with the SLAC/NMC and HKN determinations, the neutrino results clearly yield different behavior as a function of $x$, particularly in the shadowing/anti-shadowing region.   In the $\overline\nu$ case, these differences are smaller but persist in the low-x shadowing region.  They emphasize that both the charged-lepton and neutrino results come directly from global fits to the data, and there is no model involved.  They further suggest that this difference between the results in charged-lepton and neutrino DIS is reflective of the long-standing {}``tension'' between the light-target charged-lepton data and the heavy-target neutrino data in the historical global PDF fits \cite{Botts:1992yi,Lai:1994bb}. Their latest results further suggest that the tension is not only between charged-lepton \emph{light-target} data and neutrino heavy-target data, but also between neutrino and charged-lepton \emph{heavy-target} data.  In other words, a difference between charged-lepton ($\ell^{\pm}A$) and the neutrino ($\nu A$) when comparing the same A.

For both the EPS and dFS  no significant difference is found between the charged-lepton ($\ell^{\pm}A$) and the neutrino ($\nu A$) cases, and particularly no difference in the shadowing region, where the nCTEQ group finds a significant difference and where differences are actually expected theoretically, although certainly not of the magnitude as found by the nCTEQ group.

Concentrating on this interesting difference found by the nCTEQ group, if the nuclear corrections for the $\ell^{\pm}A$ and $\nu A$ processes are indeed different there are several far-reaching consequences.  Considering this, the nCTEQ group has performed a unified global analysis~\cite{arXiv:1012.0286} of the $\ell^{\pm}A$, DY, and $\nu A$ data (accounting for appropriate systematic and statistical errors) to determine if it is possible to obtain a {}``compromise'' solution including both $\ell^{\pm}A$ and $\nu A$ data. Using a hypothesis-testing criterion based on the $\chi^{2}$ distribution that can be applied to both the total $\chi^{2}$ as well as to the $\chi^{2}$ of individual data sets, they found that it was {\em not possible} to accommodate the data from $\nu A$ and $\ell^{\pm}A$ DIS by an acceptable combined fit.  That is, when investigating the results in detail, the tension between the $\ell^{\pm}Fe$ and $\nu Fe$ data sets {\em does not permit a compromise fit}, which adequately describes the neutrino DIS data along with the charged-lepton data and, consequently, $\ell^{\pm}Fe$ and $\nu Fe$ based on the NuTeV results, have different nuclear correction factors.

A compromise solution between $\nu A$ and $\ell^{\pm}A$ data can be found {\em only} if the full correlated systematic errors of the $\nu A$ data are not used and the statistical and all systematic errors are combined in quadrature, thereby neglecting the information contained in the correlation matrix.  In other words, the larger errors resulting from combining statistical and all systematic errors in quadrature reduces the discriminatory power of the fit, such that the difference between $\nu A$ and $\ell^{\pm}A$ data are no longer evident.  This conclusion underscores the fundamental difference~\cite{arXiv:1012.0286} of the nCTEQ analysis with the analyses of EPS and dFS.  

On the other hand, a difference between $\nu A$ and $\ell^{\pm}A$ is not completely unexpected, particularly in the shadowing region, and has previously been discussed in the literature \cite{Brodsky:2004qa,kp, Brodsky:2004qa,Qiu:2004qk}.  The charged-lepton processes occur (dominantly) via $\gamma$-exchange, while the neutrino-nucleon processes occur via $W^{\pm}$-exchange. The different nuclear corrections could simply be a consequence of the differing propagation of the hadronic fluctuations of the intermediate bosons (photon, $W$) through dense nuclear matter.  Furthermore, as stressed above, since the structure functions in neutrino DIS and charged lepton DIS are distinct observables with different parton model expressions, it is clear that the nuclear correction factors will not be exactly the same.  What is, however, unexpected is the degree to which the $R$ factors differ between the structure functions $F_{2}^{\nu Fe}$ and $F_{2}^{\ell Fe}$. In particular the lack of evidence for shadowing in neutrino scattering down to $x\sim0.02$ at low $Q^2$ is quite surprising.

\subsection{Evidence for Shadowing in $\nu$-A Scattering}

The ratio of $F_{2}^A(x,Q^2)$ with respect to the free-nucleon $F_{2}(x,Q^2)$ naturally reflects the general behavior of the cross section. However, instead of integrating over $Q^2$ to get a cross section in a given x interval as was done above, the x-behavior of the structure function within a specific $Q^2$ range can be examined.

For the lower $Q^2$ regions of the average $F_{2}^{Fe}(x,Q^2)$ shown already in Figure~\ref{fig:F2lowQ}, the comparison between the nCTEQ $\nu$-A fit and the nCTEQ charged-lepton fit suggests very different behavior.  There is no indication of shadowing in the lowest-x, lowest-Q regions, with the $\nu$-A fit only minimally approaching the charged-lepton fit with increasing Q.  However, even when examining the higher $Q^2$ range, as shown in Figure~\ref{fig:F2hiQ}, where the low-x shadowing region becomes kinematically inaccessible as Q increases, the slope of the $\nu$-A fit approaching the shadowing region from higher x for 1.2  $\leq Q^2 \leq $ 30  $GeV^2$  would make it difficult to mimic the shadowing evidenced in charged-lepton nucleus scattering. 
\begin{figure}[h]
\begin{center}
\includegraphics[width=0.75\textwidth]{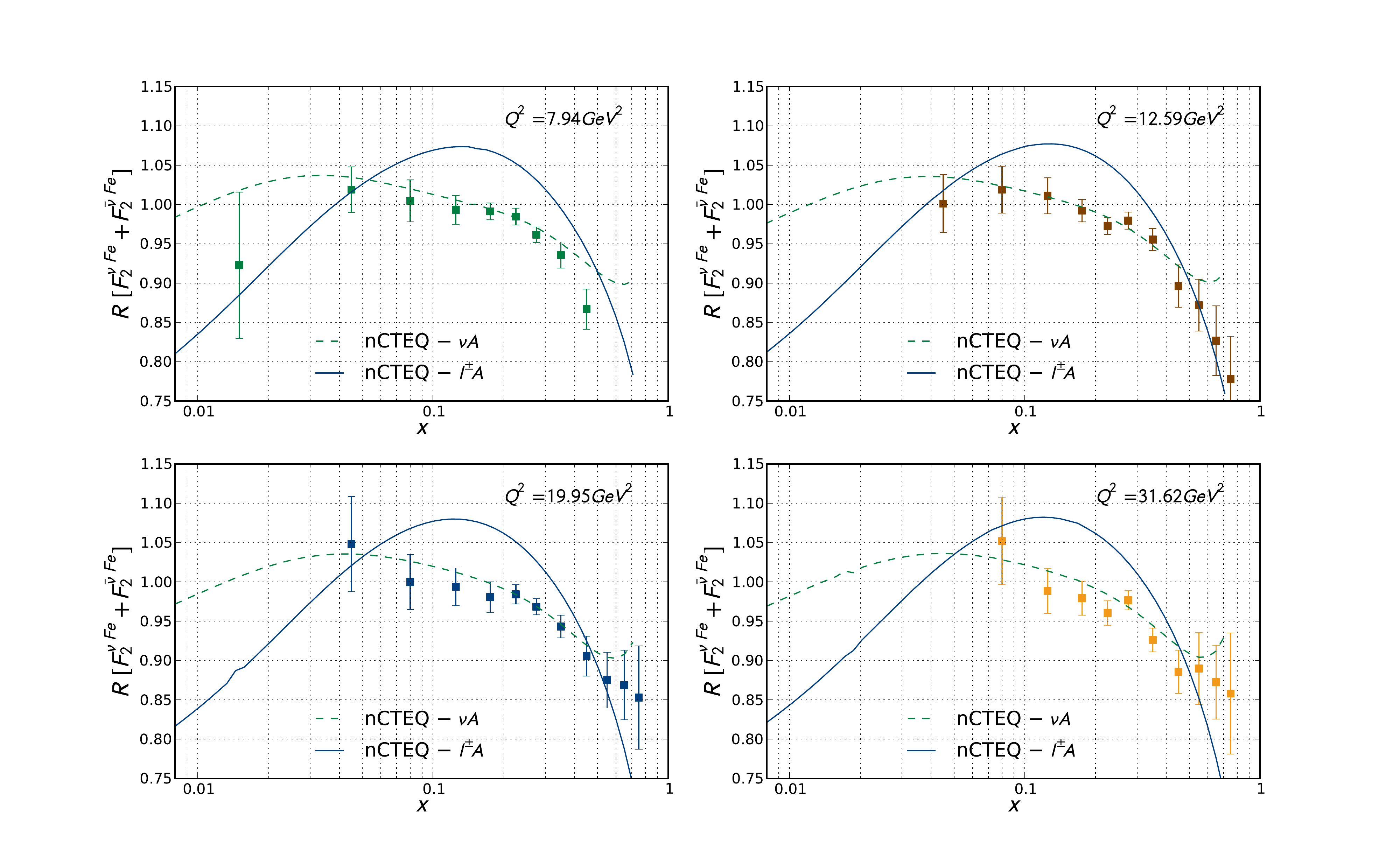}
\caption{ Nuclear correction factor $R$ for the average $F_2$ structure function in charged current $\nu Fe$ scattering at  $Q^2$ = 8.0, 12.5, 20.0 and 31.6  $GeV^2$ compared to the measured NuTeV points.  The green dashed curve shows the result of the nCTEQ analysis of $\nu$ A differential cross sections plotted in terms of the average $F_2^{Fe}$ divided by the results obtained with the reference fit (free-proton) PDFs.  For comparison, the nCTEQ fit to the charged-lepton data from Figure~\ref{fig:nucl+-} is shown by the solid blue curve.}
\label{fig:F2hiQ}
\end{center}
\end{figure}

It is important to emphasize that this apparent disagreement between $\nu$A and $\ell^{\pm }A$ shadowing behavior is based only on the NuTeV data and the use of their full covariant error matrix in the global analysis.  This disagreement was not seen by the two other analyses that, however, did not use the full discriminatory power of the NuTeV data.  Figure~\ref{fig:AvgF2All} summarizes the results of various global fits to both $\nu$A and $\ell^{\pm }A$ scattering.  The curves marked nCTEQ  $\ell^{\pm }A$, EPS09 and HKN07 are fits to  $\ell^{\pm }A$ scattering, while the nCTEQ  $\nu$A and DSSZ preliminary are fits to $\nu$A scattering.  The DSSZ preliminary fit is made to the average  $F_2$ and does not use the full NuTeV covariant error matrix, while the nCTEQ fit is to the double-differential cross sections and does use the full covariant cross section.
Before any far-reaching conclusions can be drawn, it is important to gather significantly more data from other experiments using other nuclei.


\begin{figure}
\begin{picture}(500,155)(0,0) 
 \put(0,0){\includegraphics[width=0.45\textwidth]{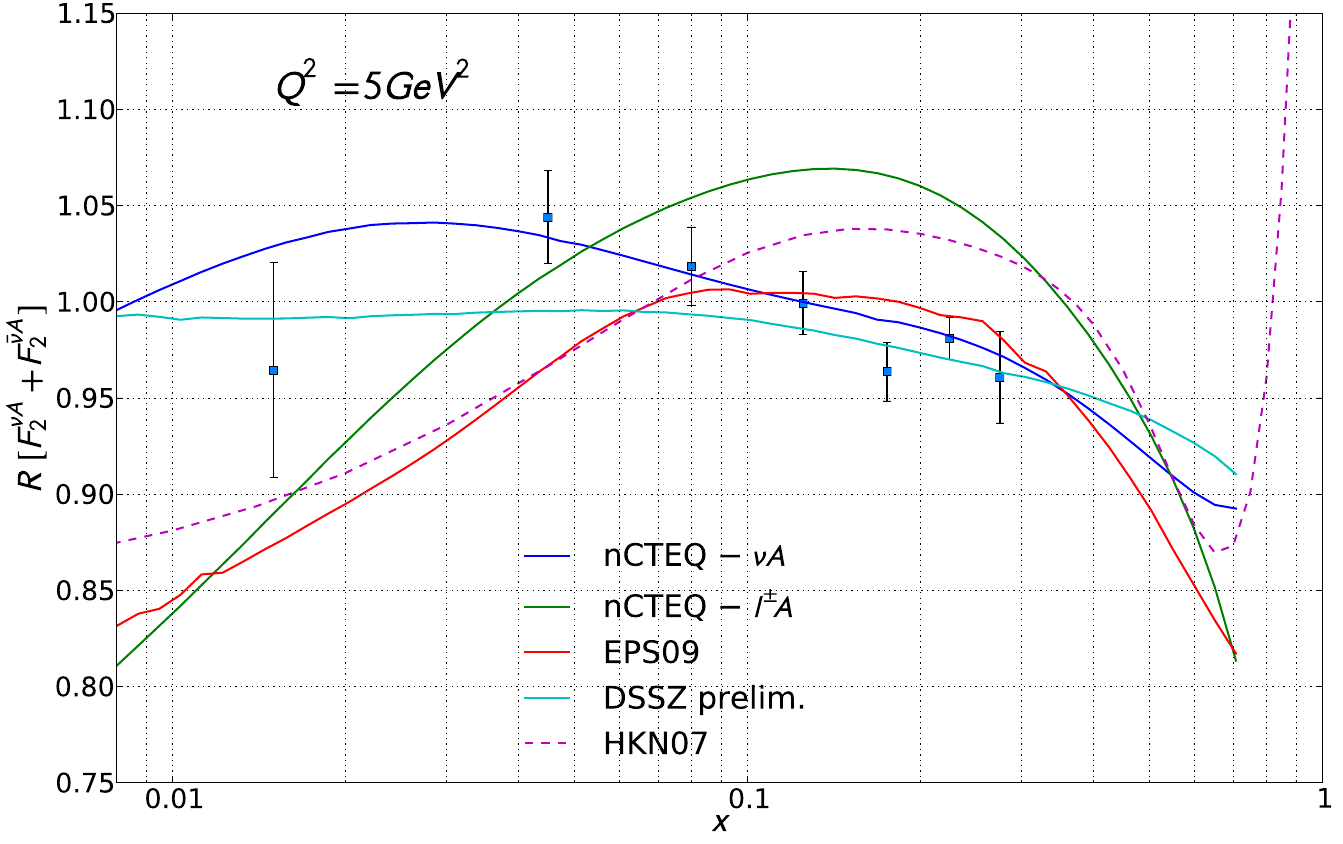}}
\put(260,0){\includegraphics[width=0.45\textwidth]{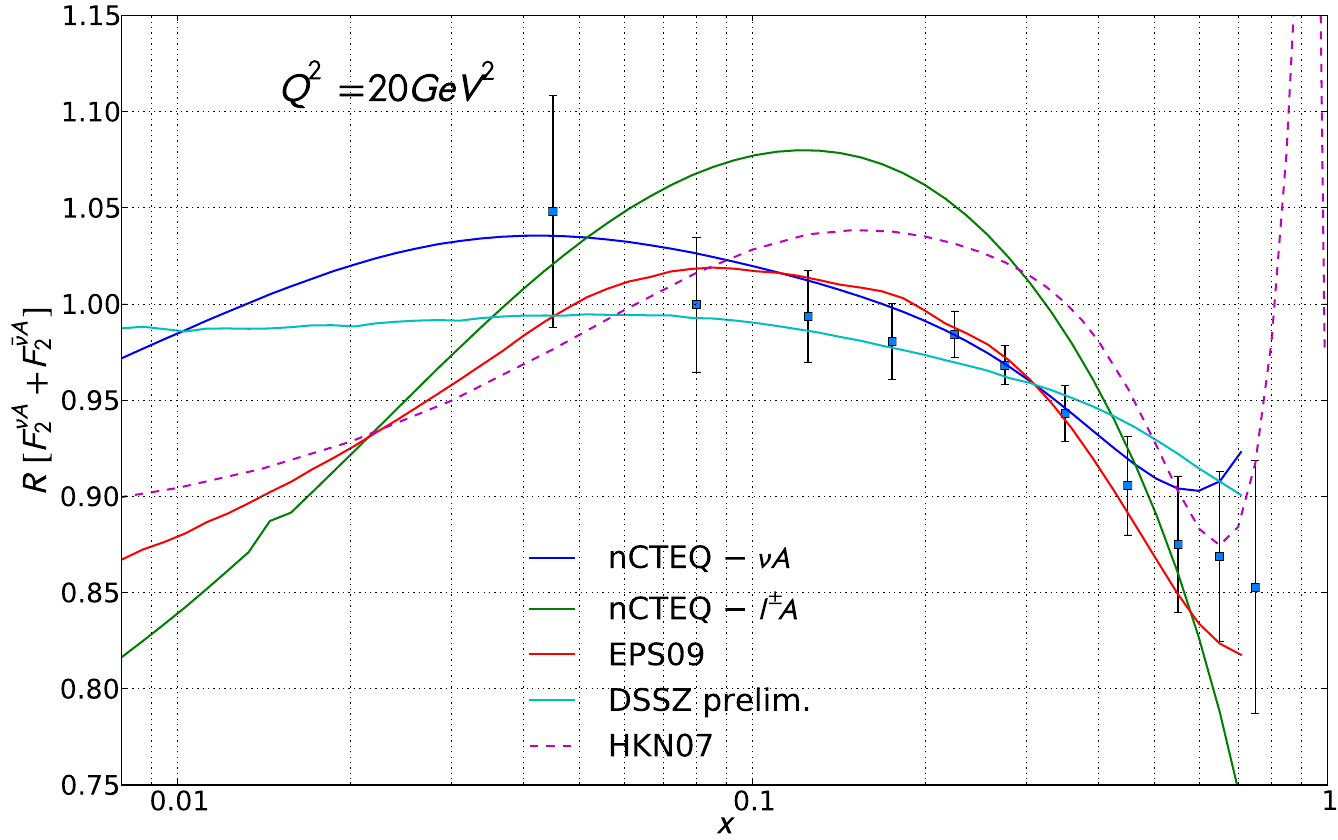}}
\put(114,0){$(a)$} \put(374,0){$(b)$} 
\end{picture} 
\caption{ Nuclear correction factor $R$ for the average $F_2$ structure function in $\nu A$ and $\ell^{\pm }A$ scattering for   a) $Q^2$ = 5 and b) 20.0  $GeV^2$ compared to the measured NuTeV average $F_2$ points shown with statistical errors.  The blue curve shows the result of the nCTEQ analysis of $\nu$ A (Fe and Pb) differential cross sections, plotted in terms of the average $F_2^{Fe}$ divided by the results obtained with the reference fit (free-proton) PDFs.  The green curve is the PRELIMINARY deFlorian et al fit to the NuTeV average structure functions.  For comparison, the nCTEQ fit, the EPS09 fit and the HKN07 fit to the charged-lepton data are also shown, to emphasize the difference in the shadowing region}
\label{fig:AvgF2All}
\end{figure}

\section{Current Experimental Studies of $\nu$-A Nuclear Effects}

\subsection{The MINER$\nu$A  Experiment at the FNAL Main Injector Neutrino Beam (NuMI)}

One such experiment that is currently taking data is the  MINER$\nu$A (Main Injector ExpeRiment: $\nu$ A) experiment~~\cite{Drakoulakos:2004qn}, a collaboration of elementary-particle and nuclear physicists.  The experiment has installed a fully active fine-grained solid  scintillator detector in the NeUtrinos from the Main Injector (NuMI) beam.  The overall goals of the experiment are to measure  absolute exclusive cross-sections, study nuclear effects in $\nu$ - A  interactions (with A varying from He to Pb),  and perform a systematic study of the resonance-DIS transition and the full DIS region.  A determination of these x-dependent nuclear effects with neutrino and antineutrino off a wide range of nuclear targets will start with the current exposure and will continue when the neutrino beam is tuned to higher energies starting in 2013.

A schematic of the MINER$\nu$A detector is shown in Fig.~\ref{minerva-side}.  The detector consists of five main regions: the fully active central detector, the upstream nuclear targets, a downstream electromagnetic and hadron calorimeter, and a surrounding electromagnetic and hadron calorimeter.  

The central fully-active detector serves as both the primary target and the tracking detector. The downstream electromagnetic calorimeter consists of alternating planes of Pb and scintillator planes, while the hadron calorimeter is similar, with planes of steel instead of Pb.  The side electromagnetic calorimeter consists of Pb plates between tracking plane in the outer region of the central detector.  The side hadron calorimeter consists of planes of steel with scintillator strips embedded.

Upstream of the central detector are planes of passive targets, with two planes 2.5 cm thick of mixed Fe/Pb, one plane with 2.5 cm thick Fe/Pb and 7.5 cm C, a solid plane of Pb 0.80 cm thick, and a mixed plane of Fe/Pb 1.30 cm thick.  The mixed Fe/Pb planes are split with part of the plane being iron and part of the plane being lead, such that the total mass is approximately equal.  Tracking planes are placed between each plane of passive targets. Recently a 15 cm water target has been installed downstream of the Fe/Pb/C target and a tank of liquid 4He about 1 m in diameter has been installed upstream of the main detector.  Figure~\ref{minerva-nuc} shows a schematic of the solid nuclear target region along with the placement of carbon, lead, and iron in the various targets.

\begin{figure}[ht]
\begin{center}
 \includegraphics[width=85mm]{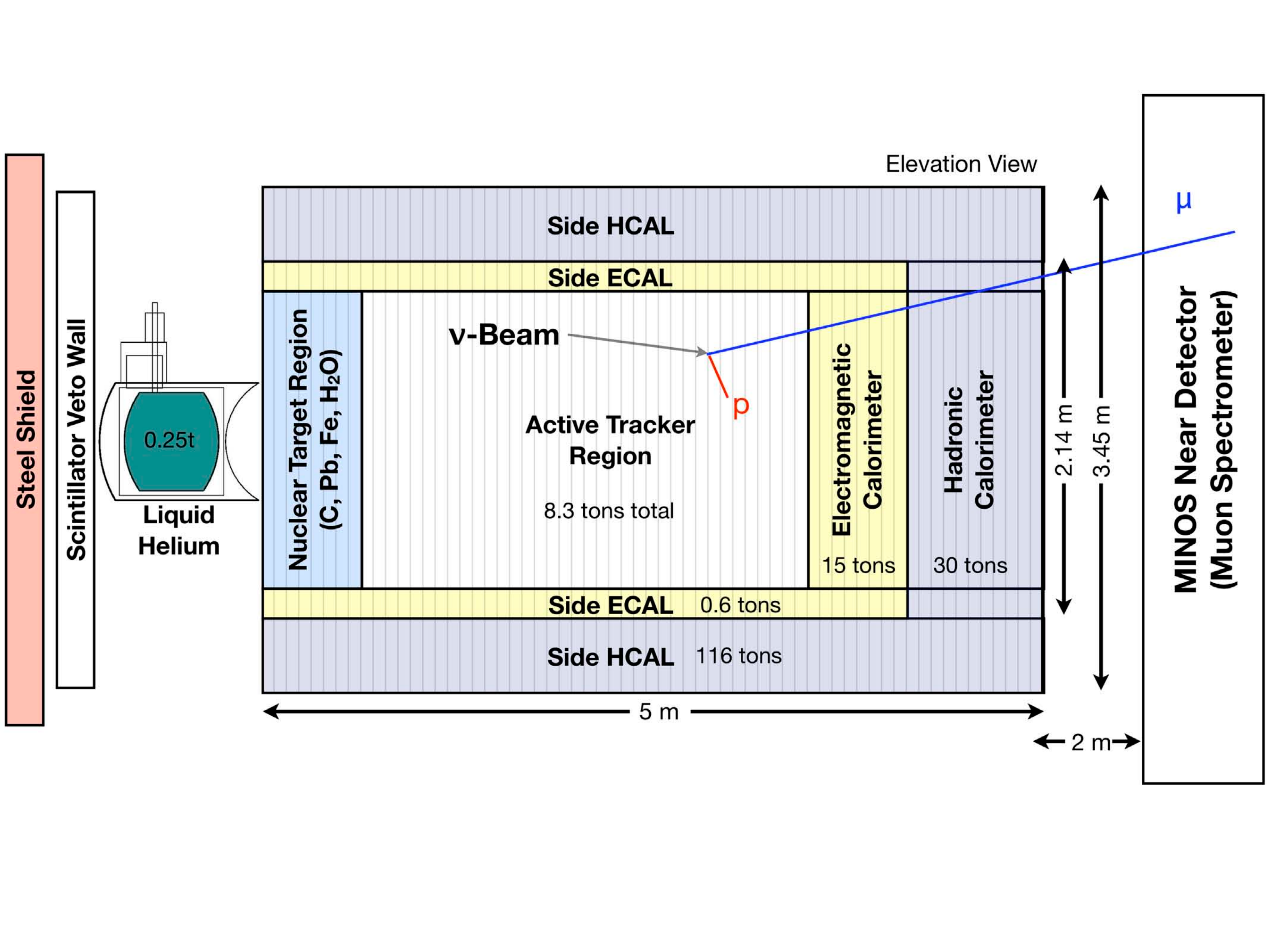}%
 \caption{Schematic side view of the MINER$\nu$A detector.}
\label{minerva-side}
\end{center}
 \end{figure}

 \begin{figure}[ht]
\begin{center}
 \includegraphics[width=100mm]{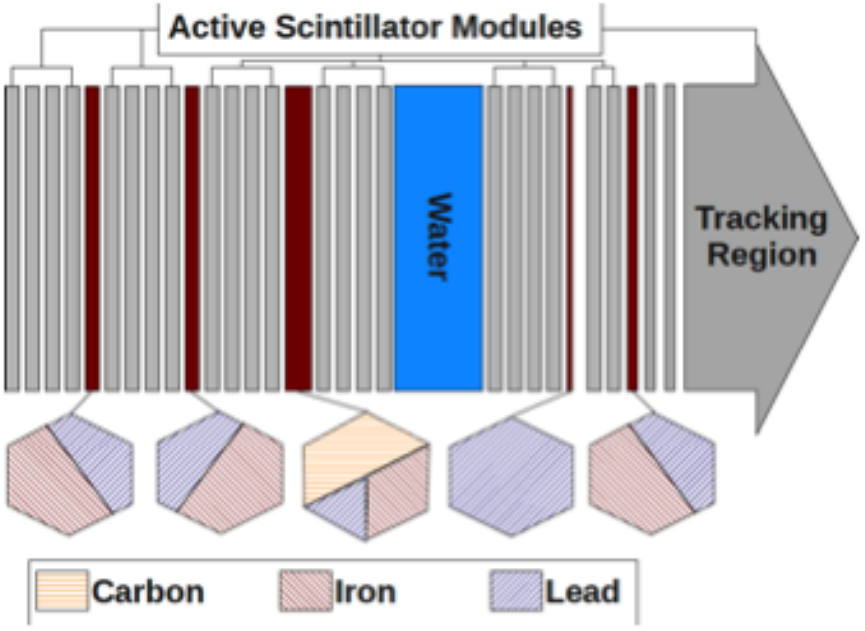}%
 \caption{Nuclear target region of the MINER$\nu$A detector.}
\label{minerva-nuc}
\end{center}
 \end{figure}

Charged current events originating in the central detector are fully contained, except for the muon, for neutrino energies of less than about 10 GeV.  The MINOS near detector, directly downstream of MINER$\nu$A, acts as a muon spectrometer and gives both muon energy and charge for forward going muons.  For particles stopping in MINER$\nu$A, particle identification can be determined from the dE/dX, but there is no charge determination.

The NuMI beam is currently being converted to the Medium Energy (ME) configuration for the NOvA experiment.  Figure~\ref{minerva-MEstat} displays the expected produced (no corrections for acceptance or reconstruction) statistics in the fiducial volume of the fully-active scintillator tracker for 6 x $10^{20}$ protons on target.  The units of the z-axis are 1k events and this represents the statistics we can expect per 1 year of running, when the Medium Energy beam configuration is running at expected power.  The statistics in the Fe and Pb targets will be about 1/3 the statistics displayed in Figure~\ref{minerva-MEstat}.  

\begin{figure}[ht]
\begin{center}
 \includegraphics[width=85mm]{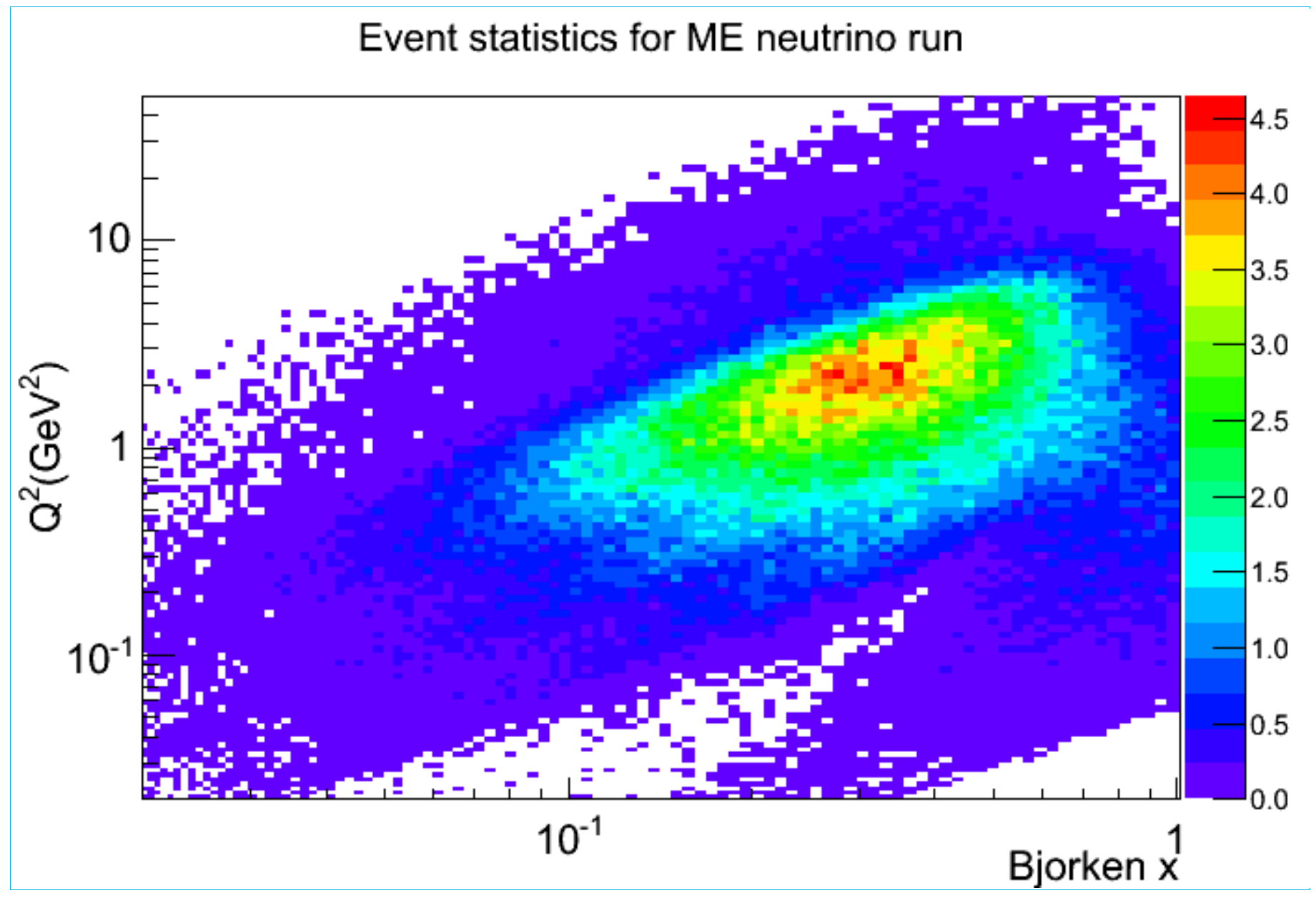}%
 \caption{Contour of the x vs $Q^2$ distribution of events produced (not corrected for acceptance or reconstruction efficiencies) in the ME beam configuration. The z-axis units are 1000 events and are for a 3-ton scintillator fiducial volume and 6 x $10^{20}$ protons on target. }
\label{minerva-MEstat}
\end{center}
 \end{figure}

Since MINER$\nu$A plans to be running for many years in the ME configuration with NOvA, it will be able to determine the nuclear dependence of the inclusive cross section, and in particular the relative ratio of cross sections off lead, iron, water and carbon to a statistical precision of better than 1\%.   MINERvA has also proposed to fill the present He cryogenic target with deuterium providing a direct ratio of A/$D_2$.
	As can be seen, even with the ME beam, MINER$\nu$A will not be able to reach deep into the shadowing region.  However, several data points integrated over $Q^2$ between x = 0.04 and 0.10 should be obtainable with reasonable statistics. This should be sufficient to determine the direction of the ratio of nuclear to nucleon cross sections as it enters the shadowing region.


\section{Conclusions: Experimental Evidence for Shadowing in Electroweak Interactions}

Shadowing with $\ell^{\pm }A$ scattering has been studied for decades and its experimental evidence is now overwhelming and conclusive.  The A-dependence of the shadowing phenomena over a wide range of nuclei has been measured with both electron and muon beams, and has been fitted to various theoretical models, as summarized in the theoretical part of this review.    

What is new in lepton nucleus scattering is the introduction of statistically significant analyses of $\nu A$ scattering.  The experimental evidence for shadowing in $\nu$ A scattering is a comparatively new study and is therefore much less settled.  In this review of experimental data the emphasis has thus been placed on these new $\nu A$ scattering results and two very different and distinct conclusions have been reached among groups studying them.  Whereas all three analyses - nCTEQ, EPS and dFS - agree with both the shape and the magnitude of the shadowing exhibited by $\ell^{\pm}A$ scattering, there is disagreement between the nCTEQ analysis and those of EPS and dFS when considering $\nu A$ scattering.  

The nCTEQ analysis suggests an apparent difference in $\nu$-A and $\ell^{\pm }A$ shadowing behavior, which is not seen by the analyses of EPS and dFS.  However EPS and dFS did not use the full NuTeV covariant error matrix, yielding the full discriminatory power of the NuTeV data in their analyses, while nCTEQ did use the full covariant error matrix. 

It is important to again emphasize that any significant conclusions on shadowing with $\nu$ and $\overline \nu$ nucleus scattering that we are able to draw here are based on {\em ONE analysis of ONE experiment using ONE nucleus}, the NuTeV $\nu$ Fe scattering data.  The other current data set used in these analyses, CHORUS $\nu$ Pb scattering, has significantly larger systematic errors than NuTeV and thus plays a much reduced role in the analysis.  Before any far-reaching conclusions can be drawn, it is important to gather significantly more data from other experiments, using other nuclei.  It is expected that the MINER$\nu$A experiment, soon to be taking data in the higher energy ME beam configuration at Fermilab, will provide data to compare with the NuTeV results.

Should subsequent experimental results confirm the difference between charged-lepton and neutrino scattering in the shadowing region at low-$Q^2$, it is interesting to speculate on the possible cause of the difference.  A recent study of EMC, BCDMS and NMC data by a Hampton University - Jefferson Laboratory collaboration~\cite{Guzey}  suggests that anti-shadowing in charged-lepton nucleus scattering may be dominated by the longitudinal structure function $F_L$.  As a by-product of this study, their figures hint that shadowing in the data of EMC, BCDMS and NMC $\mu$ A scattering was being led by the transverse cross section with the longitudinal component crossing over into the shadowing region at lower x compared to the transverse.   This is consistent with the expectations expressed in the equations and figures of ~\ref{sec:L/T}. 

In the low-$Q^2$ region, the neutrino cross section is dominated by the longitudinal structure function $F_L$ via axial-current interactions, since $F_T$ vanishes as $Q^2$ as $Q^2 \Rightarrow$ 0, similar to the behavior of charged lepton scattering.  If the results of the NuTeV analysis are verified, one contribution to the different behavior of shadowing at low-$Q^2$ demonstrated by $\nu$ A and $\ell$ A, in addition to the different hadronic fluctuations in the two interactions, could be due to the different mix of longitudinal and transverse contributions to the cross section of the two processes in this kinematic region.  

Another interesting hypothesis to explain this possible difference was recently proposed by V. Guzey~\cite{Guzey-2012}.  He suggests that at low-x and low-$Q^2$  the value of y is close to unity and the neutrino interactions primarily probe the down and strange quarks.  This is very different from the situation with charged-lepton scattering, where the contribution from down and strange quarks are suppressed by a factor of 1/4 compared to the up and charm.  If the strange quark is not subject to the same degree of shadowing as the valence quarks, then the neutrino cross section in this kinematic regime would also not exhibit strong shadowing.  On the other hand, the charged-lepton cross section, with its comparatively small strange quark content in this same kinematic regime, would still exhibit strong shadowing.   

\vspace{0.5 cm}

{\bf Acknowledgments}

\vspace{0.5 cm}

This work was supported in part by Fondecyt (Chile) grants 1090291 and 1100287.  The work of B.Z.K. was
supported also by the Alliance Program of the Helmholtz Association, contract HA216/EMMI "Extremes of Density and Temperature: Cosmic Matter in the Laboratory".
Fermilab is operated by Fermi Research Alliance, LLC under Contract No. De-AC02-07CH11359 with the United States Department of Energy.

\end{document}